\newif\ifprintfig
\printfigtrue

\documentclass{emulateapj}
\usepackage{graphicx}
\usepackage{amsmath}

\newcommand\degrees{{^\circ}}
\newcommand\pc{{\rm\,pc}}

\newcommand\kpc{{\rm\,kpc}}

\newcommand\Myr{{\rm\,Myr}}
\newcommand\Myrs{{\rm\,Myrs}}
\newcommand\Gyr{{\rm\,Gyr}}

\newcommand\kmsec{{\rm\,km\,s^{-1}}}
\newcommand\kms{\kmsec}

\newcommand\msun{{\rm\,M_\odot}}

\newcommand\numden{{\rm\,arcsec^{-2}}}
\newcommand\mags{{\rm\,mag}}
\newcommand\clock{\count0=\time \divide\count0 by 60
     \count1=\count0 \multiply\count1 by -60 \advance\count1 by \time
     \number\count0:\ifnum\count1<10{0\number\count1}\else\number\count1\fi}

\slugcomment{Ap.J., in press}
\shortauthors{Dalcanton et al.}
\shorttitle{Dust Mapping in M31}

\begin{document}

\title{The Panchromatic Hubble Andromeda Treasury VIII: A Wide-Area, High-Resolution Map of Dust Extinction in M31}

\author{Julianne J.\ Dalcanton\altaffilmark{1,2}, 
  Morgan Fouesneau\altaffilmark{1,2},
  David W. Hogg\altaffilmark{3}, 
  Dustin Lang\altaffilmark{4,5}, 
  Adam K.~Leroy\altaffilmark{6,7}, 
  Karl D.~Gordon\altaffilmark{8},
  Karin Sandstrom\altaffilmark{9,10},
  Daniel R.~Weisz\altaffilmark{1,11,12},
  Benjamin F.~Williams\altaffilmark{1}, 
  Eric F.~Bell\altaffilmark{13},
  Hui Dong\altaffilmark{14,15},
  Karoline M.~Gilbert\altaffilmark{8},
  Dimitrious A.\ Gouliermis\altaffilmark{2},
  Puragra Guhathakurta\altaffilmark{11},
  Tod R.\ Lauer\altaffilmark{14},
  Andreas Schruba\altaffilmark{16,17},
  Anil C.~Seth\altaffilmark{18},
  Evan D.~Skillman\altaffilmark{19}}

\altaffiltext{1}{Department of Astronomy, Box 351580, University of
  Washington, Seattle, WA 98195; jd@astro.washington.edu}
\altaffiltext{2}{Max Planck Institute f\"{u}r Astronomie, K\"{o}nigstuhl 17, 69117, Heidelberg, Germany}
\altaffiltext{3}{Center for Cosmology and Particle Physics, Department of Physics, New York University, 4 Washington Pl \#424, New York, NY, 10003 USA}
\altaffiltext{4}{McWilliams Center for Cosmology, Department of Physics, Carnegie Mellon University, 5000 Forbes Ave, Pittsburgh, PA 15213 USA}
\altaffiltext{5}{Department of Physics and Astronomy, University of Waterloo, 200 University Avenue West, Waterloo, Ontario N2L 3G1, Canada}
\altaffiltext{6}{National Radio Astronomy Observatory, 520 Edgemont Road, Charlottesville, VA 22903, USA}
\altaffiltext{7}{Department of Astronomy, The Ohio State University, 140 West 18th Avenue, Columbus, OH 43210}
\altaffiltext{8}{Space Telescope Science Institute, 3700 San Martin Drive, Baltimore, MD, 21218, USA}
\altaffiltext{9}{Steward Observatory, University of Arizona, 933 N Cherry Ave, Tucson, AZ, 85721 USA}
\altaffiltext{10}{Center for Astrophysics and Space Sciences, Department of Physics, University of California, San Diego, 9500 Gilman Drive, La Jolla, CA 92093, USA}
\altaffiltext{11}{Department of Astronomy and Astrophysics, University of California Santa Cruz, 1156 High Street, Santa Cruz, CA 95064 USA}
\altaffiltext{12}{Hubble Fellow}
\altaffiltext{13}{Department of Astronomy, University of Michigan, 500 Church St., Ann Arbor, MI 48109, USA}
\altaffiltext{14}{National Optical Astronomy Observatory, 950 North Cherry Avenue, Tucson, AZ 85719, USA}
\altaffiltext{15}{Instituto de Astrof́\'{i}sica de Andaluc\'{i}a (CSIC), Glorieta de la Astronom\'{a} S/N, E-18008 Granada, Spain}
\altaffiltext{16}{California Institute of Technology, Cahill Center for Astrophysics, 1200 E.\ California Blvd, Pasadena, CA 91125, USA}
\altaffiltext{17}{Max-Planck-Institut f\"{u}r extraterrestrische Physik, Giessenbachstrasse 1, 85748 Garching, Germany}
\altaffiltext{18}{University of Utah, Salt Lake City, UT, USA}
\altaffiltext{19}{Minnesota Institute for Astrophysics, University of Minnesota, 116 Church Street SE, Minneapolis, MN 55455, USA}


\begin{abstract}
  We map the distribution of dust in M31 at $25\pc$ resolution, using
  stellar photometry from the Panchromatic Hubble Andromeda Treasury
  survey.  The map is derived with a new technique that models the
  near-infrared color-magnitude diagram (CMD) of red giant branch
  (RGB) stars.  The model CMDs combine an unreddened foreground of RGB
  stars with a reddened background population viewed through a
  log-normal column density distribution of dust. Fits to the model
  constrain the median extinction, the width of the extinction
  distribution, and the fraction of reddened stars in each $25\pc$
  cell. The resulting extinction map has a factor of $\gtrsim$4 times
  better resolution than maps of dust emission, while providing a more
  direct measurement of the dust column.  There is superb
  morphological agreement between the new map and maps of the
  extinction inferred from dust emission by \citet{draine2013}.
  However, the widely-used \citet{draine2007} dust models overpredict
  the observed extinction by a factor of $\sim\!2.5,$ suggesting that
  M31's true dust mass is lower and that dust grains are significantly
  more emissive than assumed in \citet{draine2013}.  The observed
  factor of $\sim\!2.5$ discrepancy is consistent with similar
  findings in the Milky Way by \citet{planck2015}, but we find a more
  complex dependence on parameters from the \citet{draine2007} dust
  models. We also show that the discrepancy with the
  \citet{draine2013} map is lowest where the current interstellar
  radiation field has a harder spectrum than average.  We discuss
  possible improvements to the CMD dust mapping technique, and explore
  further applications in both M31 and other galaxies.
\end{abstract}

\vfill
\setcounter{footnote}{0}
\clearpage

\section{Introduction} \label{introsec}

Dust plays an increasingly important role in extragalactic
astronomy. It has long been known that dust shapes the observational
properties of disk galaxies, particularly in the optical and
ultraviolet \citep[e.g.,][]{disney1989, xilouris1999, misiriotis2001,
  pierini2004,tuffs2004, mollenhoff2006,driver2007, bianchi2008,
  popescu2011}.  However, over recent decades, dust has become a
wide-spread source of study in its own right
\citep[][]{savage1979,draine2003}, thanks in large part to
observational facilities that directly probe emission from the dust in
the mid- and far-infrared, and at sub-millimeter wavelengths.  This
emission has also become a widely used tracer of key astrophysical
quanitites, including the star formation rate and the interstellar
radiation field \citep[see review by][]{kennicutt2012}.

Dust has also emerged as one of the most effective tracers of cold,
dense gas.  Molecular gas and cold H{\sc i} are the immediate
precursors to star formation \citep[see, for
  example][]{bergin2007,mckee2007,hennebelle2012}, and the properties
of this gas is likely to be coupled to the ability of the interstellar
medium (ISM) to form stars.  Unfortunately, these cold gas components
are extremely difficult to trace. Cold molecular hydrogen is nearly
impossible to see in emission, and cold atomic H{\sc i} cannot be
reliably distinguished from warmer phases without absorption line
techniques.

The study of cold, dense gas has been fundamentally changed by the
widespread use of dust extinction as a tracer
\citep{lilley1955,dickman1978,frerking1982}.  Methods that used star
counts to identify dust extinction had been in use for many years
\citep[e.g.,][among many others]{dickman1978,cernicharo1984,
  cambresy1997, cambresy1999,arce1999a,dobashi2005}, but were
gradually supplanted by new methods taking advantage of the growing
availability of near-infrared imaging, particularly due to the all-sky
2MASS \citep{skrutskie2006} and the UKIDSS/Galactic Plane
\citep{lucas2008} surveys. Many groups have developed optimized
methodologies to use near-infrared (NIR) color-color (or
color-magnitude) diagrams \citep[e.g.,][]{lada1994, ciardi1998,
  lombardi2001, cambresy2002, lombardi2005, froebrich2006,
  lombardi2009, gonzalez2011} to identify molecular clouds in the
Milky Way and to map their structure \citep[][and many
  others]{ciardi1998,alves1998, lada1999,alves2001, cambresy2002,
  teixeira2005,lombardi2006, froebrich2007, lombardi2008,
  kainulainen2009, rowles2009, lombardi2010, romanzuniga2010,
  frieswijk2010, scandariato2011,
  dobashi2011,kainulainen2011a,kainulainen2011b,gonzalez2012,alves2014,schlafly2015},
with even more recent work using mid-infrared observations to map even
denser clouds \citep{vasyunina2009, butler2009, majewski2011,
  butler2012,nidever2012,kainulainen2013}, and other work looking to
exploit growing optical databases \citep[e.g.,][]{sale2009}.  The
resulting maps are likely to be more direct tracers of the total
column density than methods based on dust or molecular emission
\citep[e.g.,][]{goodman2009}.

The impact of extinction mapping can easily be seen in the diversity
of problems they have been used to address.  These maps have been
used: to measure the statistics of the column density distribution
\citep[e.g.,][]{lada1999, kainulainen2009, lombardi2010,
  froebrich2010, schneider2011, alves2014} and its connection to star formation
\citep[][]{lada2009, rowles2011, kainulainen2011b, stutz2015}; to explore the
origins of ``Larson's Laws'' \citep[][]{kauffmann2010, lombardi2010,
  beaumont2012, ballesterosparedes2012}; to measure the cloud
structure function and clump statistics \citep[e.g.,][]{padoan2002,
  padoan2003, kirk2006,lombardi2008, lombardi2010, kauffmann2010}; to
study individual molecular cores and globules
\citep[e.g.,][]{teixeira2005, romanzuniga2009,racca2009, schmalzl2010}; to constrain
the relationship between dust column density and emission from
molecular gas \citep[e.g.,][]{lada1994, hayakawa2001, harjunpaa2004,
  lombardi2006, kainulainen2006, pineda2010}; to distinguish among models of
turbulence \citep[][]{padoan1997, federrath2010, kainulainen2013}; to
derive the 3-dimensional structure of the ISM \citep[e.g.,][]{sale2014,schlafly2015, green2015}; and
to constrain models of the dust itself \citep[e.g.,][]{roy2013}.

Extinction mapping has also become a routine tool in the study of
external galaxies. The Magellanic Clouds are sufficiently
close to apply similar mapping techniques using NIR colors of
individual stars \citep[e.g.,][]{dobashi2008, dobashi2009}, although
their low extinctions allows optical and ultraviolet data to be used
as well \citep[e.g.,][]{harris1997,zaritsky2002, zaritsky2004}. At larger
distances, where individual stars can no longer be resolved,
extinction maps are typically based on modeling the spectral energy
distribution in individual pixels \citep[e.g.,][]{zibetti2009}, but
usually only as a way to removing the effects of dust, rather than as
a route to studying the cold ISM itself.

In this paper, we focus on using M31 to bridge these two
regimes. Andromeda is close enough that we can probe the ISM on the
scales of molecular clouds. However, thanks to the large
area covered by the Panchromatic Hubble Andromeda Treasury
\citep[PHAT;][]{dalcanton2012}, we can also cover large areas,
allowing us to view the entire cold ISM, rather than just individual
clouds.  This approach gives us our first view of the statistical
properties of molecular clouds across a large fraction of a massive
spiral galaxy.

Thanks to its proximity, M31 has been the target of many previous dust
studies, including surveys of dust emission using space-based mid- and
far-infrared (FIR) imagers on board {\emph{IRAS}} \citep{devereux1994},
{\emph{Spitzer}} \citep[][]{barmby2006,gordon2006}, and
{\emph{Herschel}} \citep{fritz2012,groves2012}.  These observations have elucidated star
formation, dust heating, the dust-to-gas ratio, and dust composition
\citep[e.g.,][]{ford2013, montalto2009, tabatabaei2010, leroy2011,
  groves2012,smith2012, draine2013}, particularly when complemented by
direct studies of the cold ISM in H{\sc i} \citep{brinks1986,
  braun2009, chemin2009} and CO \citep[][though the latter two cover
limited area]{nieten2006,rosolowsky2007,tosaki2007}.  

Some of these studies have used the dust emission and other tracers to
derive extinction maps \citep[e.g.,][]{montalto2009,draine2013}.  Most
recently, \citet{draine2013} used mid- and far-infrared data to derive
dust column densities throughout M31.  These maps use state of the art
models of dust to derive the spatial distribution of dust composition,
heating, and dust column density. The models also make specific
predictions for the extinction within M31.

In this paper we introduce a technique for probing the cold dusty
ISM. We take advantage of the Panchromatic Hubble Andromeda Treasury's
\citep[PHAT;][]{dalcanton2012} large database of NIR HST observations,
and use photometry of individual RGB stars to derive the distribution
of dust extinction on 25$\pc$ scales.  Specifically, we use the
structure of the RGB seen in NIR color-magnitude diagrams (CMDs) to
fit for the distribution of extinctions along the line of sight. The
methodology therefore gives us both the median extinction and the
width of the extinction distribution in each resolution element
(pixel).  

In Sec.\ \ref{overviewsec} we give an overview of this technique, and
explain its connection to observations of molecular clouds in the
Milky Way.  In Sec.\ \ref{cloudsec} we derive the expected
distribution of color and/or reddening for a log-normal distribution
of dust embedded in thicker stellar disk. In Sec.\ \ref{modelingsec},
we give details of how we fit the parameters of the dust$+$star model
to data from the PHAT survey, and discuss the distribution and
accuracy of the derived parameters in
Sec.\ \ref{reddingpropertiessec}. We show the global dust extinction
map and compare it to the extinction inferred from fits to dust
emission Sec.\ \ref{resultssec}. We then conclude in
Sec.\ \ref{conclusionsec}.

\section{Overview of Measuring Extinction with CMDs} \label{overviewsec}

Mapping extinction in the Milky Way requires coping with the complex
distribution of both dust and stars along the line of sight. In
external galaxies, however, the spread in distance is negligible, and all
stars can be assumed to be at the same distance. This difference
allows the full color and magnitude distribution of stars to be used
when constructing extinction maps.

\begin{figure*}
\centerline{
\includegraphics[width=7.25in]{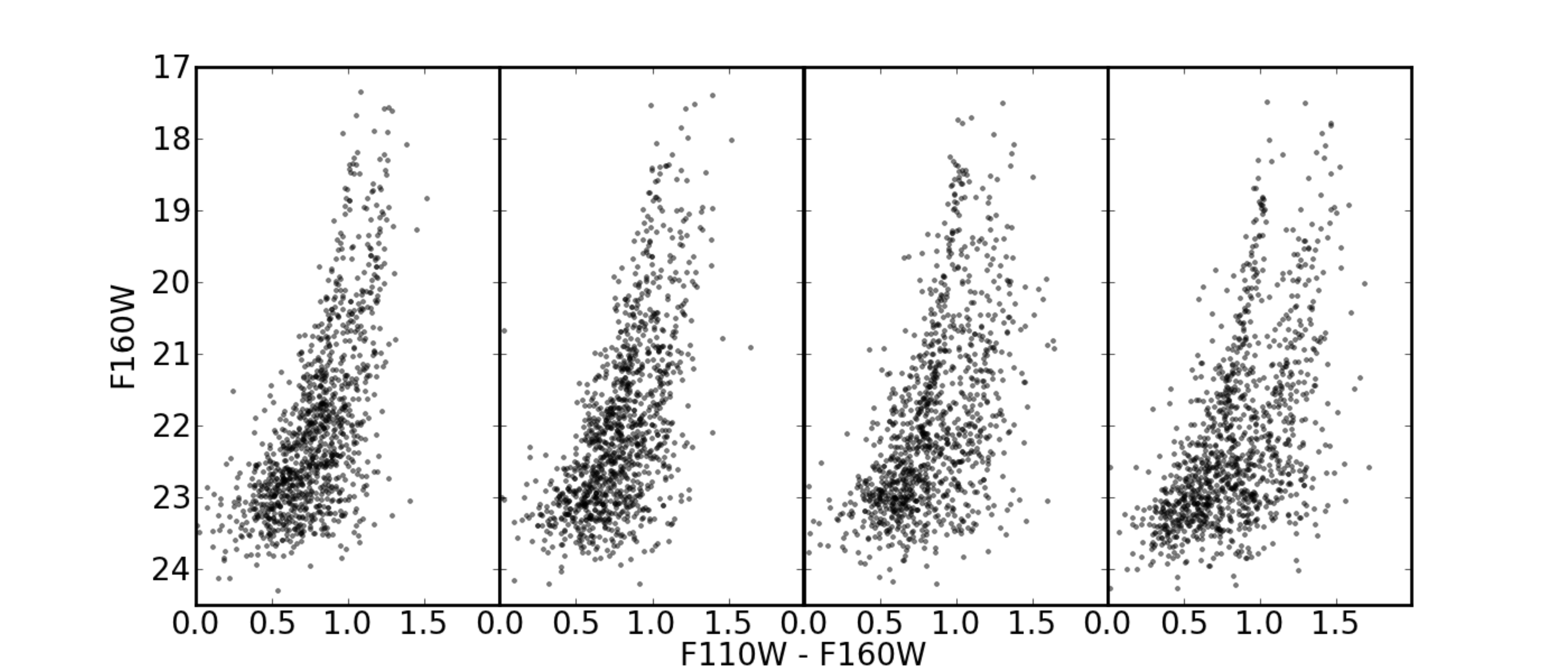}
}
\caption{The spatial variation in the NIR color magnitude diagrams,
  shown in 4 adjacent $20\arcsec$ bins within the disk of M31.  Each
  CMD is dominated by red giant branch stars. There is a narrow, bluer
  RGB sequence that is similar in all regions, but that is accompanied
  by a highly variable redder RGB sequence, which varies in both width
  and position. RGB stars trace a sufficiently old stellar population
  that they should be well mixed at these spatial scales. The spatial
  variation in the redder population is therefore best explained by
  spatial variation in the dust that is obscuring some fraction of the
  RGB stars. \label{originalcmdfig}}
\end{figure*}

The impact of dust can be most easily perceived in the NIR CMD, which
is populated almost entirely with RGB stars that occupy a small range
of intrinsic color at given magnitude.  Theoretically, the RGB should
form a narrow, nearly vertical sequence in a NIR CMD. Empirically,
however, the CMD structure of the RGB is complex and spatially
varying, as shown in Figure~\ref{originalcmdfig}. In many regions of
M31, the NIR RGB is quite thin (as seen later in
Figure~\ref{unreddenedCMDfig}), but in other regions it appears broad, and in
some cases even bimodal.  The broadening and/or bimodality
is always seen redward of the thinner, bluer component, the latter of
which always stays in essentially the same place on the CMD.  Given
that the color distribution of the bimodal RGB is not sensible in any
stellar population model, and that the unusual morphologies change
rapidly over very short distances, the natural explanation for the
effects seen in Figure~\ref{originalcmdfig} is dust.

The CMD morphology in Figure~\ref{originalcmdfig} can be explained as
follows. The thinner, bluer RGB sequence is due to old or intermediate
age stars seen in front of the cool dusty ISM.  The redder stars are
then those that lie behind the layer of dust\footnote{Based on the
  Milky Way \citep[e.g.,][]{combes1991,juric2008}, the cold ISM is
  almost certainly in a thinner layer than the RGB stars, which in M31
  are primarily in a $>0.5\kpc$ thick disk (as we show in a companion
  paper).  Thus we can safely neglect the effects of stars that are
  embedded within the dust.  This assumption would not necessarily
  hold when analyzing a much younger population of stars that is
  expected to be largely embedded within the cool ISM.}.  The cool ISM is highly
structured, and thus we see rapid small-scale spatial variations in
the behavior of the reddened stars.  Based upon this interpretation,
we developed a method to simultaneously fit the unreddened and
reddened stars to derive the distribution of dust extinction, which we
describe in detail in Secs.\ \ref{cloudsec}~\&~\ref{modelingsec}.

\begin{figure*}
%
\centerline{
\includegraphics[height=2.4in]{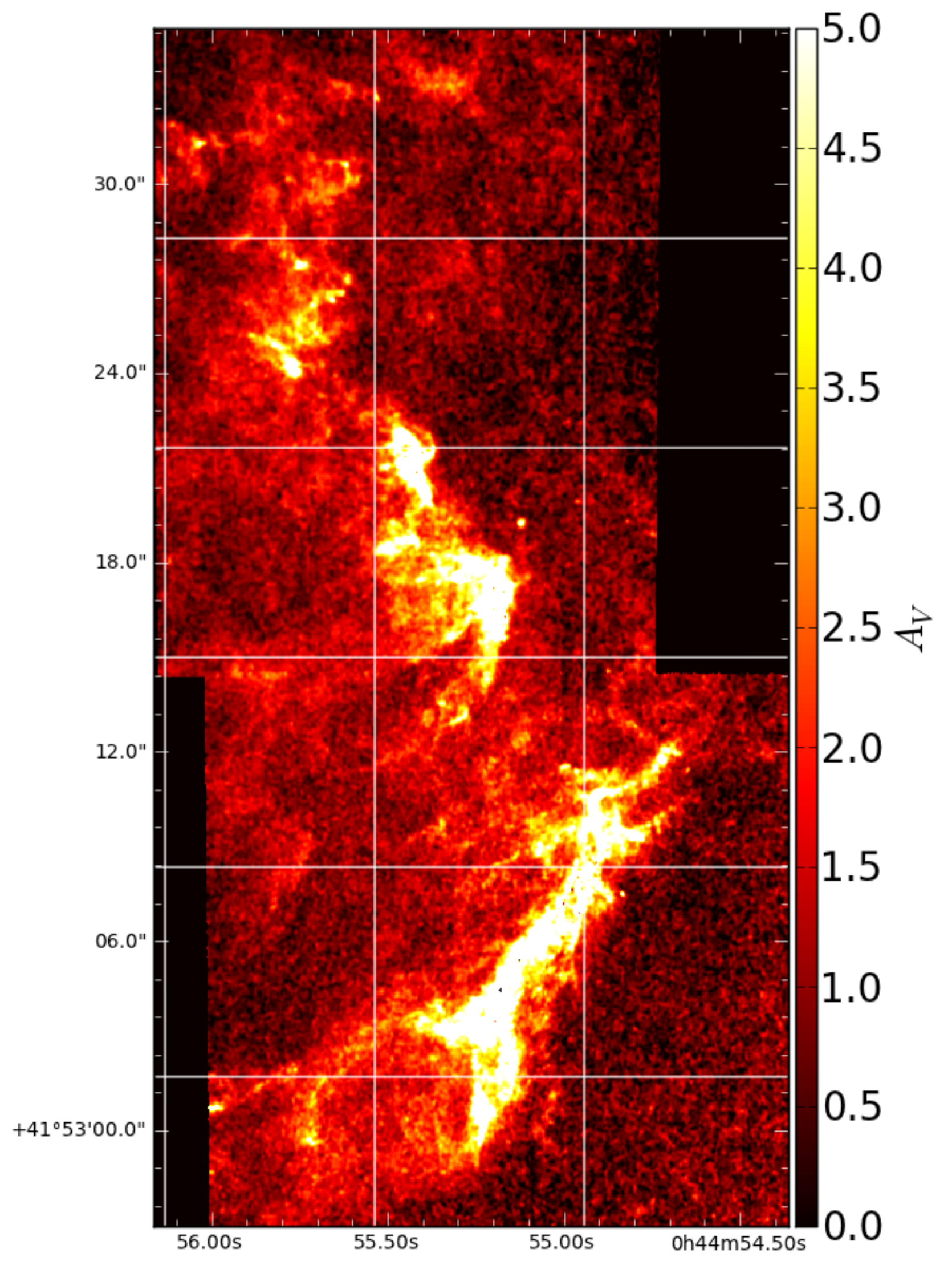}
\includegraphics[height=2.4in]{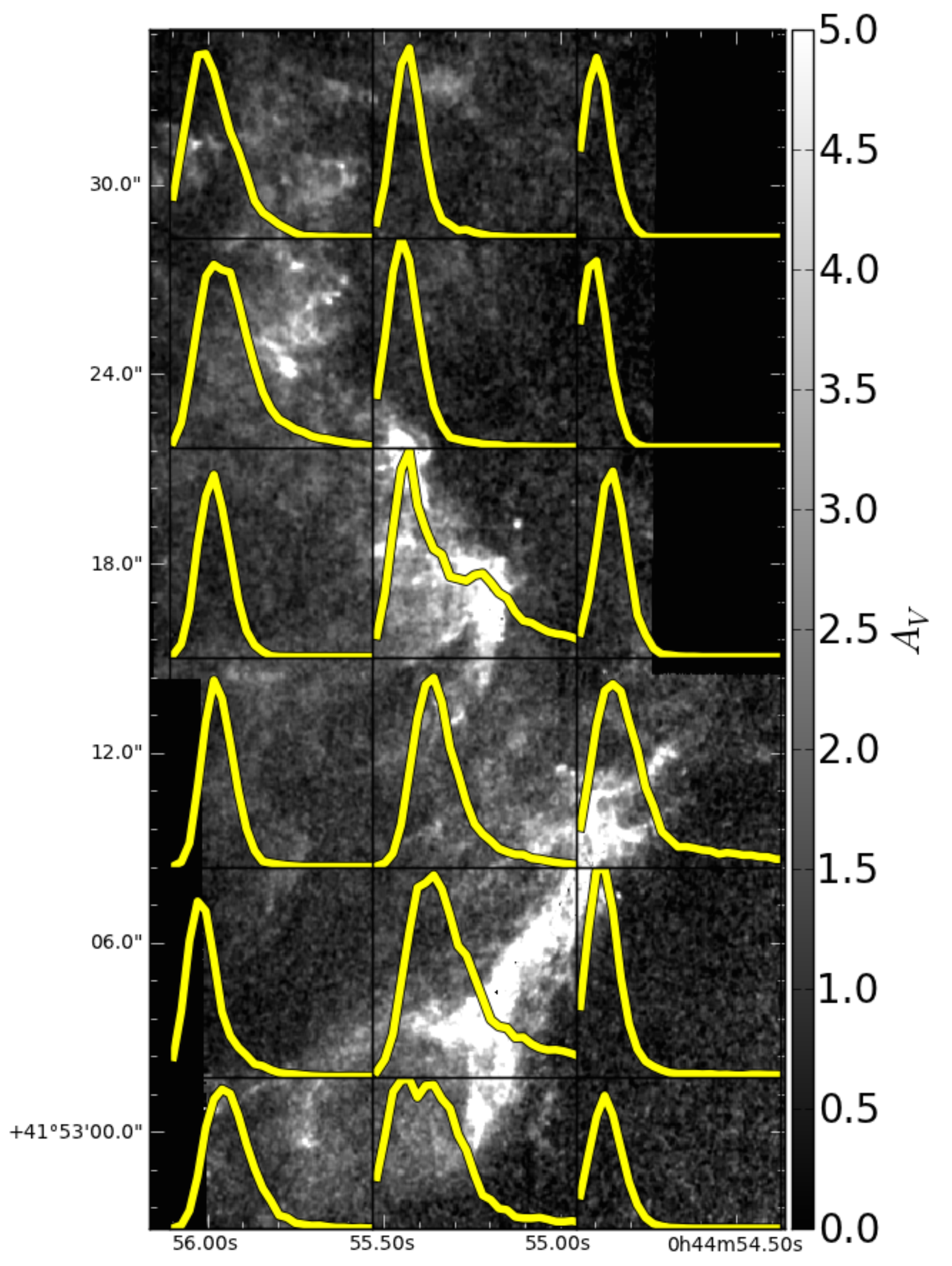}
\includegraphics[height=2.4in]{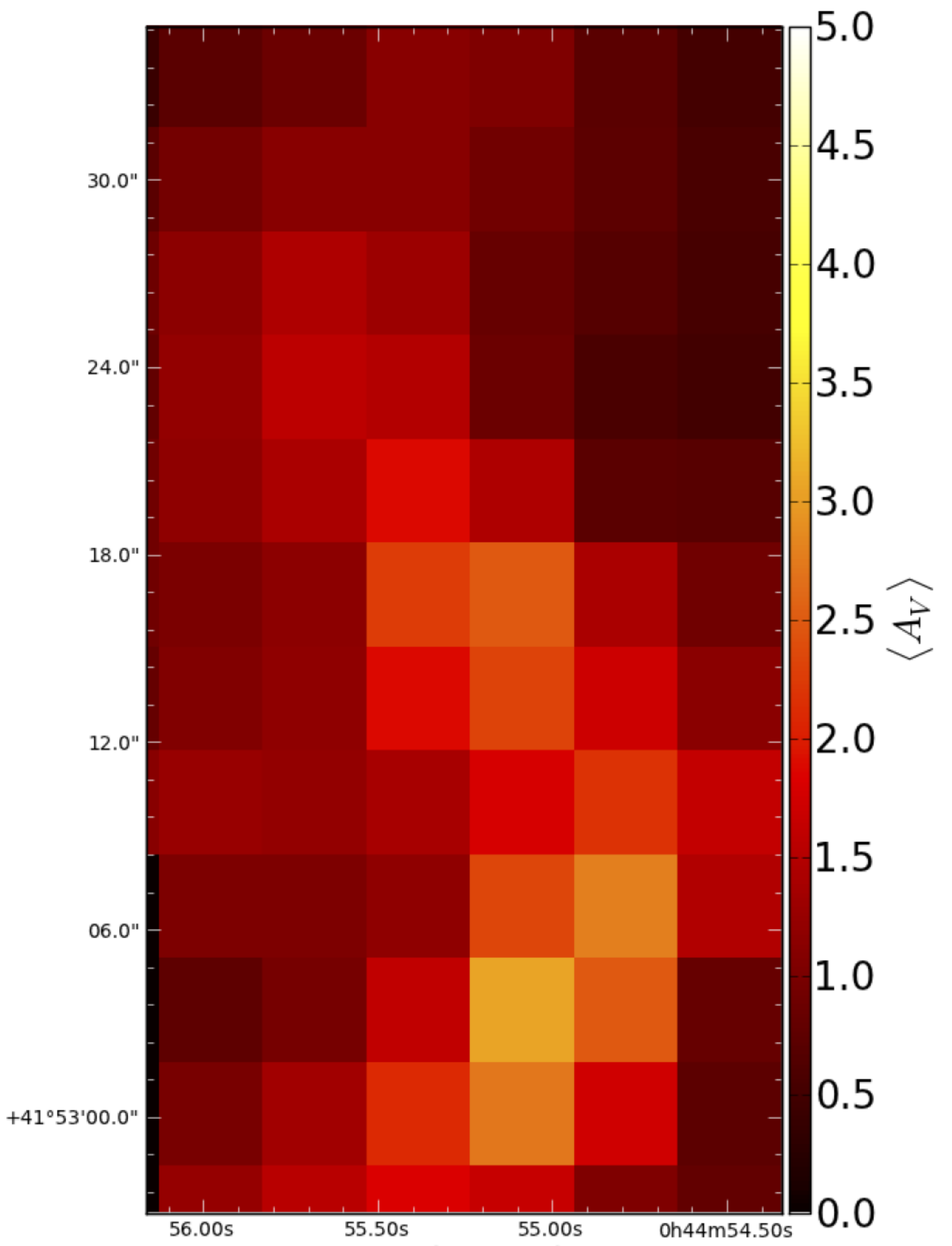}
\includegraphics[height=2.4in]{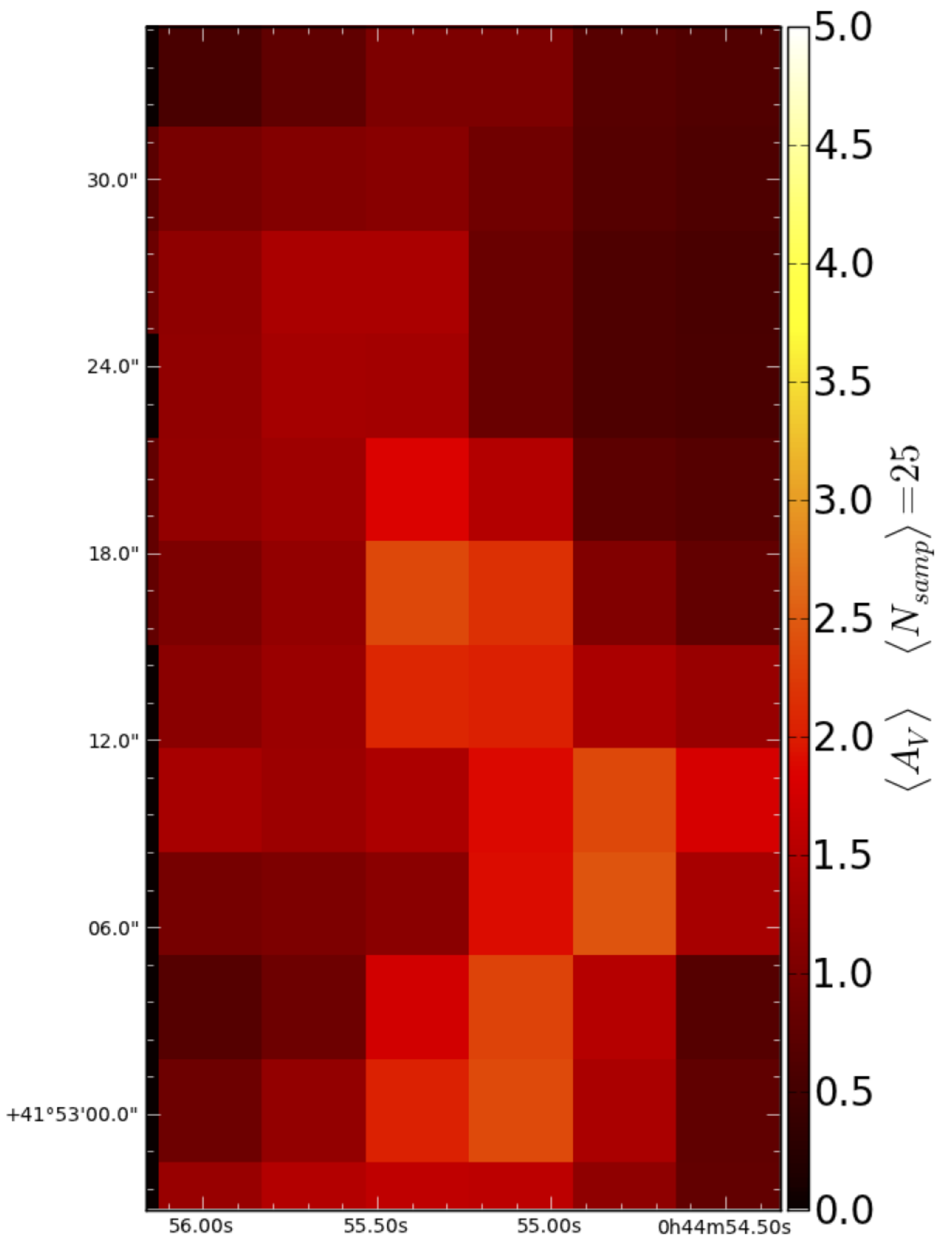}
}
\caption{Simulation of sampling the extinction distribution of the
  Orion molecular cloud at the distance of M31. (First Panel) The
  extinction map of Orion from \citet{kainulainen2009}, with the
  angular scale adjusted to place it at the distance of M31, and the
  RA and Dec shifted to a location in M31's $10\kpc$ star-forming
  ring.  Grid lines are 25$\pc$ ($=6.645\arcsec$ apart). (Second Panel)
  The distribution of extinctions within each grid region (yellow),
  superimposed on the map from the top left. The same range of
  extinction is shown in each panel ($0<A_V<5\mags$). The distribution
  of extinctions can be reasonably characterized as a log-normal
  distribution in most cases. (Third Panel) The mean extinction,
  calculated using a dithered grid of 25$\pc$, in steps of
  12.5$\pc$. In spite of the low resolution, the basic structure of
  the molecular cloud is still clearly visible. (Fourth Panel) The mean
  extinction in the same grid as in the lower left panel, but now
  calculated using only $\sim$25 randomly sampled pixels from the high
  resolution left-hand image in each 25$\pc$ bin. Even this modest
  number of samples is sufficient to capture the same morphology seen
  in the middle panel, with a slight tendency to miss some of the
  highest extinctions. \label{orionfig}}
\end{figure*}

At its heart, the method is based on recognizing that the RGB stars
are point-like ``samples'' of the column density distribution function
of the dust.  We demonstrate this effect in Figure~\ref{orionfig}.
The first panel shows a 2MASS-based extinction map of the Orion
molecular cloud, kindly provided by J. Kainulainen
\citep{kainulainen2009}\footnote{Although only a small fraction of
  M31's ISM is likely to be in the form of molecular clouds comparable
  to Orion, this example still serves as a useful illustration of the
  technique.}.  We have adjusted the angular scale of the map to be
consistent with being located at the distance of M31.
The grid on the image indicates regions $25\pc$ across, which is the
scale we use in our analysis below.  The second panel shows the
distribution of extinctions within each of these grid regions.  In the
PHAT NIR photometry, there are typically $\sim$100 RGB stars that fall
within a region this size, and 20--80\% of them lie behind the
dust. The RGB stars therefore sample the dust structure on scales
below that of the $25\pc$ grid, and even adjacent RGB stars may fall
on regions with quite different extinctions.

If we assume that any stars coincident with regions in
Figure~\ref{orionfig} are primarily found in a stellar disk that is
significantly thicker than the cool ISM \citep[consistent with the
  high velocity dispersion of $\sim\!90\kms$ seen in RGB
  stars;][]{dorman2015}, some fraction of the stars will be behind the
dust layer, with the exact fraction depending on the disk geometry.
These stars will be reddened, and will sample the local distribution
of reddening in the molecular cloud.  By analyzing the reddening
distribution, we can infer the statistical properties of the
extinction distribution of the gas layer that they lie behind, for a
given choice of extinction law. The same approach applies for any
dusty gas layer, not just the densest regions identified as molecular
clouds, as long as the amount of reddening is large enough to produce
a measurable shift in the RGB.

As an example, the third panel of Figure~\ref{orionfig} shows the mean extinction
calculated in $25\pc$ bins (i.e, the same width as the grid lines in
the upper panels), if the same region of the Orion molecular cloud
shown in the upper panels were observed in M31; this bin size matches
the fiducial bin size used in our analysis below, and provides a good
balance between ensuring a large number of stars per bin, and
maximizing resolution.  To improve the spatial sampling of the
distribution, we calculate the mean in 4 different grids, each shifted
by half of a $25\pc$ pixel\footnote{Although not an exact equivalent,
  this approach is close to that of dithering to produced a Nyquist
  sampled image --- i.e., we generate $12.5\pc$ pixels of a map with
  native $25\pc$ resolution.}, and then interleave them into a grid
with $12.5\pc$ pixels.  The resulting map of the mean extinction
clearly traces the same structure as the actual molecular cloud,
albeit with lower resolution.  The dynamic range is reduced
because we are tracing the mean, and not the extremes of the
distribution, and are convolving the data over the scale of a pixel.

The fourth panel of Figure~\ref{orionfig} shows the same calculation of the mean
extinction, but instead of averaging all the points in the original
image, we use only averages from randomly selected lines of sight,
which were laid down with an average spatial density of $\sim\!25$
samples per $25\pc$ bin (i.e., equivalent to what would be observed
for an equivalent number of stars observed through the cloud).  In
spite of the fact that the fourth panel has dramatically fewer
samples per bin than in the third panel, the map looks nearly
identical.  Thus, even a relatively coarse sampling of the extinction
distribution is sufficient to map out broad statistical properties of
the reddened distribution.

The example in Figure~\ref{orionfig} bolsters our assumption that we
can neglect depth effects within molecular clouds.  The features in
the Orion map have typical sizes that are less than $50\pc$ in their
longest extents, and that are much smaller perpendicularly.  Assuming
that these small projected sizes are also comparable to the depth of
the cloud along the line of sight, then the cloud must have little
depth compared to the overall distribution of RGB stars.  The scale height of M31's
stars is of order $\sim\!0.5-1\kpc$, which
is more than a factor of 10 times larger than the likely depth of the
cloud.  Thus, the distribution of extinctions for groups of
neighboring stars primarily reflects sampling the distribution of
extinctions across the face of individual clouds, rather than a
distribution of line-of-sight extinctions from RGB stars at different
depths. It also suggests that when modeling the extinctions of
comparable regions, we can most likely neglect any depth effects, and
treat the dust as a thin screen within the stellar disk, such that the
vast majority of stars are either in front or behind the dust layer.
This assumption is likely to also hold even if there are multiple
molecular clouds along the line of sight, given the typical thinness
of molecular gas distributions.

That said, the neglect of depth effects may be less valid in regions
where most of the gas is in the atomic phase, rather than in the dense
molecular phase that has been the typical domain of dust extinction
studies in the Milky Way.  M31 is H{\sc i} dominated, and thus in many
regions the gas layer may well be intrinsically thicker than is
characteristic for individual molecular clouds, given H{\sc i}'s
typically larger velocity disperions. In practice, however, it seems
unlikely that most of the H{\sc i} is in a layer as thick as the
stars.  The H{\sc i} dominated regions are also likely to be the lower
extinction regions where our methodology is most likely to break down
for other reasons (Sec.\ \ref{accuracysec} and \ref{systematicsec}
below).

Before proceeding, it is worth noting the similarities and differences
with other extinction mapping efforts.  By concentrating on mapping a
large area of an external galaxy, our work is similar in spirit to the
series of Harris \& Zaritsky studies of the Magellanic Clouds
\citep{harris1997,zaritsky2002,zaritsky2004}.  However, we use only 2
NIR filters rather than fitting the full spectral energy distribution
of the stars, and limit our analysis to only stars on the RGB.  These
choices reduce the data requirements and reliance on stellar models,
but at the expense of being unable to probe differences in the dust
column experienced by different kinds of stars.  In addition, our
interpretation of the reddening distribution is ``sampling of a
log-normal distribution'', whereas \citet{zaritsky2004} largely
interprets the width of the extinction distribution as being due to
stars' different depths within a thicker diffuse dust layer, although
with some clumpiness being present \citep{harris1997}. Likewise, our
approach has similarities to the Milky Way extinction mapping
described above, given that they are both based in modeling the NIR
color distribution.  However, we are able to use magnitude information
as well as color information, thanks to the lack of any significant
distance spread along the line of sight.  This allows us to use only
two filters (i.e., one color), whereas Milky Way studies typically
employ 3 (i.e., the $J$, $H$, and $K_s$ filters of the 2MASS
survey). In an appendix, we discuss possible extensions of this
technique to bluer filters (Sec.\ \ref{opticalsec}), which could in
principle be more sensitive to lower extinctions.

\section{Expected Distribution of $A_V$} \label{cloudsec}

Simulations of ISM turbulence \citep[see reviews by][and references
  therein]{ballesterosparedes2011, hennebelle2012} suggest that ISM
densities follow a log-normal distribution, both in terms of the
volume density and the projected surface density.  Observationally,
analyses of dense molecular clouds and the diffuse ISM
\citep[e.g.,][]{berkhuijsen2008, hill2008} have found that the
probability distribution function of dust and/or gas follows a broad
distribution around a peak, that in most cases appears to be well fit
by a log-normal density distribution.  Although there are recent
questions about whether the distributions are best fit by a log-normal
or by broken power laws \citep[particularly at low densities where the
  measurements are most uncertain;
  e.g.,][]{alves2014,lombardi2014,lombardi2015}, or whether deviations
from log-normal distributions are the result of integrating over large
areas that sample very different cloud conditions
\citep[e.g.,][]{brunt2015}, or whether somewhat different fitting
functions are preferred in numerical simulations
\citep[e.g.,][]{hopkins2013}, these broad, peaked distributions appear
to be a generic feature of the ISM, and are not tied to any particular
gas phase.

The only consistent deviations from a log-normal density that have
been noted are in extinction studies of the very densest regions,
where some molecular clouds -- particularly those that are star-forming --
are observed to have an additional power law tail to very high
densities \citep[i.e., to high extinctions;][]{kainulainen2011b,
  russeil2013, schneider2013, alvesdeoliveira2014, schneider2015b,
  abreu-vicente2015}, potentially due to the onset of self-gravity
\citep[e.g.,][]{tassis2010, ballesterosparedes2011, kritsuk2011}.
However, the fraction of a cloud's mass that is in this power law tail
is typically small, such that the majority of the area and/or volume
of the cloud samples the log-normal distribution.  For example, in the
map of Orion above, no more than 5\% of the area falls in the
densities associated with the power law tail, for any of the $25\pc$
analysis bins.

Given its ubiquity, we expect the log-normal distribution of ISM
densities to be imprinted on the distribution of reddenings seen in
the color-magnitude diagram.  Since these reddened background stars
are simply sampling different lines of sight, they should experience
the same distribution of ISM column densities seen in the Milky Way,
which will then be manifested as a log-normal distribution of
reddening.  The one caveat is that the extinction distributions
reported in the Milky Way are typically calculated over size scales
that may not agree with our fiducial $25\pc$ bins.  In cases where we
analyze smaller areas we might expect less averaging over ISM
properties, and thus potentially larger departures from the log-normal
assumption.  This effect can be seen in the second panel of
Figure~\ref{orionfig}, where three of the sub-regions show more
complexity in their extinction distributions than can be easily
described by a log-normal. Even in such cases, however, the log-normal
still can provide a meaningful functional form for characterizing the
typical extinction and its dispersion.

Based on the above, for the remainder of our analysis we model the
effect of dust reddening on the predicted distribution of $A_V$ as
follows.  We first assume that all the extinction in an analysis
region comes from a single layer of dusty gas along the line of sight,
across which there is a log-normal distribution of line-of-sight
values of $A_V$.  Moreover, we assume that the gas layer has
negligible thickness compared to the stellar distribution, allowing us
to assume that stars either experience all of the dust column, or none
of it.  The probability distribution $p_A$ for $A_V$ then becomes a
$\delta$-function at $A_V=0$ normalized to the fraction of stars that
lie in front of the gas, plus a log-normal function $p_{A,gas}$
describing the distribution of extinctions across the face of the gas
layer, normalized to $f_{red}$, the fraction of stars that lie behind
the layer:

\begin{equation}  \label{pAVeqn}
\begin{split}
p_A(A_V|f_{red},\widetilde{A_V},\sigma)\,dA_V = (1-f_{red})\,\delta(A_V)\,dA_V \\
 + f_{red}\,p_{A,gas}(A_V|\widetilde{A_V},\sigma)\,dA_V,
\end{split}
\end{equation}

\noindent where the extinction distribution of the dusty gas layer
$p_{A,gas}$ is

\begin{equation}  \label{pAVcloudeqn}
p_{A,gas}(A_V|\widetilde{A_V},\sigma)\,dA_V = \frac{1}{A_V\sqrt{2\pi\sigma^2}} \,
                 e^{-\frac{(\ln{(A_V/\widetilde{A_V})})^2}{2\sigma^2}}\,dA_V.
\end{equation}

\noindent The parameters of the log-normal distribution $p_{A,gas}$
are $\widetilde{A_V}$, which is the median extinction of the stars
behind the dust, and $\sigma$, which is a dimensionless parameter that
sets the width and skewness of the log-normal distribution;
smaller values of $\sigma$ produce more symmetric distributions with
less significant tails to large values.  These quantities can be
related to the mean extinction $\langle A_V \rangle$ and the standard
deviation of the extinction $\sigma_{A}$ using equations found below
in Sec.\ \ref{otherquantitiessec}. In our model, the median extinction
$\widetilde{A_V}$ is independent of the fraction of stars that are
actually reddened, and will therefore have the same value whether the
fraction of the stars behind a particular molecular cloud is 10\% or
90\%.  This treatment could be easily generalized to multiple gas
layers along the line of sight, but in practice we expect this
occurrence to be rare.

When modeling the CMD, the majority of the leverage on measuring $p_A$
does not come from measuring extinction, but instead from measuring the
reddening in the NIR $F110W-F160W$ color\footnote{While extinction
  does offer some constraints through the structure of the RGB
  luminosity function, the intrinsic narrowness of the RGB makes color
  a far more sensitive indicator of the presence of dust.  Other
  features --- such as the tip of the RGB --- have the needed intrinsic
  narrowness in luminosity, but do not contain many stars, making them
  a noisier probe of extinction.}.  We therefore transform the
extinction probability distribution into one of reddening for an
arbitrary color $X-Y$.  Once the values of $A_X/A_V$ and $A_Y/A_V$ are
fixed by adopting a specific attenuation curve $A_\lambda/A_V$, the
reddening $E(X-Y)$ becomes

\begin{equation}  \label{reddeninglaw}
E(X-Y) = A_V \,\left[\frac{A_X}{A_V} - \frac{A_Y}{A_V}\right].
\end{equation}

\noindent Changing variables from $A_V$ to the reddening ${\mathcal E}_{XY}
\equiv E(X-Y)$, the probability distribution of reddenings becomes

\begin{equation}  \label{pEXYeqn}
\begin{split}
p_{{\mathcal E}}({\mathcal E}_{XY}|f_{red},\widetilde{A_V},\sigma,A_\lambda/A_V) = 
     (1-f_{red})\,\delta({\mathcal E}_{XY}) \\
     + f_{red}\,p_{{\mathcal E},gas}({\mathcal E}_{XY}|\widetilde{A_V},\sigma,A_\lambda/A_V),
\end{split}
\end{equation}

\noindent where

\begin{equation}  \label{pEXYcloudeqn}
\begin{split}
p_{{\mathcal E}, gas}({\mathcal E}_{XY}|\widetilde{A_V},\sigma,A_\lambda/A_V) \,d{\mathcal E}_{XY} \equiv \\
     p_{A,gas}(A_V=\frac{{\mathcal E}_{XY}}{\left[\frac{A_X}{A_V} - \frac{A_Y}{A_V}\right]}|\widetilde{A_V},\sigma) \\
     \times \left[\frac{A_X}{A_V} - \frac{A_Y}{A_V}\right]\,dA_V.
\end{split}
\end{equation}

For the specific case of NIR data in the PHAT data set, we adopt
$A_{F110W}/A_V=0.3266$ and $A_{F160W}/A_V=0.2029$ from
\citet{girardi2008}'s application of a standard $R_V=3.1$
\citet{cardelli1989} attenuation law to typical RGB spectrum with a
NIR color of $F110W-F160W=0.75$. These ratios vary somewhat with the
temperature of the star, and thus are not constant for all of the
stars on the RGB.  However, fitting to the \citet{girardi2008}
isochrones for RGB stars shows that the color dependence is quite
weak, with $A_{F110W}/A_V=0.3448 - 0.0243 ({\rm F110W} - {\rm F160W})$
and $A_{F160W}/A_V=0.2061 - 0.0043 ({\rm F110W} - {\rm F160W})$. The
RGB stars in our analysis have intrinsic colors in the range
$0.5\lesssim {\rm F110W} - {\rm F160W} \lesssim 1.0$, and thus the
adopted extinctions are good to $\pm$2\% in F110W and $\pm$0.5\% in
F160W throughout the RGB, assuming $R_V=3.1$.  These scalings would be
$\sim$15\% larger if the true attenuation law is much steeper
($R_V=5$), with $A_{F110W}/A_V$ increasing by 0.051, and
$A_{F160W}/A_V$ by 0.031. This change would be smaller than our
typical precision.

We will quote results in terms of $A_V$ to simplify comparisons with
previous work and theoretical calculations, but the reader should note
that the fit for $A_V$ is driven almost entirely by $E(F110W-F160W)
= 0.124 A_V$.  In addition, our adopted ratio $E(F110W-F160W)/A_V$
is 6\% smaller than the canonical value for a solar-type star, due to
the underlying cooler temperature we have assumed. Although the exact
conversion from reddening to extinction depends on assumptions for the
underlying stellar spectrum and for the attenuation law, the former
does not vary dramatically among RGB stars, and the latter appears to
be quite stable in the NIR among different lines of sight in the Milky
Way \citep[e.g.,][]{martin1990}.  

\subsection{Other Useful Quantities} \label{otherquantitiessec}

When interpreting dust maps, it is frequently useful to use the mean
extinction $\langle A_V \rangle$, which is a better measure of the
integrated column density of gas along the line of sight for a fixed
dust-to-gas ratio. For a log-normal distribution, the mean depends on
both the median $\widetilde{A_V}$ and the dimensionless width
$\sigma$:

\begin{equation}  \label{meaneqn}
\langle A_V \rangle = \widetilde{A_V} \, e^{\sigma^2/2}.
\end{equation}

In some circumstances it may be useful to consider the actual width of
the log-normal in magnitudes of extinction (i.e. the standard
deviation $\sigma_A$), rather than the dimensionless parameter
$\sigma$.  If needed, one can relate $\sigma$ to the standard deviation
$\sigma_A$ of the distribution using either

\begin{equation}  \label{sigmaAeqn}
\sigma_A^2 = \widetilde{A_V}^2 e^{\sigma^2}(e^{\sigma^2}-1)
\end{equation}

\noindent or

\begin{equation}  \label{sigmaeqn}
\sigma^2 = \ln{\left[\frac{1 + \sqrt{1 + 4w^2}}{2}\right]}
\end{equation}

\noindent where $w \equiv \sigma_A / \widetilde{A_V}$.  Note that
$\sigma_A$ scales linearly with the median $\widetilde{A_V}$, such
that in regions of high extinction, the same value of $\sigma$
corresponds to an intrinsically wider distribution.  Because of this
dependence, we choose to fit for the dimensionless width $\sigma$,
which is more likely to be invariant across the disk.

\section{Modeling the Observed Distribution of Reddening} \label{modelingsec}

The basic outline of our fitting for the distribution of reddening
(i.e., $p_{\mathcal E}$) is as follows.  We first use NIR data from
the PHAT survey to generate high-quality CMDs, focused on the
well-populated RGB (Sec.\ \ref{nirdatasec}).  We then create an
empirical model of the unreddened foreground as a function of
position, by isolating small subregions with extremely low reddening,
as indicated by their narrow, blue RGB and low FIR emission (Sec.\
\ref{unreddenedsec}). Assuming (1) that the RGB stars are well-mixed
along the line of sight, such that the reddened stars on the far side
of the gas layer have the same intrinsic CMD as the unreddened stars;
and (2) that the distribution of extinctions is log-normal, we use the
unreddened CMD and eqn.~\ref{pAVeqn} to generate models for the NIR
CMD for arbitrary values of the median extinction $\widetilde{A_V}$,
the dimensionless width $\sigma$ of the log-normal distribution of
extinctions, and the fraction of reddened stars $f_{red}$
(Sec.\ \ref{modelcmdsec}).  The first two of these parameters
characterize the properties of the dust distribution, and the third
constrains the relative geometry of the gas and stars.  We then use a
Monte Carlo Markov-Chain (MCMC) to evaluate the relative posterior
probability that a given set of model parameters are an acceptable fit
to the observed CMD (Sec.\ \ref{fittingsec}).  Finally, we repeat this
process in a series of interleaved, oversampled pixels to generate a
map of the three parameters that describe the dust distribution, and
their associated uncertainties. We now describe the details of all the
stages in this process, for the specific data in the PHAT survey.

\subsection{NIR Data}  \label{nirdatasec}

We select stars with photometry in the WFC3/IR F110W and F160W
filters from the simultaneous six-filter photometry described in
\citet{williams2014}. We analyze individual ``bricks'', which are
large contiguous rectangular regions consisting of 3$\times$6 WFC3/IR
pointings that have been astrometrically aligned, merged in overlap
regions, and culled at the outer edges such that each star appears
only once in a ``brickwide catalog''.  Although we include Brick 1 in
fits involving the local stellar surface density to improve
continuity, we do not analyze its reddening distribution; this brick
is dominated by M31's high surface brightness bulge and the majority
of its photometry is quite shallow due to high crowding.  The stellar
population in the bulge also contains a significant high metallicity
population, leading to a very broad RGB, limiting our ability to apply
the extinction mapping technique.  Although we cannot map the dust in
the bulge using resolved stars, \citet{dong2014} has analyzed its dust
distribution using PHAT's integrated photometry in tandem with GALEX
and SWIFT observations.

We apply further culls to the publicly released photometry
to reduce the number of stars with obviously bad photometry (largely
due to unresolved blends and diffraction spikes from bright stars).
While the number of such stars is small, their errors typically place
them in the region redward of the RGB, close to the detection limit in
F110W where they can be mistaken as highly reddened stars. Thus,
even a small amount of contamination can bias the reddening
distribution to spuriously high values of the extinction.

To reduce rare instances of corrupted or highly uncertain photometry,
we adopt cuts on DOLPHOT's \citep{dolphin2000} photometric
quality parameters to ensure: (1) high signal-to-noise; (2) low
probability of significant errors in subtraction of neighboring stars;
and (3) efficient rejection of cosmic rays and of false-detections on
the diffraction spikes of bright stars.  Specifically, we require
that the signal-to-noise ratio of each star was greater than 5 in both
NIR filters, that the crowding parameter (which indicates the amount
of contamination from neighboring stars\footnote{See
  \citet{dolphin2000} for the specific definition of the photometric
  quality parameters used here.}) and ``roundness'' parameter were
small (${\tt{crowd_1}}^2+{\tt{crowd_2}}^2<2.5^2$ and
${\tt{round_1}}^2+{\tt{round_2}}^2<4.0^2$).  We also restrict the
sharpness parameter, which evaluates whether the stellar profile is
sharper or flatter than the point spread function. We fit an
ellipse to the 2-dimensional distribution of sharpness in both
filters and stars whose distribution in {\tt{sharp$_1$}} versus
{\tt{sharp$_2$}} fell within an ellipse with major axis length 0.35,
position angle $-128\deg$, and axis ratio 0.55.
These combinations eliminate clear
outliers in the space of photometric quality parameters while
eliminating very few stars with clearly good photometry (i.e., those
that fell on the CMD features occupied by the vast majority of
stars). 

\begin{figure*}
\centerline{
\includegraphics[width=3.25in]{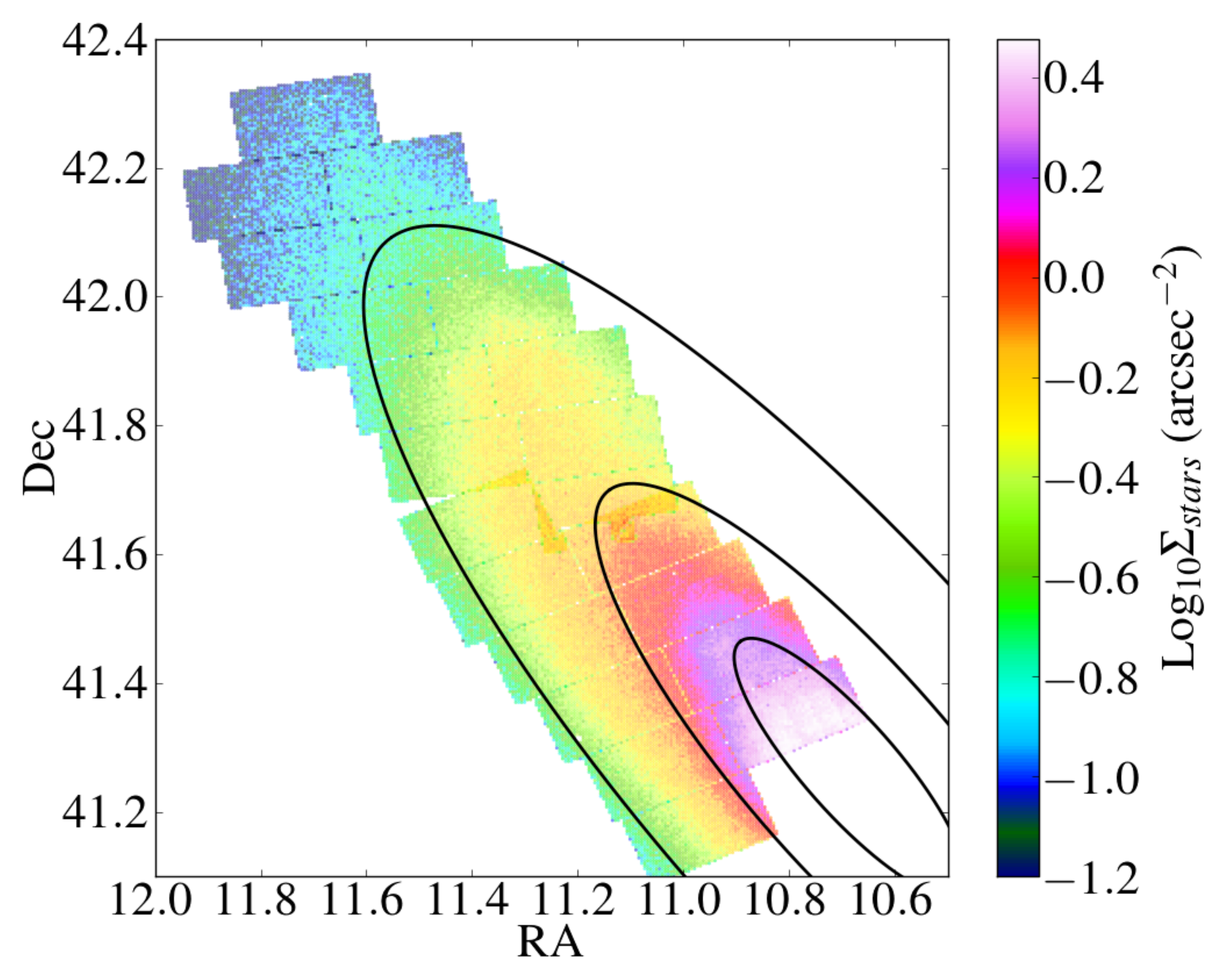}
\includegraphics[width=3.25in]{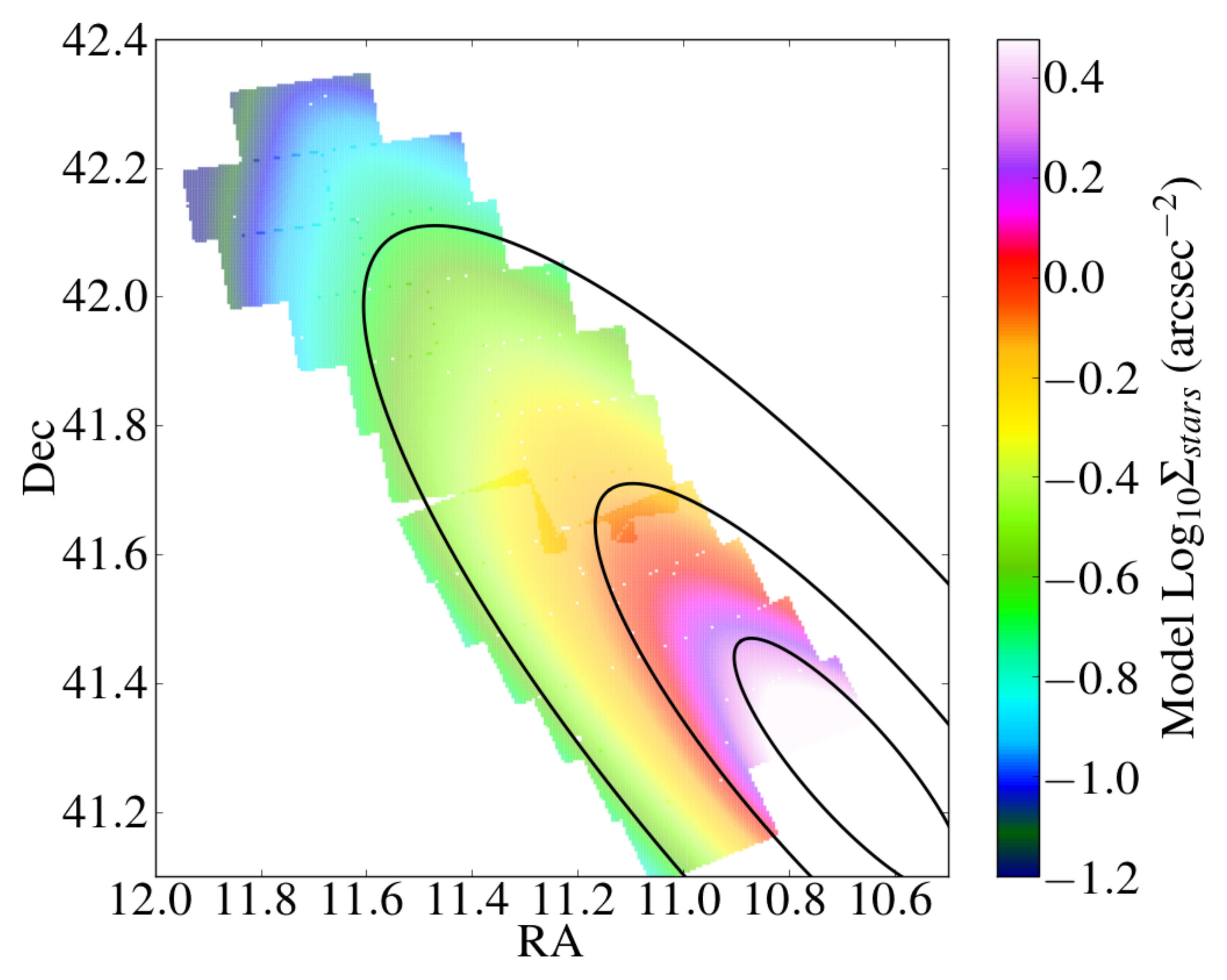}
}
\caption{Mapping the stellar surface density $\Sigma_{stars}$
  (Sec.\ \ref{nirdatasec}). (Left) Log$_{10}$ of the surface density
  of NIR-detected RGB stars (in arcsec$^{-2}$) with $18.5>F160W>21.0$
  and $F160W > 0.3$, calculated within $15\arcsec$ wide pixels.
  Outside of the dense inner bulge, the surface density falls smoothly
  with increasing distance from the center, with no obvious ``holes''
  caused by regions of high dust extinction.  For reference, we plot
  solid ellipses to represent the isophotes expected for an inclined
  disk with a position angle of 38.5$\degrees$, inclination of
  74$\degrees$, and major axis lengths of 0.25$\degrees$,
  0.55$\degrees$, and 1.05$\degrees$; compared to these locii, it is
  clear that the internal stellar density distribution of M31 is
  complex, showing multiple components. (Right) A smooth model of the
  NIR surface density, generated from the sum of 12 Gaussian radial
  surface density profiles, each with a different position angle and
  inclination. The model reproduces the observed distribution to
  within $\sim$15\% at all radii. \label{surfdensmapfig}}
\end{figure*}

We have considered whether these cuts are unfairly removing reddened
stars.  We find that there is no {\emph{a priori}} reason why the
reddened stars should be any more affected by the cuts in crowding and
sharpness than the unreddened foreground, given that the cuts remove
stars with known photometric limitations that are due to features that
are unaffected by reddening (i.e., the random probability of blending
with a star of a small projected separation, or of lying on a cosmic
ray or diffraction spike).  We also see no anti-correlation between
the number density of stars and the local extinction. Instead, the
surface density of stars is very smooth throughout the survey area
(left panel of Figure~\ref{surfdensmapfig}, where $\Sigma_{stars}$ is
defined to be the number density of stars in arcsec$^{-2}$ with
$18.5>{\rm{F160W}}>21.0$ and ${\rm{F110W}} - {\rm{F160W}} > 0.3$,
generated in $15\arcsec$ wide pixels).

\begin{figure}
\centerline{
\includegraphics[width=3.05in]{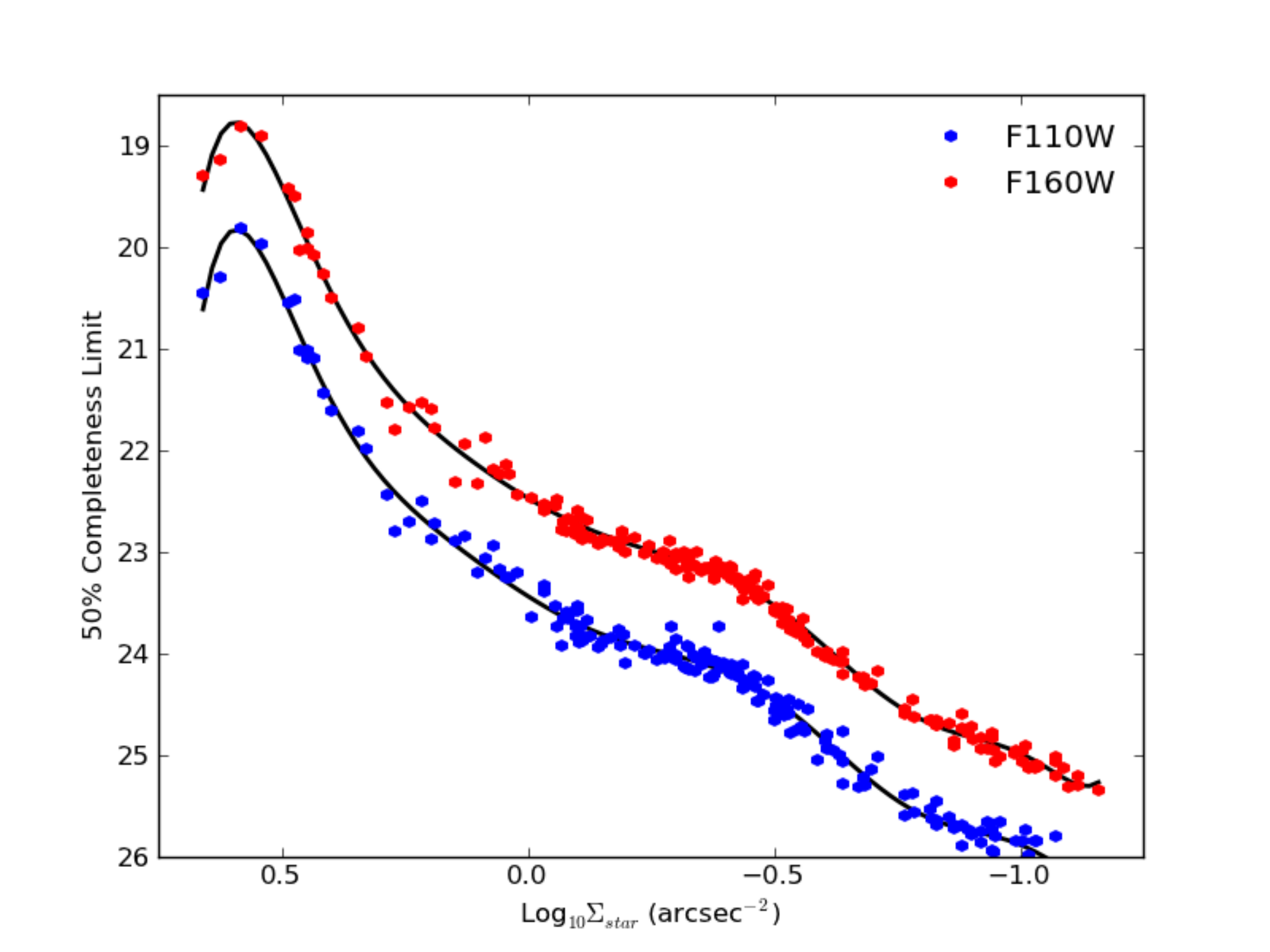}
}
\caption{The depth of the photometry as a function of radius. The blue
  and red points show the 50\% completeness limit $m_{50}$ as a
  function of log$_{10}$ of the local stellar surface density in
  bright RGB stars, for the F110W and F160W filters, respectively.
  The solid lines show 10th order polynomial fits to the data, used to
  interpolate to an appropriate limiting magnitude at each position in
  the galaxy. The PHAT photometry is crowding-limited, and thus the
  limiting magnitude is a strong function of the local stellar surface
  density.  The apparent roll-off in the inner disk (at high surface
  densities in the left side of the plot) is an artifact of the
  extraordinarily high crowding levels in the inner bulge, which lead
  to biases in the photometry; while we include these regions for
  completeness here, we do not actually analyze their $A_V$
  distributions. \label{maglimfig}}
\end{figure}

To minimize the effect of incompleteness and magnitude biases, we
impose additional magnitude cuts beyond those implicit in the
signal-to-noise cut.  We use artificial star tests to define the 50\%
completeness limit, $m_{50}$, in each filter (i.e., the magnitude for
which a star has a 50\% chance of being detected) for a large number
of fields across our survey area.  Fainter than the 50\% completeness
magnitude, the detection fraction falls dramatically while the
photometric bias rises, due to an increase in the fractional flux
contributed by blends with faint undetected stars. For each filter we
fit a polynomial to $m_{50}$ as a function of the local stellar
density, as determined by a multicomponent model fit to the log of the
stellar surface density in bright RGB stars (i.e., right panel of
Figure~\ref{surfdensmapfig}).  The data and polynomial fits are shown
in Figure~\ref{maglimfig}.  We then use the local surface density to
define a spatially-variable magnitude cut at $m_{50}$ that we 
apply to the photometry.  The effect of the 50\% completeness cut
can be seen in Figure~\ref{unreddenedCMDfig}, which shows $m_{50}$
superimposed on CMDs constructed from low extinction regions at a
range of representative local densities.  We make the F160W cut even
more stringent (by $\Delta m_{50,F160W}=0.5\mags$) than shown in
Figure~\ref{unreddenedCMDfig} to remove stars near the bottom of the
CMD, where even small amounts of reddening would move the star beyond
the magnitude limit in F110W; such stars are numerous but do not add
significant leverage in measuring the reddening distribution and thus we
wish to minimize their impact on the fit to the CMD.

In addition to culling poorly measured stars, we also make an initial
correction for spatially-dependent variations in the photometry across
the WFC3/IR chip. The version of the photometry used in
\citet{williams2014} does not include spatial variation in the PSF
model in the NIR. This can lead to systematic variations in the
photometry across the face of the camera, such that stars in one
position on the chip will be measured systematically brighter or
fainter than in another.  Errors in the large scale flat-field can
also lead to similar issues.  While these effects are partially
corrected for by DOLPHOT's perturbations of the PSF model, small
differences on the scale of a hundredth of a magnitude persist. Our
sensitivity to dust depends on small shifts in the NIR color, and thus
failure to correct for these effects will limit the sensitivity we
have to low levels of attenuation.

We develop a position-dependent correction map for the color by
analyzing fields in the outer galaxy that have low extinction on the
\citet{draine2013} emission-based dust map. We use 54 individual
fields in Bricks 2, 6, 8, 10, 11, 12, 14, 18, 19, 20, 22, and 23, and
calculate each star's color relative to the RGB locus.  We bin the
stars as a function of position on the NIR chip in a 40$\times$40
grid, keeping only bright stars that fall in positions where the extinction
$A_{V,emission}$ predicted by \citet{draine2013} is less than
$0.15\mags$, and then calculate the median of the color
offset.  The resulting map shows coherent structure, with redder
colors found for the largest values of $Y$, peaking at the middle of
the range in $X$. Bluer colors are found for the lowest values of $Y$,
peaking for large values of $X$. The amplitude of the variation is
small, with peak-to-peak variations of roughly $\pm0.01\mags$, comparable
to the amplitude seen by \citet[][; see their Fig. 24]{williams2014}. The
structure appears robust with respect to different choices of
magnitude range, binning scheme, mean vs. median, field choice,
etc. We find similar patterns when using colors derived from measuring
the magnitude of the red clump, although the resulting maps are less
accurate due to the red clumps larger intrinsic width.  

We apply this correction to all of the photometry before beginning
analysis, choosing to modify the F160W magnitude to produce the
required color shift. Although the pattern in the map of color shifts
undoubtedly reflects differences in photometry between F110W and F160W,
we cannot deduce the magnitude shifts from the RGB color alone. In
practice, the effect of shifting magnitudes by $\pm0.01\mags$ is
minimal, since these shifts are significantly smaller than the
magnitude error of the size of the magnitude bins we use when fitting
the observed RGB.

\subsection{Generating a Model for the Unreddened RGB} \label{unreddenedsec}

Building a model of the unreddened RGB that includes
spatial variations is essential to mapping the extinction.  Our
extinction mapping technique relies on detecting stars that are
``redder than expected'', and thus spatial variations in the expected
stellar colors can be misinterpreted as reddening if the underlying
RGB model is incorrect. We therefore must build a model for the unreddened
RGB and red clump that captures as much of these variations as possible.

Unfortunately, stellar population gradients, density-dependent
photometric quality, and projection effects make the CMD structure of
the RGB and red clump complex and difficult to model.  Generating a
theoretical model for the unreddened RGB would require solving for the
detailed star formation history across the disk to account for
gradients in age and metallicity, while also building a model for
projection effects and photometric errors.  Such a model would
be subject to errors in the disk model, the characterization of
photometric errors throughout the disk, and the star formation history
itself, as well as any errors in the underlying stellar isochrones and
atmospheric models, which can be particularly significant for cool
stars in the NIR.

Instead of trying to simulate these effects, we adopt an empirical
approach.  We isolate low-extinction regions throughout the galaxy, and then
bin the stars in these regions to generate a model CMD.  We create
these models as a function of the local stellar surface density, which
captures variations in both the underlying stellar populations and in
the photometric errors. 

Here we discuss the process used to generate the spatially-dependent
model of the unreddened RGB, the properties of the RGB itself, and the
associated uncertainties.

\subsubsection{Spatial Variations in the NIR RGB}

Almost all of the stars in the PHAT NIR photometry lie on the RGB or in
the ``red clump'' found at the red end of the horizontal branch.
These stars span a range of ages ($\gtrsim1\Gyr$) and metallicities,
depending on the exact star formation and chemical enrichment history
of M31.  The colors of these stars depend on their age and
metallicity, with older and more metal rich stars favoring redder
colors\footnote{M31's older, more metal poor RGB stars have similar
  NIR colors to its younger, more metal rich population (i.e., the
  bluer colors of younger RGB stars are partially cancelled by their
  being more metal rich), and thus the NIR RGB is typically
  narrow. This narrowness allows even modest amounts of reddening to
  be detectable.}. The RGB's mean color and width can therefore change
as a function of position if the past star formation history and
enrichment varies spatially.  Metallicity and age gradients are common
in disk galaxies, and distinct structures such as M31's spheroid and
bar may likewise have star formation histories that differ from those
in the disk. The net result is that the structure of the RGB is likely
to vary with position within the galaxy. While the RGB is sufficiently
old that dynamical evolution will have smoothed out any sharp
features, failure to account for the anticipated smooth variations
will lead to errors when trying to model the color distribution of RGB
stars.

The spatial variations in RGB and red clump structure can also be
subtly altered by viewing angle.  The RGB and red clump are dominated
by older stars, many of which were born in or dynamically heated into a
thicker disk. When this disk is inclined, a range of radii will be
present along a given line of sight.  The stars at a given projected
radius are therefore a blend of the stars from a range of radii and
scale heights.  The degree of this ``projection mixing'' will depend
on distance from the major axis. Stars on the major axis come from a
range of scale heights and azimuthal angles, but will all be at the
same radius. Away from the major axis, however, projection
effects will mix together stars from a range in radii, not just scale
height.  This projection mixing will leave a signature on the
structure of the RGB and red clump as a function of distance from the
major axis, if the age and metallicity vary with radius and if the
disk is thick compared to the scale length over which the intrinsic
RGB properties vary.  We currently do not try to capture the impact
of projection mixing, which requires knowledge not just of the radial
gradients in the stellar populations, but on the amplitude of such
gradients with scale height; we expect that projection mixing will be
second-order compared to the radial gradient and photometric errors.

Finally, in addition to spatial variations due to projection effects
and in the intrinsic properties of the RGB and red clump, we expect
there to be spatial variations in the quality of the photometry.  The
NIR observations of M31 are highly crowded, which affects the behavior
of photometric errors, biases, and depth.  In general, all of these
quantities are adversely affected when the stellar density increases,
such that photometric errors and biases are larger at a given
magnitude, and the 50\% completeness limit is much brighter
\citep[see][]{williams2014}. We therefore expect the RGB to be broader
and shallower in regions of the highest stellar density.

\begin{figure}
\centerline{
\includegraphics[width=3.05in]{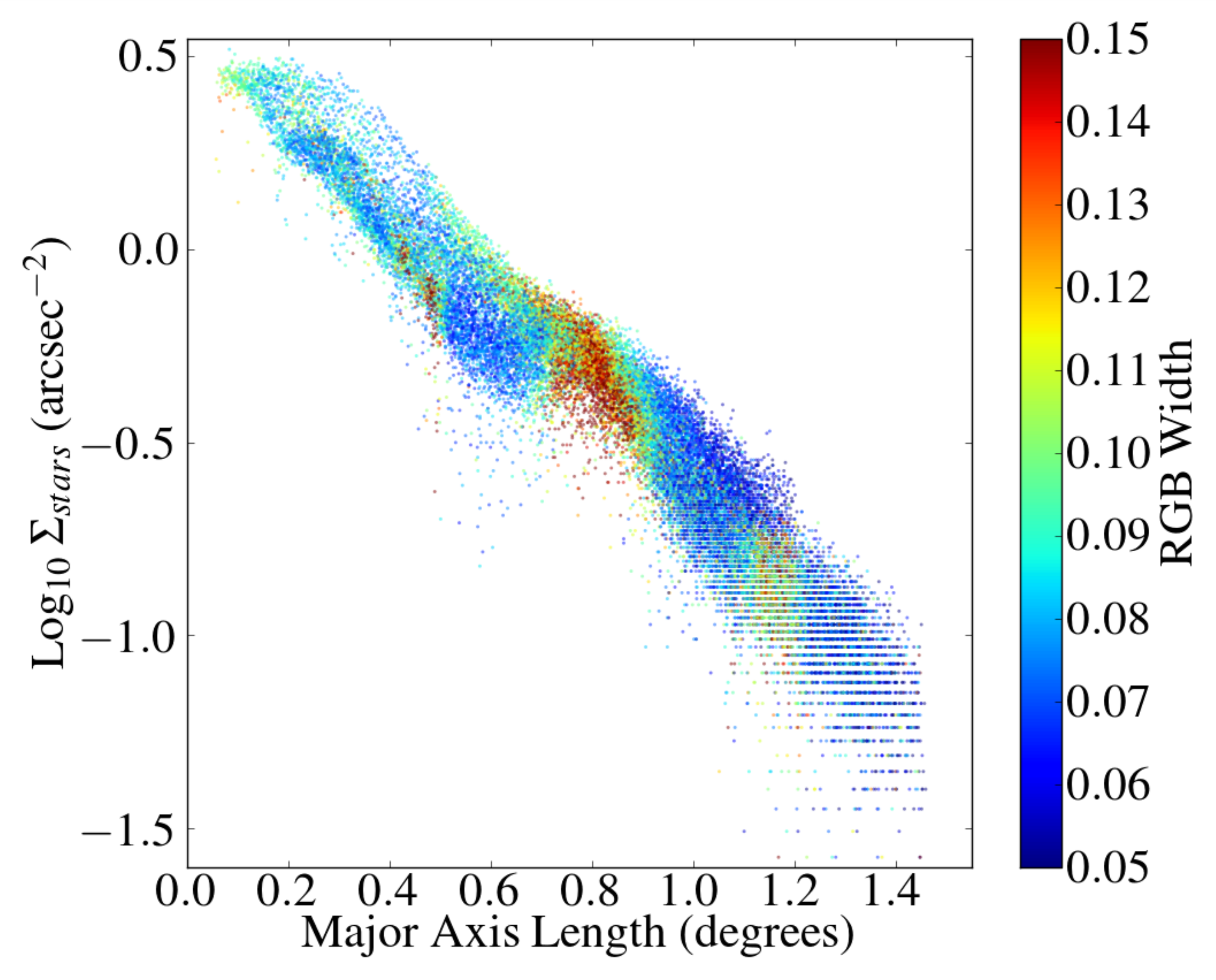}
}
\caption{The mapping between Log$_{10}$ of the surface density of
  bright RGB stars and the distance along the major axis (see
  discussion in Sec.\ \ref{unreddenedsec}). Points are color coded by
  the standard deviation of the upper RGB, relative to a fiducial
  locus, as in Figure~\ref{RGBwidthcolormapfig}.  The color coding
  indicates the position of the major star-forming ring and the more
  minor outer star-forming arm, at $\sim0.75\degrees$ and
  $\sim1.15\degrees$ from the center, respectively. The outer disk
  declines approximately as an exponential disk, but the inner disk is
  multivalued, because the single inclined disk model used to
  calculate the major axis length is not appropriate for the complex
  inner disk.  The falloff towards the center ($\lesssim\!0.2^\circ$)
  is due solely to increasing incompleteness in the highly crowded
  inner bulge.  \label{radialsurfdensfig}}
\end{figure}

To characterize the unreddend RGB as a function of position, we use
the stellar surface density $\Sigma_{stars}$ to track position, rather
than projected radii.  We choose to work in the space of stellar
surface density (rather than radius), because the local surface
density is the dominant factor setting the quality and depth of
photometry for crowding-limited images.  This variation in photometric
quality has far more of an effect on the CMD morphology of the
unreddened RGB than do the radial gradients in the underlying stellar
population (which are small, as we show below). Moreover, the local
density is an equally good proxy for ``distance from the center of
M31'', allowing it to track stellar population gradients without
needing to be tied to a specific model for the 3-dimensional stellar
distribution; a single disk model is not appropriate where the
structure of the galaxy is far more complex (e.g., the inner bulge and
bar \citep[e.g.,][]{beaton2007} and the outer warp
\citep{innanen1982}). The decrease in the stellar density with radius,
and the structurally complex inner region can be seen in
Figure~\ref{radialsurfdensfig}, where we plot the log of the surface
density in bright RGB stars (from Figure~\ref{surfdensmapfig}) as a
function of the projected major axis for a fiducial inclined disk
model\footnote{We refer to the projected major axis length as
  ``radius'' throughout this paper. We adopt a fiducial inclined disk
  when converting position into radius, assuming PA=38.5$\degrees$,
  inclination=74$\degrees$, and center ($\alpha_0$,$\delta_0$) =
  (10.6847929$\degrees$, 41.2690650$\degrees$) in J2000, which
  approximates the disk's isodensity contours in the Spitzer 3.6$\mu$m
  image of M31. However, M31 has significant structural complexity,
  and thus this projected radius is not necessarily an accurate
  measure of the distance to the center of M31.}.

\subsubsection{Identifying Candidate Low Extinction Regions} \label{candidatelowAVsec}

We isolate low-extinction regions by identifying areas where the RGB
is narrow and unreddened.  We first divide the stars within each PHAT
``brick'' into bins $15\arcsec$ on a side ($\sim\!57\pc$) and then
make a broad cut in color and magnitude to isolate stars on the upper
RGB ($19.0\!<\!{\rm F160W}\!<\!22$).  These regions are larger than
our fiducial $25\pc$ analysis regions to ensure there are enough stars
on the upper RGB to provide reliable measurements of the RGB's width
and color. We then calculate mean color of the RGB relative to a
fiducial RGB locus (approximated as ${\rm F110W}-{\rm F160W} = a_0 +
a_1({\rm F160W}-m_0) + a_2({\rm F160W}-m_0)^2$ where $m_0=22.0$,
$a_0=0.752$, $a_1=-0.1$, $a_2=-0.006$).  We also calculate the
standard deviation of the distribution of color differences between
RGB stars and the fiducial locus; we refer to this standard deviation
as the ``width'' of the RGB in subsequent plots.  This definition of
``width'' is somewhat sensitive to differences in slope between the
fiducial and the actual RGB.  It therefore does not solely indicate
broadening, given that an RGB with a different slope than the fiducial
could also have a significant width by this measure, even if it were
intrinsically narrow. However, such slope differences are modest and
vary smoothly with radius (Fig.\ \ref{RGBgradientfig}; left panel),
and thus the ``width'' will still identify a locally broader
RGB.

\begin{figure*}
\centerline{
\includegraphics[width=3.25in]{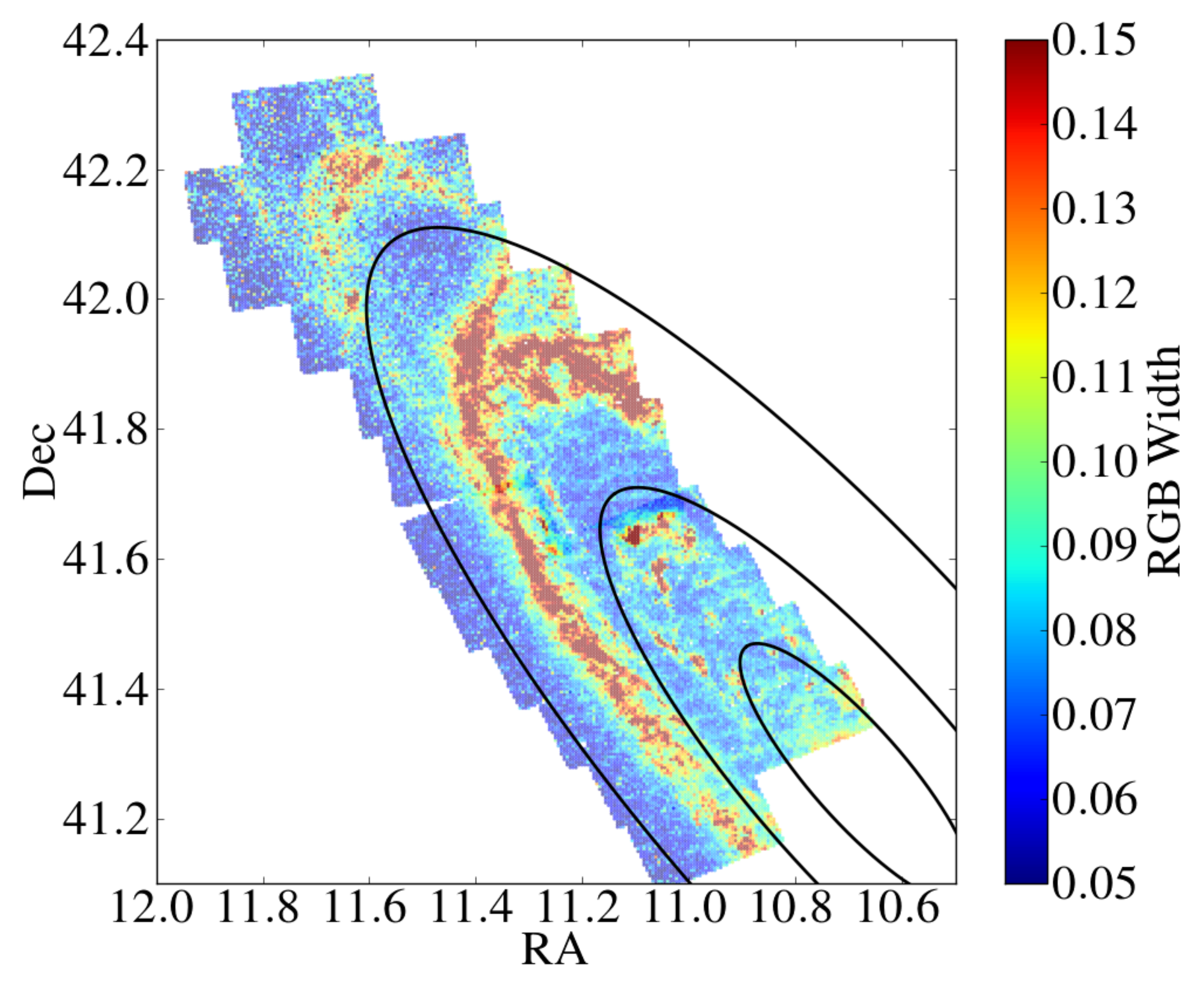}
\includegraphics[width=3.25in]{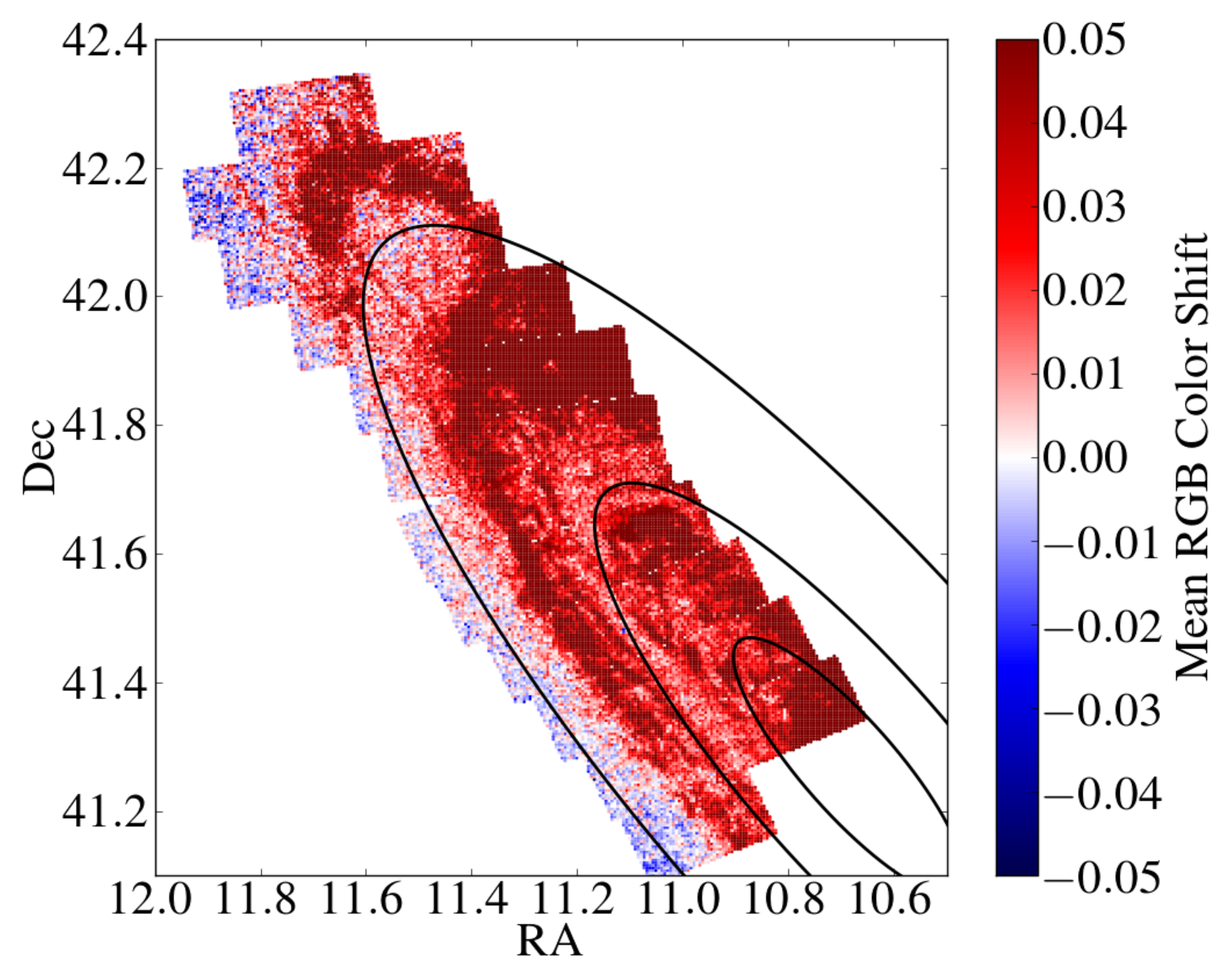}
}
\caption{Maps of the width and color of the upper RGB
  (Sec.\ \ref{candidatelowAVsec}). (Left) Standard deviation of bright
  ($19.0 < F160W < 22$) stars around a fiducial RGB locus, as a
  function of position. The RGB is broader in regions of higher
  extinction.  The typical width of the RGB also increases towards the
  inner galaxy due to larger photometric errors from stellar crowding
  and undetected blends, and potentially due to a broader range of
  intrinsic properties in the underlying stellar population.  (Right)
  The mean color of bright ($19.0 < F160W < 22$) stars relative to a
  fiducial RGB locus, as a function of position.  The mean colors are
  redder in regions of high extinction, as expected.  There are also
  radial trends towards bluer relative colors in the inner
  disk. However, this does not indicate a systematic bluing of the
  population, and instead reflects increasing mismatch between the RGB
  and the fiducial locus. Note also that there is some repeating
  structure in the mean RGB color that mimics the positions of the
  WFC3/IR fields.  This structure is likely to be due to residual
  $<0.015\,{\rm mag}$ errors in the WFC3/IR flat fields, or in the
  spatially-varying PSF model. Ellipses are at the same location as in
  Figures~\ref{surfdensmapfig}. \label{RGBwidthcolormapfig}}
\end{figure*}

Figure~\ref{RGBwidthcolormapfig} shows the resulting maps of the width
of the RGB and the mean color shift of the stars with respect to the
fiducial RGB locus. The main gas-rich star-forming rings clearly show
up as regions that have larger RGB widths and redder colors than is
typical for their location in the galaxy. The shift in mean color is
also largest eastward of the major axis, which is the near side of
M31.  In this region, a much larger fraction of stars are reddened, as
is obvious from optical images of the galaxy.

We isolated the low-extinction regions from these maps as follows.
First, we assembled a list of candidate low-extinction regions for
each brick.  Within each brick, we tagged regions by their projected
major axis length for the fiducial inclined disk model, and sorted
these subregions into 20 bins of radius within each brick. Then,
within each of those radial bins, we identify candidate low $A_V$
regions by tagging the 20\% of subregions that have the narrowest
RGBs, as characterized by the standard deviation in the color of their
RGB relative to the fiducial RGB locus.  This process guarantees that
candidate regions are identified uniformly throughout the brick. We
repeat this process for every brick and then merge the candidate
low-extinction regions into a list that is guaranteed to sample the
entire area covered by the PHAT data.

\subsubsection{Producing an RGB Model as a Function of Local Surface Density} \label{modelRGBsec}

After merging the candidate low-extinction regions from all bricks, we
analyze the resulting list to find the lowest exinction subregions
associated with a given stellar surface density.  

\begin{figure*}
\centerline{
\includegraphics[width=3.25in]{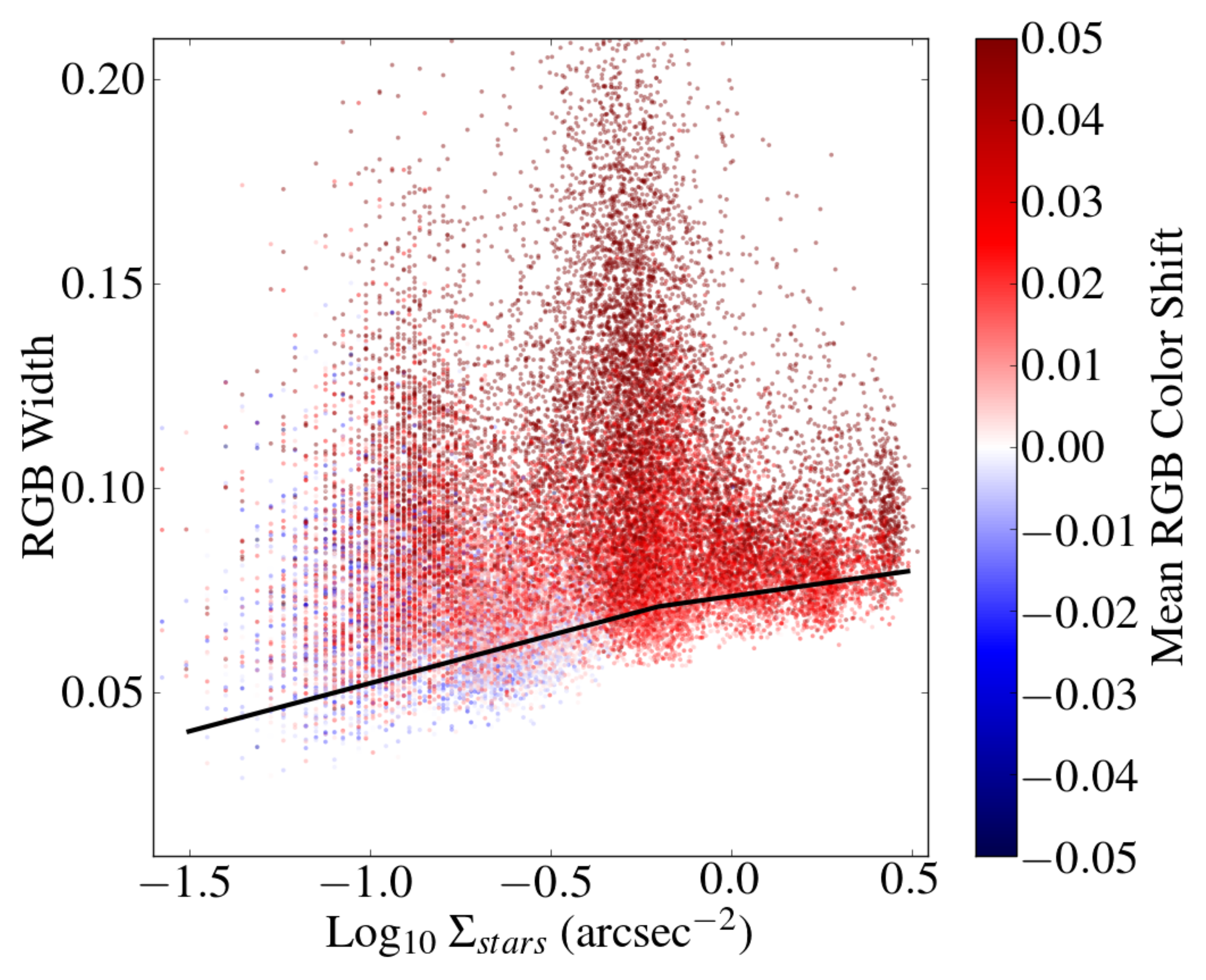}
\includegraphics[width=3.25in]{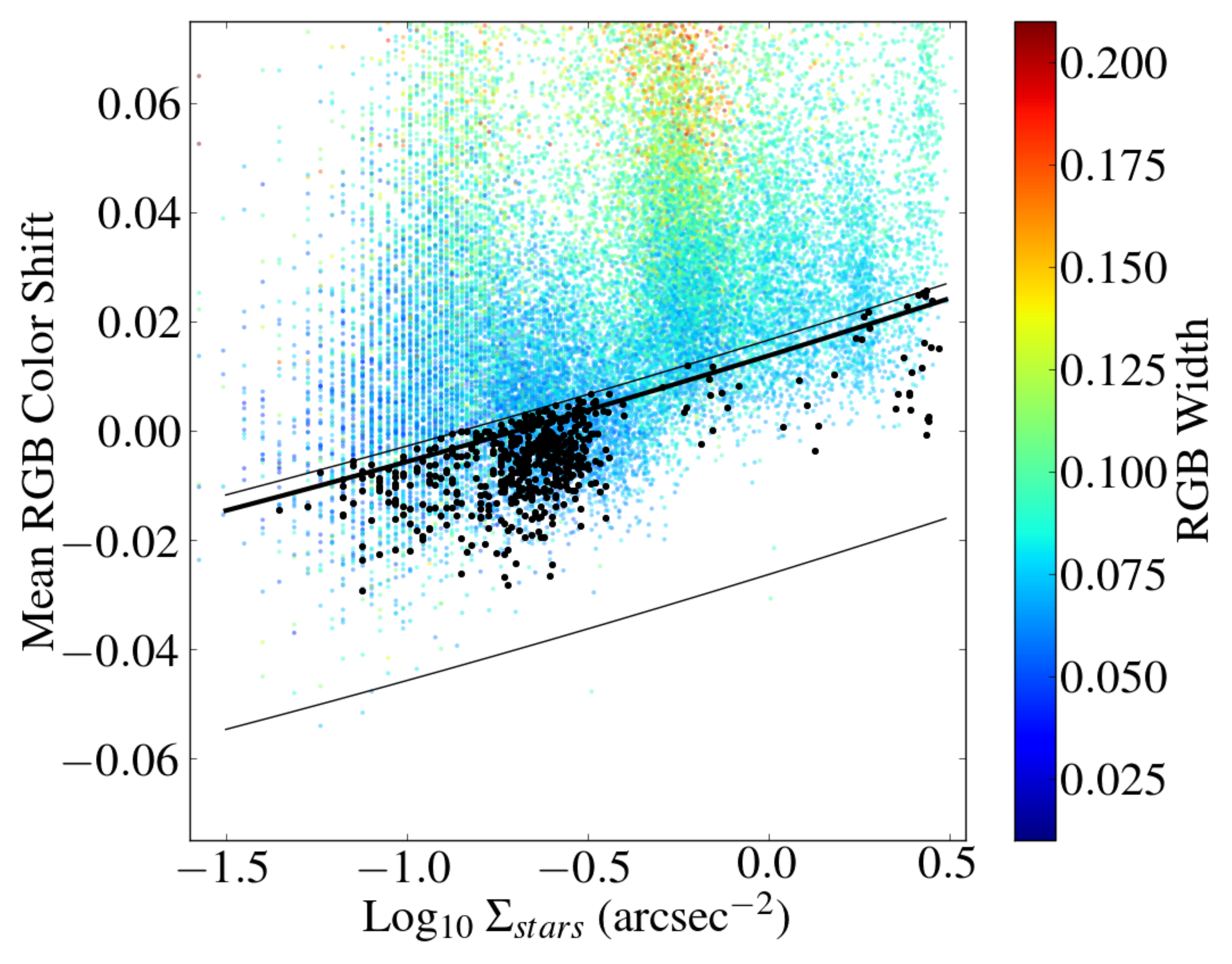}
}
\caption{Identifying low exinction regions using the width and mean
  color of the RGB, as defined in
  Sec.\ \ref{candidatelowAVsec}. (Left) The standard deviation of the upper RGB,
  relative to a fiducial RGB locus, as a function of the log of the
  local stellar surface density of bright RGB stars, color coded by
  the mean color of the RGB relative to the same fiducial RGB
  locus. Solid lines show the selection limit below which we identify
  the lowest extinction regions at each stellar surface
  density. (Right) The mean color of the RGB relative to the fiducial
  RGB locus, as a function of the log of the local stellar surface
  density of bright RGB stars, color-coded by the width of the upper
  RGB (defined as the standard deviation, relative to the same
  fiducial RGB locus).  Light solid lines indicate the range in color
  used to exclude regions with possible extinction (upper curve) or
  high contamination from RHeB and/or AGB stars (lower curve) from the
  selection region identified in the left panel. Heavy solid points
  indicate regions that survive both cuts, and which also have low
  extinction in the \citet{draine2013} dust map. The location of these
  points on the disk are shown in Figure~\ref{lowAVmapfig}.
  \label{surfdensrgbwidthfig}}
\end{figure*}

From the list of low extinction candidates, we select the regions that
have the narrowest RGBs as a function of local surface density, by
including any region that falls below the heavy black lines in the
left panel of Figure~\ref{surfdensrgbwidthfig} ($\sigma_{RGB}=0.071 +
0.0125(\log_{10}{\Sigma_{stars}} + 0.2)$ and $\sigma_{RGB}=0.071 +
0.0235(\log_{10}{\Sigma_{stars}} + 0.2)$, for the high and low density
regions, respectively).  We then impose an additional cut on the color
of these regions by iteratively fitting a fourth order polynomial to
the mean RGB color relative to the RGB fiducial, and selecting the
subset of regions with the narrowest RGB that are also no more than 0.5
standard deviations redder or 7 standard deviations bluer than the
polynomial fit\footnote{The width of this selection region is
  $\sim0.04$ magnitudes in color, which is larger than the 
  systematic magnitude variation that we see across the face of the
  WFC3/IR chips (right panel of Figure~\ref{RGBwidthcolormapfig});
  correcting the WFC3/IR calibration would allow us to narrow this
  selection region further, potentially improving the measurement of
  $A_V$ at low extinctions if the true population width is in fact
  narrower.}; the selected color range is indicated with light solid
lines in the right panel of Figure~\ref{surfdensrgbwidthfig}.  The red
selection limit culls any potentially reddened regions and the blue
selection limit reduces fields with large contamination on the blue
side of the RGB from asymptotic giant branch or red core Helium
burning (RHeB) stars.  Finally, we compare the surviving regions to
the dust map recently published by \citet{draine2013}, and reject any
regions that fall where the implied extinction is greater than
$A_V>0.175$.  The stars that fall in the remaining regions are then
tagged with the local stellar surface density, and then used to
generate model unreddened CMDs.

\begin{figure}
\centerline{
\includegraphics[width=3.05in]{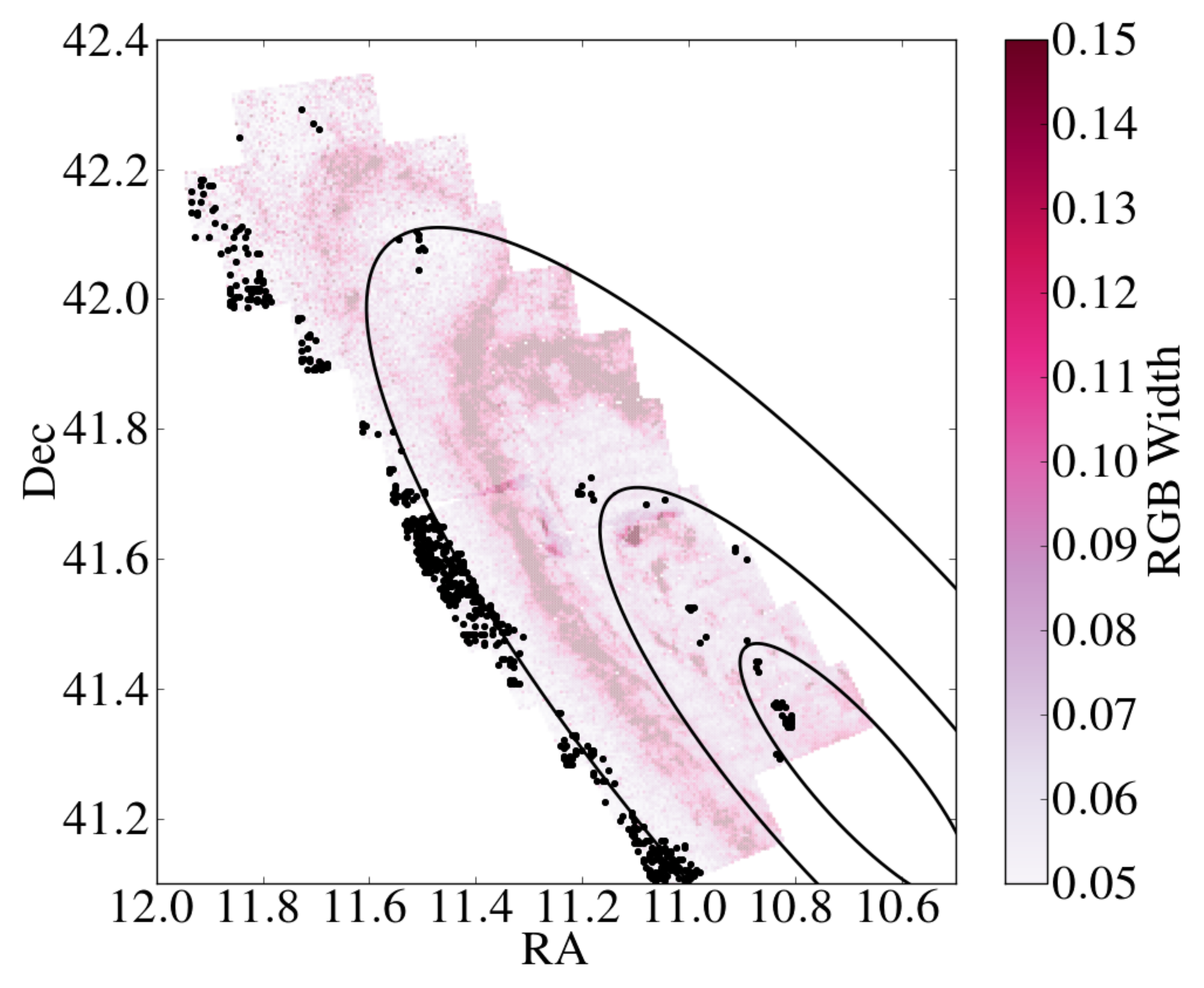}
}
\caption{Location of the lowest extinction regions (blue dots),
  superimposed on a map of the width of the RGB (e.g., left panel of
  Figure~\ref{RGBwidthcolormapfig}; see
  Sec.\ \ref{candidatelowAVsec}).  Ellipses are at the same location
  as in Figures~\ref{surfdensmapfig} and
  \ref{surfdensrgbwidthfig}. \label{lowAVmapfig}}
\end{figure}

The regions that survive the cuts are highlighted in dark blue in the
right panel of Figure~\ref{surfdensrgbwidthfig}. There are
$\sim153,000$ stars total in these regions, sampling the full range in
local surface density, although not uniformly; for example, there are
no low extinction lines of sight identified in the dusty $10\kpc$ star
forming ring ($-0.5\lesssim \log_{10}{\Sigma_{stars}} \lesssim-0.2$),
and there are many more low extinction regions that sample the diffuse
outer disk. The spatial distribution of these regions is shown in
Figure~\ref{lowAVmapfig}, superimposed on a map of the width of the
upper RGB.  By design, these regions lie in the lowest extinction
regions of the recently published \citet{draine2013} dust mass map,
derived from models of dust emission.  We note, however, that our
selection does not ensure that these regions truly have zero
extinction.  The RGB has a finite width due to photometric
uncertainties and stellar population effects, and we cannot detect
extinction variations that broaden the RGB by amounts smaller than the
intrinsic width.

To generate the final model CMD for unreddened stars as a function of
local surface density, we group the unreddened stars into bins of
$\log_{10}{\Sigma_{stars}}$.  We first rank the regions by their local
surface density, and then use a sliding, adjustable bin to generate
CMDs which each contain at least 2500 bright RGB stars in the range
$19\!<\!{\rm F160W}\!<\!21.5$). Each of the resulting set of bins
contain comparable numbers of stars, but do not sample equal ranges of
local surface density; in the low surface density outer disk, one must
merge stars in a wider range of $\log_{10}{\Sigma_{stars}}$ to reach
the same total number of stars.  We also allow sequential bins to
overlap by one third, to provide a smoother interpolation over local
density. To avoid having unnecessarily high numbers of nearly
identical adjacent bins, we then merge neighboring density bins
together until there is a step of at least $\Delta
\log_{10}{\Sigma_{stars}}=0.05$ between each bin. We then keep a
record of the mean and the range of $\log_{10}{\Sigma_{stars}}$
associated with each bin.  When fitting data, we adopt the model CMD
from the bin with the closest mean surface density to that of the
region being analyzed.

\subsubsection{Properties of the RGB in Low Extinction Regions} \label{RGBpropertysec}

\begin{figure*}
\centerline{
\includegraphics[width=3.25in]{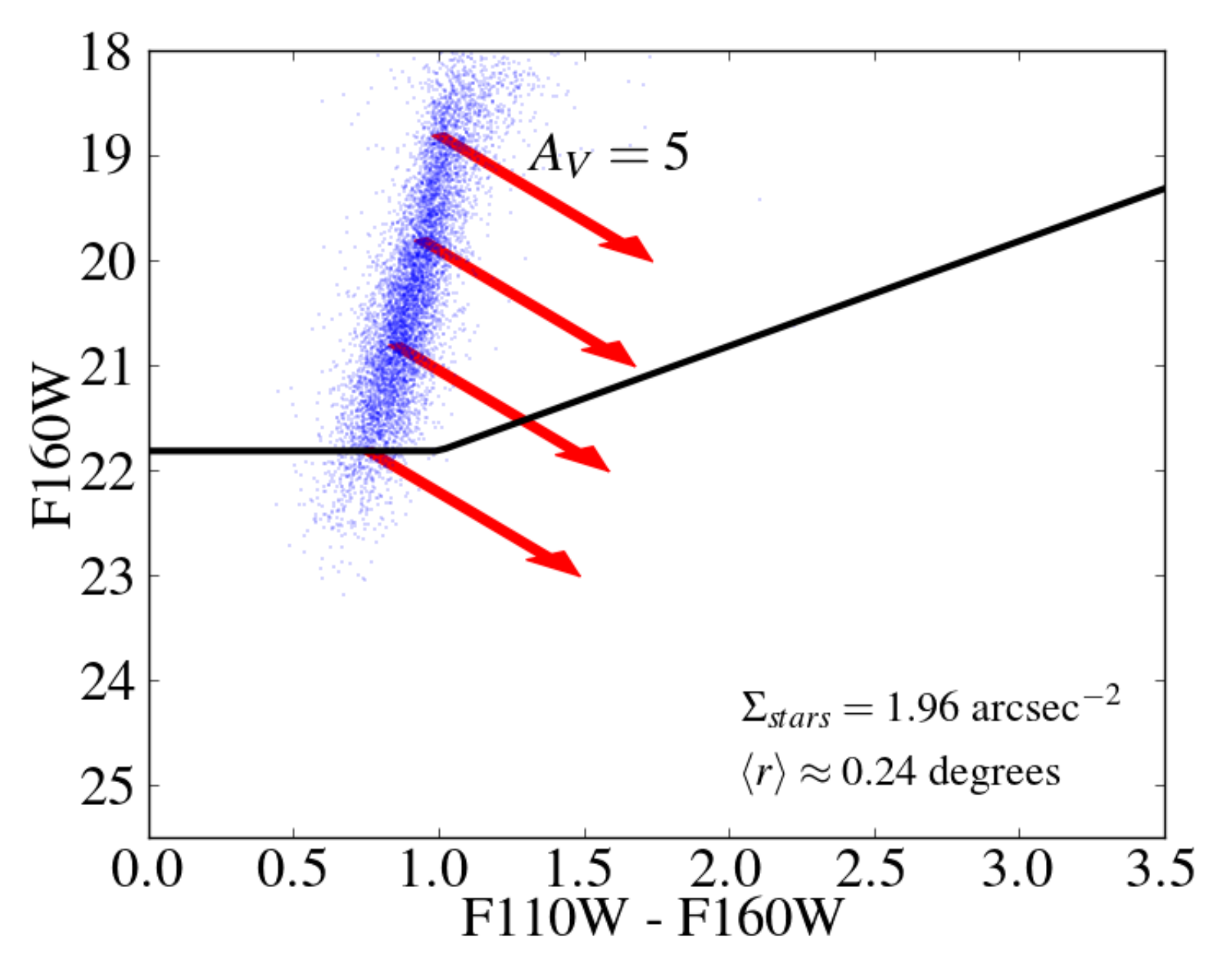}
\includegraphics[width=3.25in]{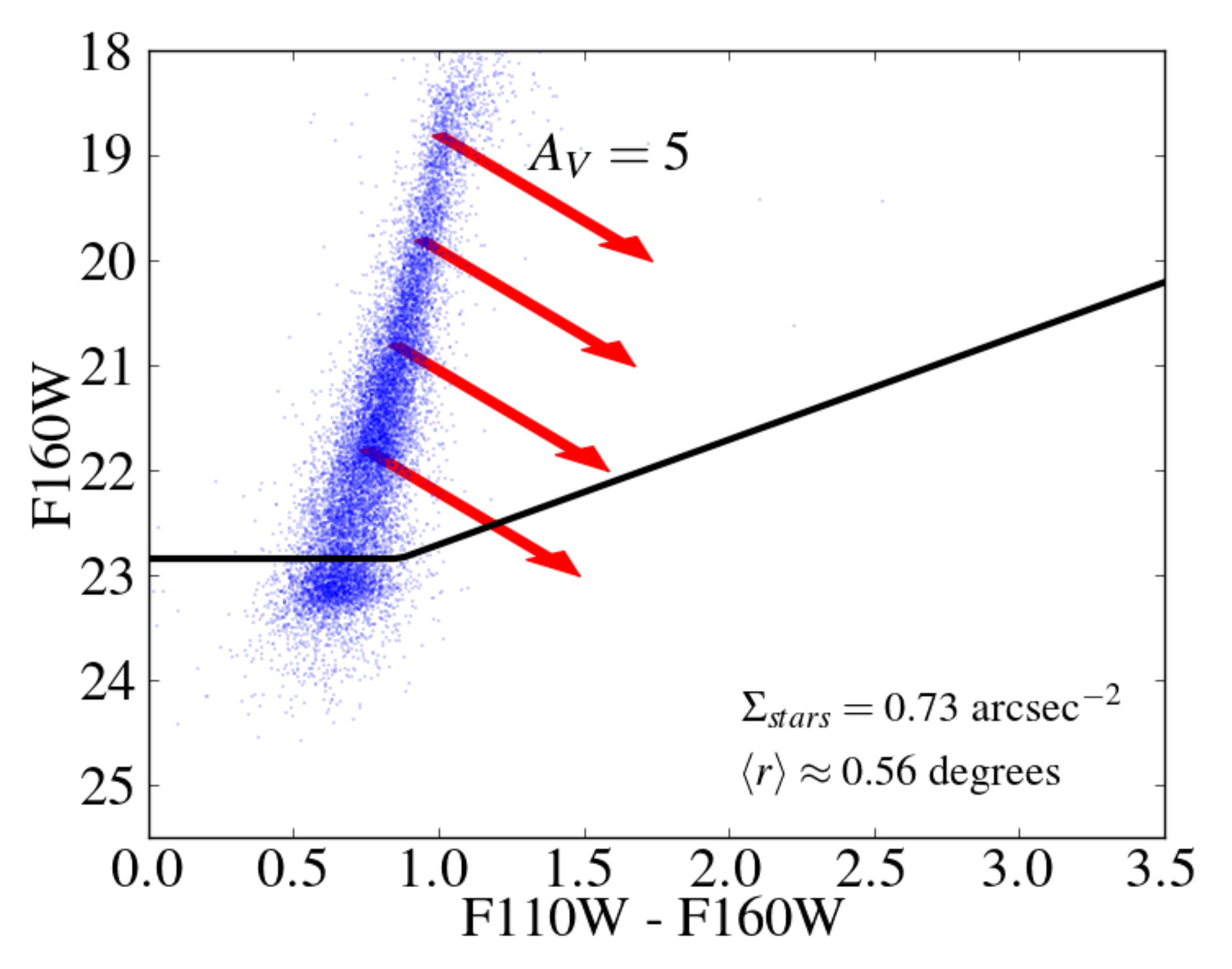}
}
\centerline{
\includegraphics[width=3.25in]{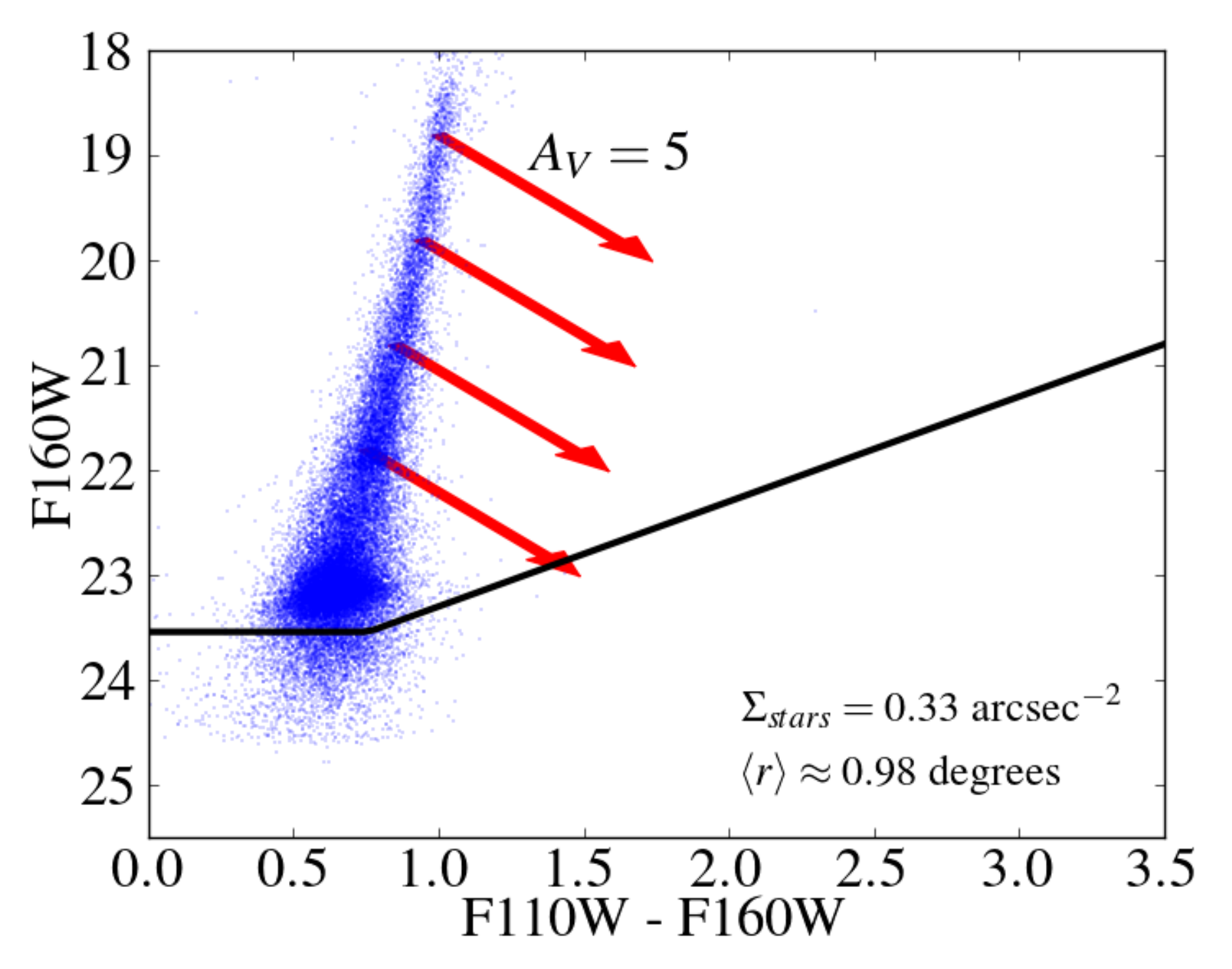}
\includegraphics[width=3.25in]{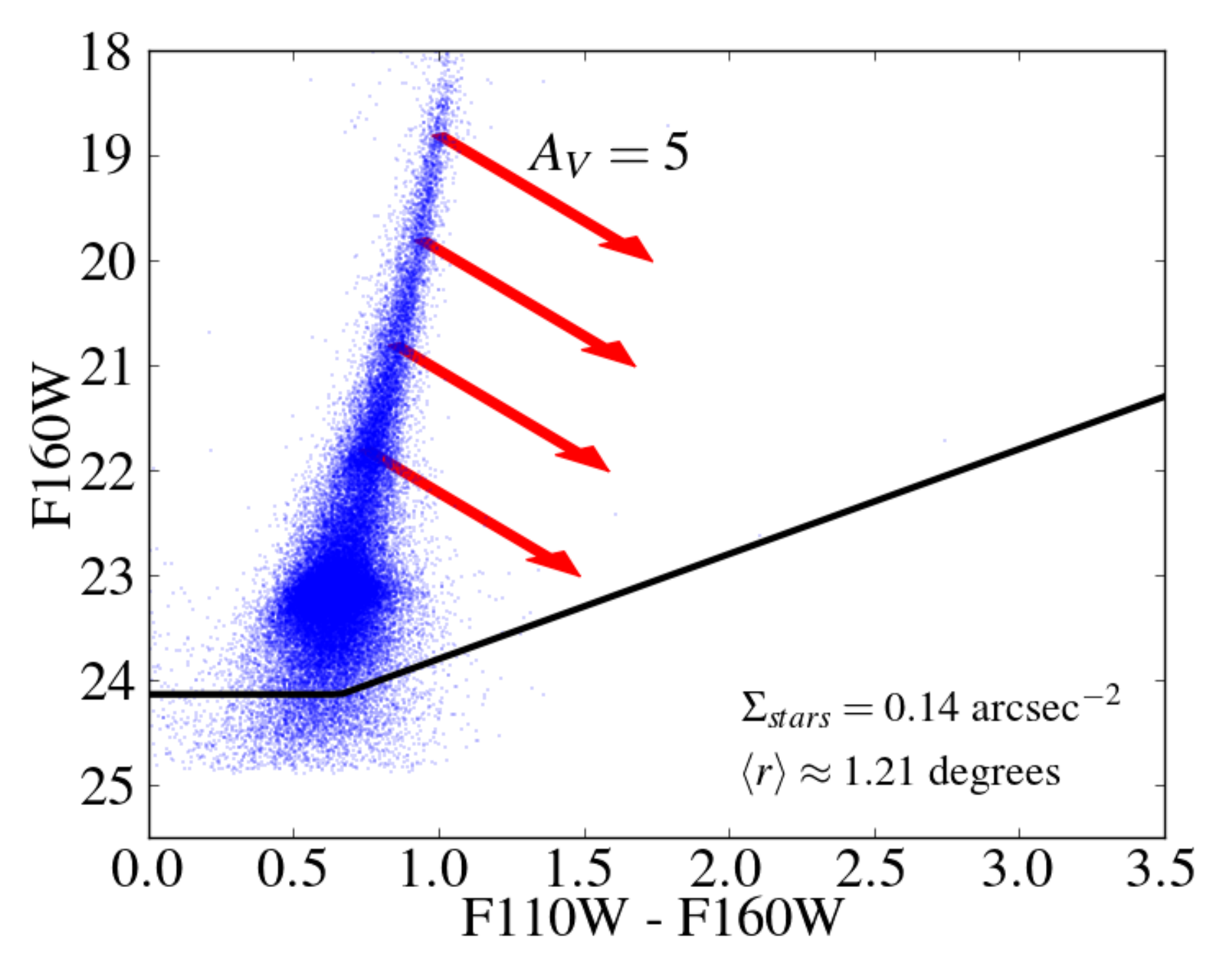}
}
\caption{CMDs of stars in 4 unreddened regions, spanning a range of
  local RGB surface density $\Sigma_{stars}$
  (Sec.\ \ref{nirdatasec}) and/or deprojected semi-major axis distance
  $\langle r \rangle$ (Sec.\ \ref{unreddenedsec}).  The black lines
  indicate the 50\% completeness limits adopted at that stellar
  surface density. Note the near absence of stars redward of the CMD,
  indicating the quality of the photometric cuts and of the
  identification of low extinction regions. Red arrows show the
  expected effect of 5 magnitudes of extinction in $V$.  Extinction
  will cause fainter stars to drop below the completeness limit at
  lower $A_V$ than for brighter stars.  The effect of the magnitude
  cuts on the range of observable $A_V$ is more severe in the inner
  disk, where the cuts are closer to the tip of the RGB.  The RGB in
  this region is also intrinsically wider, making it more difficult to
  measure small offsets due to reddening.
\label{unreddenedCMDfig}}
\end{figure*}

Four representative CMDs of the low extinction regions, in narrow
ranges of local surface density, are shown in
Figure~\ref{unreddenedCMDfig}.  These CMDs are extremely clean,
showing only a narrow RGB sequence. The CMDs also demonstrate the
quality of the photometric cuts, as measured by the nearly total
absence of spurious detections near the magnitude limit.

\begin{figure*}
\centerline{
\includegraphics[height=2.75in]{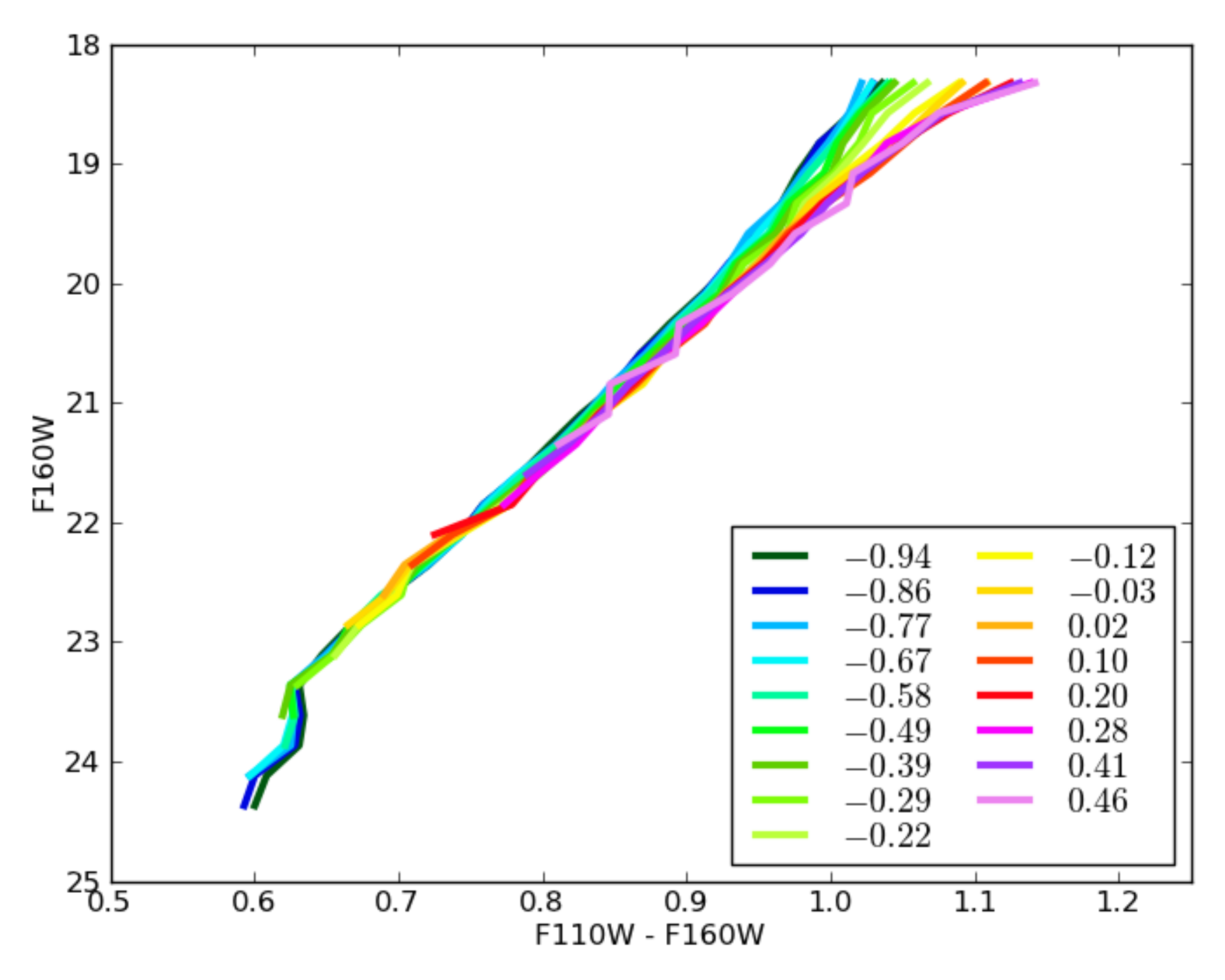}
\includegraphics[height=2.75in]{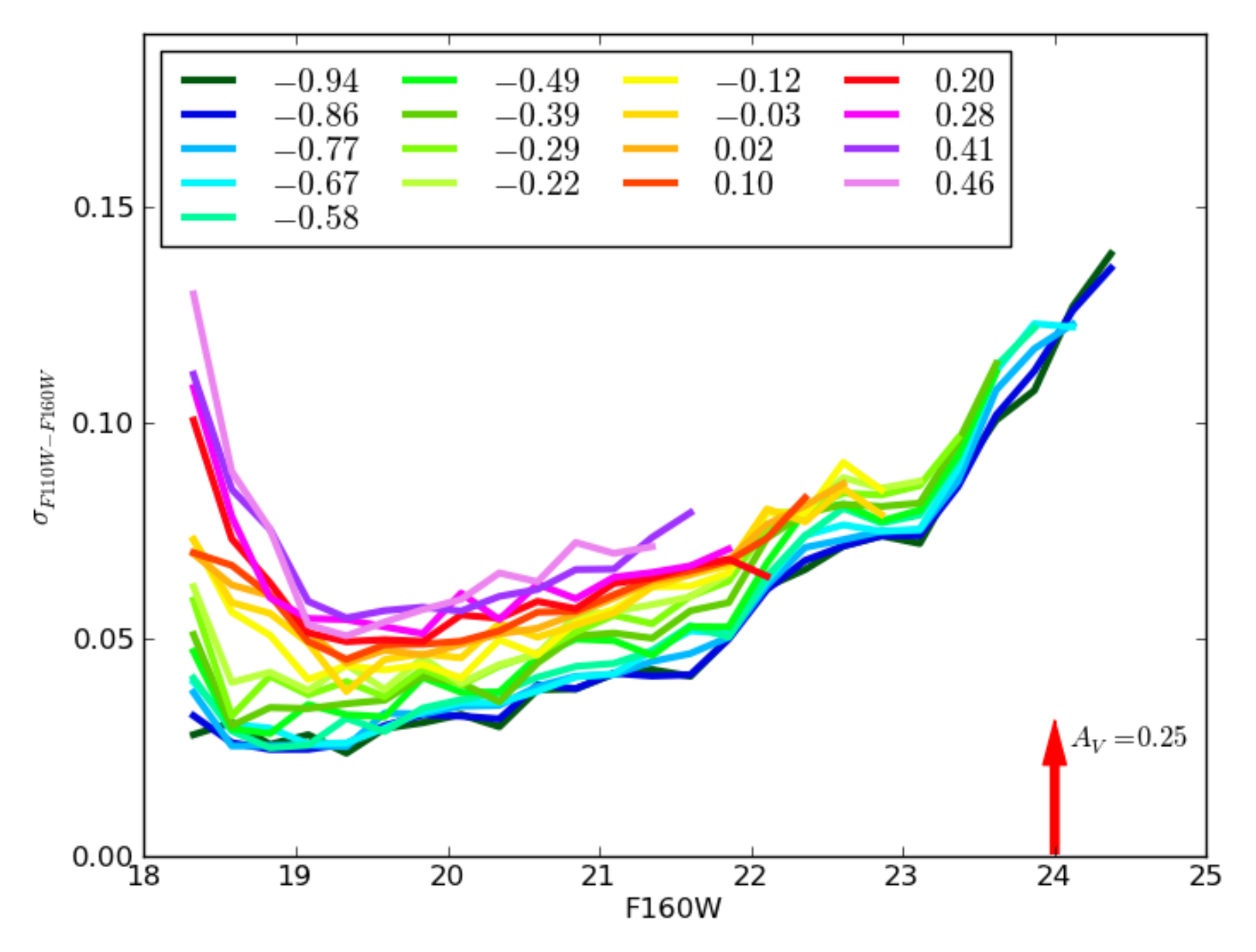}
}
\caption{The morphology and width of the unreddened RGB as a function
  of position within M31, as characterized by the stellar surface
  density $\Sigma_{stars}$ . (Left) Color locus of the unreddened RGB
  as a function of magnitude. Each line corresponds to a bin of
  $\log_{10}{\Sigma_{stars}}$ (in units of arcsec$^{-2}$), which can
  be mapped to major-axis length using
  Figure~\ref{radialsurfdensfig}).  The mean color was calculated in
  bins of 0.15$\,{\rm mag}$, above the limiting 50\% completeness
  magnitude shown in Figure~\ref{maglimfig}. The morphology of the RGB
  varies slowly and smoothly from the inner regions (pink) to the
  outer disk (dark blue). The RGB can be traced to fainter magnitudes
  in the less crowded outer regions, due to the smaller photometric
  errors and fainter crowding limit. (Right) Standard deviation of
  stars around the mean color locus as a function of magnitude,
  calculated in the same bins as in the left panel. The arrow
  indicates the expected shift in color for an extinction of
  $A_V=0.25$. See Sec.\ \ref{RGBpropertysec} for a discussion of
  this figure.\label{RGBgradientfig}}
\end{figure*}

In Figure~\ref{RGBgradientfig}, we show the properties of the
unreddened RGB as a function of magnitude (calculated in bins of
$0.15\,{\rm mag}$), in each of the resulting bins of local surface
density. The mean color of the RGB (left panel) is systematically
bluer and the RGB slope is steeper towards low local surface densities
(i.e., towards the outer disk), as would be characteristic of an
increasingly metal poor population \citep[see][]{gregersen2015}.
However, the shift in RGB morphology is extremely small between
adjacent bins, such that our choice to bin in
$\log_{10}{\Sigma_{stars}}$ leaves no detectable imprint on the
extinction maps that are eventually derived from the unreddened CMDs.

The right panel of Figure~\ref{RGBgradientfig} shows the standard
deviation of the color of the RGB as a function of magnitude.  The RGB
becomes increasingly narrow at bright magnitudes and towards the outer
disk, as would be expected for decreasing photometric
uncertainties\footnote{ The only exception is right near the RGB tip
  at ${\rm F160W}\sim18.5$, where the decreasing effective temperature
  of the stars leads to strong metal-dependent effects on the stellar
  photosphere, broadening the distribution of colors at the coolest
  temperatures.}, and a narrower mixture of stellar populations.
There is also a clear floor at $\sim0.025\,{\rm mag}$, which is due to
a combination of the instrinsic width of the RGB, photometric
uncertaintes, and systematic calibration errors in the WFC3/IR chip
($\pm0.01\,{\rm mag}$).  The minimum width corresponds to roughly $0.3$
mag of extinction, setting a threshold below which we are
significantly less sensitive to broadening due to reddening.  We
could potentially improve our sensitivity to low-but-not-zero
extinction with improved WFC3/IR calibrations if systematics dominate,
but may not be able to if this floor is the intrinsic width of the
underlying population.

\subsubsection{Limitations in Building an Unreddened RGB} \label{limitationsec}

There are a number of unavoidable limitations in our procedure for
generating a model of the unreddened distribution.  While these
effects are quite small in terms of their effect on the structure of
the RGB, they can potentially affect the inferred
extinctions.  In addition to their impact on the model
of the unreddened RGB, these effects will also impact measurements
of the redding at low extinction. We now discuss each of these in turn.

\smallskip
\centerline{\it{The Finite Width of the RGB}}
\smallskip

The width of the RGB (Figure~\ref{RGBgradientfig}) creates a limit on
the lowest extinction we can reliably identify at a given location.
If the color change produced by a given $A_V$ is much smaller than the
typical width of the RGB, then small redward shifts of the CMD will
not be detectable.  The finite width of the NIR RGB thus directly
affects our ability to accurately identify the lowest extinction lines
of sight.

Although we have taken great pains to identify the lowest reddening
regions throughout the galaxy, there is no guarantee that these are
truly zero reddening lines of sight, due to the fact that low level of
diffuse dust may not produce measurable shifts in the NIR colors
and/or width of the RGB. We have attempted to mitigate for this effect
by using the \citet{draine2013} dust emission maps to place more
stringent limits on the local extinction than are possible with the
NIR CMD alone. However, due to background subtraction, the
\citet{draine2013} maps have uncertainties at low extinction as well,
such that some stars with $A_V>0$ may still remain in our ``zero
reddening'' model CMD.  The resulting model CMD would be biased
slightly to the red, and would lead to underestimating $A_V$ when
modeling the observed CMD.

We can estimate the amount of reddening that could be present in our
``unreddened'' CMD by inspecting the widths of the RGB at bright
magnitudes, shown in the right panel of Figure~\ref{RGBgradientfig}.
If the width of the RGB is $\sigma_{F110W-F160W}\lesssim 0.03$ over a
reasonably well populated part of the RGB\footnote{This plot also
  suggests that the well populated red clump at $F160W\approx22.6$ is
  of somewhat limited utility for detecting small extinctions, because
  the intrinsic width of the RGB$+$red clump feature is much broader
  than the RGB alone.}, then one could expect to identify extinction
at the level of $A_V=0.25$, indicated by the vertical arrow. We
therefore expect that the model of the unreddened CMD could be
``contaminated'' with reddened regions with no more than this level of
extinction.  Inspection of the RGB widths in
Figure~\ref{RGBwidthcolormapfig} suggests that identifying low
reddening regions will be harder in the inner galaxy, where the RGB
appears to be wider, both due to higher photometric errors from
crowding, and from the wider range of metallicities present as one
nears M31's bulge.  However, the overall level of extinction in the
inner galaxy appears to be low \citep{draine2013}, which may
perhaps mitigate against this effect.

Including bluer filters in our analysis could potentially improve our
ability to correctly identify low extinction regions, while also
giving more sensitivity in measuring $A_V$.  In Appendix
\ref{opticalsec} we discuss the trade-off between the intrinsic width
of the RGB and the sensitivity of RGB color to extinction in other
PHAT filter combinations.

\smallskip
\centerline{\it{Stellar Population Gradients}}
\smallskip

Spatial variation in M31's underlying stellar population pose another
possible obstacle to creating a realistic model of the unreddened CMD.
We construct models from stars in individual low reddening regions,
but frequently, these are far from the individual pixel being
analyzed.  Thus, if there are spatial variations in the stellar
population, we may sometimes use a model CMD that does not correctly
reflect the local population.  To first order, this should not present
a severe problem because we match the local stellar density of the
model to the pixel being analyzed, which isolates regions at
comparable deprojected radii, and thus of similar underlying stellar
population.  Empirically, Figure~\ref{RGBgradientfig} shows that the
RGB color and width is a very weak function of local stellar density,
particularly where $\log_{10}\Sigma_{stars} < -0.2\numden$.  Thus,
systematic offsets in the structure of the RGB are likely to involve
color differences of only a few hundredths of a magnitude, which would
produce a very small bias in the modeled RGB.  The only place where we
anticipate that stellar population gradients might significantly
affect the model unreddened CMD is in the inner regions of the galaxy
where the stellar populations change more rapidly with radius, and
where the structure of the galaxy is complex (due to M31's bar, and
multi-component bulge; see Figure~\ref{radialsurfdensfig} and
\citet{gregersen2015}).

A potentially more significant impact of stellar population gradients
comes through their effects when viewing an inclined disk (as in M31).
Projection effects lead to pixels along the minor axis having stars
from a wider range of radii than along the major axis. The net result
is that while matching of the stellar densities correctly matches the
photometric properties of the RGB, it may not successfully match the
underlying color and shape of the RGB, due to a mismatch in the
stellar populations between the model and the pixel being analyzed.

\smallskip
\centerline{\it{Foreground Extinction from the Milky Way}}
\smallskip

The final limitation when constructing a model for the unreddened CMD
is that there may be foreground dust extinction from the Milky Way or
from a diffuse dust halo in M31 itself.  This foreground would not affect
our results if it were uniform, because our measurement is
fundamentally relative; if both the reddened and
``unreddened'' stars are affected equally, the measurement of the reddening
relative to the unreddened model is still sound even if the
``unreddened'' model has been affected by dust in the Milky Way
foreground. If the foreground dust is not uniform (which indeed is
most likely), then it broadens the RGB in the model for the
unreddened CMD, due to including lines of sight with different
reddenings. The net result will be to reduce sensitivity to low levels
of reddening within M31.  Unfortunately, one cannot use maps of the
Milky Way dust \citep[e.g.,][]{schlegel1998,schlafly2011} to correct
for the foreground dust, since these maps fail when there is a bright
background source of FIR emission (such as M31) that dominates the
emission from the Milky Way.

While the list above describes the principal limitations in
constructing the unreddened CMD, there are additional uncertainties
associated with the model fitting itself. We discuss these below in
Sec.\ \ref{systematicsec}, and make empirical tests of systematics
in our final maps in Sec.\ \ref{spatialsystematicssec}.

\subsection{Generating the Model Reddened CMD}  \label{modelcmdsec}

Our data is set of F160W magnitudes $m_i$ and F110W$-$F160W colors
$c_i$.  We therefore need to translate our adopted model for the
expected distribution of extinctions $p_A$ and/or reddenings
$p_{\mathcal E}$ into a model of the expected color magnitude diagram
for a given set of model parameters $\vec{\theta}=\{\widetilde{A_V},
\sigma, f_{red}, \delta_c\}$, given the empirical model for the
distributions of colors and magnitudes of unreddened RGB stars that we
derived in Sec.\ \ref{modelRGBsec}. In other words, we wish to derive
$p_{CMD}(c, m |\vec{\theta})$, the probability of finding
a star at color $c$ and magnitude $m$, given $\vec{\theta}$. This
model will depend on the local surface density $\Sigma_{stars}$, but
we do not make this surface density dependence explicit, for
notational compactness.

Before proceding with the calculation of $p_{CMD}$, we take two steps
to increase the computational efficiency.  First, we bin
both the data and the models into bins of color and magnitude.  We
use a grid spacing of 0.015$\,{\rm mag}$ in color and 0.2$\,{\rm
  mag}$ in magnitude. This choice guarantees that there are at least 2
pixels across the 1$\sigma$ width of the narrowest RGB, effectively
Nyquist sampling the width. Experiments with smaller bins produced no
noticeable changes in the resulting extinction maps, while
dramatically increasing the computation time.

Second, we translate all F160W magnitudes into ``reddening-free''
magnitudes $q$, defined as

\begin{equation}  \label{qeqn}
\begin{split}
q \equiv {\rm F160W} - (({\rm F110W} - {\rm F160W}) - c_0)\\
    \times\frac{A_{F160W} / A_V}{(A_{F110W} / A_V) - (A_{F160W} / A_V)}
\end{split}
\end{equation}

\noindent where $c_0$ is an arbitrary color for which $q\equiv {\rm
  F160W}$.  With this transformation, increases in extinction will
cause a star's color to redden, while leaving the reddening-free
magnitude unchanged.  With this skewed version of the CMD, changes in
the extinction move stars horizontally in the grid, rather than
diagonally (i.e., the reddening vector becomes horizontal).  We can
then do 1-dimensional convolutions when calculating the reddened CMD,
rather than more computationally expensive 2-dimensional convolutions.
We will therefore calculate $p_{CMD}(c, q |\vec{\theta})$ rather
than $p_{CMD}(c, m |\vec{\theta})$.

The next step in generating $p_{CMD}$ is to define what region of the
CMD will be fit.  We restrict our analysis to regions occupied by
RGB stars by generating a mask. The blue boundary of the mask is defined to
be 2.5$\sigma$ blueward of the RGB (Fig.\ \ref{RGBgradientfig}) to
reduce contributions from AGB and RHeB stars. The upper boundary is
fixed at $q=18.5$, which is approximately the tip of the RGB.
The lower faint boundary and the right diagonal boundaries were set by
the adopted F160W magnitude limit ($m_{50,F160W}-0.5$) and the 50\%
completeness limit in F110W, respectively, both translated into the
reddening-free CMD.  All normalizations and comparisons with
data are restricted to be within the unmasked regions.

Within each bin of local stellar surface density, we grid the
unreddened stars into a CMD where the reddening-free magnitude
$q$ has replaced F160W, eliminating stars that fall in the
masked region.  We then normalize the binned CMD so that it
integrate to one over the unmasked area.  We refer to the final binned,
masked, normalized probability distribution for unreddened stars as
$p_{unreddened}(c, q)$.

We then generated the probability distribution for the reddened stars,
$p_{reddened}(c, q | \widetilde{A_V}, \sigma)$, by convolving the
binned unreddened CMD with a 1-dimensional log-normal kernel, where
the properties of the kernel are set by $\widetilde{A_V}$ and
$\sigma$.  We then added the reddened and unreddened model CMDs
together to generate the model for the combined CMD $p_{unreddened +
  reddened}(c, q | \vec{\theta})$ after weighting them by appropriate fraction of
reddened stars:

\begin{equation}
\begin{split}
  p_{unreddened + reddened}(c, q | \vec{\theta}) =  (1-f_{red})\,p_{unreddened}(c, q)\\
                                  + f_{red}\,p_{reddened}(c, q | \widetilde{A_V}, \sigma).
\end{split}
\end{equation}

\noindent We then reapply the mask to $p_{unreddened + reddened}$ and renormalize.

We also include a third component to model the noise from potentially
bad photometry.  Occasionally, spurious sources will be detected with
colors and magnitudes consistent with being reddened RGB stars.  These
sources are rare, but not impossible, and thus we need to include an
additional component in our model that allows there to be a small
chance of an individual source being spurious, rather than requiring
every star redward of the RGB to be due to dust reddening.  To
model the CMD contribution from noise, $p_{noise}$,
we identify stars that are more than 3.5$\sigma$ redward of the
mean RGB color in regions that were identified as being
``unreddened''.  These red stars are unlikely to be due to intervening
dust, and instead are due to occasional photometric errors when doing
crowded field photometry.  These anomalous red stars are then
gridded into a ``noise CMD'', $p_{noise}(c, q)$, and
then re-added into the sum of the unreddened$+$reddened models in the
proportion with which they were found in the original unreddened CMD.
Letting $f_{noise}$ be the fraction of stars in the catalog of unreddened
stars were classified as noise,

\begin{equation}
\begin{split}
  p_{CMD}(c, q | \vec{\theta})=(1-f_{noise})\,p_{unreddened+reddened}(c, q | \vec{\theta})\\
                             + f_{noise}\,p_{noise}(c, q)
\end{split}
\end{equation}

\noindent The fraction of pixels in the noise model was never
more than 1.5\% in any bin of local stellar surface density, and was
less than 0.5\% in 3/4 of the bins.

Finally, we introduce a fourth parameter $\delta_c$ that allows the
color of the unreddened CMD to shift by a few hundredths of a
magnitude, with the goal of absorbing any residual issues with the
spatially-dependent WFC3/IR calibration/photometry correction
discussed in Sec.\ \ref{nirdatasec}. Because the photometric correction we
applied in Sec.\ \ref{nirdatasec} was calculated on a coarse grid of chip position, it does not
accurately trace rapid changes with position, particularly those near
the edges of the chip; these residual issues can be seen as slight
bands in Figure~\ref{RGBwidthcolormapfig} that trace the edges of the
NIR chip in adjacent observations. In addition, a handful of fields show
small global shifts in RGB color, most likely due to a small error in the
aperture correction that was applied to the entire field. Both of
these residual systematic photometry errors translate directly into
features on the resulting extinction map if not corrected for.

\begin{figure*}
\centerline{
\includegraphics[width=3.75in]{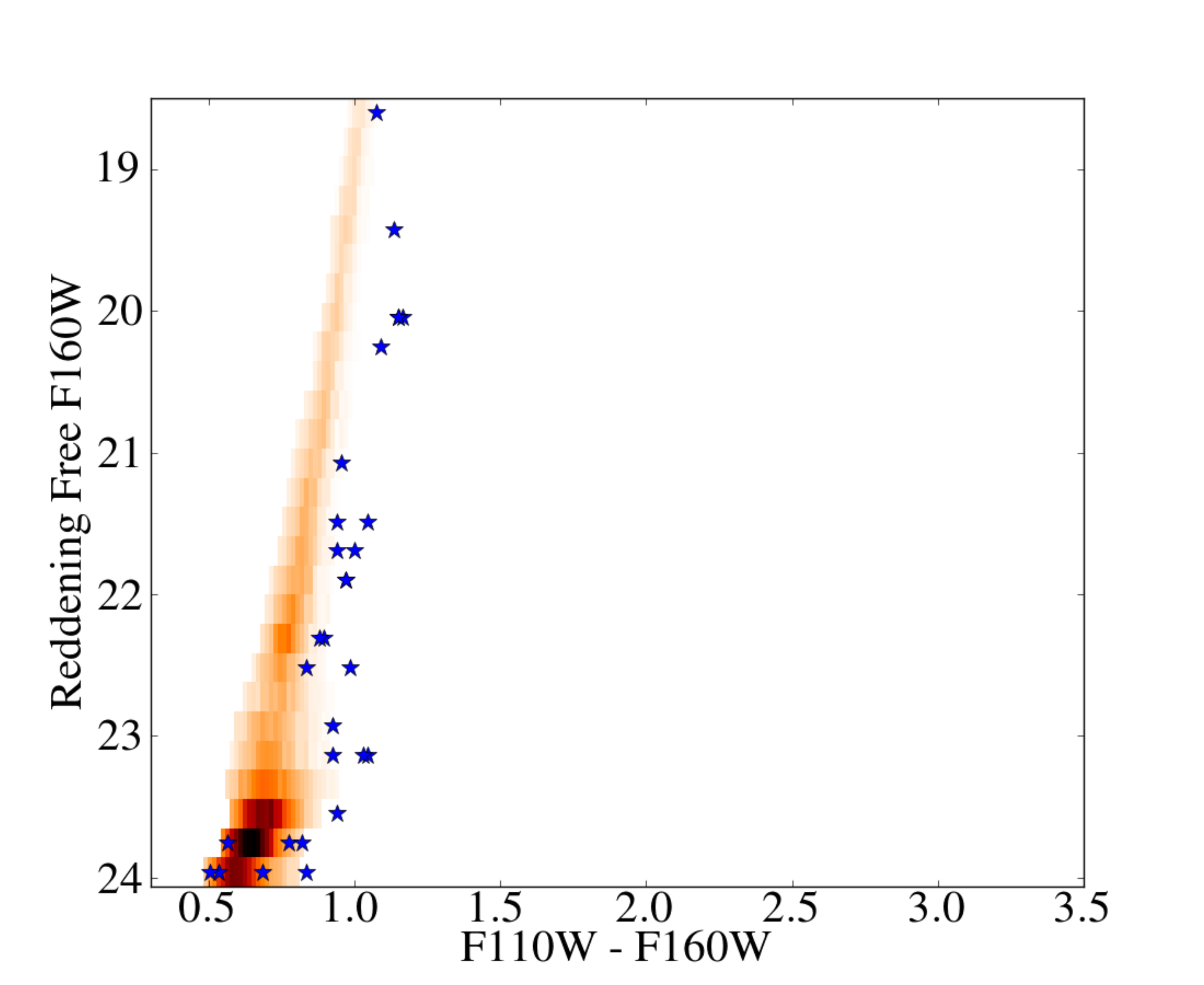}
\includegraphics[width=3.75in]{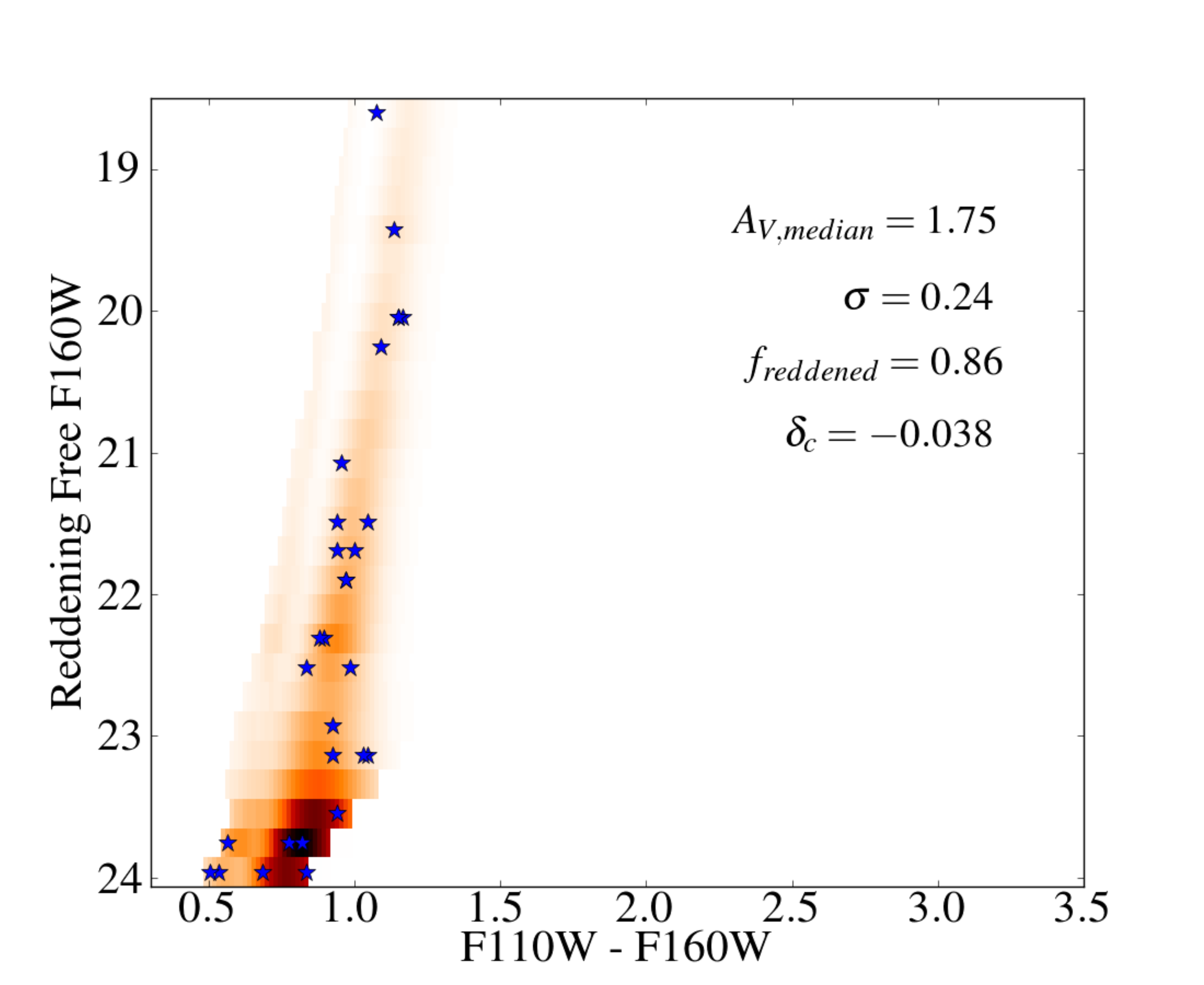}
}
\caption{Comparing a binned unreddened CMD (left) to a reddened model including
  noise (right). Star symbols show the location of stars found in a
  $25\pc$ pixel in Brick 15. The best fit parameters for that pixel
  are used in the reddened model. Note that the vertical axis is not
  F160W magnitude, but instead is the ``reddening-free'' F160W
  magnitude $q$ (eqn.\ \ref{qeqn}), such that changes in the
  amount of extinction/reddening moves stars
  horizontally. \label{modelcmdfig}}
\end{figure*}

An example of an unreddened and reddened CMD are shown in
Figure~\ref{modelcmdfig}.

\subsection{Fitting the Reddened CMD}  \label{fittingsec}

The model CMD shown in Figure~\ref{modelcmdfig} has three adjustable
parameters --- the median $\widetilde{A_V}$ of the log-normal
distribution of extinctions, the dimensionless width $\sigma$ of the
same distribution, and the fraction of reddened stars $f_{red}$
(eqns.\ \ref{pAVeqn}~\&~\ref{pAVcloudeqn}) -- in addition to one
parameter $\delta_c$ to help absorb any residual issues in spatially
variable photometry.  We solve for these parameters and their
uncertainties within small spatial regions of the PHAT imaging data,
using MCMC fitting to find the most probable fit and the associated
uncertainties, subject to sensible prior probability distributions
(``priors'') on the values of the parameters. The choice of priors is
discussed in Sec.\ \ref{priorsec} below.

Specifically, we use $p_{CMD}(c, q | \vec{\theta})$ derived in the
previous section to calculate the probability of finding a star at
color $c$ and reddening-free magnitude $q$, for a given set of
parameters $\vec{\theta}=\{\widetilde{A_V}, \sigma, f_{red},
\delta_c\}$, assuming a model for an unreddened CMD at the local surface
density $\Sigma_{stars}$.

The total probability distribution function $p_{total}$ of measuring a
set of $N$ independent and identically drawn observations $\{c_i,
q_i\}$ is then

\begin{equation}  \label{likelihoodeqn}
p_{data}(\{c_i, q_i\} | \vec{\theta}) = 
       \prod_{i=0}^{N} p_{CMD}(c_i, q_i | \vec{\theta}).
\end{equation}

\noindent We do not concern ourselves with the variation in the number
of stars per CMD point, given that any variations due to the loss of
highly extincted stars is expected to be extremely small, adding little
useful information to the fit.

We are interested not in the probability distribution function (PDF)
of a given set of observations given the parameters of the adopted
reddening model, but instead are interested in the posterior
probability distribution function for the reddening parameters
$\vec{\theta}$, given the set of observations $\{c_i, q_i\}$.  We
therefore use Bayes theorem to derive

\begin{eqnarray}  \label{parampdfeqn}
p_{param}(\vec{\theta} | \{c_i, q_i\}) &\propto& 
          p_{data}(\{c_i, q_i\} | \vec{\theta}) \, p_{prior}(\vec{\theta}) \\
       &\propto& \prod_{i=0}^{N} p_{CMD}(c_i, q_i | \vec{\theta}) \, 
                                p_{prior}(\vec{\theta}),
\end{eqnarray}

\noindent where $p_{prior}(\vec{\theta})$ encodes any prior
information on the values of the parameters.

\subsubsection{Choice of Prior Probability Distributions} \label{priorsec}

We chose specific forms for $p_{prior}$ to limit the parameters
$\vec{\theta}$ to physically plausible values.  We chose the prior in
$\widetilde{A_V}$ to require that the extinction be greater to or
equal to zero.  For $\sigma$, we adopt a wide log-normal distribution
where the most probable value of $\sigma$ is 0.3, which is
characteristic of extinction distributions seen in the Milky Way
\citep[e.g.,][]{kainulainen2009}, and the dimensionless log-normal
width of the prior on $\sigma$ is 0.3.  This choice also restricts the
value of $\sigma$ to be greater than zero.  We adopt a narrow,
flat-topped prior for $\delta_c$, since we expect the photometric
systematics to be small (i.e., as the same order as the large scale
systematics corrected for in Sec.\ \ref{nirdatasec}), but do not
have a strong prior on the exact value. We adopt a functional form
similar to a gaussian with $\sigma_\delta=0.035$, but with the
exponent raised to the fourth power, rather than squared; this change
produces a peaked distribution with a flatter top and faster fall-off
than a gaussian.

When setting the prior for $f_{red}$, we take a more sophisticated
approach than for $\widetilde{A_V}$ and $\sigma$.  The value of
$f_{red}$ is naturally bounded by zero and one, but rather than having
a hard limit at each of these values, we instead adopt a prior that
increasingly penalizes values of $f_{red}$ that approach these limits.
We do so by ``regularizing'' $f_{red}$ by switching to a closely related
parameter that more easily penalizes the reddening fraction becoming
zero or 1, while maximizing the prior probability at an arbitrary mean
value of $\langle f_{red} \rangle$.  Specifically, we adopt a new
parameter $x$ such that

\begin{equation}  \label{xeqn}
f_{red} = \left( \frac{e^x}{1 + e^x} \right)^{(\ln{\langle f_{red} \rangle}) / (\ln{0.5})}
\end{equation}

\noindent With this change of variables, $f_{red}=\langle f_{red}
\rangle$ when $x=0$, and $f_{red}$ goes to zero or 1 only when
$x\rightarrow \pm \infty$.  We then adopt a prior that has a mean
$\langle x \rangle = 0$ and that declines to both positive and
negative $x$, which maximizes the prior probability on $f_{red}$ to
have the proper mean, while increasingly (but smoothly) penalizing
values of $f_{red}$ that approach the extremes of its permitted values.

\begin{figure*}
\centerline{
\includegraphics[width=2.25in]{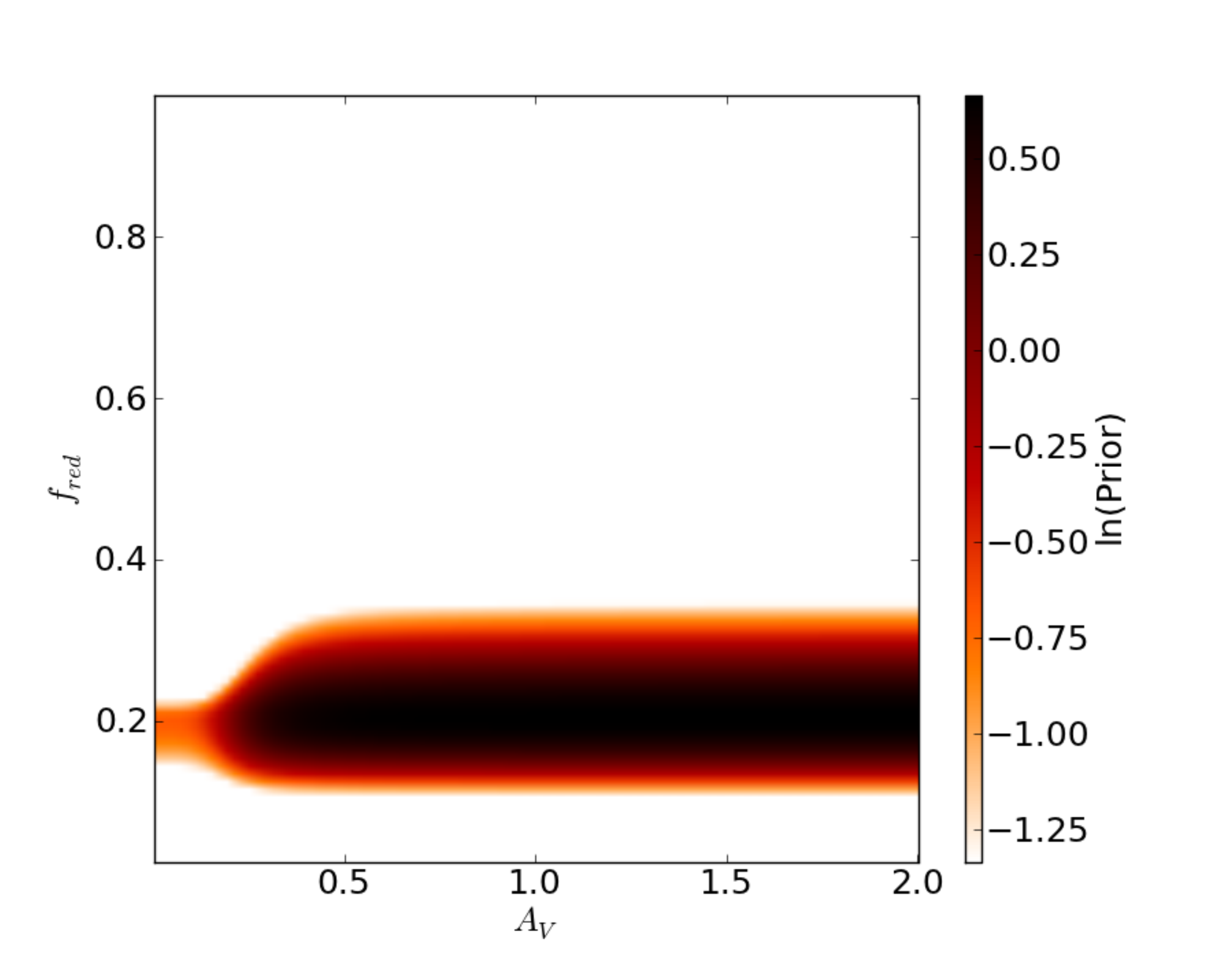}
\includegraphics[width=2.25in]{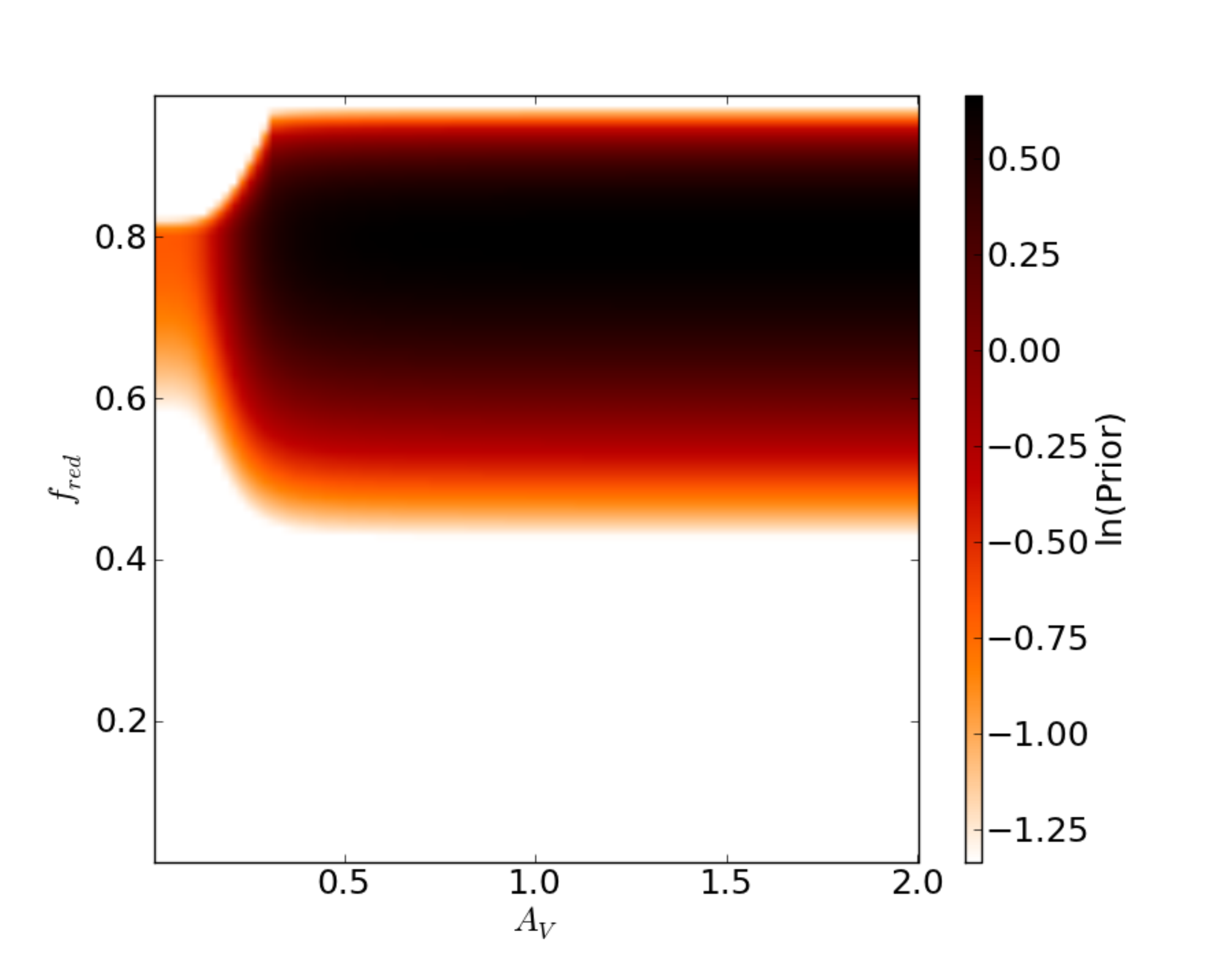}
\includegraphics[width=2.25in]{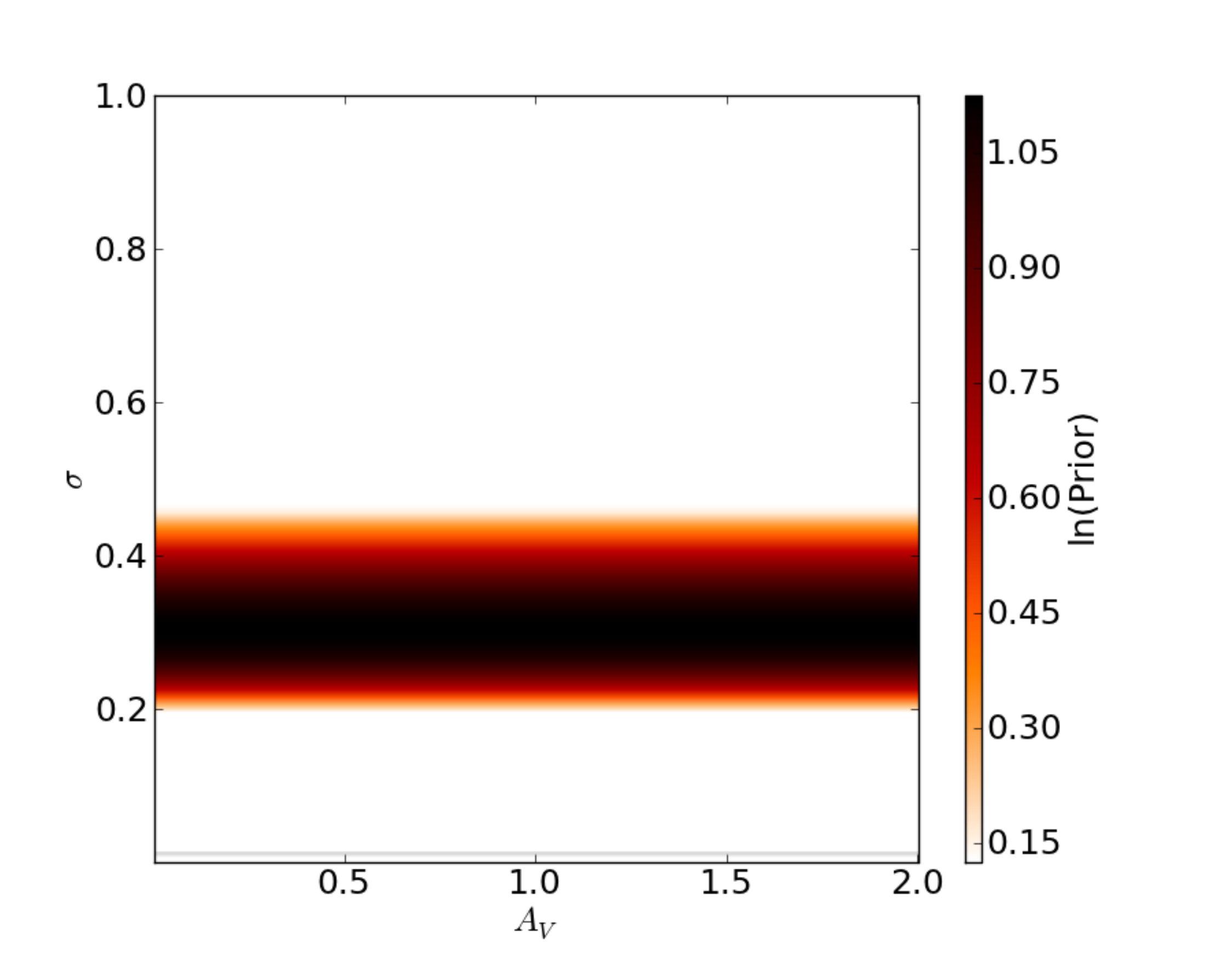}
}
\caption{The log of the priors in $f_{red}$ as a function of
  $\widetilde{A_V}$, for an assumed mean reddening fraction of
  $\langle f_{red}\rangle=0.2$ (left) and $\langle f_{red}\rangle=0.8$
  (center). The right hand plot shows the log-normal prior in
  $\sigma$, for which the width of the reddening distribution is
  assumed to be independent of extinction. These priors can be
  compared to the distribution of recovered values shown in the upper
  two panels of Figures~\ref{brickscatter16fig}
  and~\ref{brickscatter15fig}.
\label{priorfig}}
\end{figure*}

Eqn.\ \ref{xeqn} requires chosing a value for the expected mean value
of $f_{red}$.  Naively one might expect $\langle f_{red} \rangle=0.5$,
but this is generically only true along the major axis of an inclined
disk. For a thickened stellar disk, projection effects lead the stars
on the near side of the disk to come from slightly different radii
than the stars on the far side of the disk.  The number of stars
declines approximately exponentially with radius, and thus the
fraction of reddened stars at some projected radius can be much higher
or lower than 0.5, depending on whether the stars in front of the dust
layer are coming from inside or outside the nominal projected radius.
As a result, for a moderately thick inclined stellar disk like M31,
the fraction of reddened stars should vary significantly from the near
side to the far side\footnote{As indeed has long been evident in
  optical images of M31.}, by an amount that depends on the
inclination and on the thickness of the stellar disk relative to the
disk's exponential scale length.  As such, the appropriate value of
$\langle f_{red} \rangle$ should vary with position to avoid biasing
the derived parameters.

To create a prior that looks as much as possible like the actual data,
we developed a simple geometric model for the reddening fraction of a
thick, inclined disk. We adjust the properties of this model
iteratively, using the observed spatial distribution of $f_{red}$ in
regions where the measurement of $f_{red}$ has small reported errors,
and thus is essentially unaffected by the prior. Between iterations,
the only significant change in the maps of $f_{red}$ are in regions of
very low extinction where $f_{red}$ is largely unconstrained by the
data, and instead gravitates to the value of $\langle f_{red} \rangle$
set by the prior.  The fit converged after only 2 iterations
(verifying our assumption that the measured values of $f_{red}$ in low
uncertainty regions were unaffected by the choice of prior).  The
resulting model for $\langle f_{red} \rangle$ has a position angle of
$37^\circ$, inclination of $78^\circ$, and a ratio of vertical to
horizontal exponential scale heights of $h_z/h_r=0.15$.  The residuals
from this simple model are approximately $\Delta f_{red} = \pm0.1$.

After setting the most probable value of $f_{red}$ as a function of
position and extinction, we set the width of the prior probability
distribution by using an asymmetric split normal distribution in $x$.
We keep the priors broad, to avoid biases in regions where the simple
geometric model for $f_{red}$ is not ideal, and to account for our
more uncertain knowledge of the filling factor of molecular clouds
$f_{fill}\equiv \mathcal{A}_{cloud}/ \mathcal{A}_{pixel}$, where
$\mathcal{A}_{cloud}$ and $\mathcal{A}_{pixel}$ are the areas of the
gas cloud and the analysis pixel, respectively.  We narrow the prior
for low extinctions ($\widetilde{A_V}<0.3$), where the RGB is only
slightly broadened, rather than split into distinct peaks, since these
regions will not have sufficient information to constrain $f_{red}$
reliably without the prior.  Examples of the resulting priors for
regions with $\langle f_{red} \rangle = 0.2$ and 0.8 are shown
in Fig.\ \ref{priorfig}.

\subsubsection{Fitting Parameters in a Pixelized Map}

Using these priors and the likelihood (eqn.\ \ref{likelihoodeqn}), we
use a MCMC sampler to characterize the posterior
probability distribution for $\vec{\theta}$ (eqn.\ \ref{parampdfeqn}).
We use {\tt emcee} \citep{foremanmackey2013}, which implements an
affine-invariant ensemble sampler to efficiently sample the
distribution using a set of coupled MCMC chains
\citep[``walkers'';][]{goodman2010}.

We run the MCMC sampler within pixels defined by a dense spatial grid
covering the survey area.  Our fiducial grid uses square
$6.645\arcsec$ pixels, which correspond to $25\pc$ at the distance of
M31 ($776\kpc$).  M31 is highly inclined, and thus projection effects
may make the effective spatial resolution of the survey significantly
worse than $25\pc$ in the direction paralleling the minor axis.
However, although this degredation in resolution is generically true
for disk structures, there is no strong evidence in the Milky Way that
molecular clouds are flattened in the same sense as the global
galactic disk, rather than being 3-dimensional objects with no
preferred direction to their structure. In this latter case,
projection does not actually affect the physical resolution; as an
extreme example, if molecular clouds were spherical, they would appear
to have the exact same size when viewed at any angle.  We therefore
adopt $25\pc$ as the effective resolution of the dust map, in spite of
the inclination of M31's disk, but recognize that the exact physical
resolution could be worse in the direction perpendicular to the major
axis.

The adopted $25\pc$ grid size balances the accuracy in
$\vec{\theta}$ (which decreases when there are few stars per pixel, as
discussed below in Sec.\ \ref{accuracysec}) and the spatial
resolution of the resulting map. We have found that we can
produce reliable maps at significantly higher resolution, but only in
regions where a large fraction of the stars are reddened.

We oversample the $25\pc$ grid by rerunning the fitting procedure at 4
dithered locations, shifted by 0.5 pixel. We then interleave the
resulting 4 dithered maps, producing pixels of $12.5\pc$ on a
side. This sub-sampling is roughly analogous to Nyquist sampling a map
with $25\pc$ resolution.  Each of the interleaved pixels shares 50\%
of the stars with the adjacent pixels with which it shares a border,
and 25\% of the stars with the pixels with which it shares a corner.
Thus, pixels are not independent of their immediate neighbors, but are
completely independent of all other pixels.

Within each $25\pc$ spatial pixel, we use 50 sets of coupled MCMC
chains, and take 150 steps to localize the chains near the
parameter values that maximize the posterior probability.  We then run
the sampler for a final 15 steps (i.e., $50\times15$ samples) to
sample the posterior distribution.  To deal with memory limitations and
allow parallelization, we analyze the pixels in batches, defined by
the area of individual survey bricks; when a pixel overlaps adjacent
bricks, we use the results for the pixel that contained the largest
number of stars. Running the samper for all 22 bricks at all 4 dither
positions takes over a week when running on 50 cores.

\begin{figure*}
\centerline{
\includegraphics[width=3.25in]{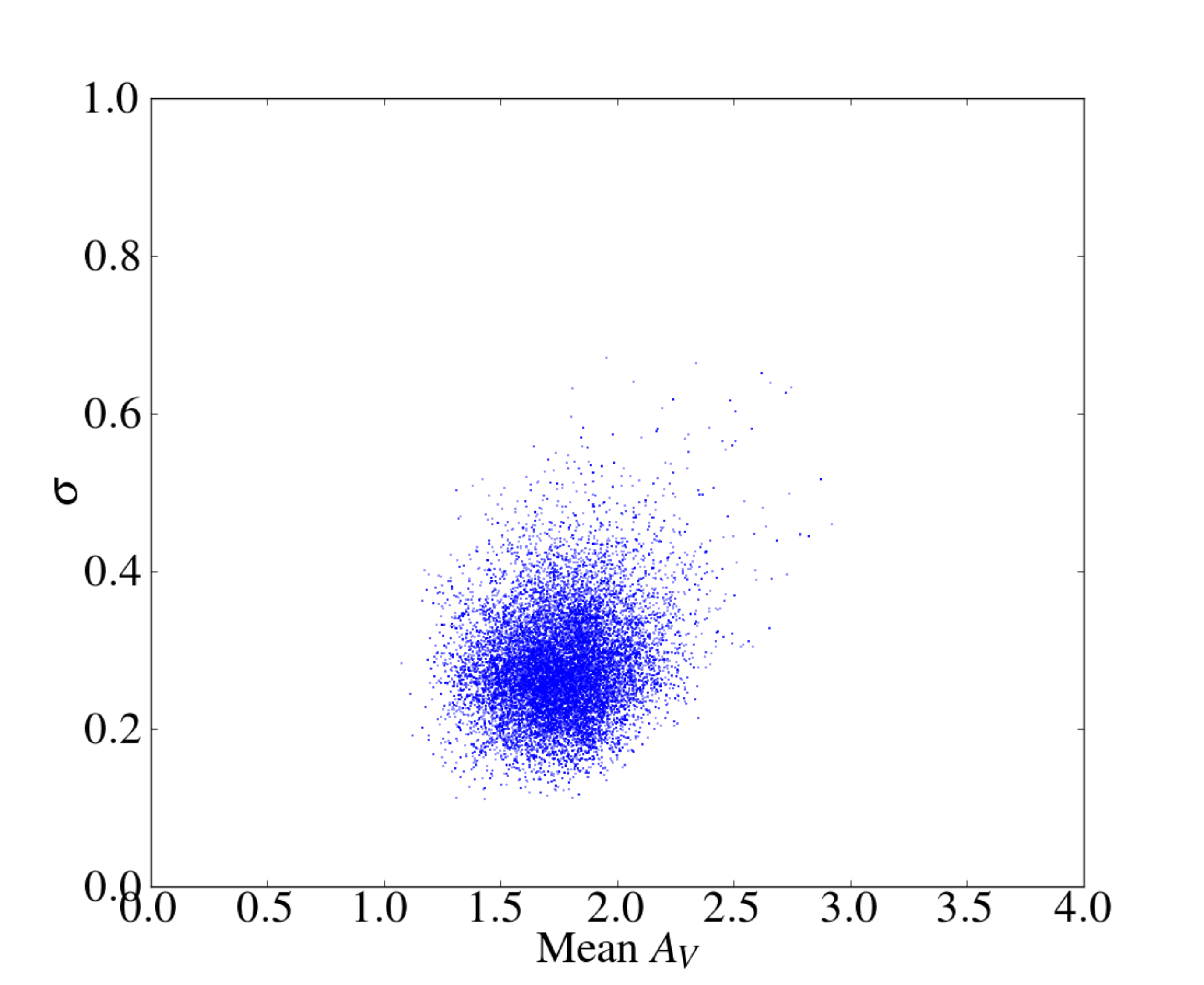}
\includegraphics[width=3.25in]{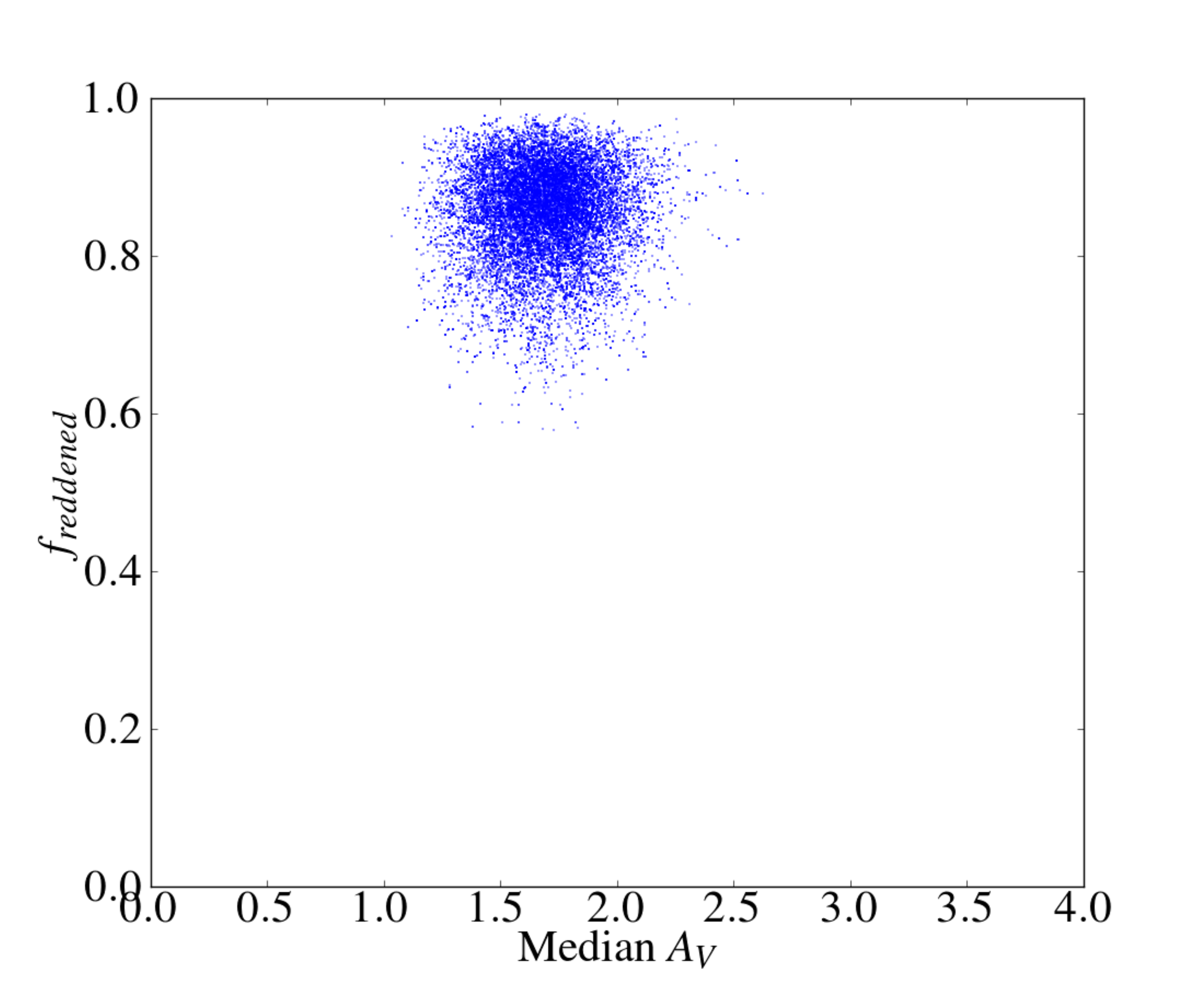}
}
\caption{Samples of the posterior probability distribution function
  for the width of the log-normal distribution $\sigma$ (left) and the
  reddening fraction $f_{red}$ (right) as a function of the mean $A_V$
  (left) and the median $A_V$ (right).  Samples were drawn for the
  same $25\pc$ pixel used in Figure~\ref{modelcmdfig}. Bestfit values
  and the 16\%-84\% range are $f_{red}=0.84_{-0.07}^{+0.05}$, $\langle
  A_V \rangle=1.80_{-0.24}^{+0.17}$, and $\sigma=0.24_{-0.06}^{+0.07}$.
\label{PDFfig}}
\end{figure*}

A representative example of the resulting distribution is shown in
Figure~\ref{PDFfig}, for the same model shown in
Figure~\ref{modelcmdfig}.  The distribution is smooth and unimodal, as
has been the case for all distributions that we have inspected
individually.  Because of the simplicity of these distributions, we do
not save the full probability distribution at each grid
point, but instead record the values of the parameters that maximize
the posterior probability.  We also marginalize the distributions for
each parameter, and record the values corresponding to 16\%, 50\%, and
84\% of the parameter distribution (i.e., the median and the $\pm 1
\sigma$ points for a Gaussian distribution).  When quoting a single
value for the uncertainty in each parameter (rather than a range), we
define the uncertainty in the $i$-th parameter to be $\Delta\theta_i =
(\theta_{i,84}-\theta_{i,16})/2$.

We also record the equivalent percentile points for quantities that we
derive from the best-fit parameters during subsequent analysis.  For
example, the mean $A_V$ depends on both $\widetilde{A_V}$ and $\sigma$
(eqn.\ \ref{meaneqn}), so we calculate the marginalized distribution
of $\langle A_V \rangle=\widetilde{A_V} \exp{(\sigma^2/2)}$ and record
the resulting 16\%, 50\%, and 84\% percentile values, rather than
relying on formulas that assume propagation of Gaussian errors in
$\widetilde{A_V}$ and $\sigma$.  The same deterministic translation of
uncertainties is used when converting from the regularized variable
$x$ back to the reddening fraction $f_{red}$ (eqn.\ \ref{xeqn}).

\section{Properties of the Reddening Parameters} \label{reddingpropertiessec}

Before presenting final maps, we use the results of fitting individual
PHAT bricks to demonstrate the overall accuracy of the method. We show
results for two independent regions, and discuss the model
parameters' spatial distribution (Sec.\ \ref{distributionsec}),
correlations (Sec.\ \ref{correlationsec}), and uncertainties
(Sec.\ \ref{accuracysec}). We then discuss the method's overall
susceptibility to systematic errors (Sec.\ \ref{systematicsec}), a
point we return to when comparing to other dust tracers in
Sec.\ \ref{spatialsystematicssec} below.

\subsection{The Spatial Distribution of Derived Reddening Parameters}  \label{distributionsec}

\begin{figure*}
\centerline{
\includegraphics[width=3.75in]{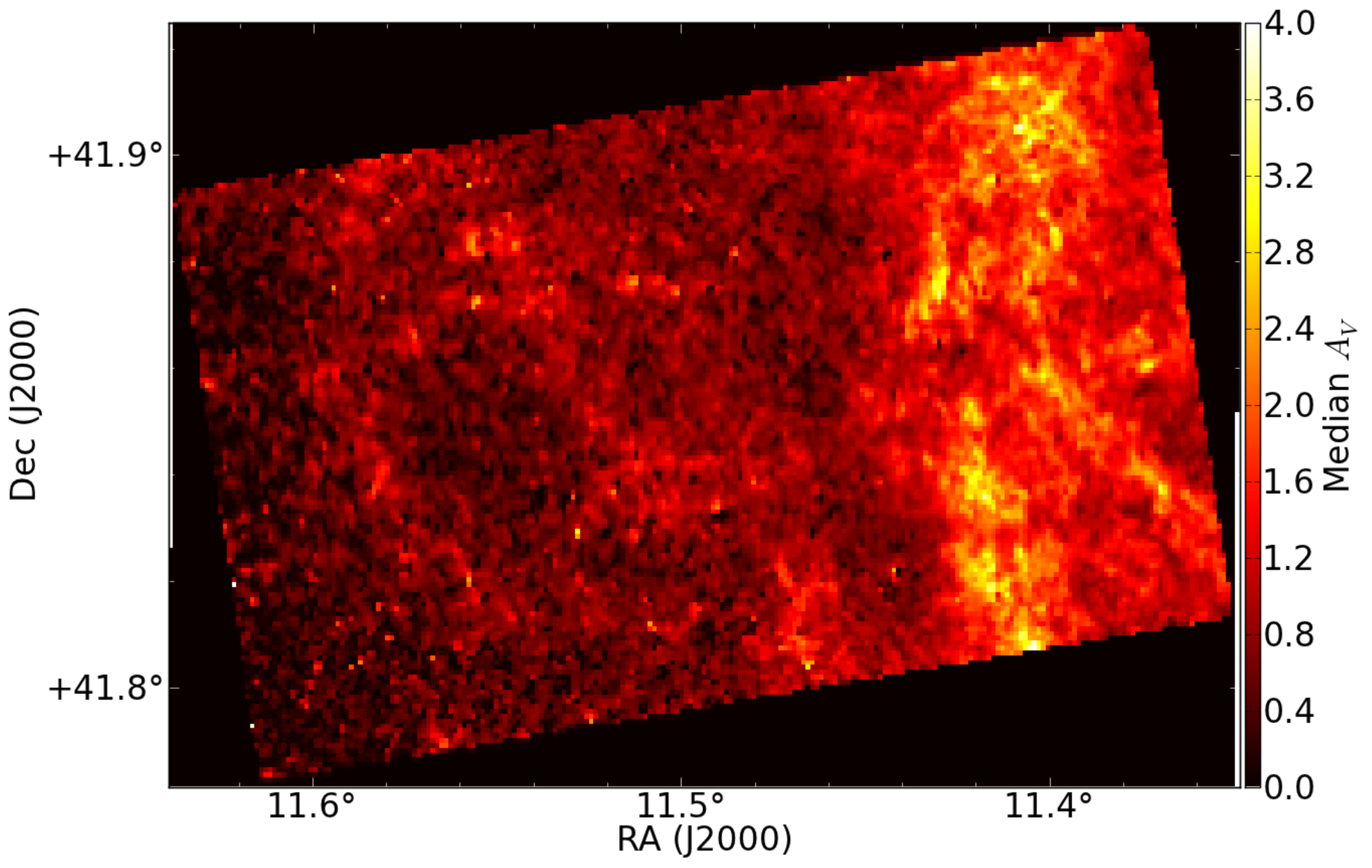}
\includegraphics[width=3.75in]{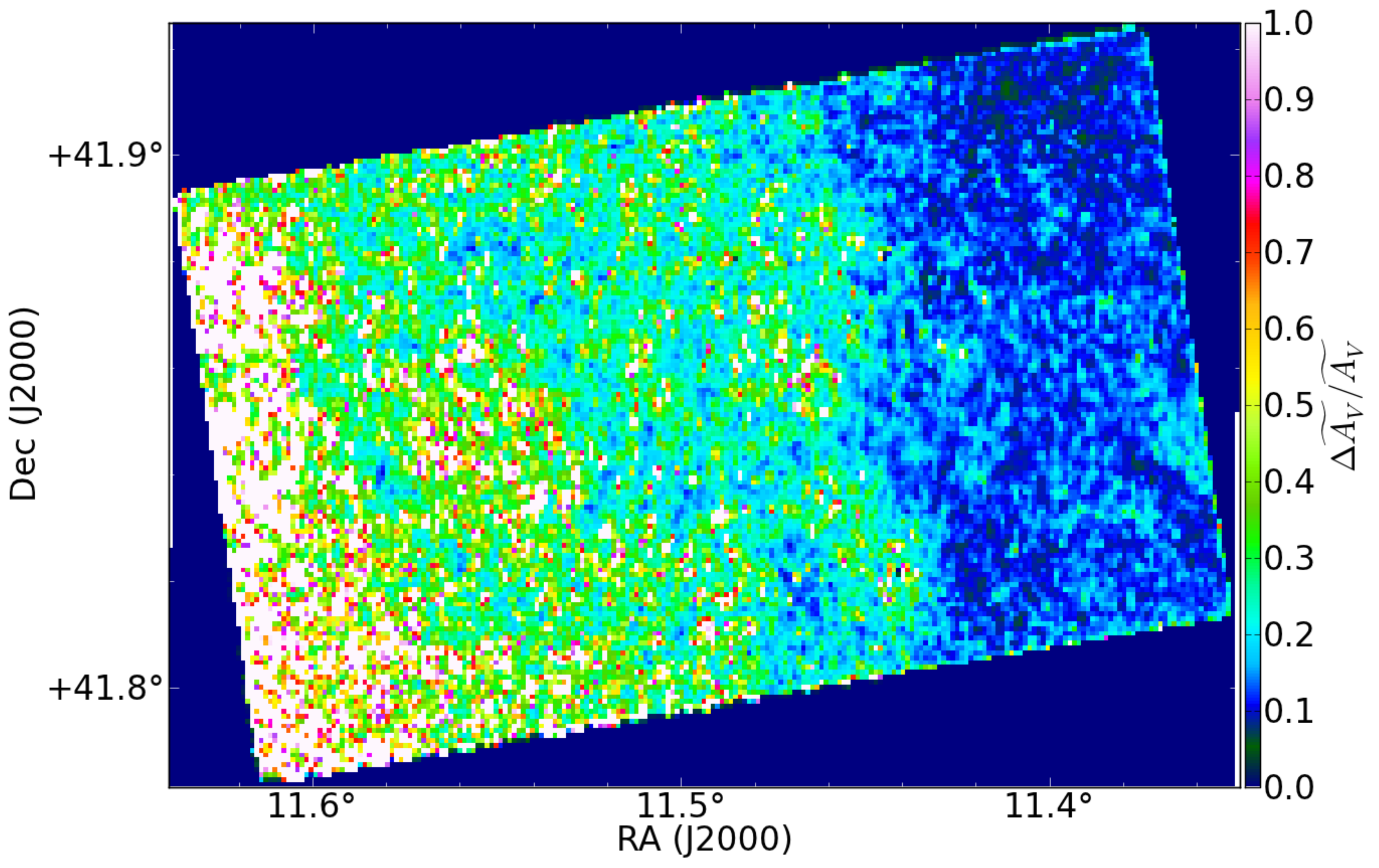}
}
\centerline{
\includegraphics[width=3.75in]{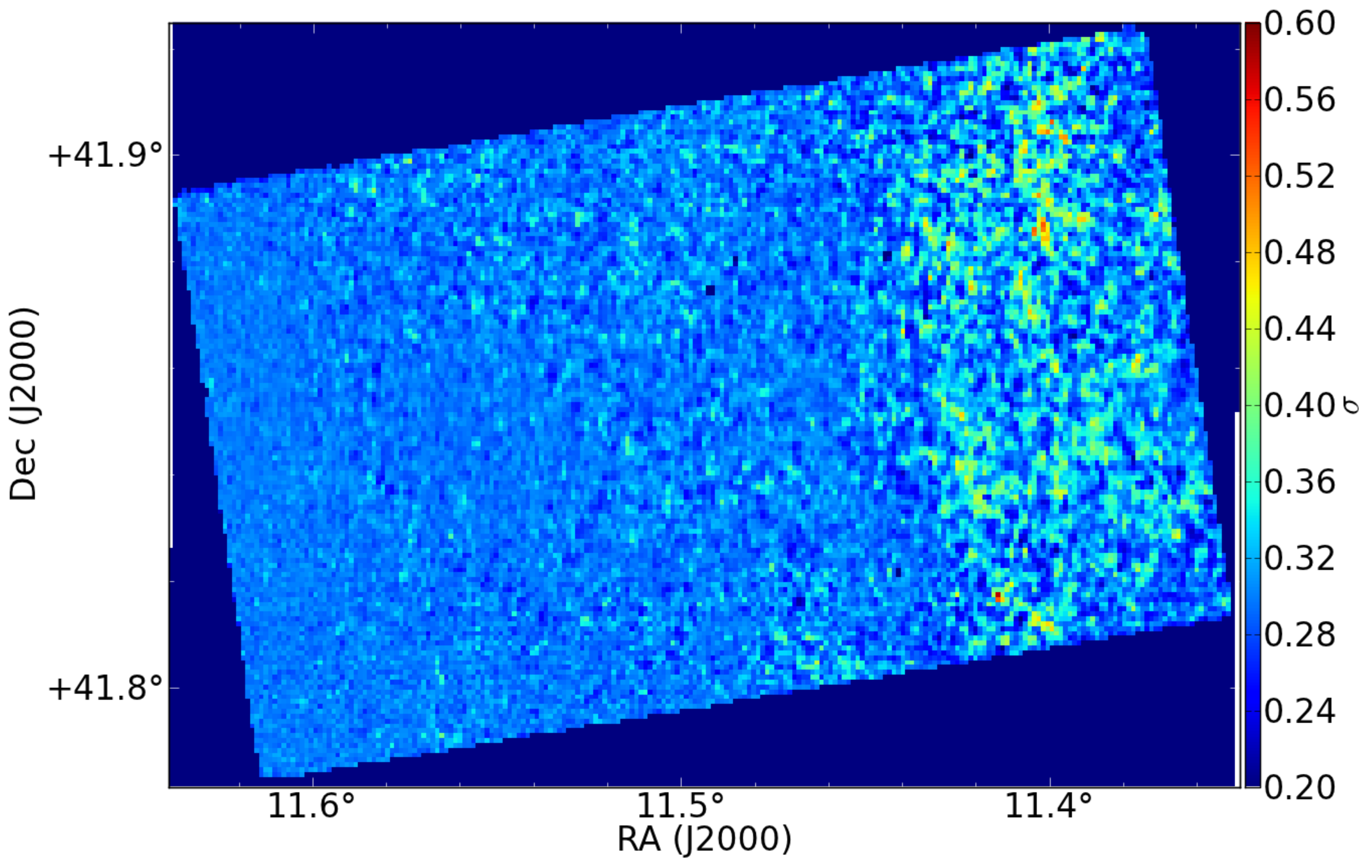}
\includegraphics[width=3.75in]{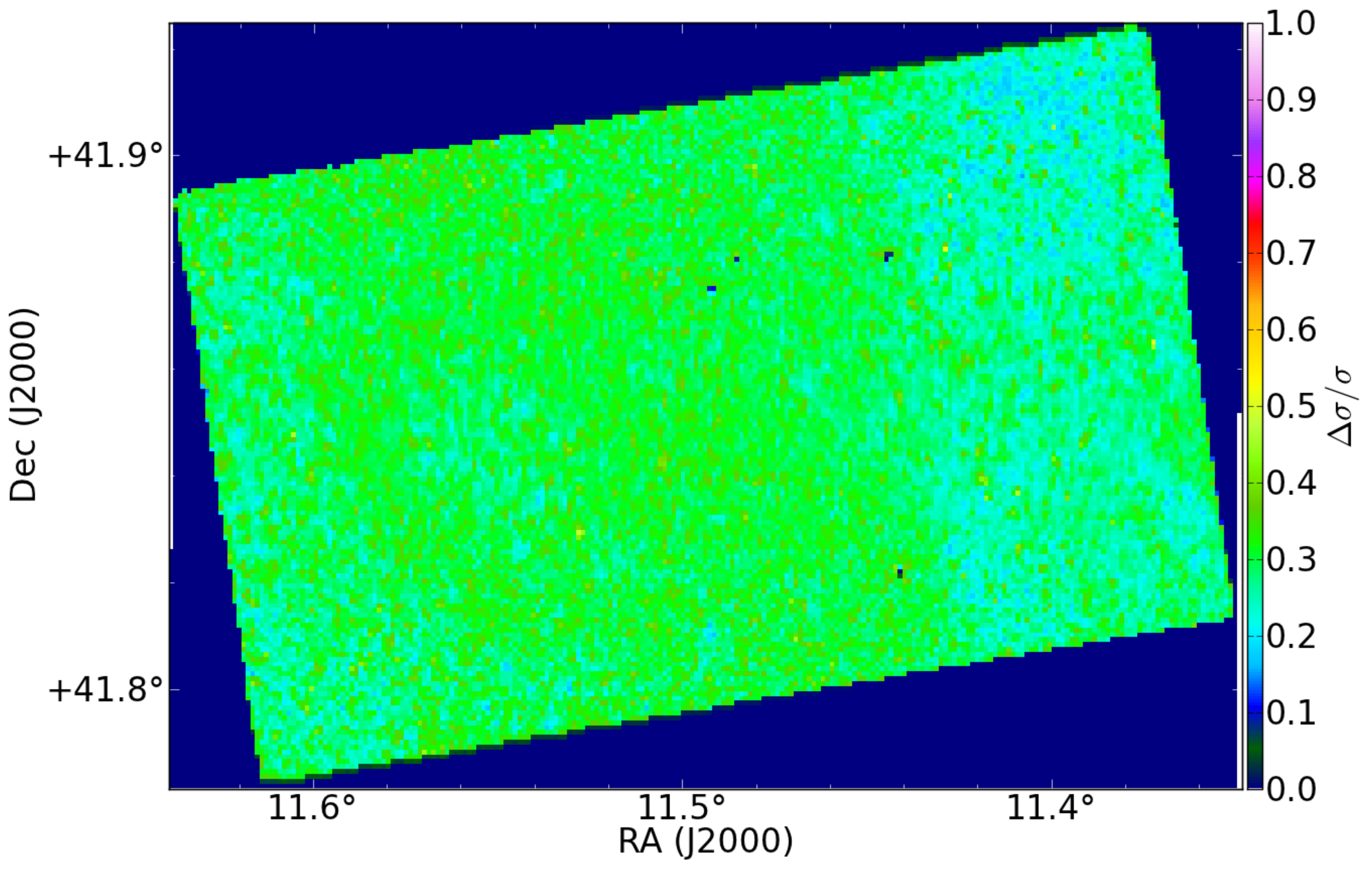}
}
\centerline{
\includegraphics[width=3.75in]{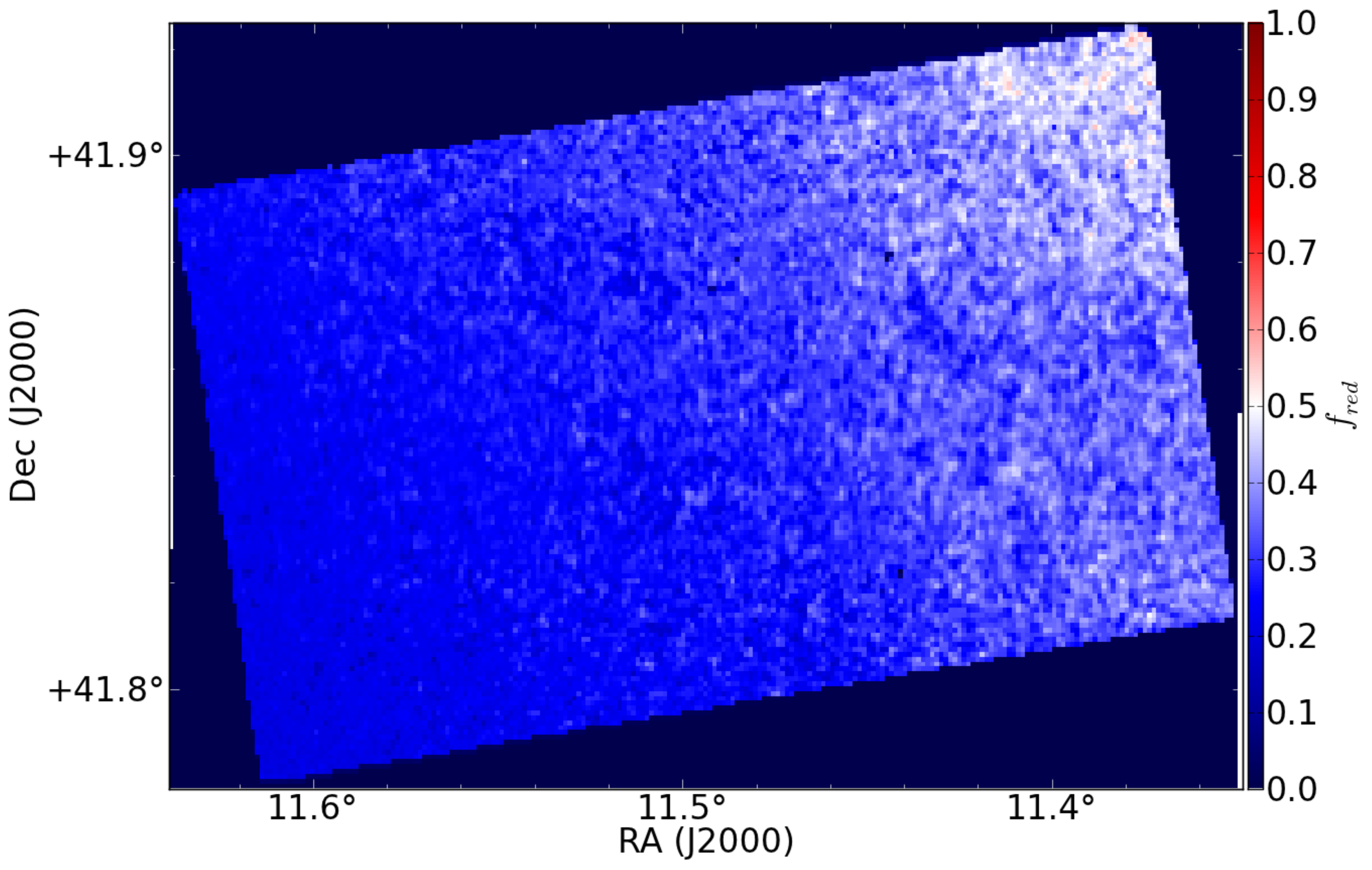}
\includegraphics[width=3.75in]{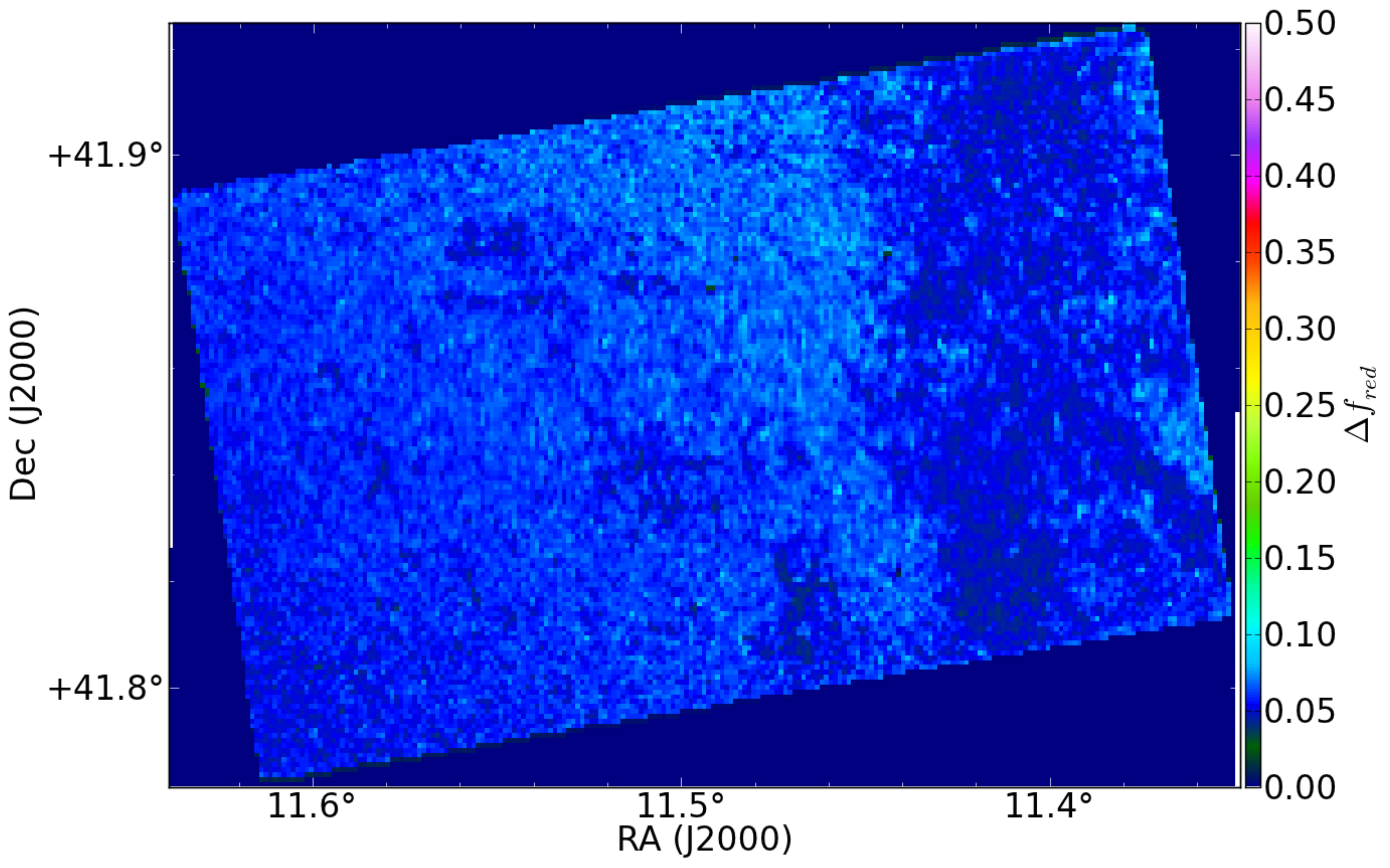}
}
\caption{Left: Maps of the median extinction $\widetilde{A_V}$ (top),
log-normal width $\sigma$ (middle), and reddened fraction $f_{red}$
(bottom) for Brick 16.  In regions of low extinction, the values of
$f_{red}$ and $\sigma$ are driven to the most likely value in their
priors ($=0.4$). Right: Maps of the fractional uncertainty in
extinction $\Delta \widetilde{A_V}/ \widetilde{A_V}$ (top) and the
log-normal width $\Delta\sigma / \sigma$ (middle), and the absolute
uncertainty in the reddening fraction $\Delta f_{red}$
(bottom). Uncertainties are smallest in the regions of highest
extinction. In general, the fractional errors in $\widetilde{A_V}$ are
much smaller than in $\sigma$.
\label{brickmap16fig}}
\end{figure*}

\begin{figure*}
\centerline{
\includegraphics[width=3.75in]{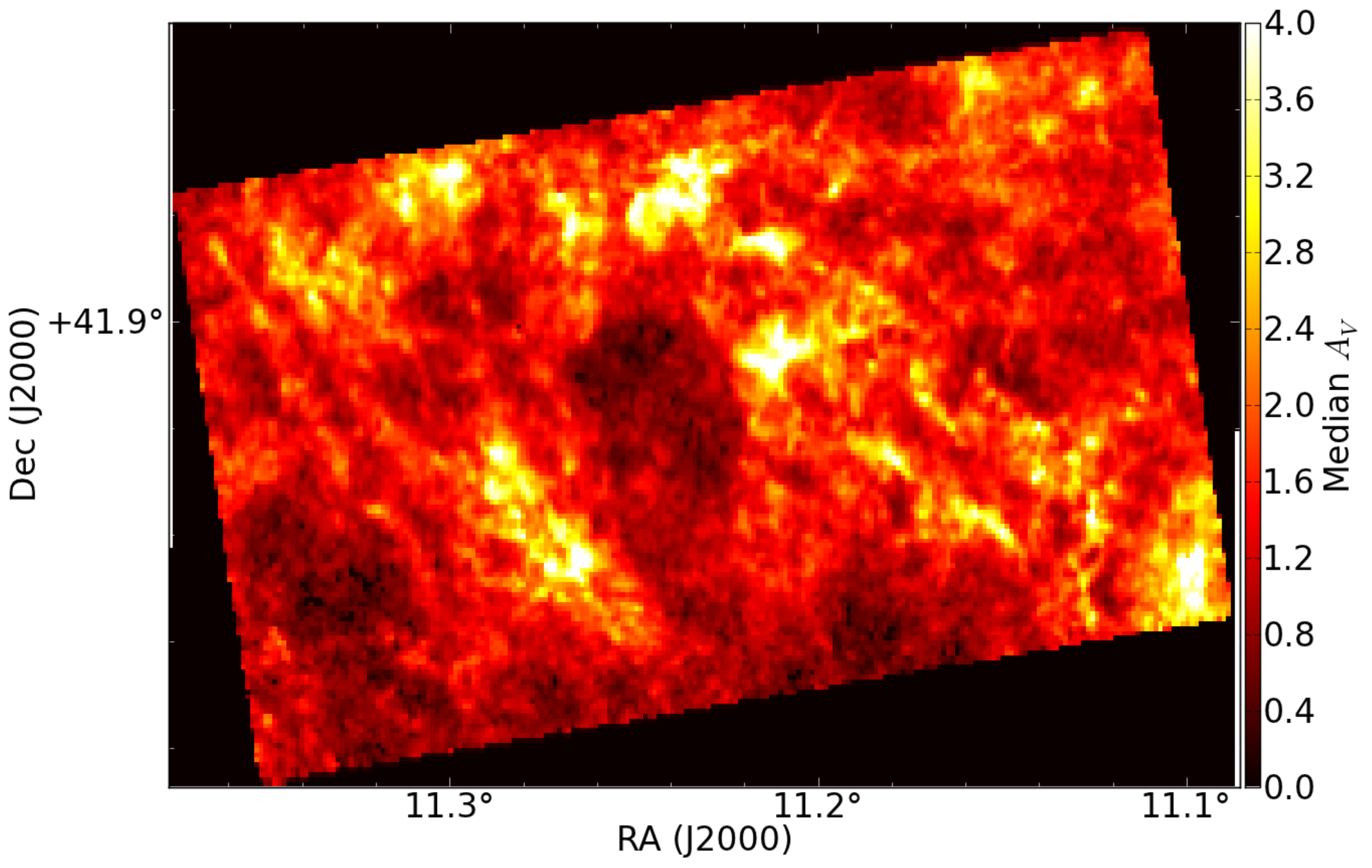}
\includegraphics[width=3.75in]{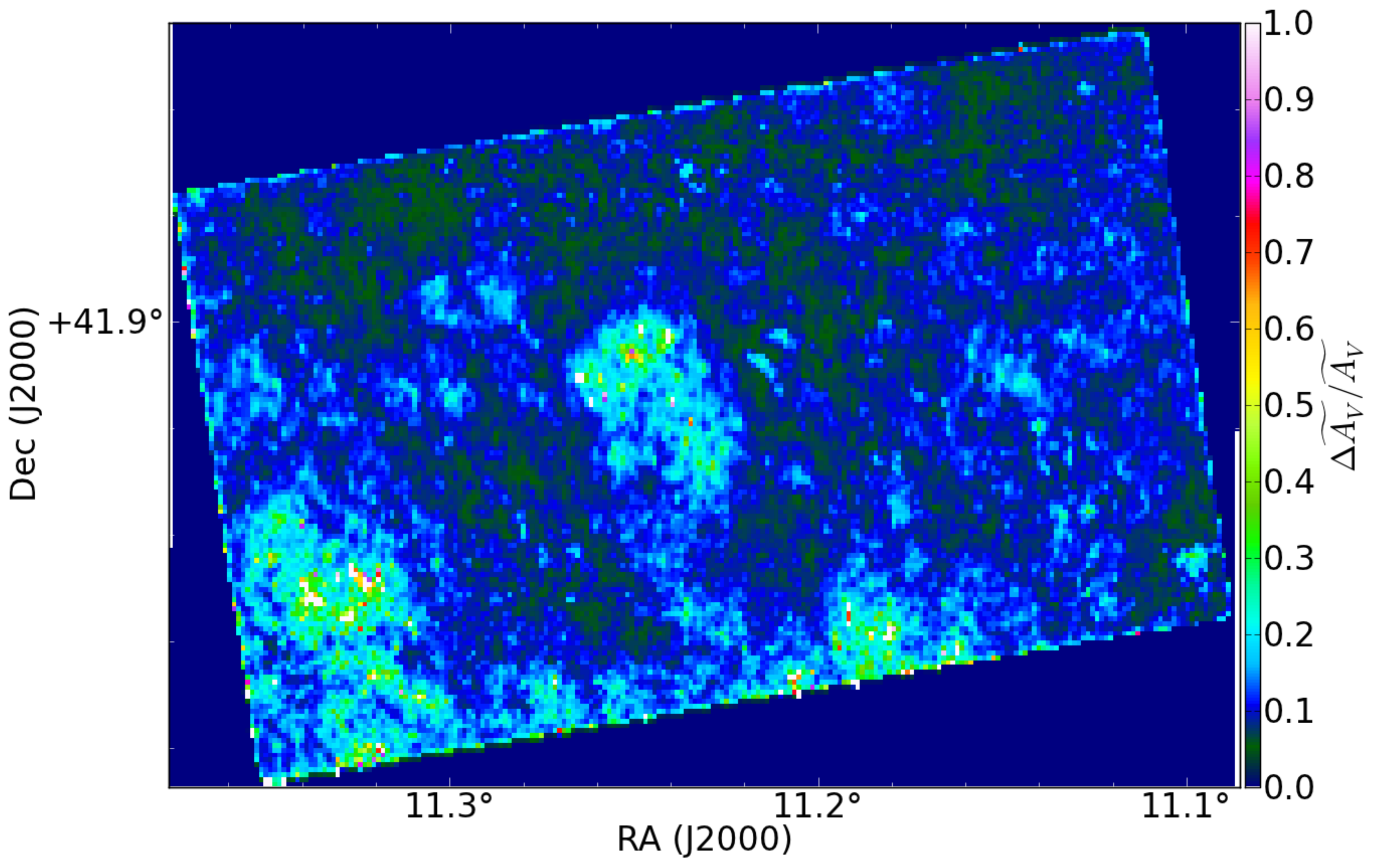}
}
\centerline{
\includegraphics[width=3.75in]{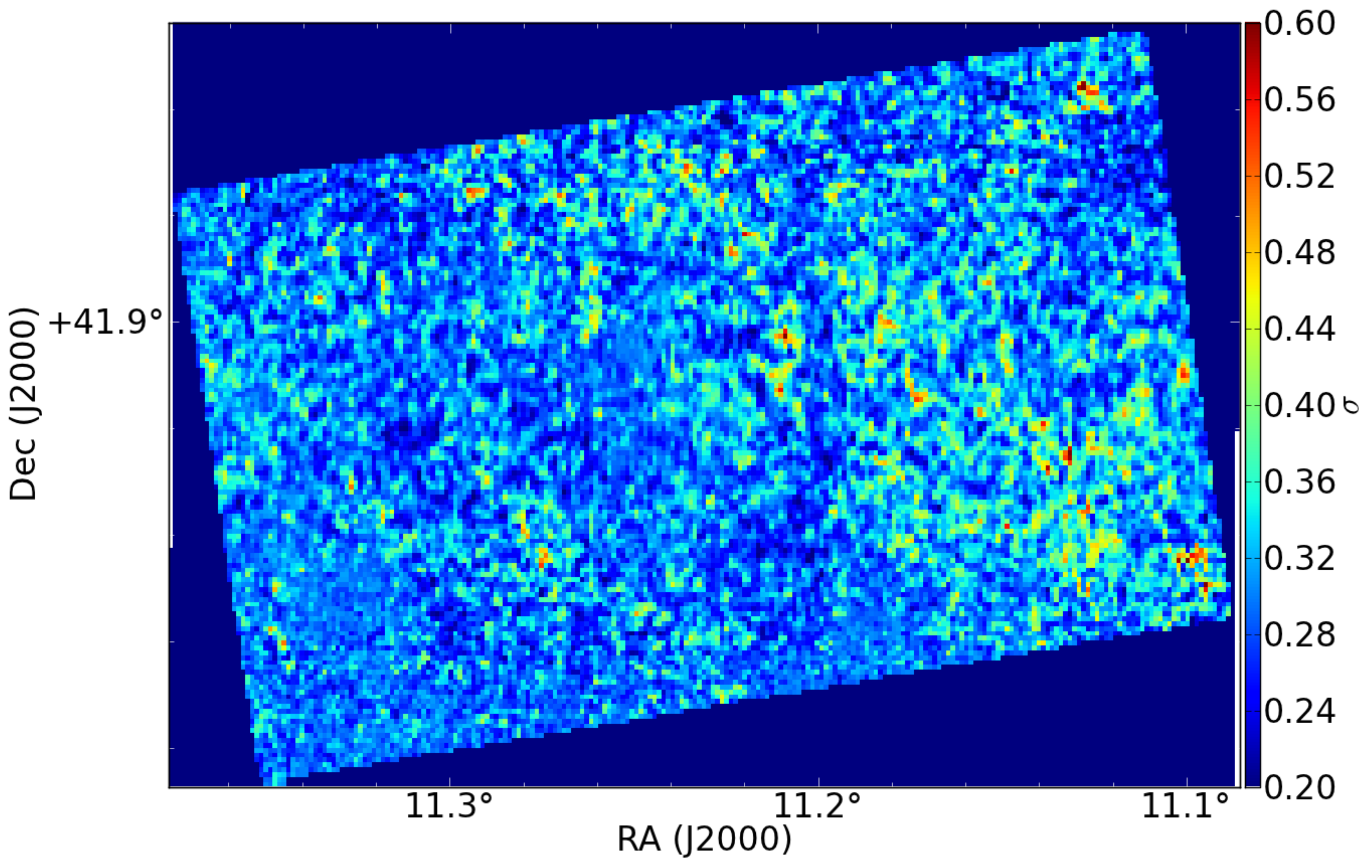}
\includegraphics[width=3.75in]{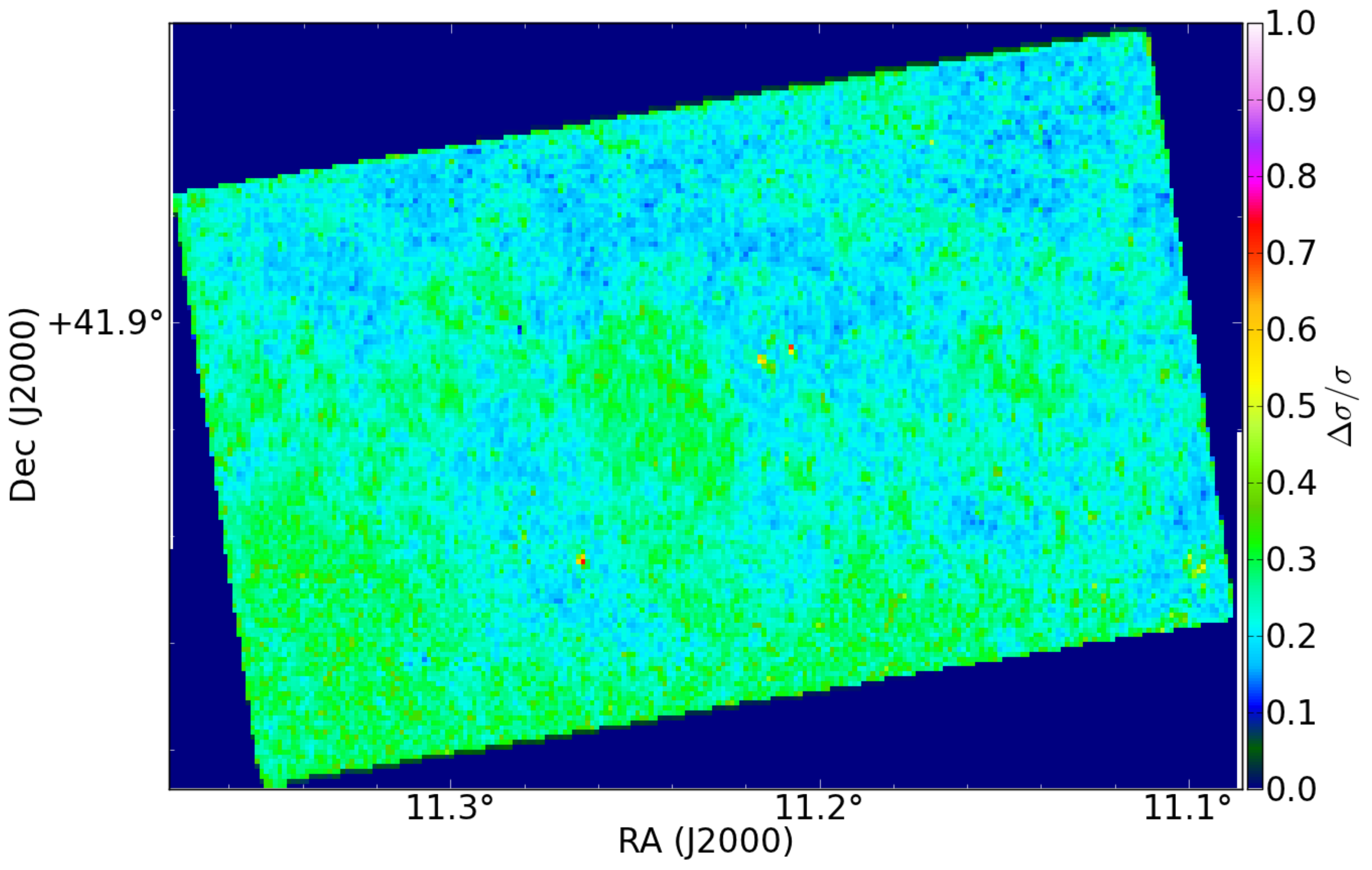}
}
\centerline{
\includegraphics[width=3.75in]{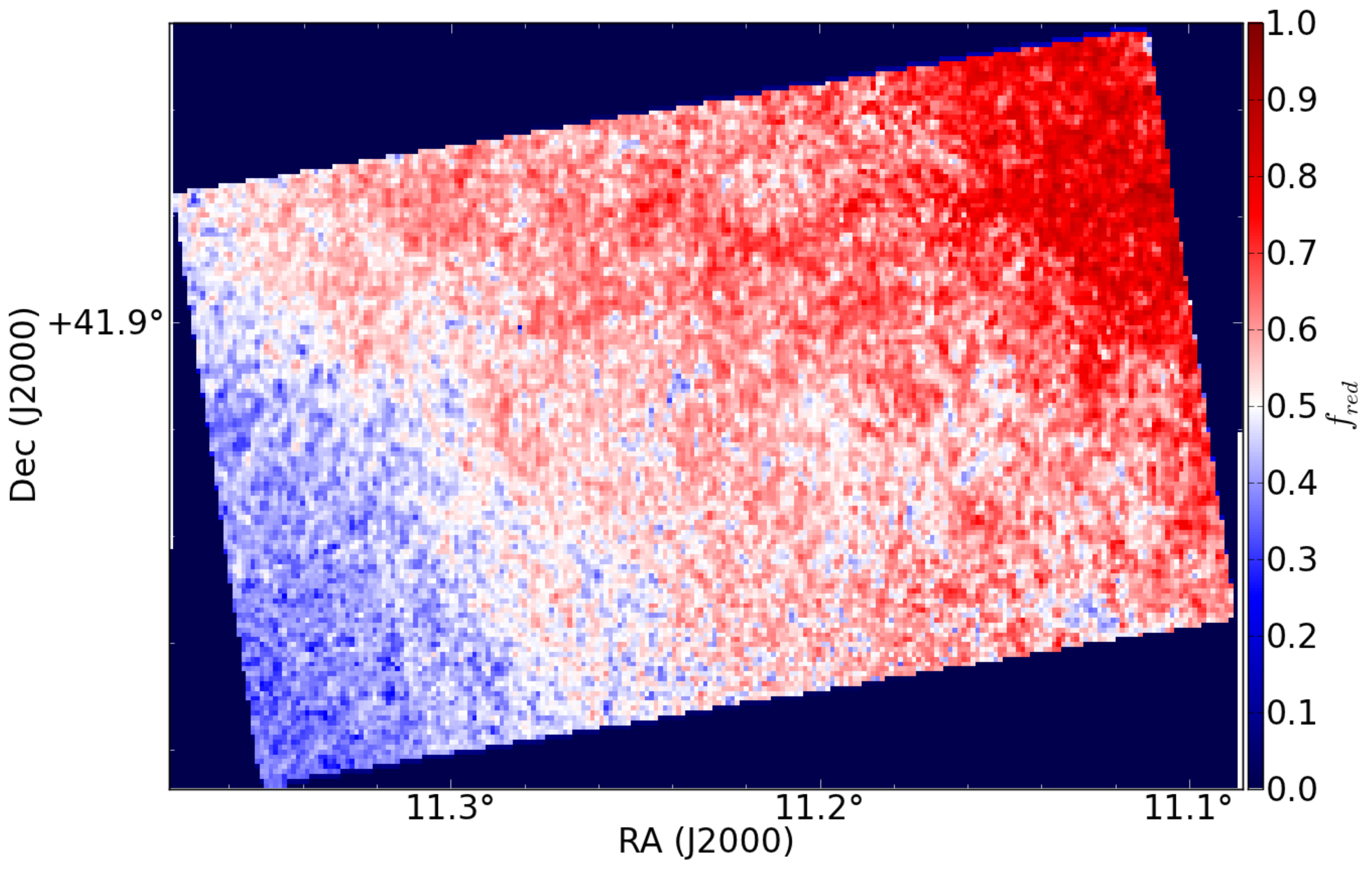}
\includegraphics[width=3.75in]{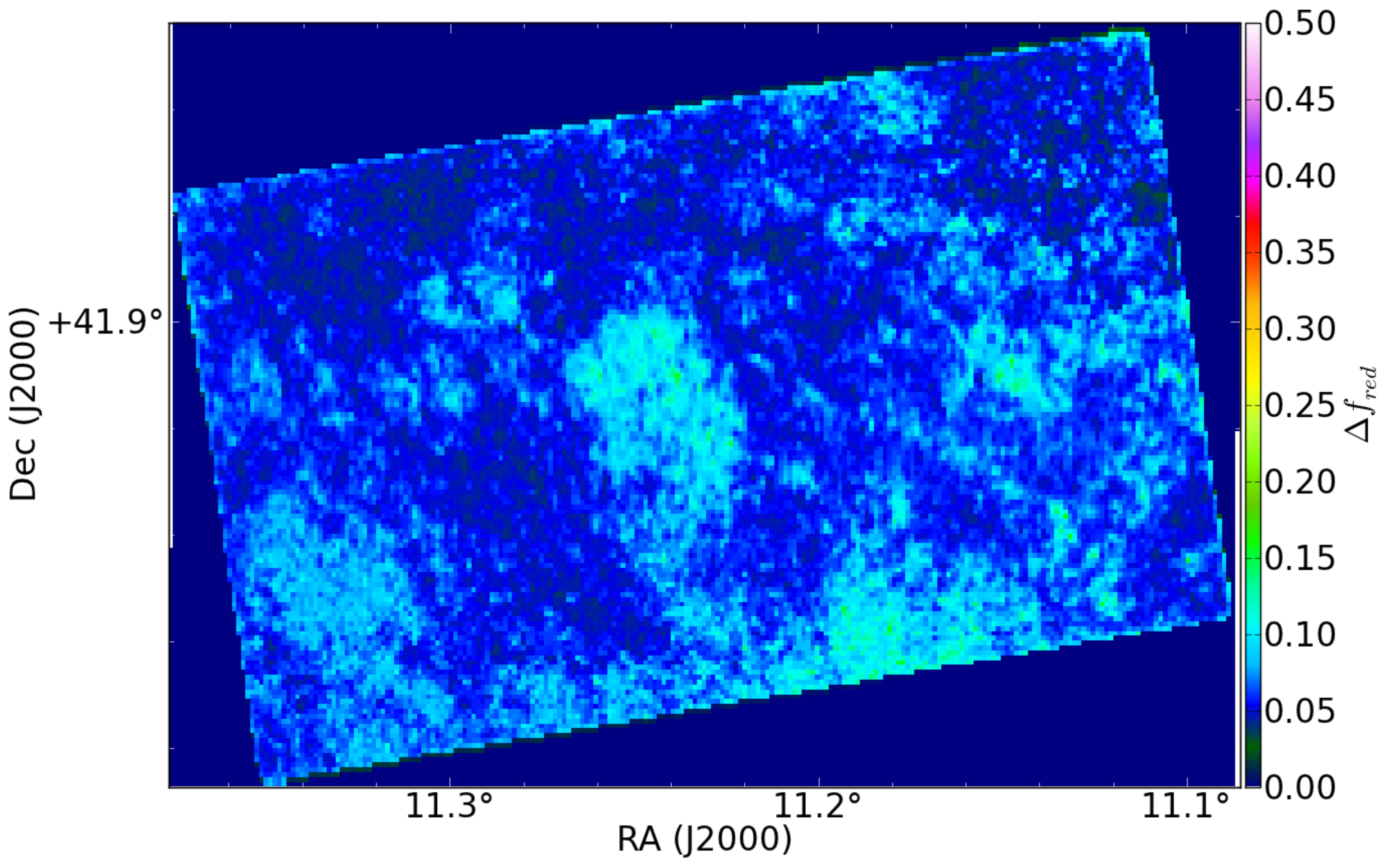}
}
\caption{Same as Figure~\ref{brickmap16fig},
  but for Brick 15, which falls on the $10\kpc$ star-forming ring on
  the major axis, and thus samples the highest star formation of any
  of the PHAT bricks. Left: Maps of the median extinction
  $\widetilde{A_V}$ (top), log-normal width $\sigma$ (middle), and
  reddened fraction $f_{red}$ (bottom).  Right: Maps of the fractional
  uncertainty in extinction $\Delta \widetilde{A_V}/ \widetilde{A_V}$
  (top) and the log-normal width $\Delta\sigma / \sigma$ (middle), and
  the absolute uncertainty in the reddening fraction $\Delta f_{red}$
  (bottom). Given that uncertainties are smallest in the regions of
  highest extinction, Brick 15 has low uncertainties
  overall. \label{brickmap15fig}}
\end{figure*}

The left panels of Figures~\ref{brickmap16fig} and
\ref{brickmap15fig} show the spatial distribution of
$\widetilde{A_V}$, $\sigma$, and $f_{red}$ for two of the PHAT
bricks that sample regions with different star formation
intensities (Bricks 16 and 15, the latter of which contains some of
the most intense star formation in the PHAT survey area).  Each brick
covers an area of roughly $1.5\kpc \times 3\kpc$; see
\citet{dalcanton2012} for details.

Although the reddening parameters are derived independently for each
spatial pixel, there are clear coherent features in all the derived
parameters across the area, on scales larger than the one pixel
coherence expected solely from oversampling the $25\pc$ pixel
grid. The distribution of dust extinction shows clear filamentary
structure throughout the kiloparsec-scale regions, and is reminiscent
of large scale maps within the Milky Way
\citep[e.g.,][]{schlegel1998,froebrich2007, kohyama2013,lombardi2011,
  schlafly2011, nidever2012}.  There is also obvious large scale
coherence in the spatial distribution of the reddening fraction
$f_{red}$.  The width of the reddening distribution $\sigma$ also is
spatially coherent and follows the structure in the reddening
distribution, but with a much smaller dynamic range.

\subsection{Correlations Among the Reddening Parameters}  \label{correlationsec}

\begin{figure*}
\centerline{
\includegraphics[width=3.75in]{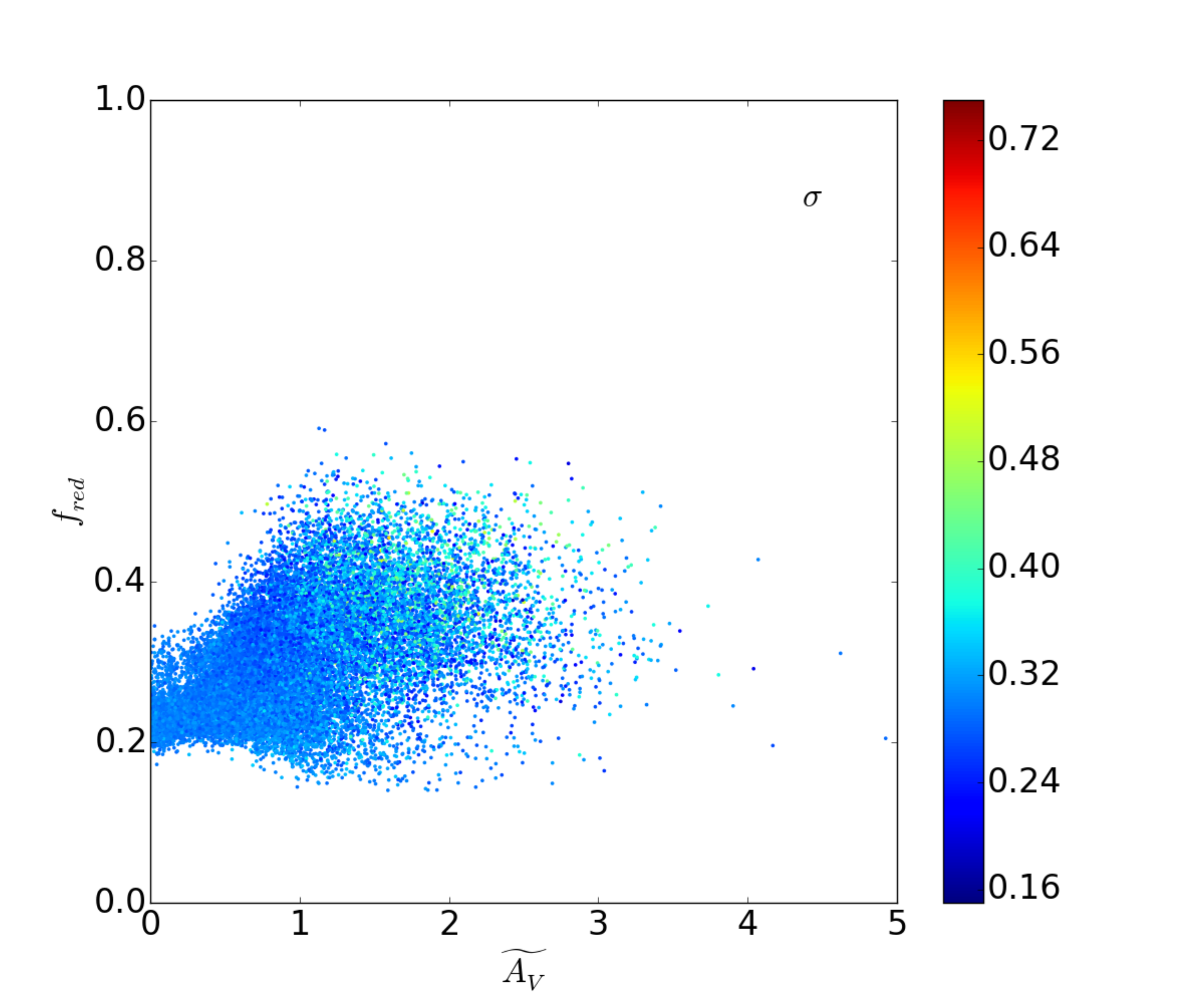}
\includegraphics[width=3.75in]{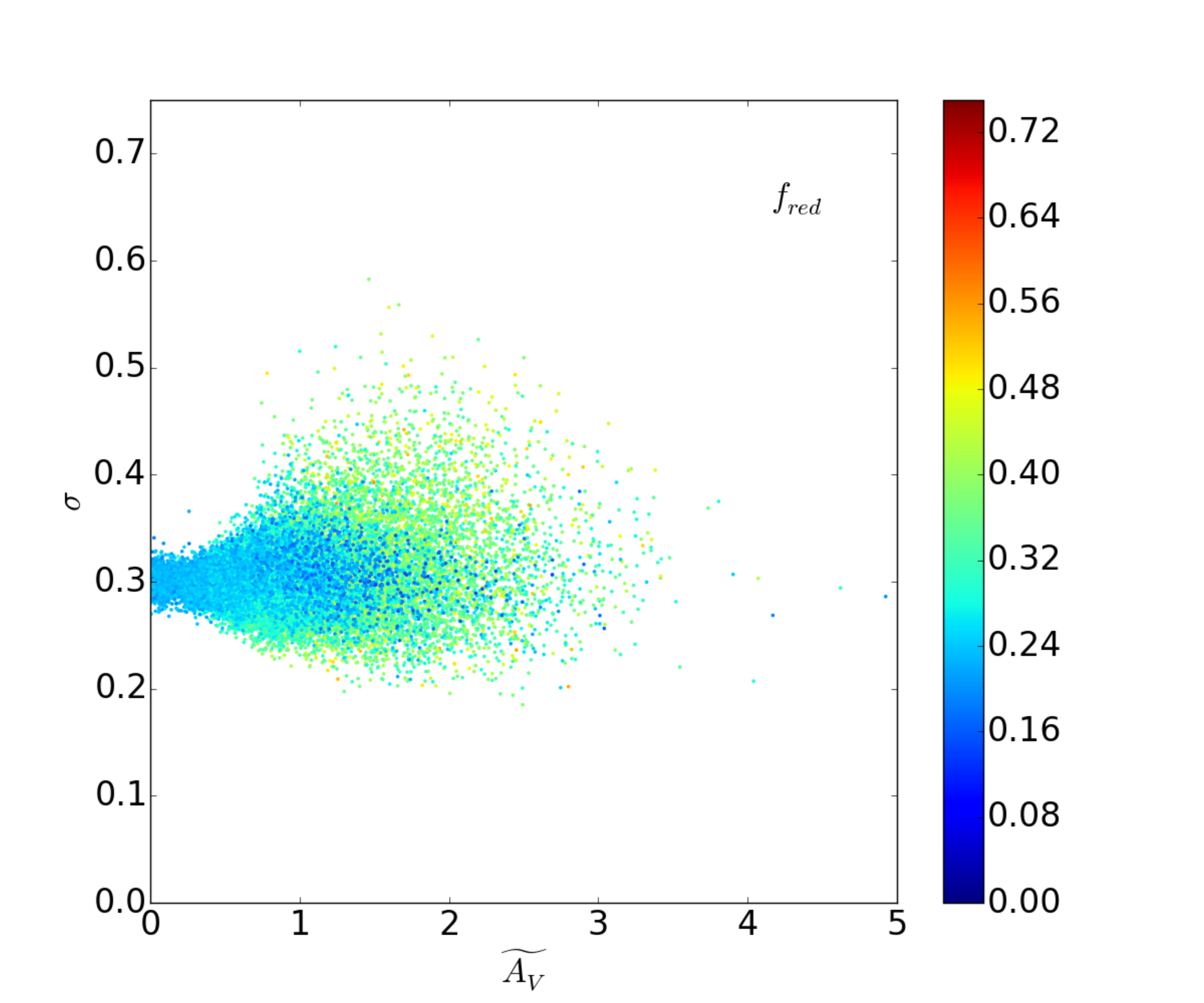}
}
\centerline{
\includegraphics[width=3.75in]{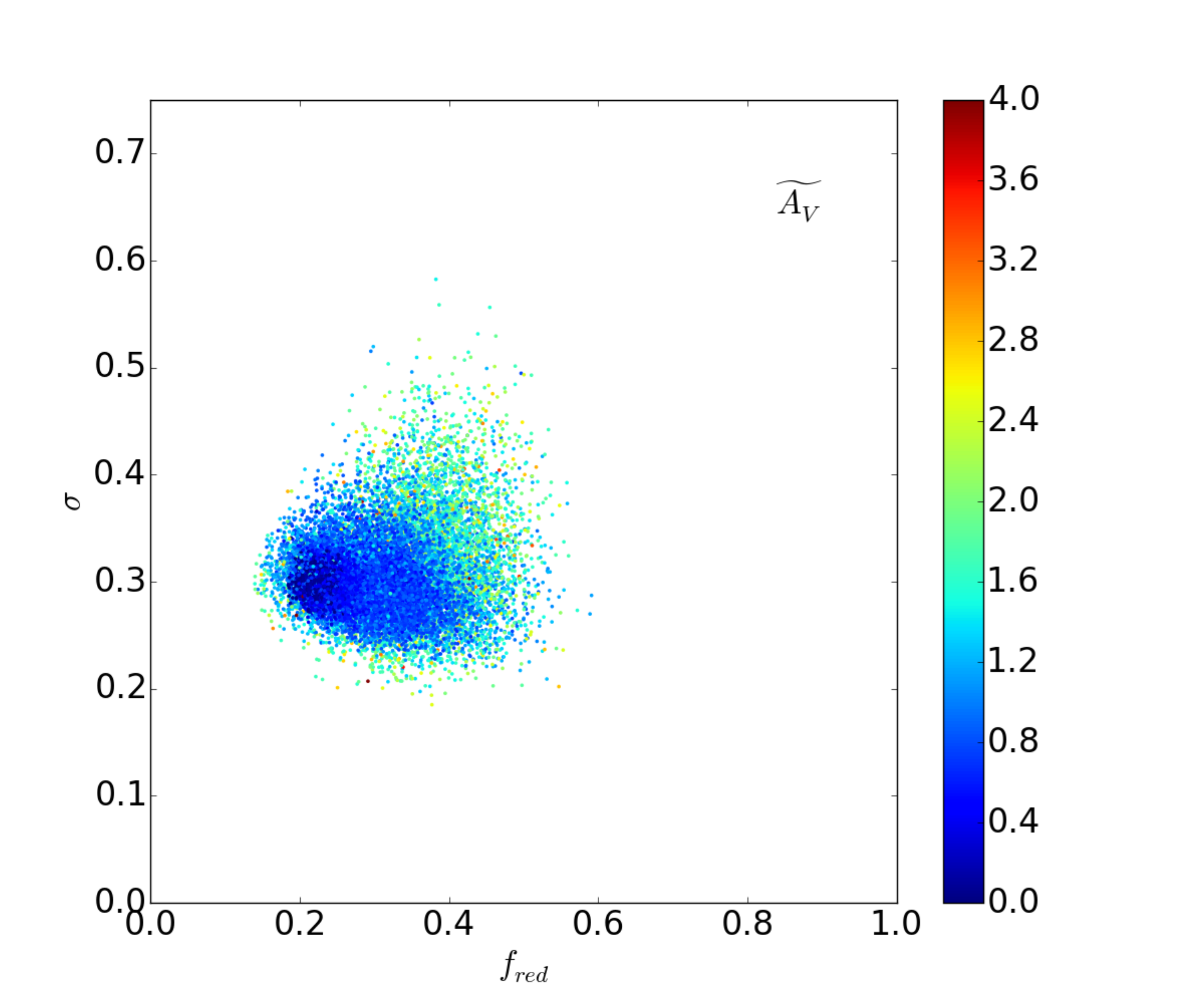}
\includegraphics[width=3.75in]{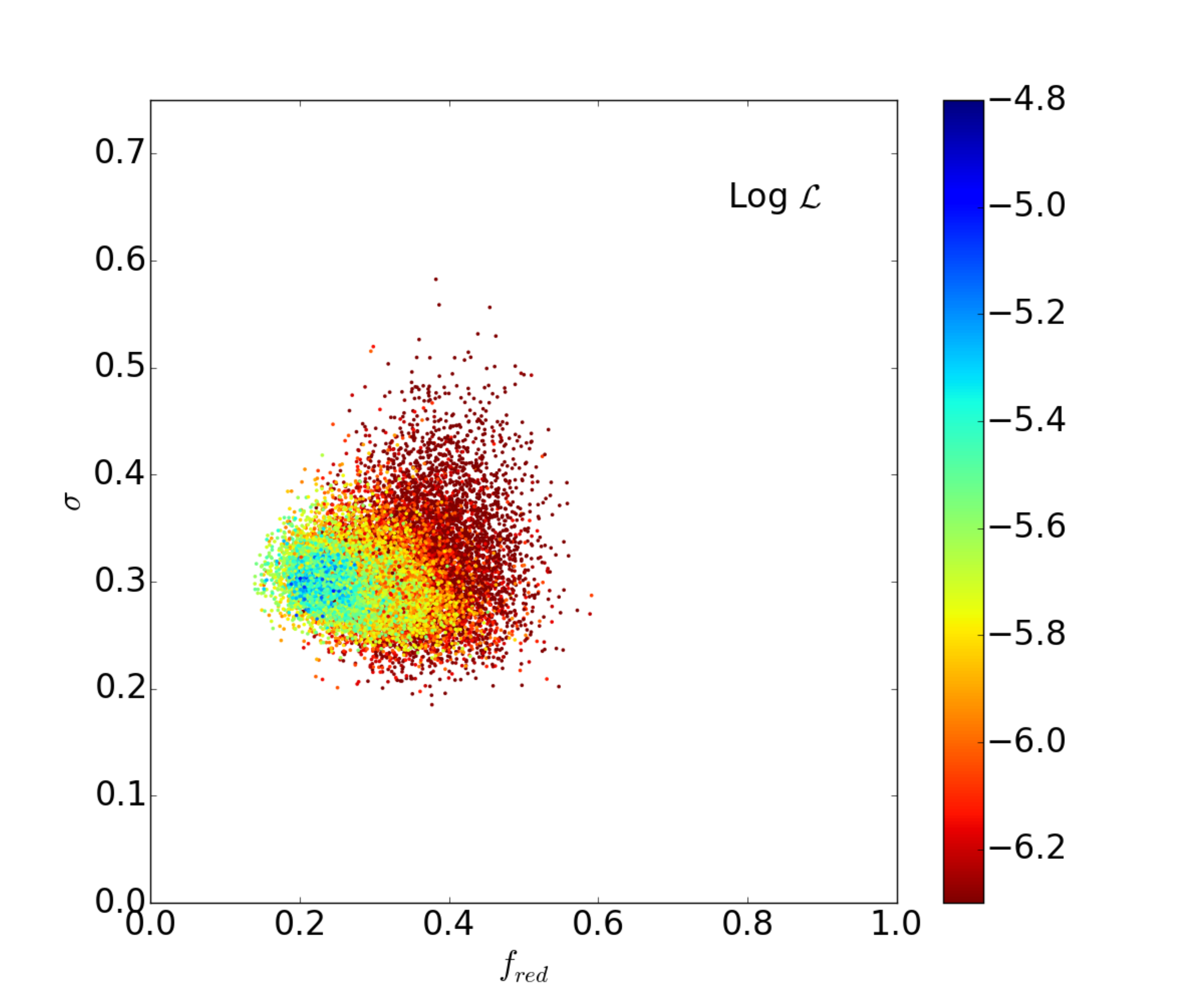}
}
\caption{Joint distributions of derived parameters ($\widetilde{A_V}$
  vs $f_{red}$, $\widetilde{A_V}$ vs $\sigma$, and $f_{red}$ vs
  $\sigma$, clockwise from top left) for each pixel in Brick 16,
  color-coded by the parameter labeling the figure; {\cal L} is the
  likelihood calculated for the fit.  At low $\widetilde{A_V}$
  ($\lesssim0.5$), the values of $f_{red}$ and $\sigma$ are clearly
  pulled towards the maximum probability of the assigned priors
  ($=0.4$).
  The bottom right shows
  $f_{red}$ vs $\sigma$ again, but now color-coded by the log
  likelihood of the model fit.
  \label{brickscatter16fig}}
\end{figure*}

\begin{figure*}
\centerline{
\includegraphics[width=3.75in]{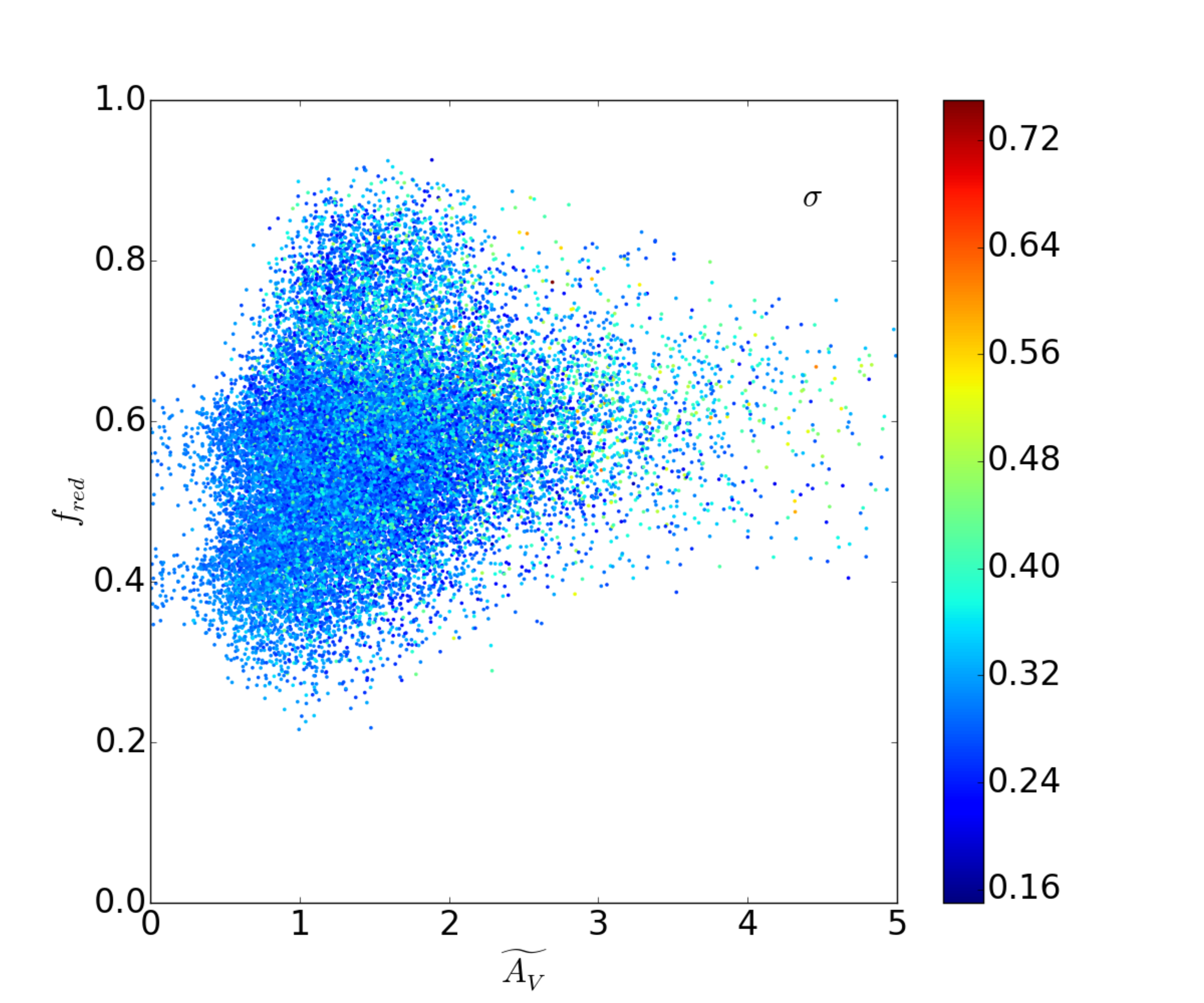}
\includegraphics[width=3.75in]{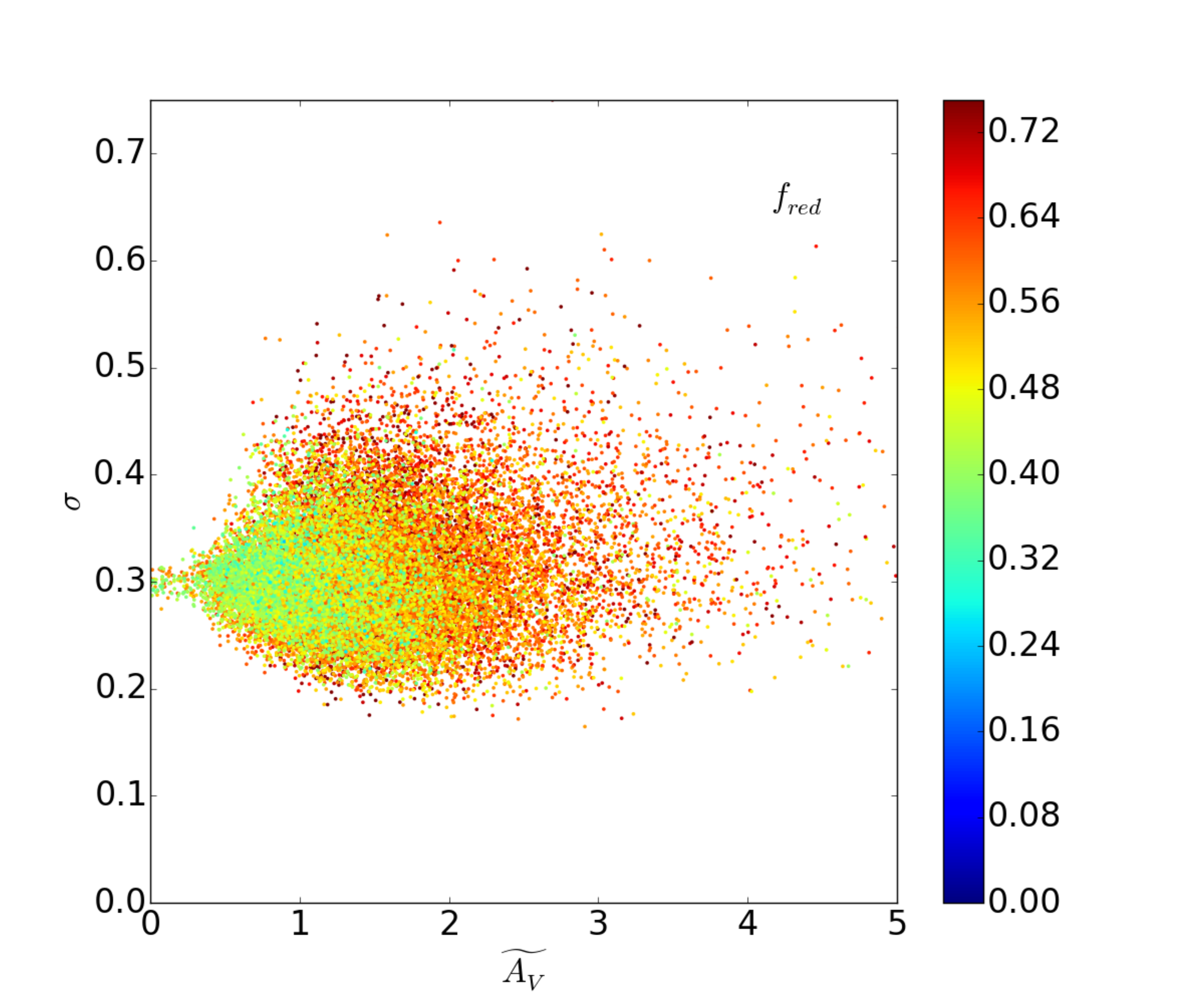}
}
\centerline{
\includegraphics[width=3.75in]{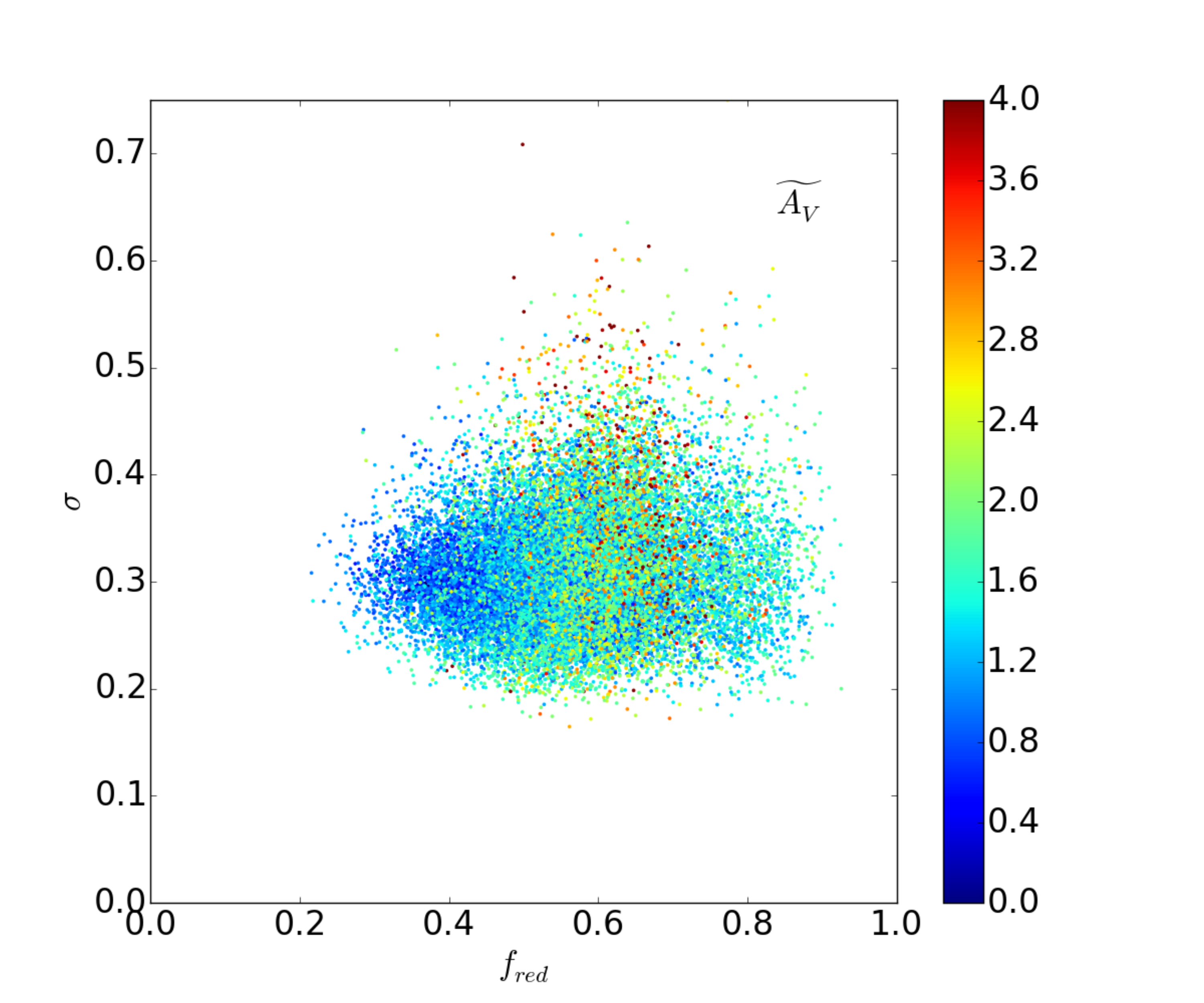}
\includegraphics[width=3.75in]{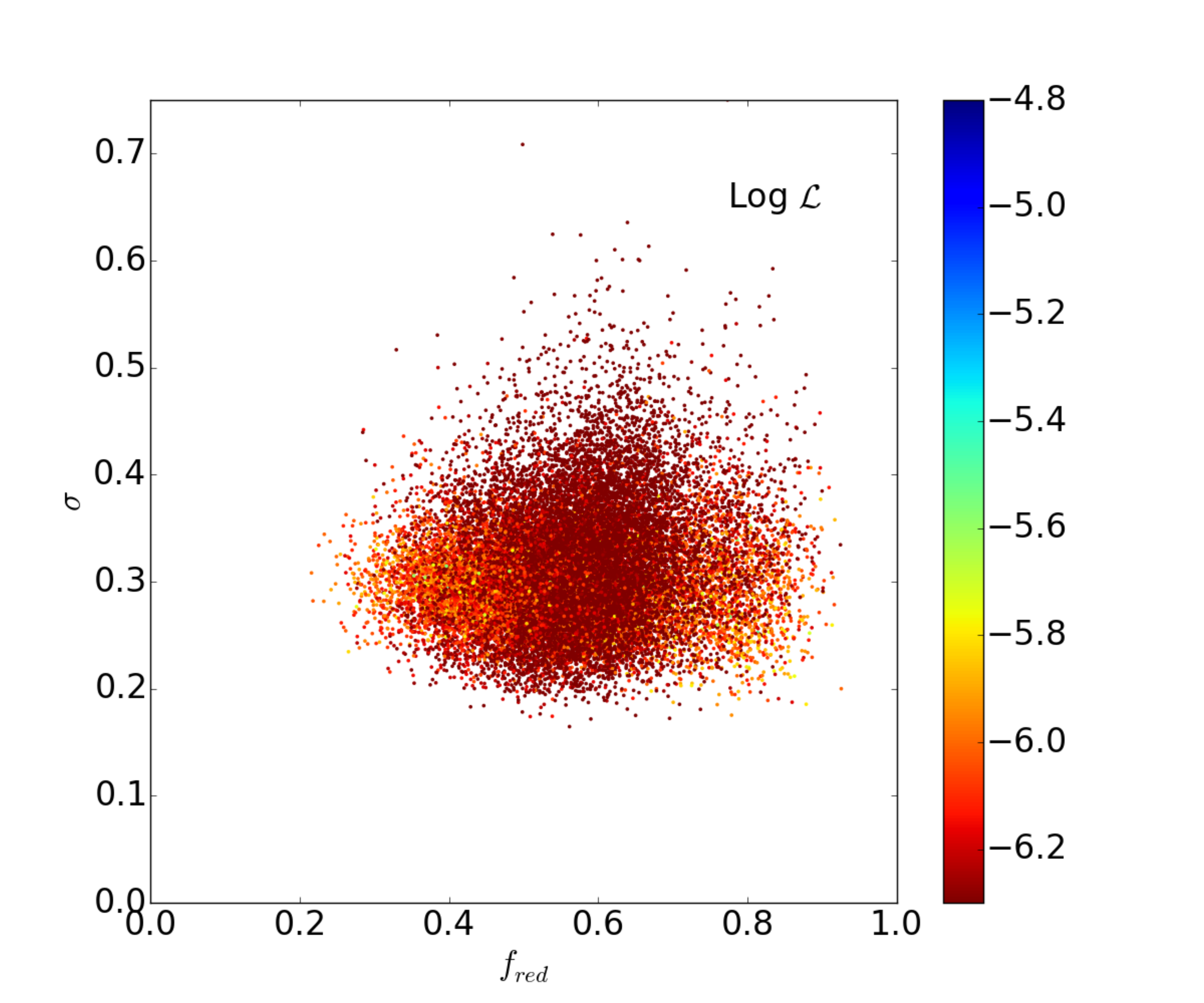}
}
\caption{Same as Figure~\ref{brickscatter16fig}, but for the high
star-formation region covered by Brick 15. Joint distributions of
derived parameters ($\widetilde{A_V}$ vs $f_{red}$, $\widetilde{A_V}$
vs $\sigma$, and $f_{red}$ vs $\sigma$, clockwise from top left) for
each pixel in Brick 15, color-coded by the unplotted parameter
(denoted by the legend). The bottom right shows $f_{red}$ vs $\sigma$ again,
but now color-coded by the log likelihood of the model fit.
  \label{brickscatter15fig}}
\end{figure*}

Figures~\ref{brickscatter16fig} and~\ref{brickscatter15fig} plot
correlations among the different reddening parameters, for the maps in
Figures~\ref{brickmap16fig} and~\ref{brickmap15fig}, respectively.

The plots of $f_{red}$ as a function of $\widetilde{A_V}$ (upper left)
reveal a number of features that are due to either the dust
distribution or to the fitting process, or both.  First, at high extinction
($\widetilde{A_V}\gtrsim 1\mags$), the mean value of $f_{red}$ 
shows no obvious trend with increasing extinction; this behavior is
most noticeable in Figure~\ref{brickscatter15fig}, where the
extinction and $f_{red}$ are highest. The simplest interpretation of
this behavior is that dusty gas largely fills
the entire $25\pc$ pixel when the extinction is high, such that the
value of $f_{red}$ reflects only geometrical effects (i.e., the
relative position of the stars and the gas along the line
of sight). In this high extinction regime, the scatter in $f_{red}$ is
due both to the uncertainty in measuring $f_{red}$
(Sec.\ \ref{accuracysec} below) and to the intrinsic variation in the relative
position of the gas with respect to the stars.

At lower extinctions ($0.3 \lesssim \widetilde{A_V} \lesssim 1\mags$),
the typical value of $f_{red}$ tends to fall with decreasing
extinction (albeit with large scatter, particularly in Brick 15,
where the value of $f_{red}$ varies significantly across the brick).  It is unlikely that
the relative placement of gas and stars along the line of sight would
``know'' about the overall extinction of the gas.  Instead, this
fall-off may reflect that dusty gas has an increasingly
smaller filling factor as the median extinction decreases (see, for
example, Figure~\ref{orionfig}).  In this regime $f_{red}$ equals the
product of the ``geometric'' value of $f_{red,geom}$ that one would
measure if the gas filled the entire pixel and of the filling factor
$f_{fill}$.  Thus, if the area covered by gas is smaller for lower
extinctions, then one expects exactly the fall-off we observe.  Note
that in this range of $\widetilde{A_V}$ we have not built the
decreasing filling factor into the prior (see Figure~\ref{priorfig}),
and thus the decrease effectively reflects the properties of the gas
itself.  On the other hand, some of the reduced scatter may be due
to the increasing role that the prior plays as the information content
of the data decreases. At lower extinctions, reddening does not cleanly
separate the reddened and unreddened RGB, making measurements of $f_{red}$
more challenging, which increases the weight given to the prior and pulls
the measured value of $f_{red}$ closer to $\langle f_{red} \rangle$.

At the lowest extinctions ($\widetilde{A_V}\lesssim 0.3\mags$), the
relationship between $f_{red}$ as a function of $\widetilde{A_V}$
changes again. In this regime, the width of the NIR RGB is comparable
to or less than the broadening expected due to reddening.  As such,
the observed CMD is barely changed by the dust, and the constraints on
the reddening parameters becomes extremely weak. As such, the most
probable value of $f_{red}$ is driven almost entirely by its prior.
In other words, when the posterior probability distribution is very
broad, the peak is set primarily by the prior probability distribution
function, which in this case drives $f_{red}$ towards $\langle f_{red}
\rangle$ as the extinction decreases below $\widetilde{A_V}\lesssim
0.3\mags$.

Unlike the clear trends between $\widetilde{A_V}$ and $f_{red}$, any
trends of $\sigma$ with the other reddening parameters are far weaker.
Below $\widetilde{A_V}\lesssim 0.75\mags$, the distribution of
$\sigma$ is clearly dominated by the prior.  Large values of $\sigma$
are only found at higher extinctions ($\widetilde{A_V}\gtrsim 1\mags$,
most prevalent in Figure~\ref{brickscatter15fig}).  Visual inspection
of Figures~\ref{brickmap16fig} and \ref{brickmap15fig} confirms that
larger values of $\sigma$ are associated with regions of high
reddening, but Figures~\ref{brickscatter16fig}
and~\ref{brickscatter15fig} suggest this is largely the effect of the
fits being able to move away from the prior when the signal from
reddening is strongest.  However, while the distribution of $\sigma$
broadens in high extinction regions, it is clearly much more likely to
scatter to higher values than lower ones, suggesting that the
distribution of reddenings can become much broader, but rarely becomes
narrower, as seen by the floor at $\sigma\approx0.25$. These high
extinction regions with broader reddening distributions may reflect
differences in the underlying turbulent structure, a breakdown in the
log-normal assumption for the reddening distribution, or an increased
likelihood of multiple gas layers falling in a single analysis pixel
when the column density is high.

\subsection{Uncertainties in the Reddening Parameters} \label{accuracysec}

In the right panels of
Figures~\ref{brickmap16fig}
and~\ref{brickmap15fig}, we plot the spatial distribution of the
uncertainties in $\widetilde{A_V}$, $\sigma$, and $f_{red}$
(i.e. $\Delta\theta = (\theta_{84}-\theta_{16})/2$).  The
uncertainties are clearly not spatially uniform, and instead mimic
structures seen in the distribution of $\widetilde{A_V}$ and
$f_{red}$.  This correspondence reflects the fact that the accuracy of
the derived parameters depends on how cleanly the extincted stars
separate from the unreddened RGB.  Thus, the uncertainties are lowest
when the median extinction $\widetilde{A_V}$ is high (as in Brick 15;
Figure~\ref{brickmap15fig}).

The accuracy of the measured parameters also depends on the number of
reddened stars in the pixel.  In Brick 15
(Figure~\ref{brickmap15fig}), the uncertainties on the extinction are
low in the upper right corner, where the fraction of reddened stars
$f_{red}$ is highest, in spite of the fact that the extinction is not
particularly high in that region.  Brick 16
(Figure~\ref{brickmap16fig}) shows a similar gradient in accuracy, but
at essentially constant $f_{red}$; here, the fall off in accuracy is
due to the lower number of stars per pixel overall, since the stellar
density falls significantly from right to left (not shown).
Uncertainties are also high for edge pixels, which may not be fully
covered by the data. These pixels disappear when we merge the bricks
into a contiguous map, because the edges of the bricks overlap
significantly.  In cases of overlaps, we adopt the fits from whichever
overlapping pixel had measurements with the smallest uncertainties.

\begin{figure*}
\centerline{
\includegraphics[width=3.75in]{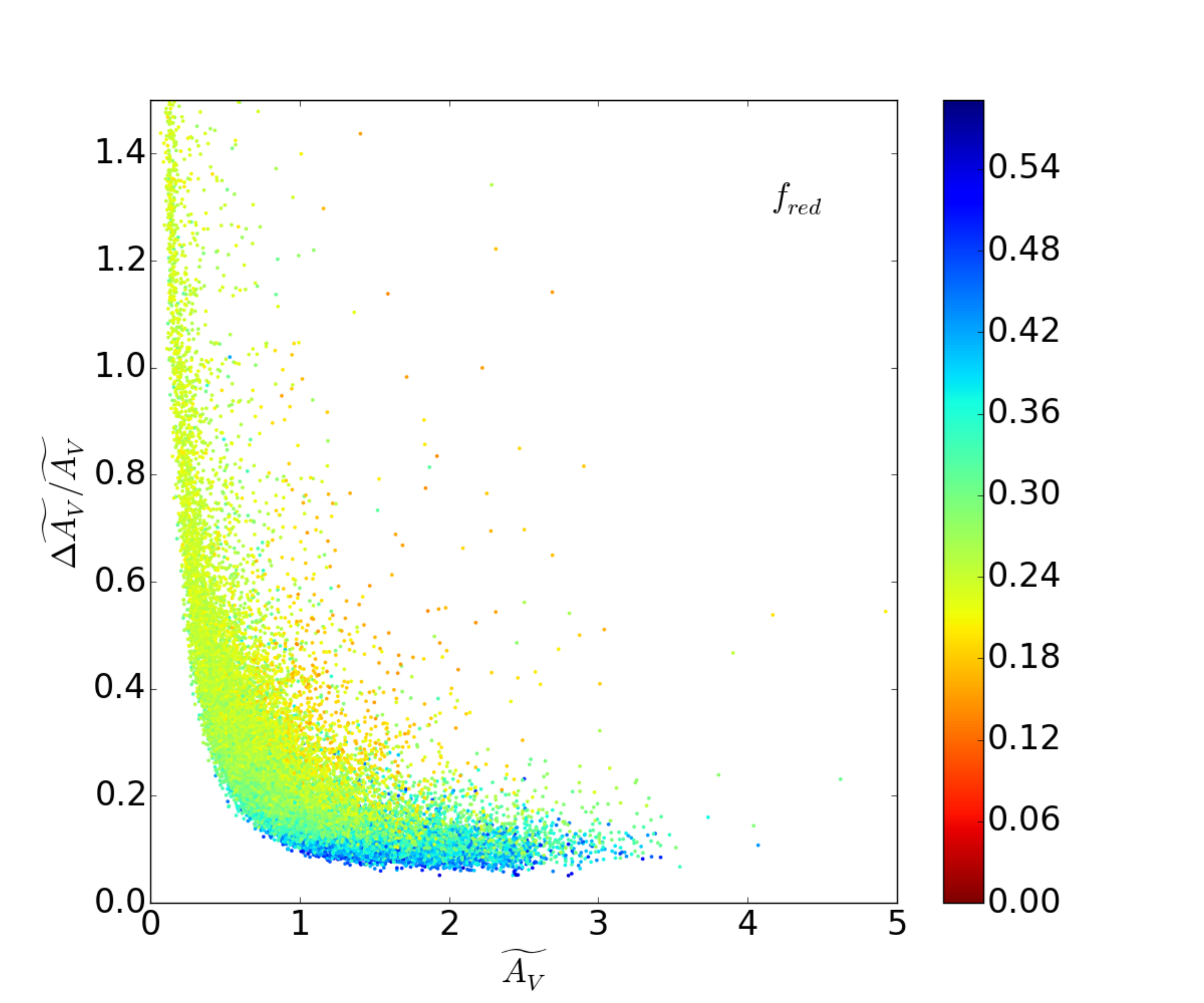}
\includegraphics[width=3.75in]{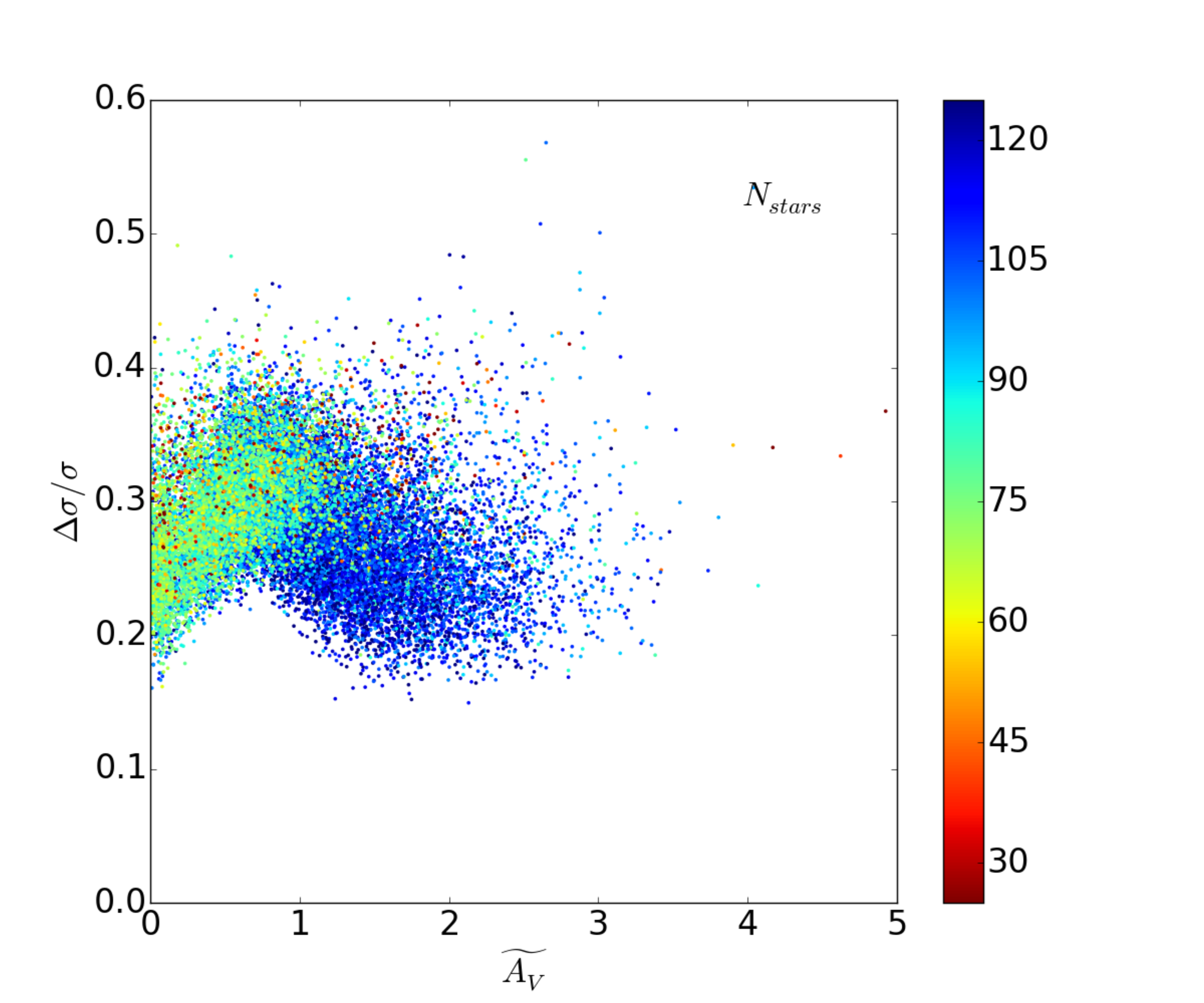}
}
\centerline{
\includegraphics[width=3.75in]{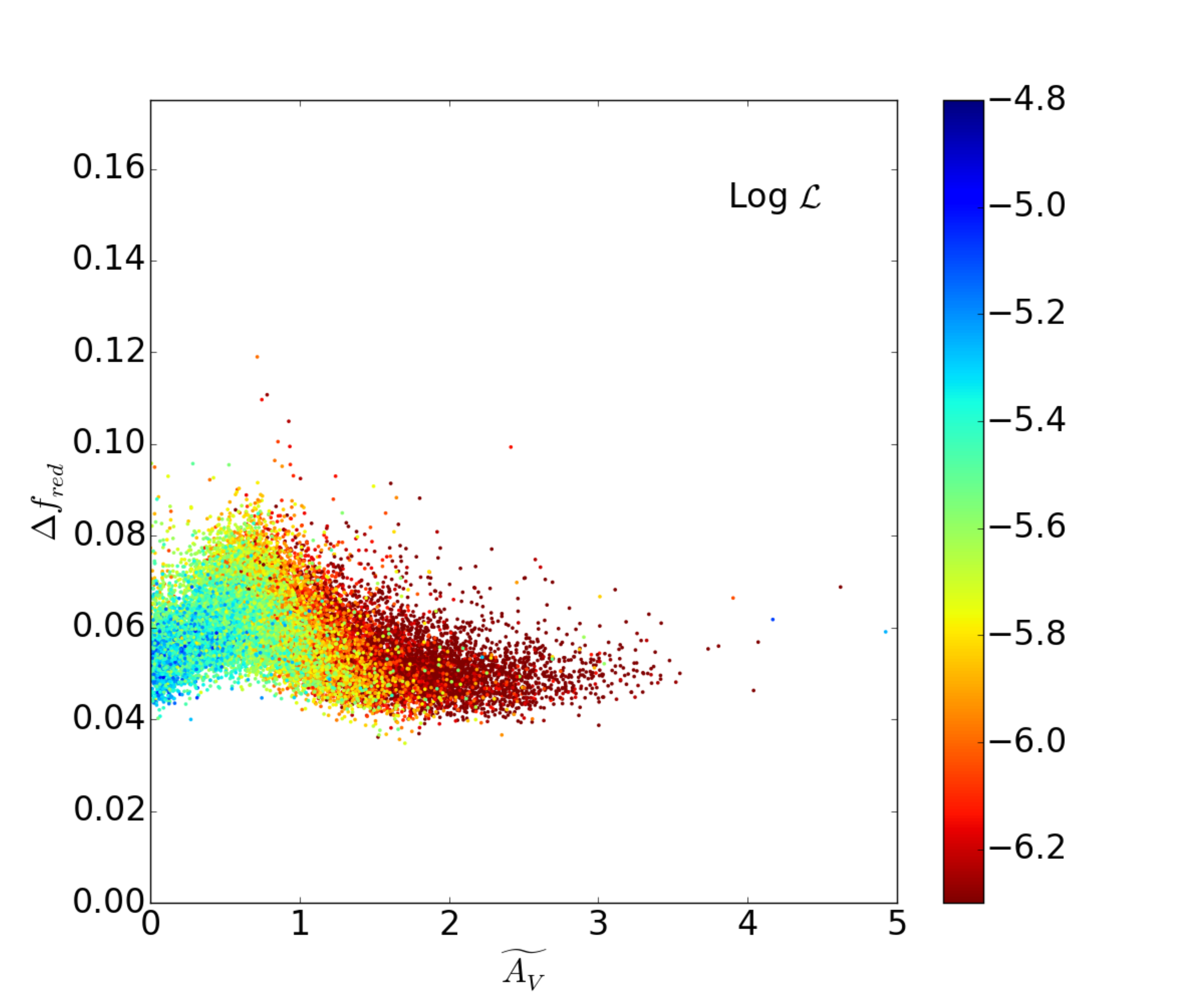}
\includegraphics[width=3.75in]{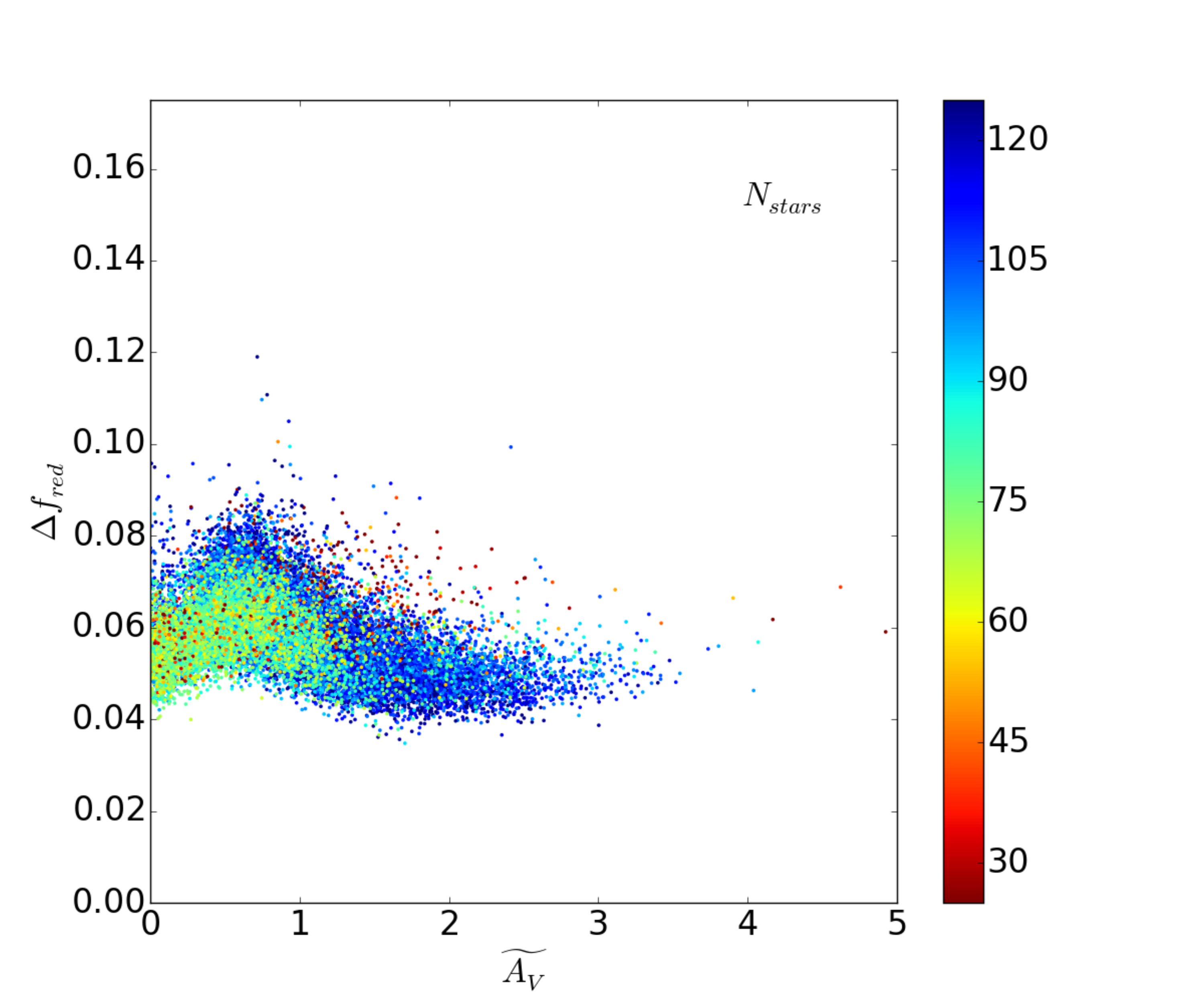}
}
\caption{Uncertainties $\Delta\widetilde{A_V} / \widetilde{A_V}$,
  $\Delta\sigma / \sigma$, and $\Delta f_{red}$ (clockwise from top
  left) as a function of median extinction $\widetilde{A_V}$, for each
  pixel in Brick 16, color-coded by the reddening fraction (upper
  left), number of stars fit per pixel (right panels) and the log
  likelihood of the model fit (lower left).  As expected from visual
  inspection of Figure~\ref{brickmap16fig}, uncertainties increase
  strongly at low extinctions, where the reddened RGB is not cleanly
  separated from the unreddened RGB. In regions of modest or high
  extinction ($>1\mags$), fractional uncertainties are typically lower
  for the median extinction than for the width of the log-normal. The
  log likelihood of the best fit decreases steadily with increasing
  $\widetilde{A_V}$ because as the constraining power of the data
  becomes stronger (as reflected in the smaller uncertainties), any
  deviations from the assumed log-normal distribution become apparent.
  The lower right panel shows $\Delta f_{red}$ again, but now
  color-coded by the number of stars in each pixel. Not unexpectedly,
  at fixed extinction, the uncertainties are higher when the number of
  stars is lower. \label{brickscattererr16fig}}
\end{figure*}

\begin{figure*}
\centerline{
\includegraphics[width=3.75in]{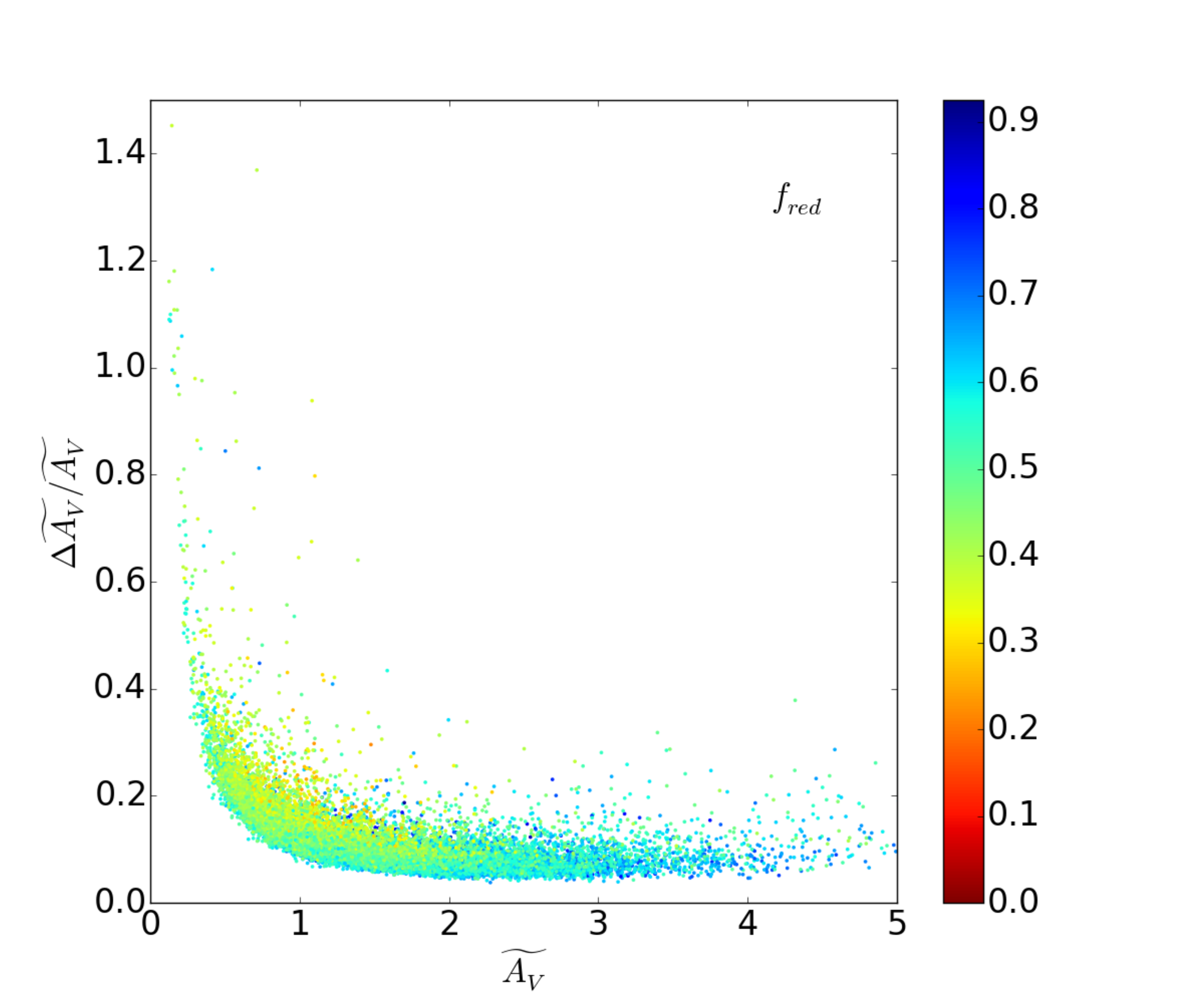}
\includegraphics[width=3.75in]{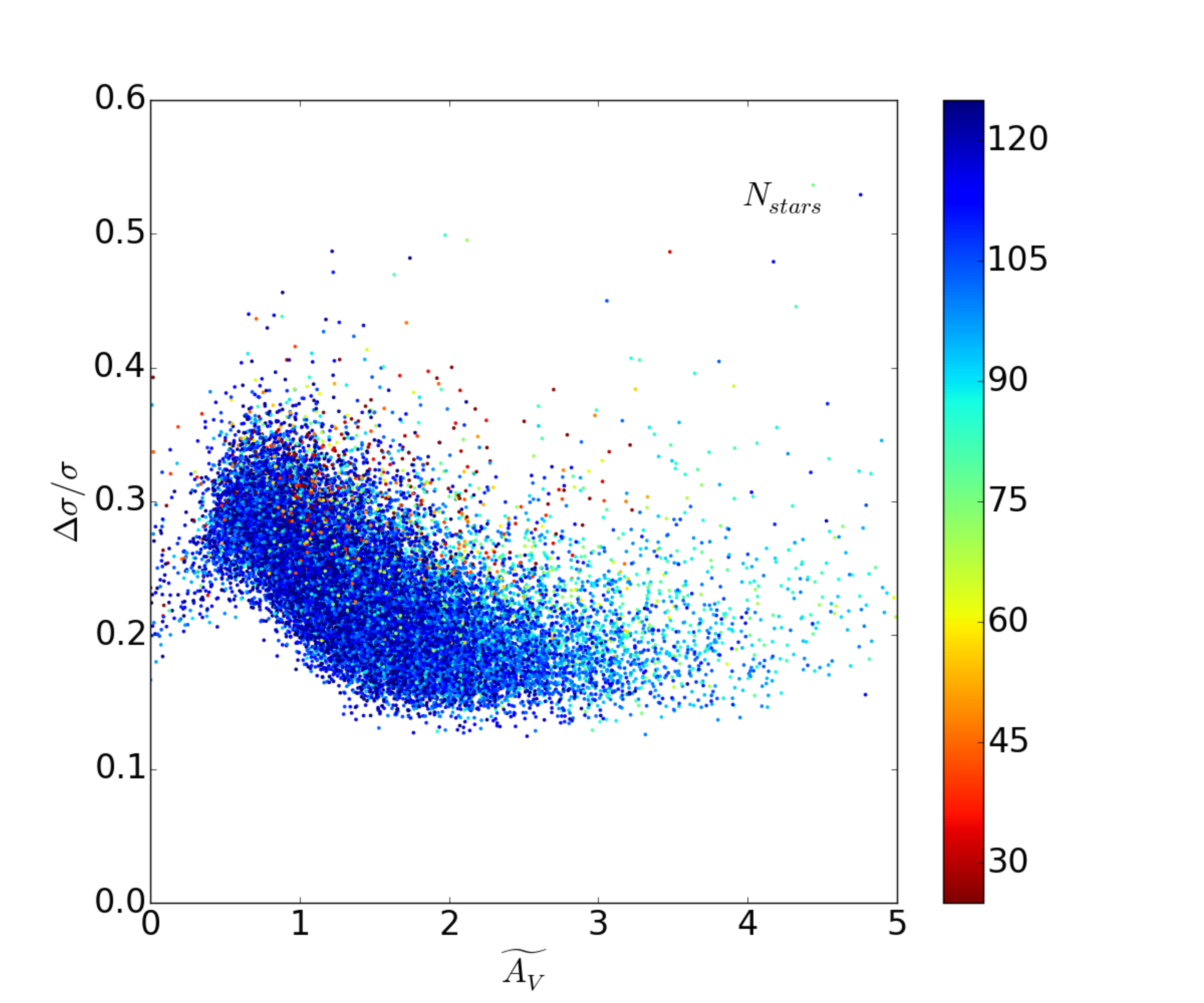}
}
\centerline{
\includegraphics[width=3.75in]{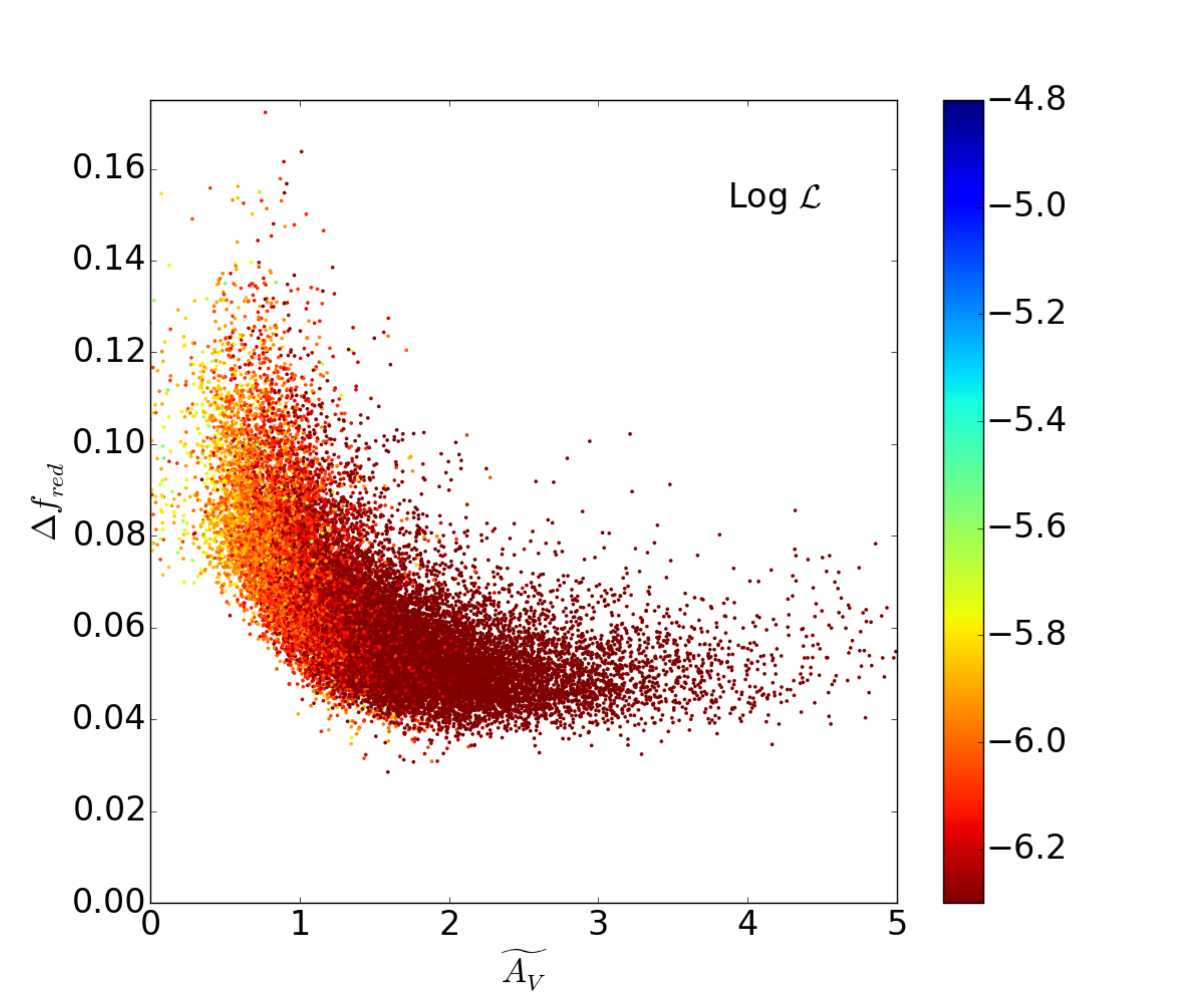}
\includegraphics[width=3.75in]{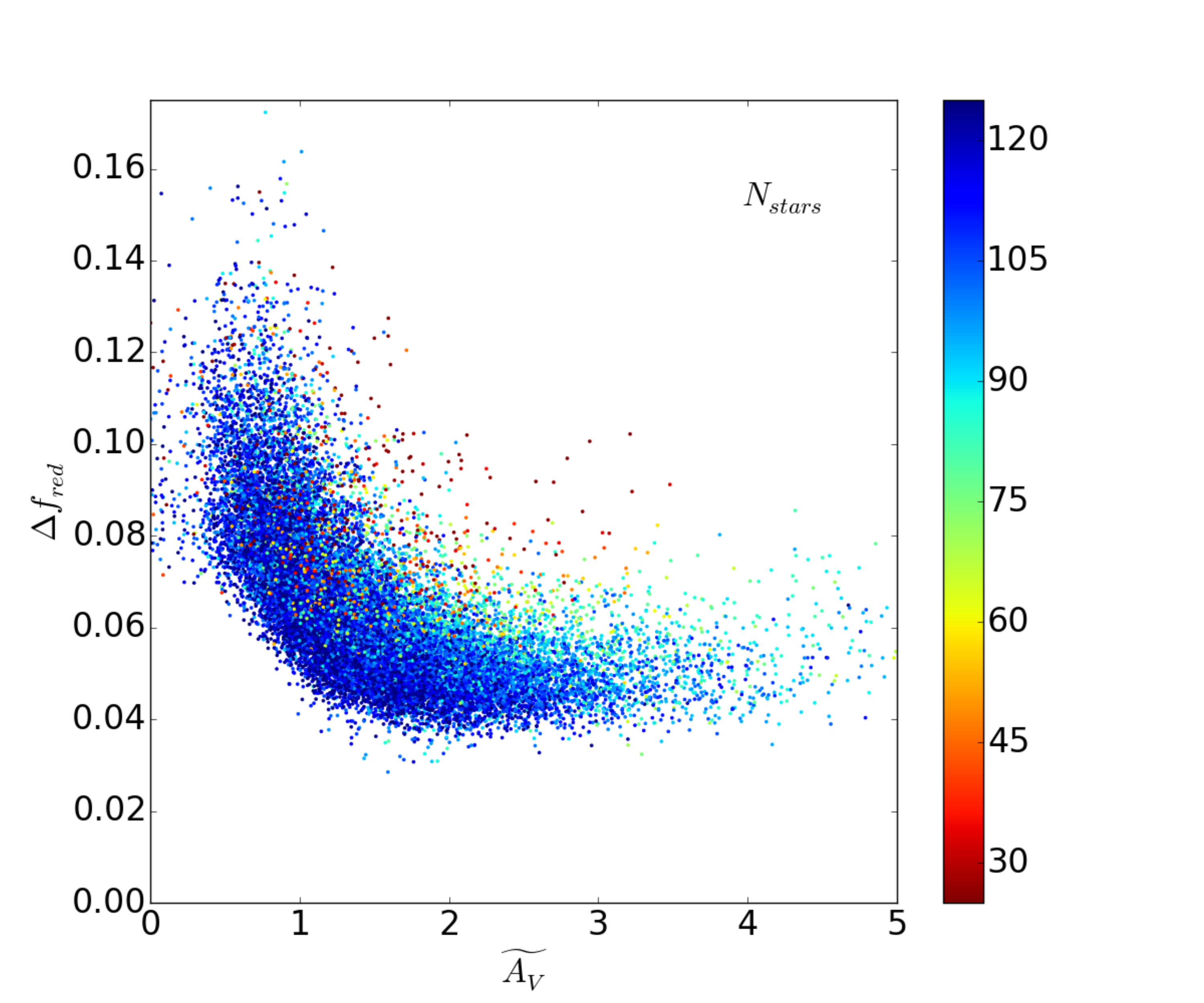}
}
\caption{Same as Figure~\ref{brickscattererr16fig}, but for the high
  extinction star-forming region in Brick 15. Uncertainties
  $\Delta\widetilde{A_V} / \widetilde{A_V}$, $\Delta\sigma / \sigma$,
  and $\Delta f_{red}$ (clockwise from top left) are as a function of
  median extinction $\widetilde{A_V}$, for each pixel in Brick 15,
  color-coded by the parameter indicated in the upper right of each
  panel.  Given the overall high extinction in this region, the
  associated uncertainties on the parameters are much smaller than in
  Figure~\ref{brickscattererr16fig}
\label{brickscattererr15fig}}
\end{figure*}

Because the accuracy depends most strongly on
the extinction of the reddened component, in
Figures~\ref{brickscattererr16fig} and~\ref{brickscattererr15fig} we
plot the fractional uncertainties of
$\Delta\widetilde{A_V}/\widetilde{A_V}$ and $\Delta\sigma / \sigma$,
and the absolute uncertainty $\Delta f_{red}$ as a function of
$\widetilde{A_V}$, for all pixels in Bricks 16 and 15 (whose parameter
distributions were shown in Figures~\ref{brickscatter16fig}
and~\ref{brickscatter15fig}).  

As expected, there is a clear trend towards better precision at higher
extinctions.  For large $\widetilde{A_V}$, the reddened distribution of
RGB stars becomes completely distinct from the unreddened RGB in the
CMD, and the distribution typically becomes broader, leading to
stronger constraints on the reddening distribution.  At high
extinctions, the typical uncertainties are smaller than 20\% in
$\widetilde{A_V}$, and smaller than $\pm0.1$ in $f_{red}$.
The uncertainty in $\sigma$ is comparatively high, being rarely
smaller than 20\%.  As with $\widetilde{A_V}$, the fractional
uncertainty in $\sigma$ drops somewhat at high extinction, because of
the better separation of the unreddened and reddened RGB
sequences.

There are rare outliers at high $\widetilde{A_V}$, where the
fractional uncertainties are much larger than typical. These outliers
usually correspond to cases where the posterior probability
distribution on $\widetilde{A_V}$ is very broad due to a very small
number of reddened stars, as may occur when the total number of stars
is small, when the reddening fraction is very low, and/or when there
is a region with bad photometry (typically due to sources detected
around saturated stars).

At lower extinctions ($\widetilde{A_V}\!\approx\!1\mags$), the
uncertainties in $A_V$ are more typically 20-30\% in regions where
$f_{red}>0.3$, and reach 50\% by $\widetilde{A_V}\lesssim
0.3\mags$. The typical uncertainties are larger in regions with even
lower reddening fractions, due to the smaller number of stars in the
reddened peak. These regions also have more spurious high
$\widetilde{A_V}$ points with large uncertainties, because the
presence of very few red stars can appear to mimic high extinction
when the prior favors small values of $f_{red}$.  The uncertainties in
$f_{red}$ and $\sigma$ decrease towards low extinction, because of the
narrowing of the prior rathan than because of increasingly
constraining data.

A comparison of Figures~\ref{brickscattererr16fig}
and~\ref{brickscattererr15fig} also shows the effect of $f_{red}$ on
the uncertainties.  The reddened fraction in Brick 16
(Figure~\ref{brickscattererr16fig}) is much lower than the typical
reddening fraction in Brick 15 (Figure~\ref{brickscattererr15fig}),
leading to lower precision at the same median extinction.  At
$\widetilde{A_V}=2\mags$, the fractional accuracy in the extinction is
$\sim12\%$ in Brick 16, whereas it is better than 6\% in Brick 15,
which has a reddening fraction that is nearly twice as high (see
Figures~\ref{brickscatter16fig} and~\ref{brickscatter15fig}).  This
difference reflects the larger number of stars in the reddened peak
when $f_{red}$ is large.

To better show the impact of the stellar density on the uncertainties,
the right panels of Figures~\ref{brickscattererr16fig}
and~\ref{brickscattererr15fig} show $\Delta f_{red}$ and $\Delta
\sigma/\sigma$ as a function of the median extinction, color coded by
the number of stars in the pixel.  There is a clear tendency for
pixels with fewer stars to have broader posterior distributions and
thus larger uncertainties.  For a fixed pixel scale, this trend
indicates that the uncertainties will be larger in the outer disk,
where the stellar surface density is low.  This trend also suggests
that better spatial resolution could be achieved in regions with high
reddening fractions, where we can reduce the number of stars per pixel
in exchange for increased resolution, while still preserving
scientifically useful results. We have not chosen to adopt
spatially-variable resolution elements here, to allow easier
comparisons to global emission maps of M31's ISM.  However, we will be
pursuing higher resolution analyses of specific high-extinction
subregions in future work.

Finally, we stress that the above uncertainties are random, and not
systematic.  However, in the low extinction regime where systematic
biases could potentially become important, the random uncertainties
are large.  Thus, while we expect biases at low extinctions (e.g.,
Sec.\ \ref{limitationsec}), we also know not to over-interpret these
extinctions because of their significant random uncertainties.

\clearpage

\subsubsection{Failures of Model Fitting}
In addition to the normal scatter due to random uncertainties, there
are individual pixels where the model fitting has obviously failed.
These cases are most noticeable in low extinction regions, where we
find occasional pixels that have extremely high values of
$\widetilde{A_V}$, in a region where there is no other evidence for
significant dust.  There are two major failure modes we have
identified in these regions.

\paragraph{Spurious Photometry \& the Noise Model} The first of these
cases corresponds to failures of our adopted noise model.  If an
individual pixel has unusually high contamination from bad photometry,
then there will be a larger than expected number of spurious
detections. When these sources are not well-represented by our noise
model the fitting code will adjust the log-normal parameters to
increase the likelihood of finding stars far redward of the RGB, where
such spurious sources typically lie (i.e., close to the photometric
limit). As a result, the best fit value of $\widetilde{A_V}$ and/or
$\sigma$ will be driven to large values in an attempt to produce an
extremely broad log-normal component, increasing the probability of
finding very red sources. This failure represents a complete breakdown of our
model.  The effect is especially pernicious in regions with low
reddening fractions, where one expects few stars redward of the
unreddened RGB. It is further compounded in the outer disk, where the
absolute number of stars per pixel is small. Luckily, however, these
same failure modes are also marked by much higher than average
uncertainties in $\widetilde{A_V}$, allowing them to be easily
identified.  An adaptive pixel size would help alleviate this issue,
at the expense of complicating comparisons with other dust tracers by
introducing spatial variations in the resolution.

Although the noise-model failures sometimes affect only a single
isolated pixel, they also sometimes affect multiple adjacent
pixels (i.e., spanning $\gtrsim13\arcsec$). The most common reasons
for such highly clustered spurious sources are (1) improper rejection
of sources on diffraction spikes around bright stars; and (2)
subregions of background galaxies.  To minimize the impact of these
bad fits in the following analysis, we use unsharp masking to identify
pixels that are wildly different from their neighbors, and then
replace those pixels with the local median of the neighboring pixels.
Our rejection threshold is extremely conservative, and therefore some
problem areas clearly remain even in the cleaned maps.  All subsequent
analysis will use these ``cleaned'' maps.

\paragraph{WFC3 Calibration Issues} The second failure of the models
in low extinction regions is due to the WFC3/IR photometric
calibration errors discussed in Sec.\ \ref{nirdatasec}.  Inspection of
the right hand plot in Figure~\ref{RGBwidthcolormapfig} shows that
there are repeating structures in the map of mean RGB color, even
after applying the first order correction in Sec.\ \ref{nirdatasec}.
These structures follow a square pattern that coincides with the
locations of the individual WFC3/IR frames that make up the survey
data.  This pattern suggests that some of the width of the RGB is due
to stars in one part of the detector consistently appearing redder or
bluer than stars in another area. As discussed in
Sec.\ \ref{nirdatasec}, such an offset could be due to different
accuracies in the flat fields used for the F110W and/or the F160W
filters, or (more likely) due to systematic differences in the
accuracy of the spatially-varying model point spread functions used
for photometry. Attempts to track down the exact calibration issue are
on-going. We have reduced the effect of these structures on the
eventual extinction map by introducing a fitting parameter $\delta_c$
to allow very small shifts in the color, but the correction is noisy,
particularly when the fraction of unreddened stars is small.  Thus,
while it helps, residual issues remain.

The pattern is created because our procedure for creating a model of
the unreddened stars identifies stars in regions where the RGB is
narrow and blue. Regions where the RGB is redder and/or broadened by
systematic errors in the photometry will therefore be best fit by
models with higher extinction. When these systematically redder and
bluer regions are consistently found in certain parts of the WFC3/IR
footprint, they imprint a pattern of apparently higher than normal
extinction on the resulting map. However, because these photometric
issues are at the level $<\pm0.02\mags$ nearly everywhere, their
effects are only noticeable when the extinction is small. We can
estimate the amplitude of this effect on the $A_V$ map by considering
the most extreme case, where all the stars in the model unreddened CMD
would come from portions of the WFC3/IR footprint with systematically
blue colors. In this case, unreddened regions that lie in the
systematically redder parts of the WFC3/IR footprint would by redder
by $\lesssim0.04\mags$, corresponding to an apparent, spurious
extinction of $A_V\lesssim 0.3\mags$. This amplitude is comparable to
the typical random errors at low extinction
(Figures~\ref{brickscattererr16fig} and~\ref{brickscattererr15fig}).
We expect these features to eventually be eliminated in the final
version of the PHAT photometry, which will use spatially-variable PSF
models in the NIR.

\subsubsection{Possible Systematic Uncertainties} \label{systematicsec}

The uncertainties derived above are tied specifically to our chosen
model. If this model is incorrect, systematic
uncertainties can be introduced into our analysis without being
reflected in the uncertainties derived from the MCMC fit.  We have
already addressed several of the systematics inherent in constructing
a model of the unreddened CMD (Sec.\ \ref{limitationsec}; i.e.,
failure to identify true zero reddening stars, biases from population
gradients and projection effects, foreground reddening, etc), but
address a few additional points here. 

\smallskip
\centerline{\it{Variations in the Reddening Law} }
\smallskip
We have assumed that there is a single conversion between reddening
and extinction that applies throughout M31 (see Sec.\ \ref{cloudsec}).
This assumption may not hold if there are local departures from a
standard $R_V=3.1$ extinction law.  In such cases we will convert the
observed reddening into the wrong extinction. However, the conversion
in the NIR is relatively insensitive to the value of $R_V$.  Even an
extreme value of $R_V=5$ leads to an inferred extinction that is only
15\% different than in the $R_V=3.1$ case, which is less than our
typical precision.  We therefore do not anticipate this to be a
significant source of error in the extinction map.

\smallskip
\centerline{\it{The Assumption of a Log-Normal $A_V$ Distribution} }
\smallskip
Our model assumes that the distribution of extinction in each
spatial pixel can be approximated as a single log-normal.  If 
the true distribution is more complex, then the
specific values of parameters are harder to interpret and their
uncertainties may not be correct.

The log-normal appears to be an adequate representation for the
majority of sightlines in typical Milky Way molecular clouds and the
diffuse ISM, but the densest star-forming clouds are usually found to
have an additional power-law at high-extinctions \citep[e.g.,][; see
  the more extensive discussion in
  Sec.\ \ref{cloudsec}]{kainulainen2009}.  In practice,
deviations from log-normal distributions are not a large concern,
given that the power-law tail is only found in the most actively star
forming clouds, and contains only a small fraction of the cloud mass
when present.  The highest column densities occupy a very small volume
of any molecular cloud, and thus random sampling of the column density
distribution using background RGB stars will tend to miss any features
that occupy only a negligible fraction of the pixel area.  The only
notable effect may be a tendency of fits to favor a somewhat broader
distribution (i.e., larger $\sigma$), particularly in regions with
large median extinctions. Inspection of
Figures~\ref{brickscatter16fig} and \ref{brickscatter15fig} suggest
that this combination of parameter values is not common.

Another violation of the log-normal assumption may occur when the gas
distribution is sufficiently complex that there are multiple gas
layers along the line of sight.  For example, two molecular clouds
that are spatially coincident when viewed in projection could have
different densities and depths along the line of sight, and thus would
produce a more complex reddening distribution than we have assumed.
However, even when multiple gas overdensities are superimposed along
the line of sight, they are unlikely to all contribute equally to the
column density distribution, such that the total reddening
distribution is likely to be dominated by the subcomponent with the
highest column density with the largest fraction of stars behind it,
potentially leaving the log-normal as an adequate fit.

As a potential test of the suitability of the log-normal model, we have
color-coded the values of the uncertainties by the log of the
posterior probability of the best fit model in
Figures~\ref{brickscattererr16fig} and~\ref{brickscattererr15fig}.
There is an obvious trend such that the log-normal model becomes
systematically less likely at higher extinction (although only by
$\sim1.5$ dex).  This behavior is not alarming, because our data
clearly have less constraining power at low extinctions; when the
ability to constrain parameters is low, making it impossible
to rule against a given adopted model. Any model is therefore acceptable in
terms of the posterior probability of the best fit.  While we could
engage in more exhaustive testing of whether a log-normal is the
optimal functional form, we proceed with an understanding that the
log-normal parameters are best interpreted as a good, but approximate,
first-order characterization of the distribution of column densities
within the pixel.

\smallskip
\centerline{\it{Biases from Reddened Young Stars} }
\smallskip

Our model assumes that all of the stars that we fit in the CMD are
reddened RGB stars.  However, in regions of significant reddening,
some of the younger stars that are blueward of our analysis region can
be reddened into the portion of the CMD that we are analyzing.  These
younger stars are typically found in high extinction regions
\citep[e.g.,][]{harris1997,zaritsky2002,zaritsky2004}, making this
mechanism a possible source of bias.

Unfortunately, the amplitude of the bias is difficult to quantify
because the young stellar populations have much more dramatic spatial
variations \citep{lewis2015} than the older RGB stars, which have been
well-mixed over many dynamical times ($\sim250\Myrs$ in M31's main
star-forming ring).  Thus, the exact degree to which highly reddenened
young stars contaminate the RGB analysis region will depend on exactly
how numerous the young stars are relative to RGB stars.  The degree of
contamination will also depend on the ages of the young stars, since
the core Helium-burning sequences (roughly 25--$400\Myr$) are redder
than the main sequence, and thus can more easily be reddened into our
analysis regions. Finally, the degree of contamination will also
depend on the depth of the observations.  At brighter magnitudes, core
Helium-burning stars are well-separated from the RGB, but at fainter
magnitudes, they merge into the red clump where their colors are
indistinguishable from the main population of RGB stars. In this
latter case, reddening of these stars would produce no bias, because
they would be included in the unreddened model.

Although the exact amount of bias is difficult to compute, we can make
a first order estimate by calculating the colors of stars at different
ages, and their relative numbers compared to the RGB. We adopt
photometric errors comparable to those in PHAT, and assume a constant
star formation rate.  Overall, this choice will tend to overestimate
the contribution of young stars, given that M31's star formation rate has
declined since a burst 2--$3\Gyr$ ago \citep{williams2015}. However,
Figure~13 of \citet{lewis2015} shows that there are some star-forming
regions that have recent star formation rates (averaged over $100\Myr$
and $\sim\!0.4\Gyr$ timescales) that are of the order of the past average
star formation rate or higher. In these regions, the potential bias will be
somewhat higher than we estimate here.

\begin{figure}
\centerline{
\includegraphics[width=3.5in]{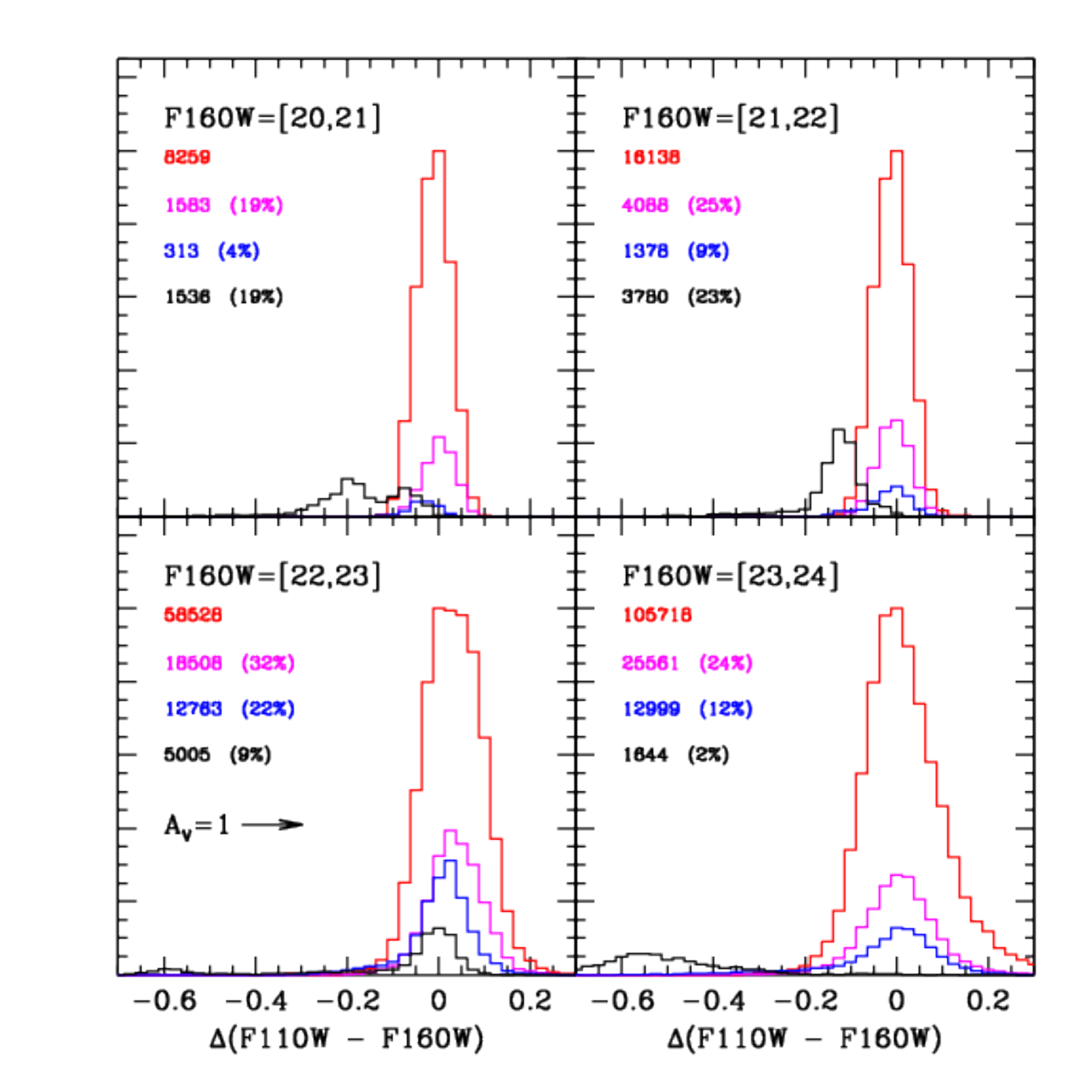}
}
\caption{The distribution of colors for stars in 4 different age
  ranges (100--$500\Myr$, $500\Myr$--$1\Gyr$, $1$--$2\Gyr$, and
  $>\!2\Gyr$; black, blue, magenta, and red, respectively), assuming a
  modest degree of metallicity evolution ([Fe/H]$=\!0$ for the two most
  recent bins, -0.25 for the 1--$2\Gyr$ bin, and $-$0.75 in the oldest
  age bin), for a simulated CMD assuming a constant star formation
  history and typical PHAT photometric errors. Each panel shows the
  distribution of colors within a $1\mags$ interval of F160W
  magnitude. To assess the possible contribution of reddened young
  stars to the RGB-dominated portion of the CMD, all colors are given
  relative to the peak of the stars in the oldest age range; the arrow
  in the bottom left panel shows the reddenening expected for
  $A_V=1\mags$ of extinction. The numbers listed in each panel give the
  number of stars in each histogram; percentages are given relative to
  the number of stars in the oldest peak. Stars older than
  $\sim\!500\Myr$ have NIR colors that are indistinguishable from the
  main RGB peak. Stars in the youngest bin can have bluer colors at young
  ages, but are not numerous and require large reddening before they
  could be mistaken as RGB stars with modest reddening.
  \label{agecolorfig}}
\end{figure}

In Figure~\ref{agecolorfig} we plot the resulting distribution of
colors for 4 different age ranges (100--$500\Myr$, $500\Myr$--$1\Gyr$,
$1$--$2\Gyr$, and $>\!2\Gyr$), assuming a modest degree of metallicity
evolution ([Fe/H] = 0 for the two most recent bins, -0.25 for the
1--$2\Gyr$ bin, and $-$0.75 in the oldest age bin); the first age
ranges encompass the ages probed in \citet{lewis2015}'s spatially
resolved maps of M31's recent star formation history.  We do not include
results for ages younger than $100\Myr$, since these stars are too few
and too blue to make any significant contribution. We plot the
distributions relative to the color of the oldest stellar populations
(i.e., relative to the peak of the RGB), for four different magnitude
ranges.  Each plot lists the numbers of stars redder
than $F110W-F160W=0.2$, both as an absolute number and as a percentage
of the stars in the oldest RGB peak; these numbers give some sense of
the relative size of the young population that might potentially
contaminate the RGB when reddened.

Figure~\ref{agecolorfig} suggests that the impact of contamination
from young reddened stars is likely to be small.  Stellar populations
older than $\sim\!0.5\Gyr$ have colors that are indistinguishable from
the RGB. While their presence or absence may affect the total number of
stars in a given pixel, our model fits only relative distribution of
stars on the CMD, and is thus insensitive to their presence.
Stars in the 100--$500\Myr$ age range will likewise have no impact at
faint magnitudes (F160W$>22$) where their fractional contribution is
small ($\le5$\% of the number of RGB stars), and their colors are
either indistinguishable from the RGB ($22<F160W<23$), or too blue to
make any contribution except at the very highest extinctions
($23<F160W<24$).

At brighter magnitudes ($22<$F160W), the 100--$500\Myr$ stars are
proportionally more important. They contain up to 20-25\% of the
number of stars found in the RGB (assuming a constant star formation
rate), and could be reddenened to the color of the RGB peak with
1--$1.5\mags$ of visual extinction.  If the young stars experienced
this amount of extinction, they would merge with the peak of
unreddened RGB stars. The model fits would then assume that the
fraction of reddened stars was smaller than it actually was. To affect
the measurement of $A_V$, however, the stars would need to experience
even more reddening, such that they moved redwards of the RGB peak.
In this case, the presence of the stars would tend to bias the
measurement of $A_V$ to lower values.  However, given that (1) RGB
stars would still outnumber the young stars 3:1 at bright magnitudes;
(2) that the fainter stars where the bias is low are factors of 4-6
times more numerous; and (3) that the inclusion of the noise model can
potentially ``absorb'' some fraction of stars that appear abberant
compared to the pure RGB fit, it is not clear by how much the best fit
model would be biased in practice.  We therefore conclude that biases
due to reddening from young $<0.5\Gyr$ old stars is likely to be
negligible in most cases, and would only potentially be detectable at
the highest extinctions ($A_V>2$), at anomalously high SFRs in the
100--$500\Myr$ age range, and where the data was shallowest (i.e.,
in the inner disk).

\smallskip
\centerline{\it{Loss of Stars at High $A_V$} }
\smallskip

Another possible source of bias would be if highly reddened stars are
systematically missing from our data in high extinction regions, due
to their suffering enough extinction that they fall below our
detection limit.  Given the range of extinction found in our analysis,
rarely is the extinction high enough to completely remove large
numbers of moderately bright RGB stars (e.g.,
Figure~\ref{unreddenedCMDfig}).  This is not to say that no individual
stars have been reddened out of the sample. Intrinsically faint RGB
stars are certainly ``missing'', and undoubtedly occasional bright RGB
stars may be extinguished as well if they happen to fall behind the
very smallest, densest core of a molecular cloud.  However, it appears
highly unlikely that either effect is not being accounted for in the
model fitting, as long as the mode of the log-normal distribution is
detectable for some of the RGB.

\smallskip
\centerline{\it{Summary of Systematic Biases} }
\smallskip

All of the potential systematic biases identified above and in
Sec.\ \ref{limitationsec} are modest ($A_V\lesssim\!0.3\mags$) or
confined to very rare regions. Systematic errors are therefore likely
to only be noticeable at low extinctions. However, the random errors
at these low extinctions are large, so that the systematic errors may
not actually dominate the error budget for any individual
pixel. However, the effects of systematics can be noticeable when
looking at the cumulative behavior of the extinction over large, low
extinction regions of the galaxy, which essentially averages over many
pixels and thus reduces the impact of random uncertainties.  We look
for evidence of these effects in Sec.\ \ref{spatialsystematicssec}
below, after presenting our final dust maps and comparing them with
other measurements of dust extinction.

\section{Results} \label{resultssec}

\subsection{A Galaxy Wide Map of M31's Cool ISM} \label{globalmapsec}

\begin{figure*}
\centerline{
\includegraphics[width=7.25in]{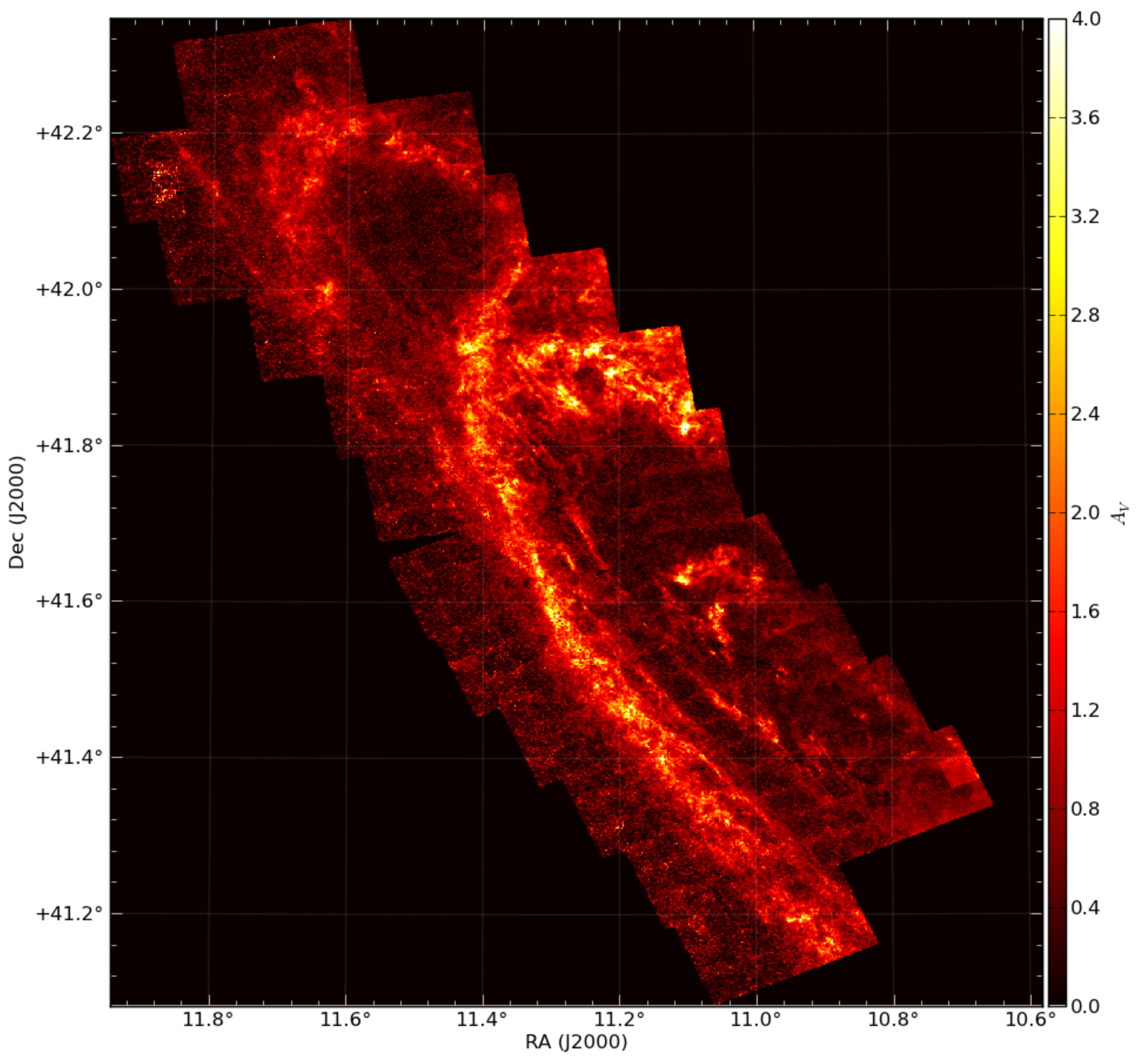}
}
\caption{Map of the median extinction $\widetilde{A_V}$ for all analyzed 
regions. Some very
low-level ``grid'' patterns are visible in the map, due to
position-dependent calibration issues with the WFC3/IR chip
(Sections~\ref{nirdatasec}~\&~\ref{accuracysec}).
\label{AVmapfig}}
\end{figure*}

\begin{figure*}
\centerline{
\includegraphics[width=7.25in]{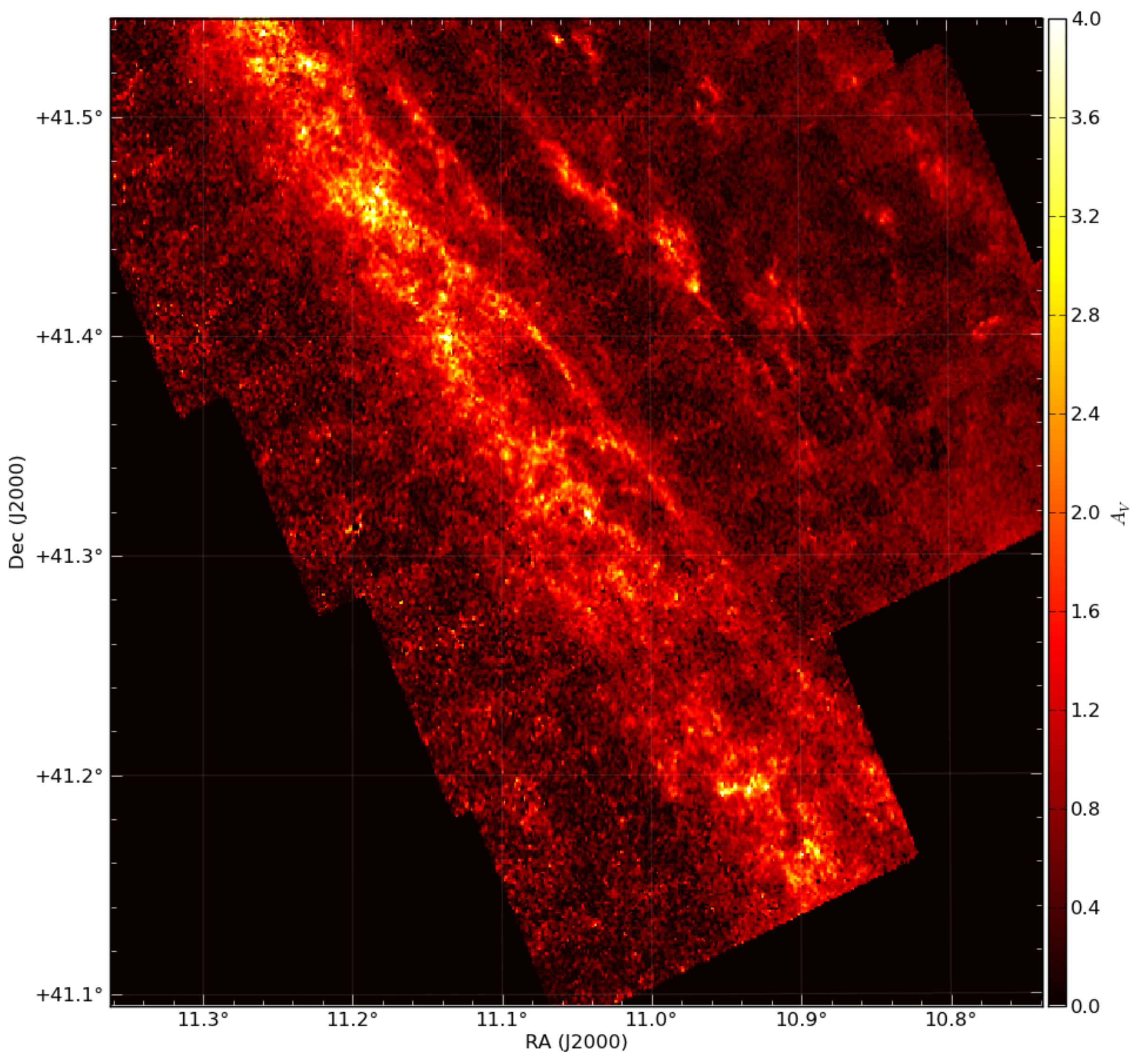}
}
\caption{Map of the median extinction $\widetilde{A_V}$ for 
the inner regions containing Bricks 2, 4, 6, and 8 along
the $10\kpc$ star-forming ring (bottom to top), and the major axis field
Brick 5, which samples a smaller, less intense star-forming ring. Note the
regular ripples or spurs along the inside edge of the star-forming ring.
\label{AVmapregion1fig}}
\end{figure*}

\begin{figure*}
\centerline{
\includegraphics[width=7.25in]{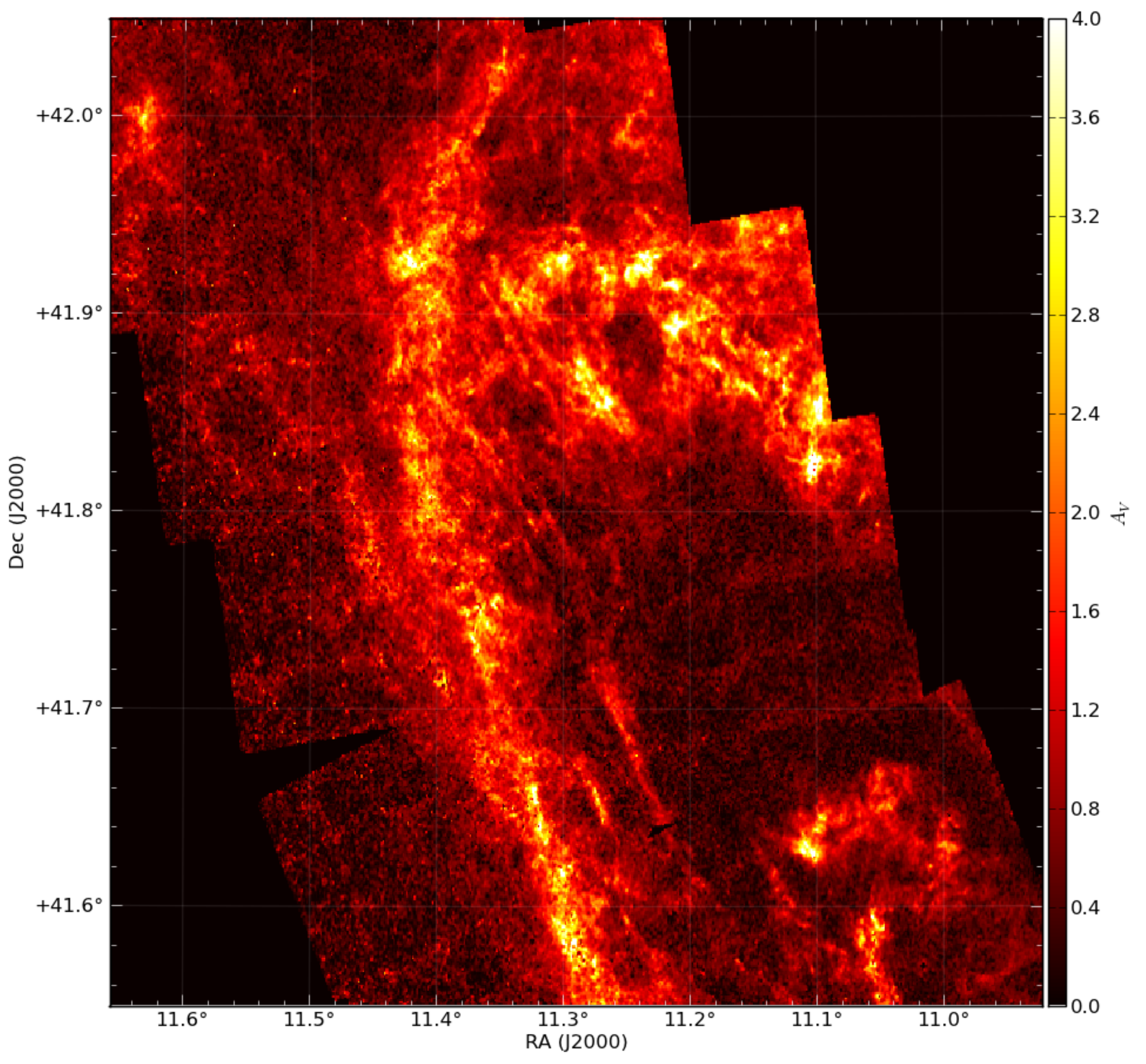}
}
\caption{Map of the median extinction $\widetilde{A_V}$ for: (1) the
major axis of the the $10\kpc$ star-forming ring (Bricks 12, 14, and
16 along the left, and Bricks 15 and 17 along the major axis, from
bottom to top); (2) the major axis field Brick 9, which samples the
second of the 2 inner star-forming rings; and (3) the star of the
outermost star-forming ring (Brick 18, in the upper left).  The
ripples/spurs noted in Figure~\ref{AVmapregion1fig} continue into
Brick 12 on this map.
\label{AVmapregion2fig}}
\end{figure*}

\begin{figure*}
\centerline{
\includegraphics[width=7.25in]{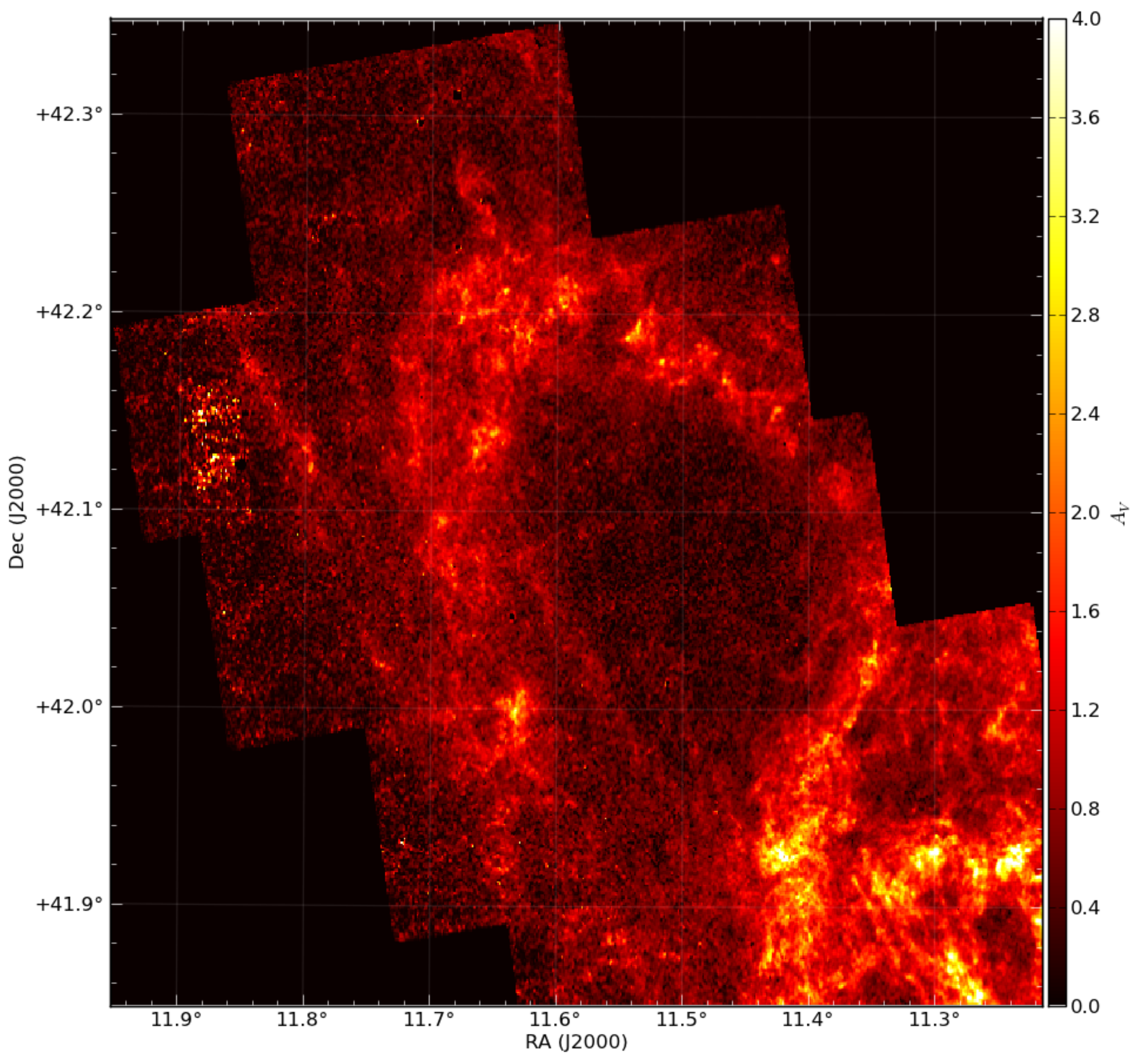}
}
\caption{Map of the median extinction $\widetilde{A_V}$ in the outer
disk, including Bricks 17, 19, 21, and 23 along the major axis,
and Bricks 18 and 22 on the left of the image. Two of the fields in
Brick 22 (near $\alpha=11.86^\circ$, $\delta=42.14^\circ$) have somewhat
higher-than-expected photometry errors, which leads to a large fraction
of erroneous pixels in the dust map.
\label{AVmapregion3fig}}
\end{figure*}

In Figure~\ref{AVmapfig} we present the full maps of M31's median
extinction $\widetilde{A_V}$.
Figures~\ref{AVmapregion1fig}--\ref{AVmapregion3fig} show a series of
zoomed in regions of the same maps, from the inner to the outer disk
(i.e., moving from bottom to top in Figures~\ref{AVmapfig}).  All
images show the full interleaved map (12.5$\pc$ pixel sampling with
25$\pc$ resolution).  We present only the cleaned maps; some remaining
spurious features due to high numbers of false detections near bright
stars are still visible, particularly in the outer
disk\footnote{ There are also some slight problems visible in Field 8
  of Brick 22 \citep[for nomenclature see][]{dalcanton2012} that can
  be traced to unresolved photometric issues. Thankfully, these are in
  very low extinction regions and thus the scientific impact of these
  problems is low. There is also an issue in Field 12 of Brick 3,
  wherein the photometry is systematically too red.}.  There are also
very low level ``grid'' structures due to the $\pm0.01\mags$
structured systematics in the photometry across the WFC3/IR chip (as
discussed in Sec.\ \ref{nirdatasec} and \ref{systematicsec}).

The maps in Figures~\ref{AVmapfig}---\ref{AVmapregion3fig} reveal
extremely rich details. Assuming that the extinction is proportional
to the column density of cold gas, the maps of $\widetilde{A_V}$ offer
one of the highest resolution views of the ISM of a spiral disk,
outside our own Milky Way. The structure of the extinction shows
networks of filaments
(Figures~\ref{AVmapregion2fig}~\&~\ref{AVmapregion3fig}), holes
(Figure~\ref{AVmapregion2fig}), ripples/feathers
(Figure~\ref{AVmapregion1fig}; seen as repeating hooked structures on the inside edge of major
star-forming ring), and spurs (Figure~\ref{AVmapregion3fig}; linear
features extending radially from the major star-forming rings and
spiral arms).  These latter two features are thought to be direct
results of hydrodynamic instabilities
\citep[e.g.,][]{kim2002,chakrabarti2003, shetty2006,kim2006,
  dobbs2006,lee2012}, making the maps a superb resource for
understanding these common features of spiral disks
\citep[e.g.,][]{elmegreen1980,lavigne2006}. For example, the regular
pattern of spurs evident in the inside edge of M31's star-forming ring
(Figure~\ref{AVmapregion1fig}) has a particularly small opening angle,
and quickly bends around to form almost a complete inner ring. In
contrast, simulations of spur formation typically find much more open
patterns of spurs than are seen here.

Inspection of Figure~\ref{AVmapfig} shows that typical extinctions are
of order $\widetilde{A_V}\!\sim\!1$, with very little surface area
having $\widetilde{A_V}\!>\!3$.  If we limit the map to only those
regions with well-measured values of $\widetilde{A_V}$ (fractional
errors $\Delta \widetilde{A_V} / \widetilde{A_V}\! \lesssim \!0.3$),
the median of $\widetilde{A_V}$ is $1.0\mags$, with only 10\% of the
pixels having $\widetilde{A_V}\!>\!1.8$, and 1\% having
$\widetilde{A_V}\!>\!2.8$.  For comparison, the Milky Way sample of 23
molecular clouds studied by \citet{kainulainen2009} has a median of
$\widetilde{A_V}\!=\!1.2$, and only 2 clouds have
$\widetilde{A_V}\!>\!3$. Thus, although one cannot draw firm
statistical conclusions based on the heterogeneous
\citet{kainulainen2009} sample, the median extinctions in M31 are not
obviously atypical for what is seen locally in the Milky Way.

One way to assess the veracity of the extinction maps is to compare to
other tracers of extinction.  A straightfoward test is to
compare the dust morphology to that seen in optical imaging in blue filters.  At
these wavelengths, dusty structures can appear as dark features
against the smooth background of unresolved stars, albeit with
confusion from foreground stars.  Optical images also have the highest
spatial resolution of various extinction tracers.

\begin{figure*}
\centerline{
\includegraphics[height=3.45in]{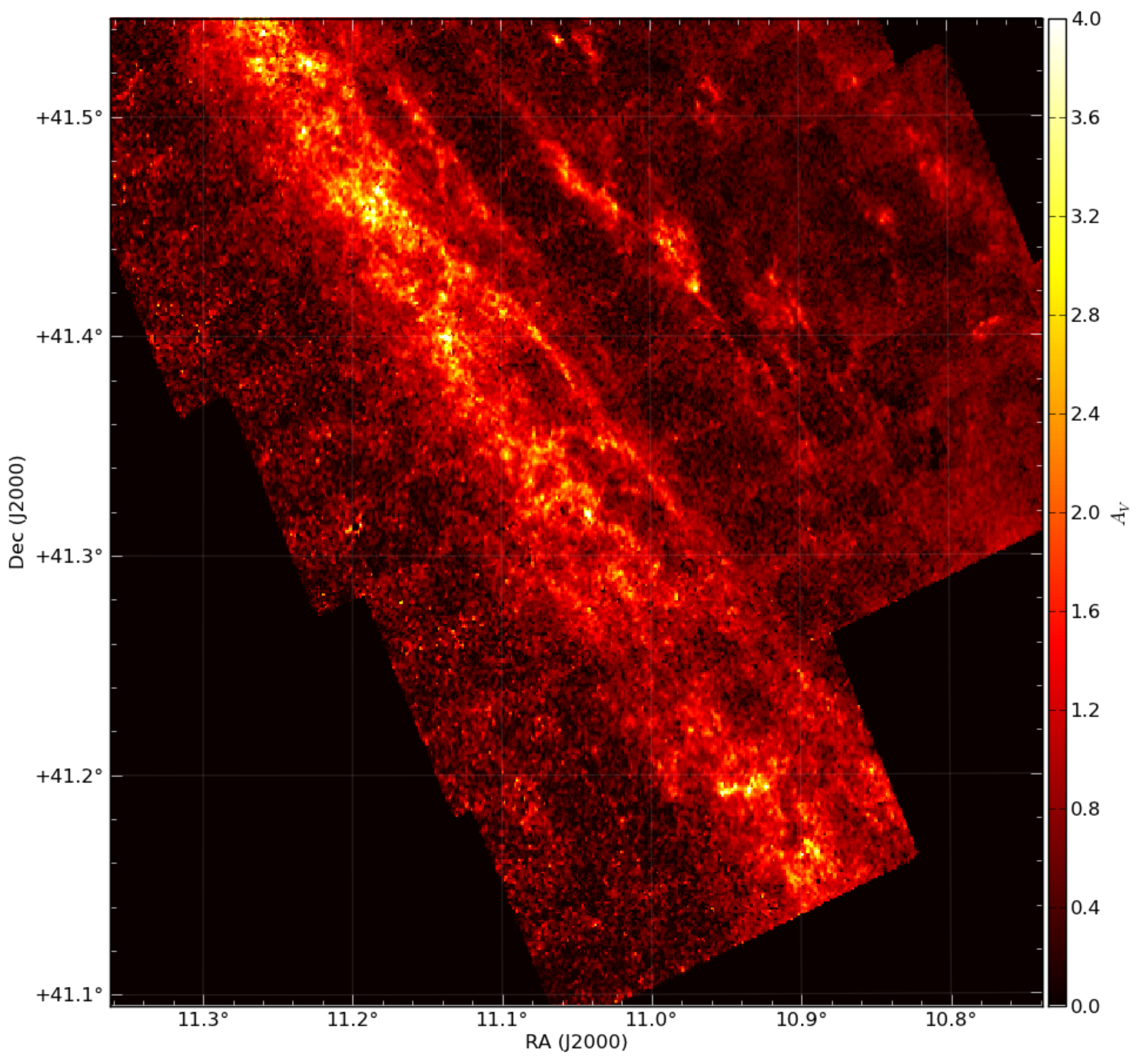}
\includegraphics[height=3.45in]{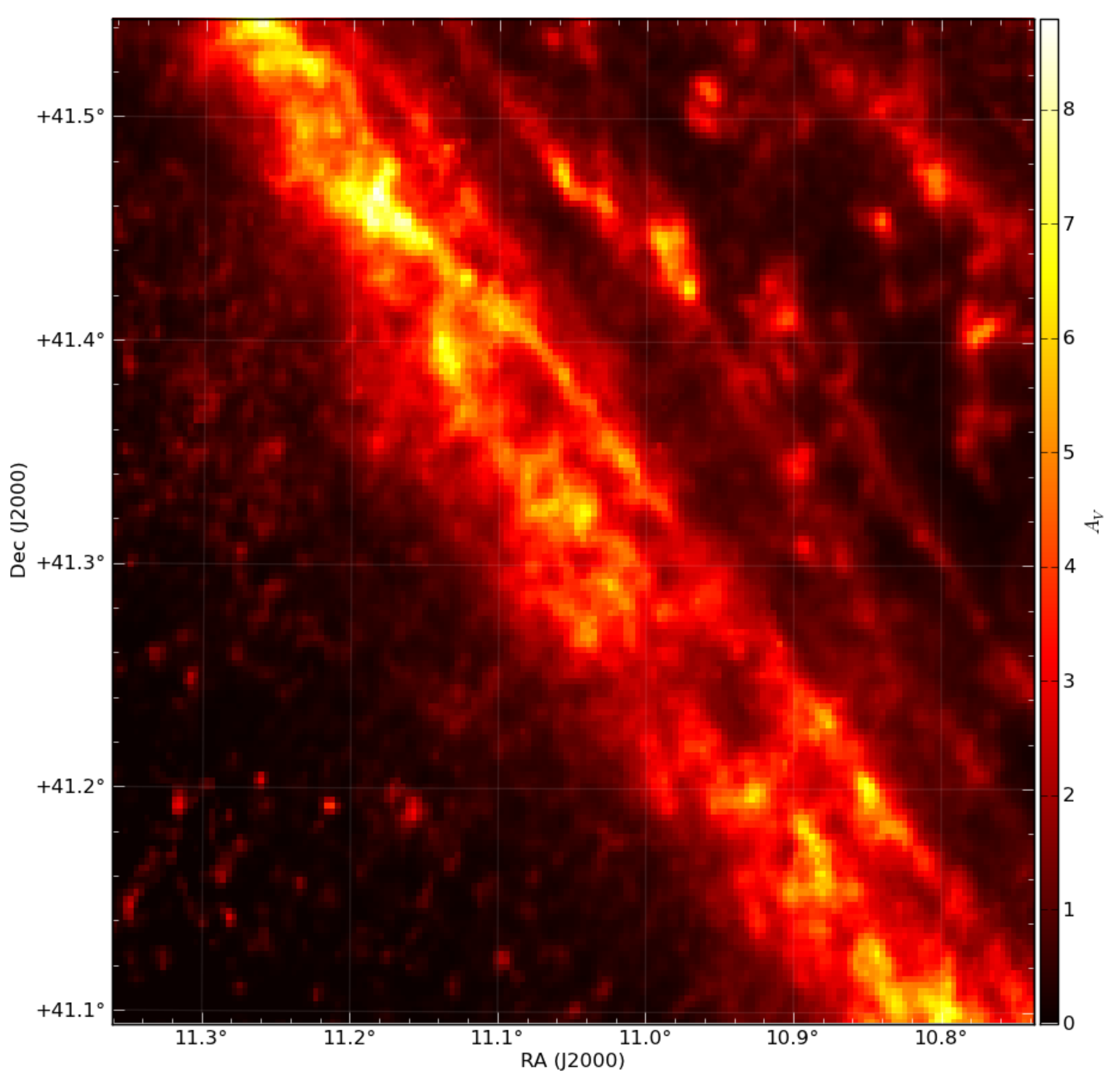}
}
\centerline{
\includegraphics[height=3.5in]{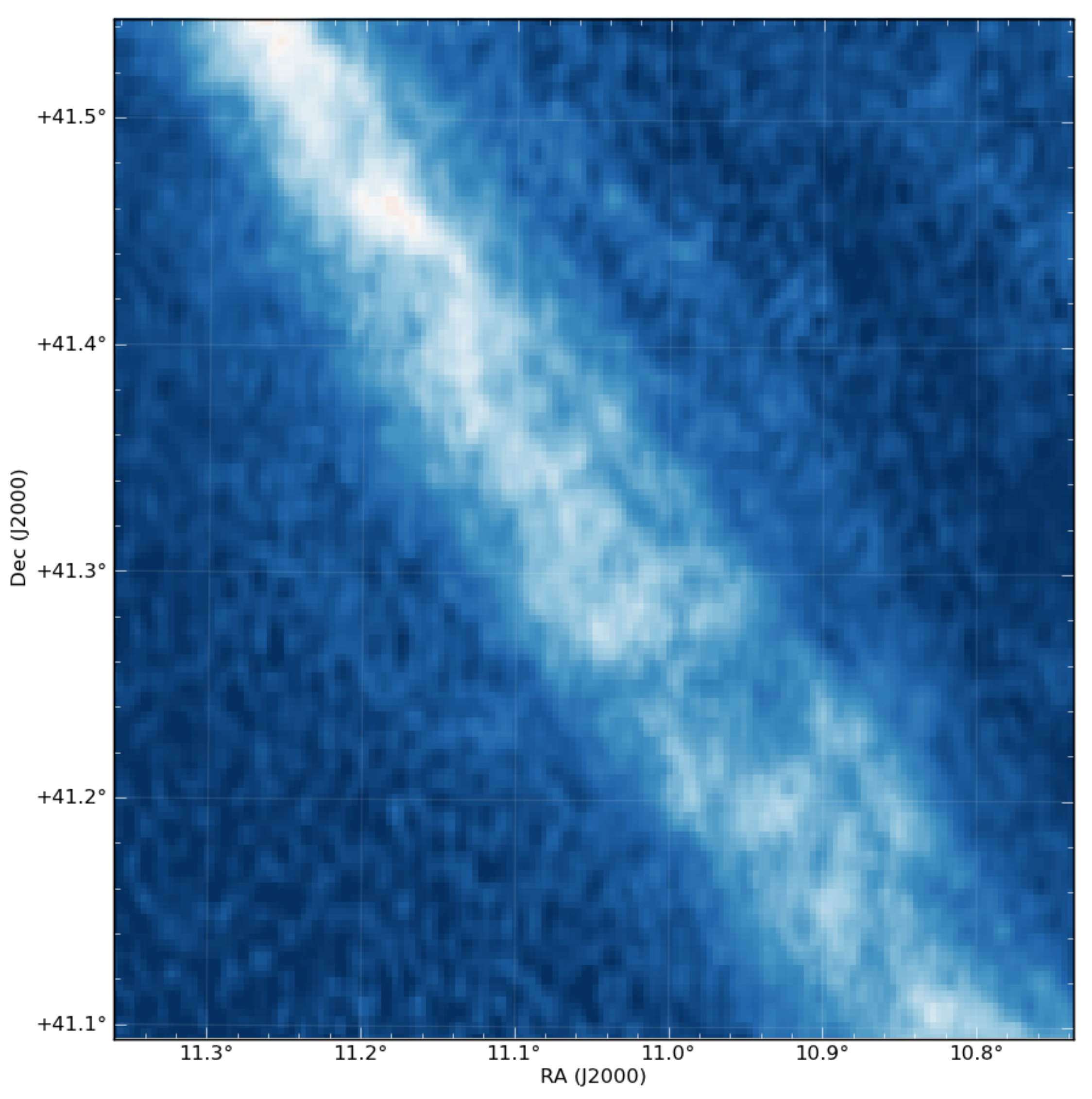}
\includegraphics[height=3.5in]{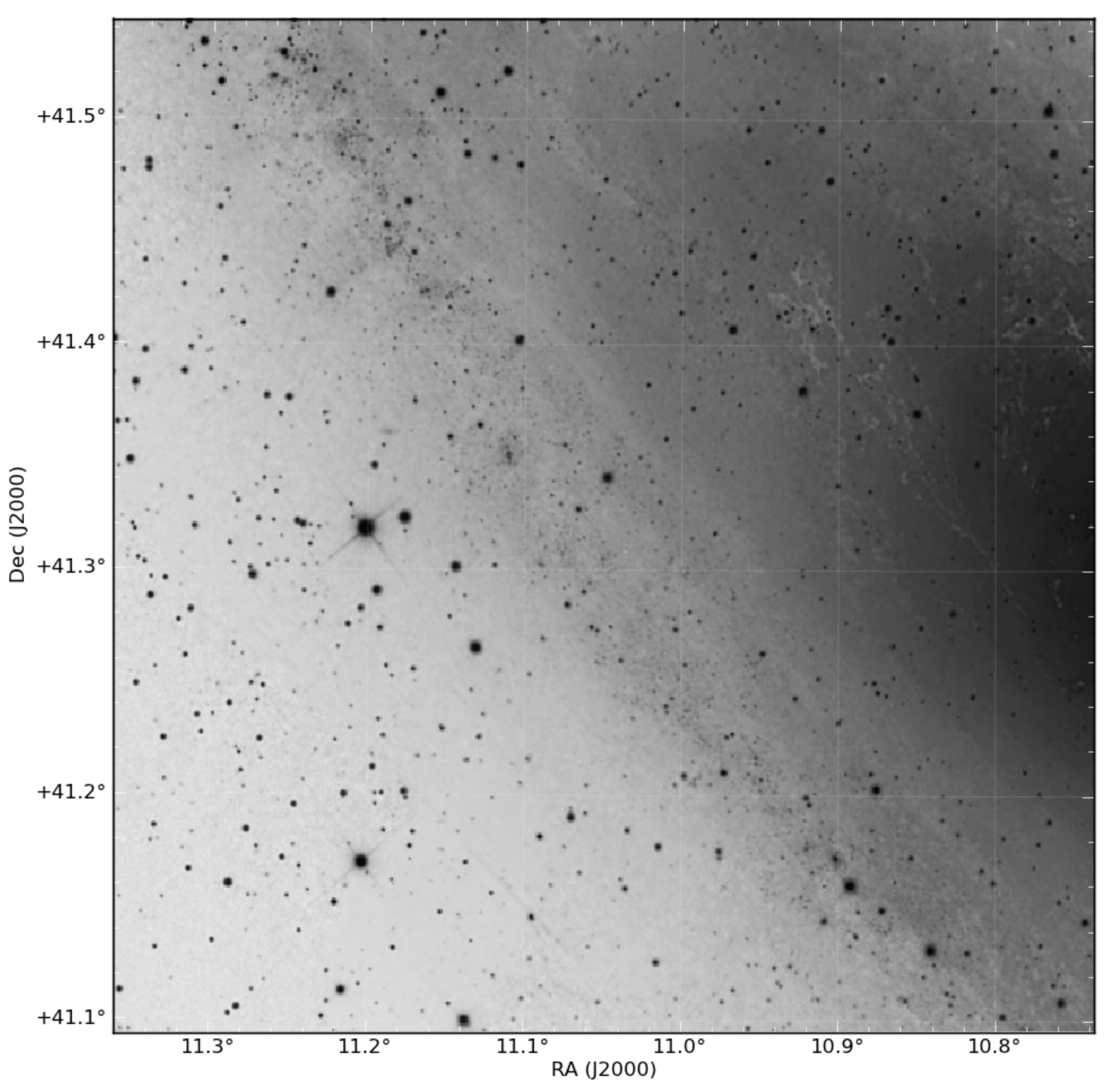}
}
\caption{Comparison between the new M31 extinction map
  ($\widetilde{A_V}$; upper left), the extinction inferred from the
  emission-based Draine et al dust map (upper right), an H{\sc i} map
  (lower left), and an optical image (lower right). These maps
  reproduce the area shown in Figure~\ref{AVmapregion1fig}.
\label{compareregion1fig}}
\end{figure*}

\begin{figure*}
\centerline{
\includegraphics[height=3.45in]{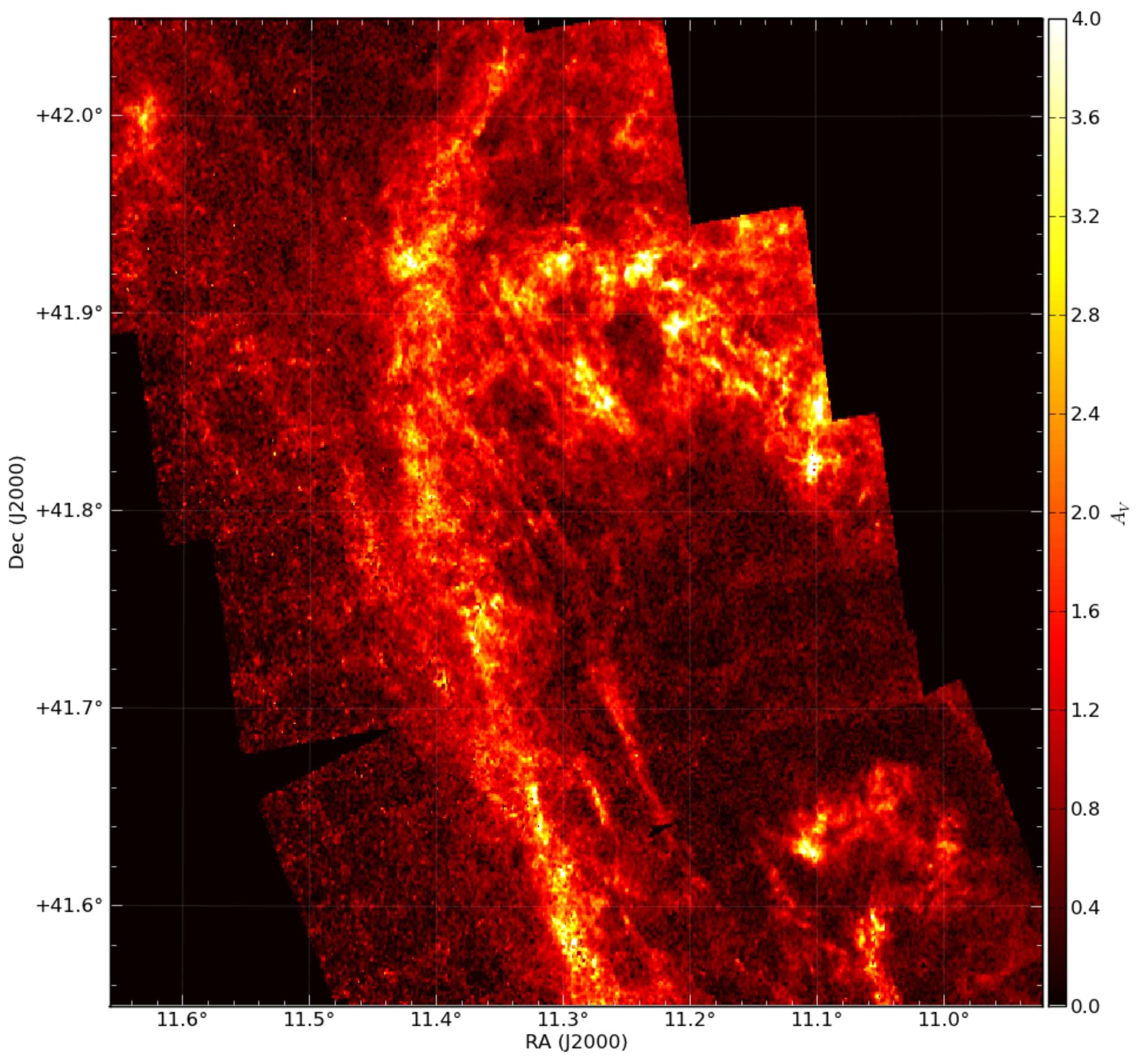}
\includegraphics[height=3.45in]{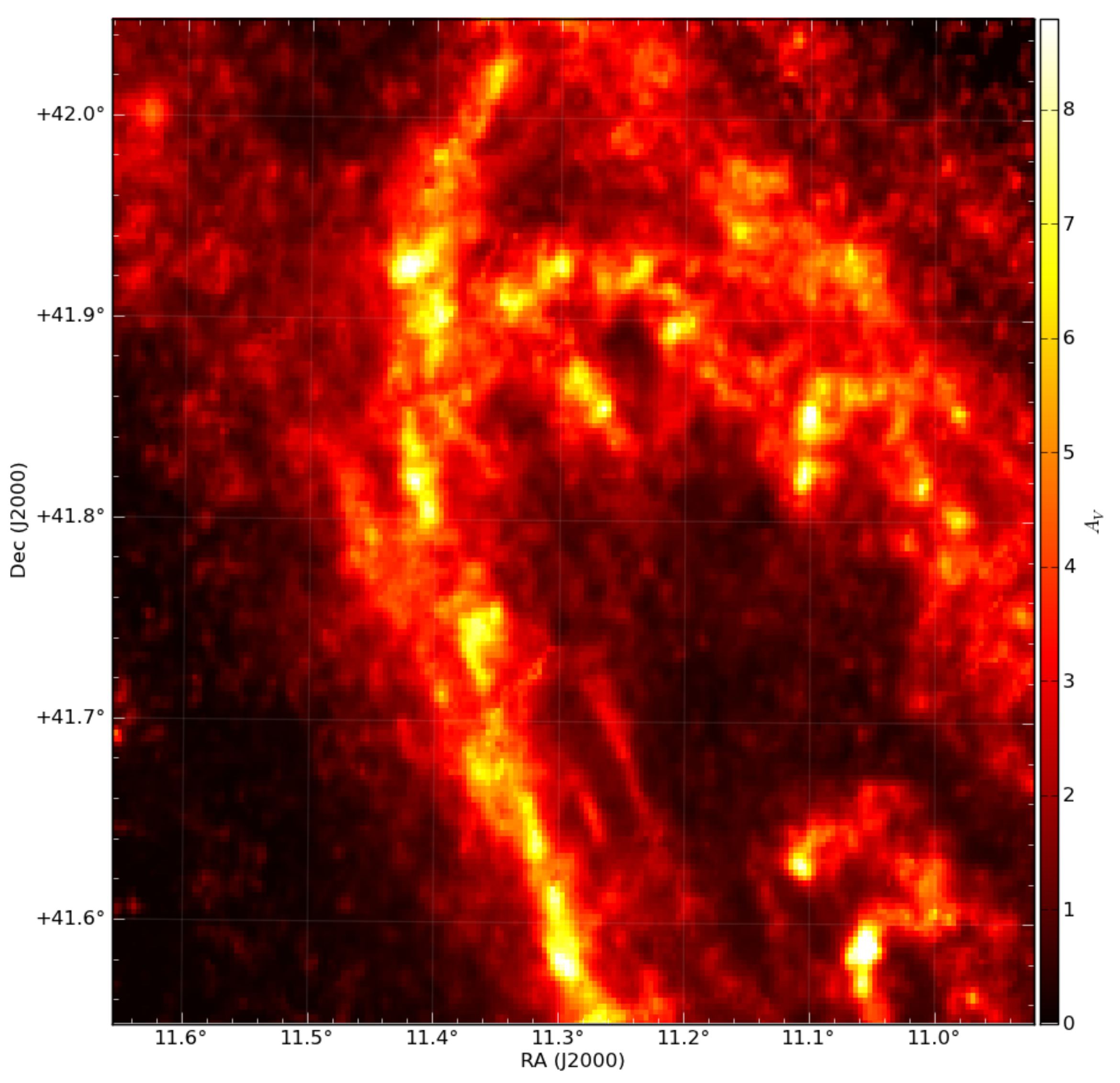}
}
\centerline{
\includegraphics[height=3.5in]{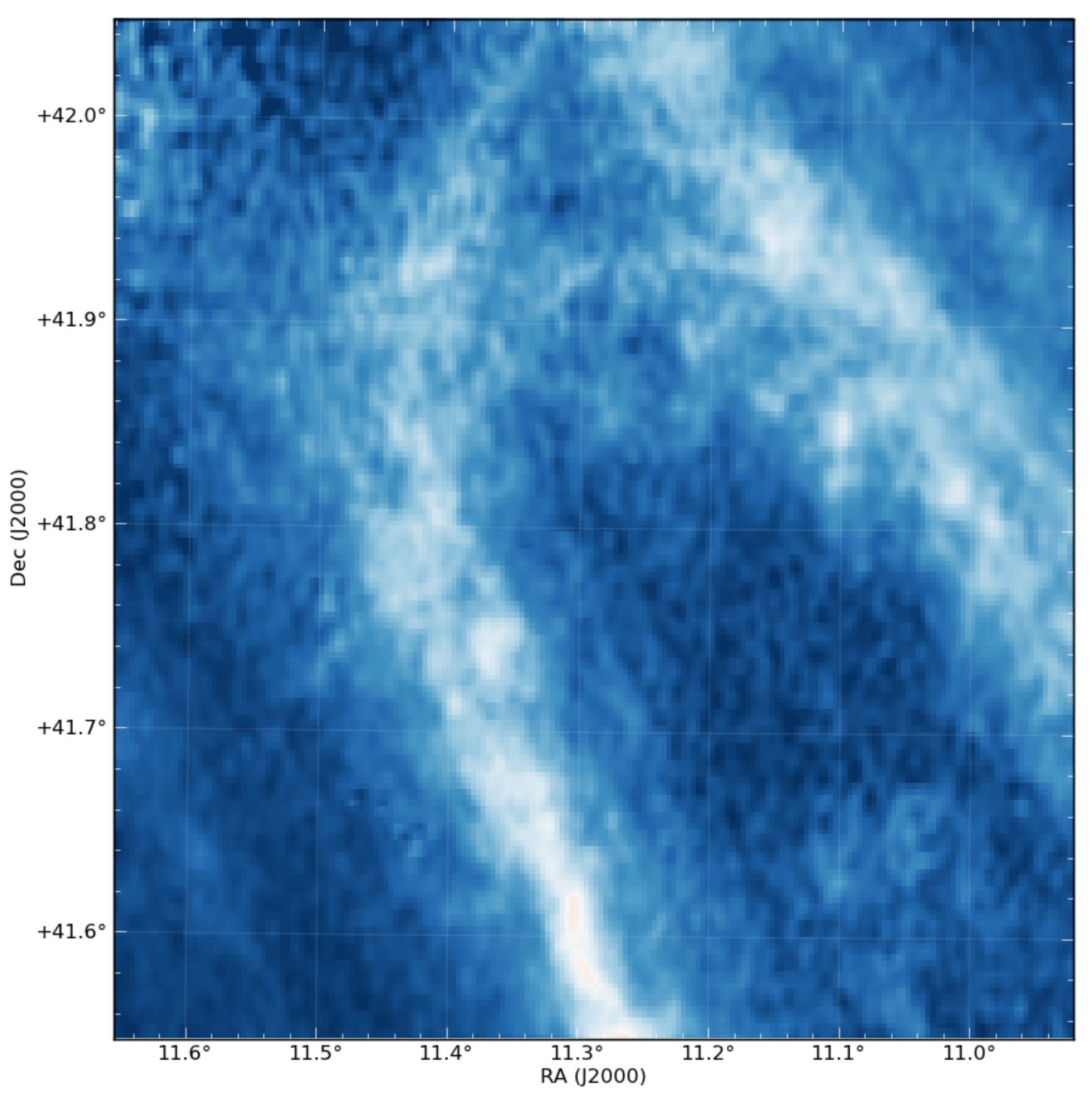}
\includegraphics[height=3.5in]{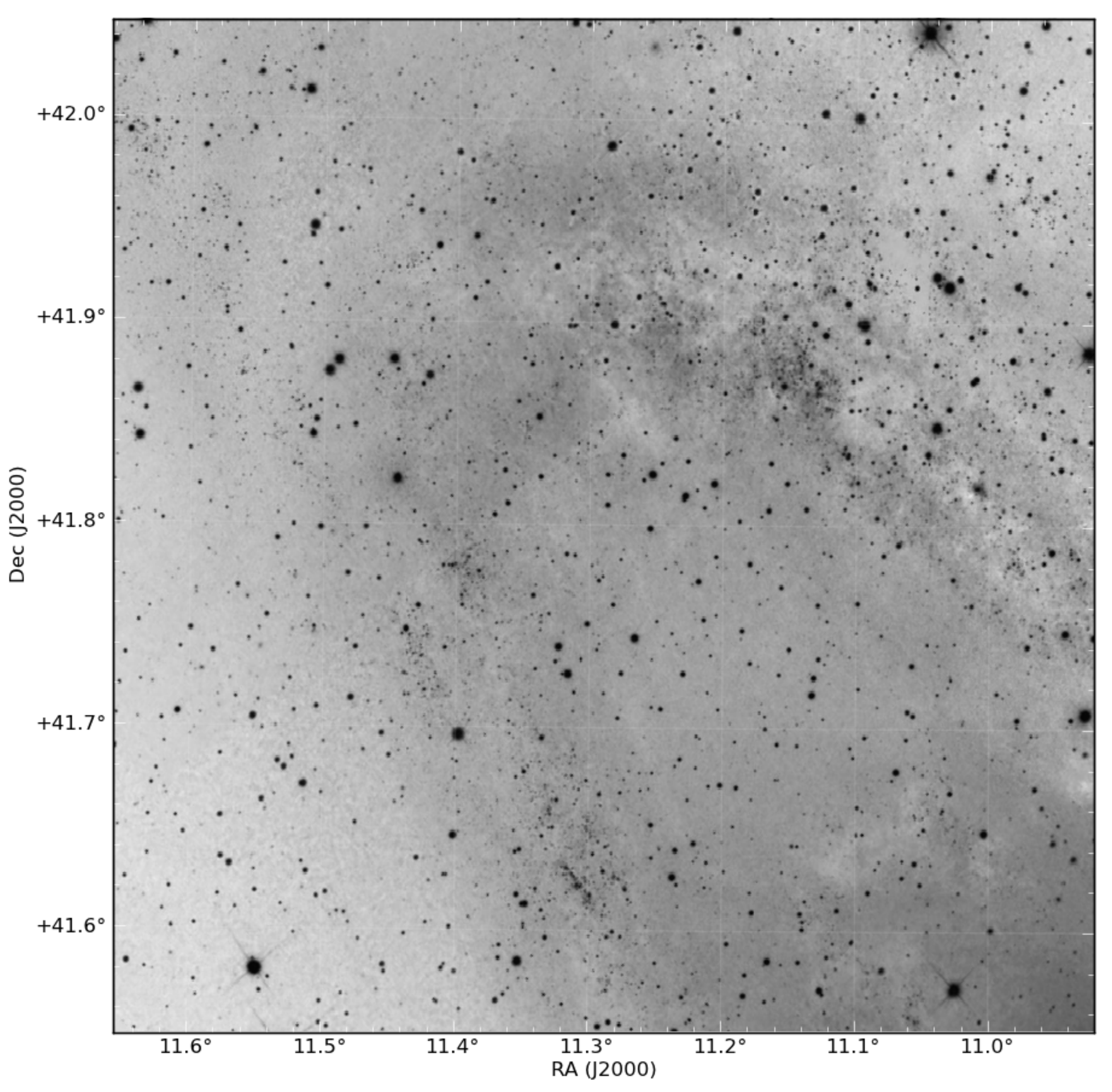}
}
\caption{Comparison between the new M31 extinction map
  ($\widetilde{A_V}$; upper left), the extinction inferred from the
  emission-based Draine et al dust map (upper right), an H{\sc i} map
  (lower left), and an optical image (lower right). These maps
  reproduce the area shown in Figure~\ref{AVmapregion2fig}.
\label{compareregion2fig}}
\end{figure*}

\begin{figure*}
\centerline{
\includegraphics[height=3.45in]{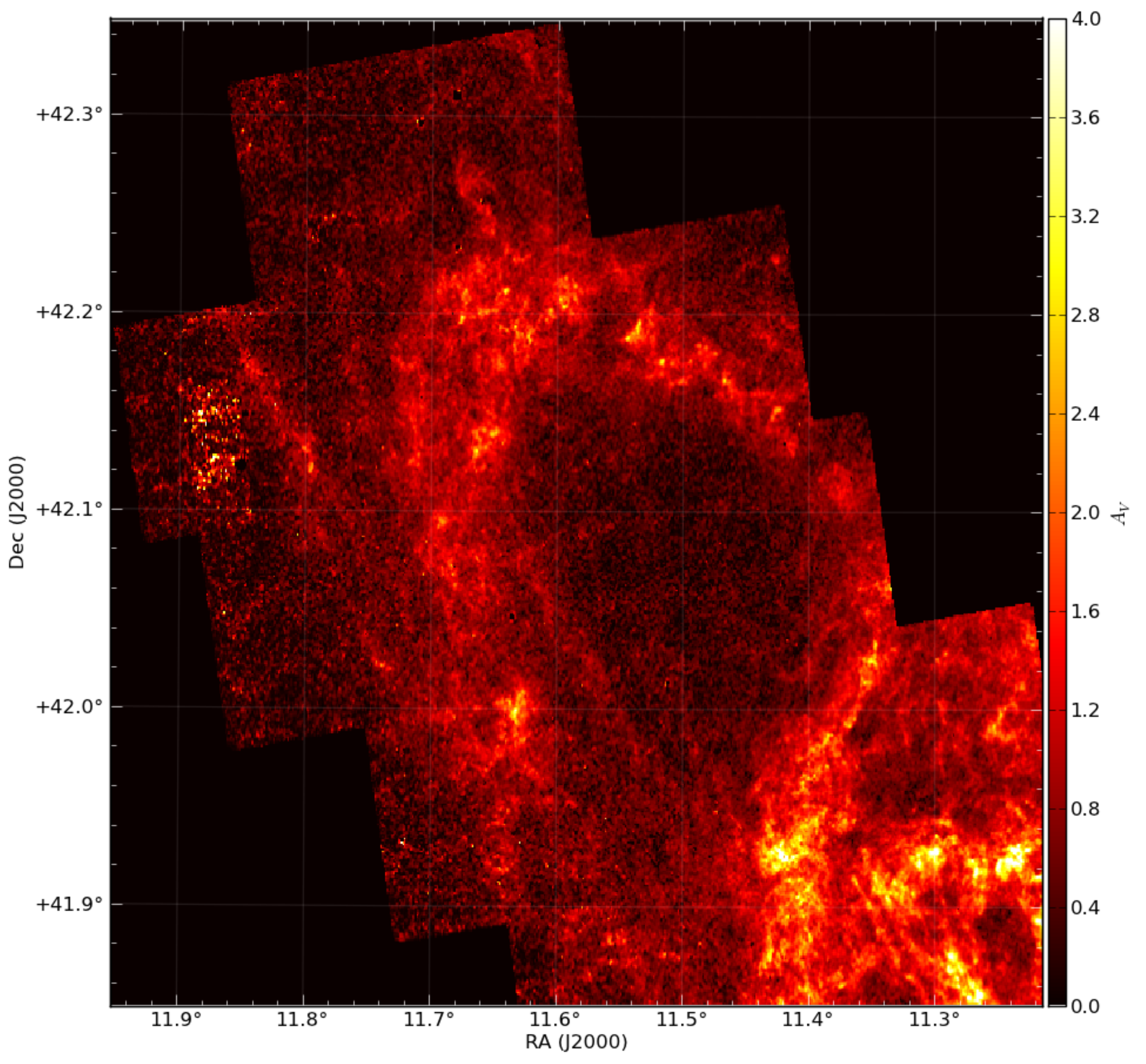}
\includegraphics[height=3.45in]{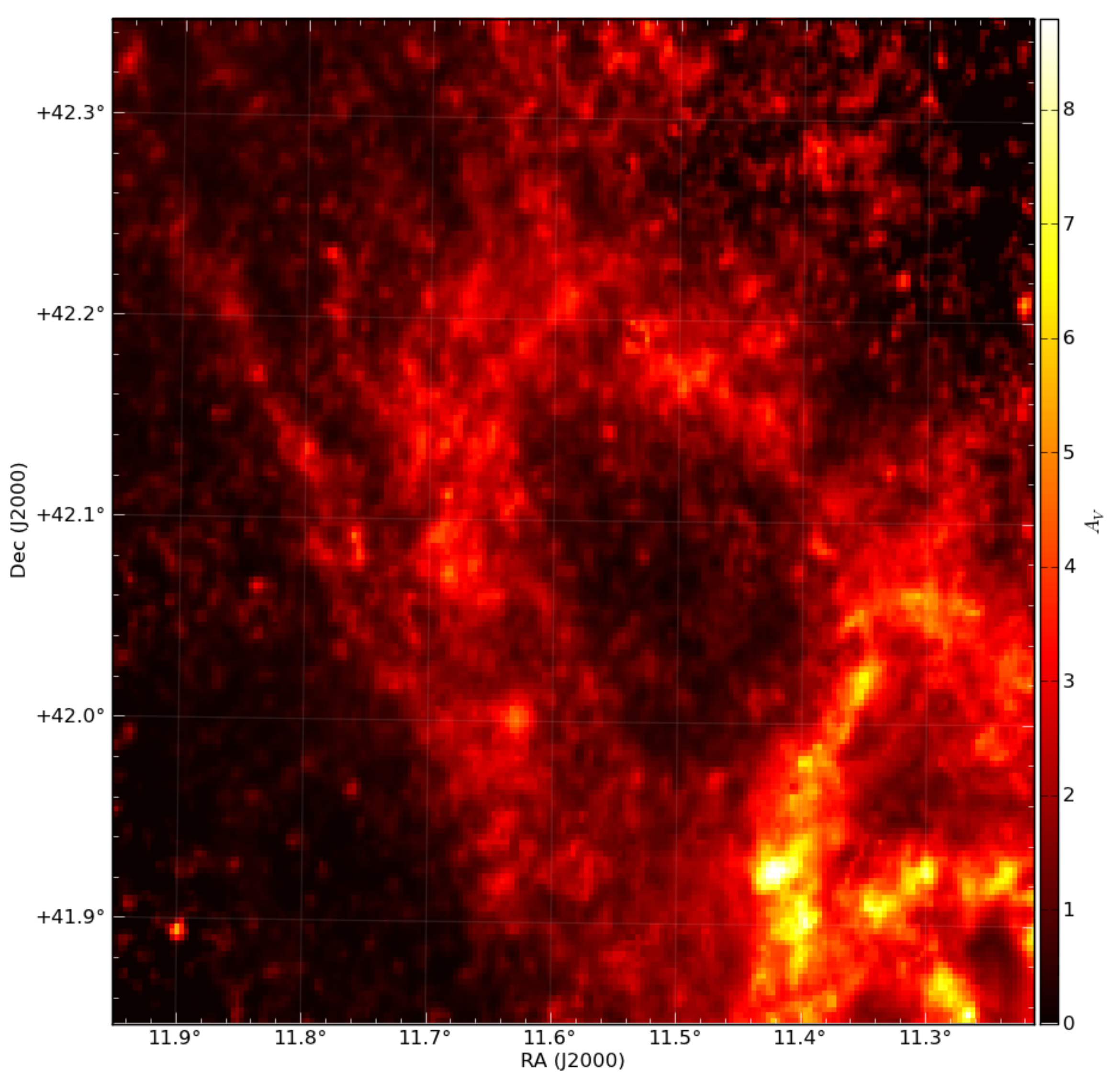}
}
\centerline{
\includegraphics[height=3.5in]{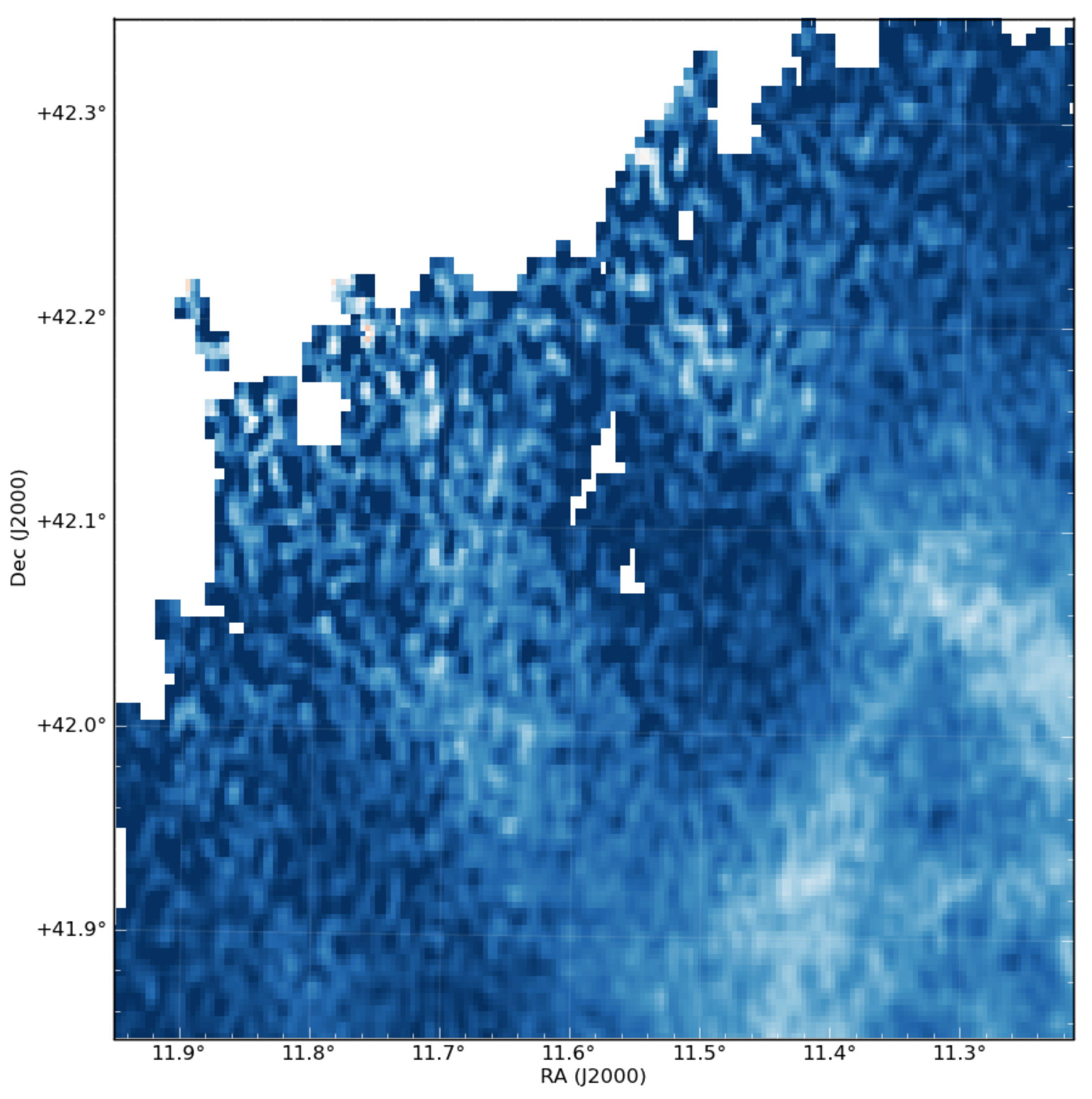}
\includegraphics[height=3.5in]{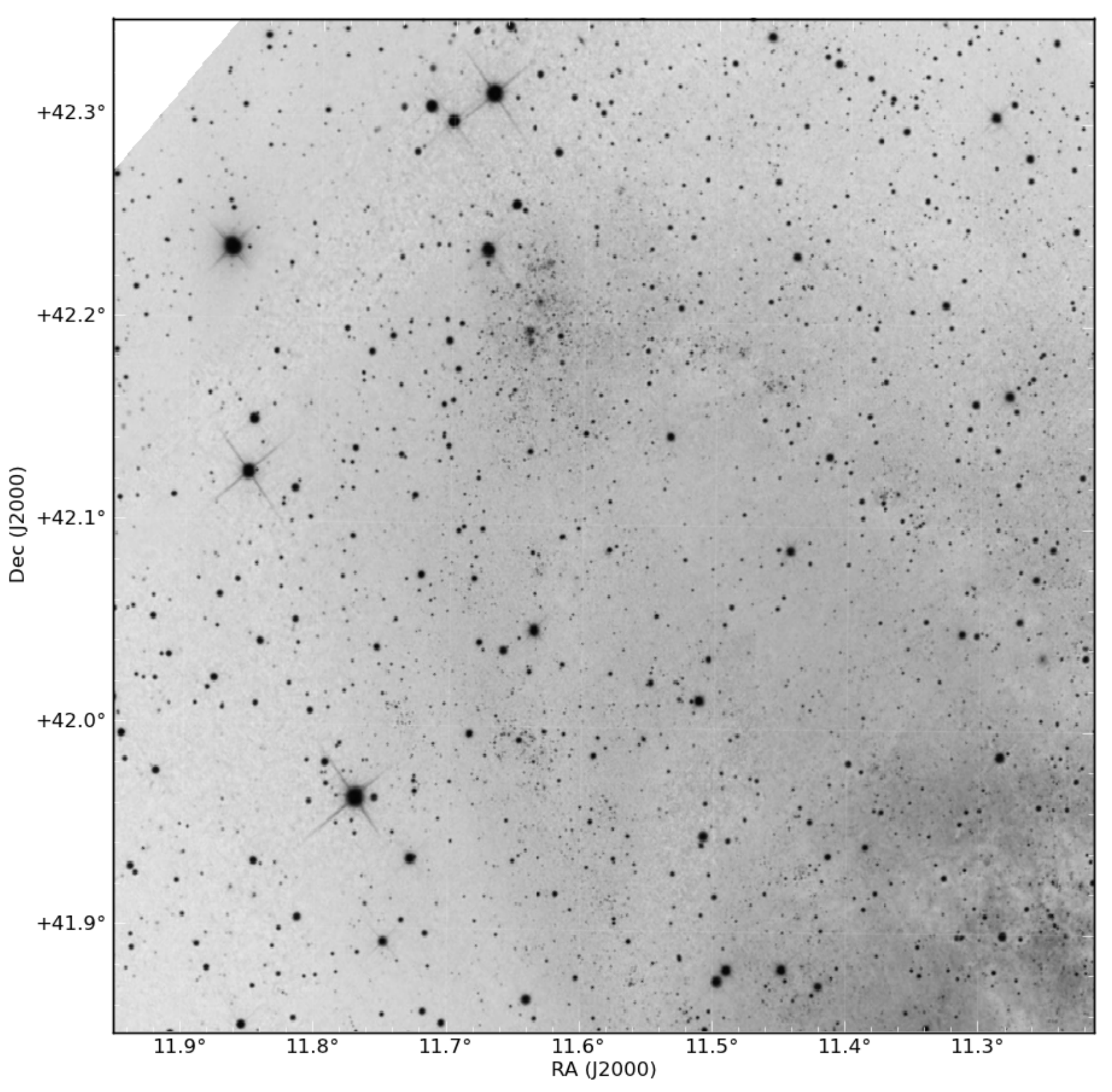}
}
\caption{Comparison between the new M31 extinction map
  ($\widetilde{A_V}$; upper left), the extinction inferred from the
  emission-based Draine et al dust map (upper right), an H{\sc i} map
  (lower left), and an optical image (lower right). These maps
  reproduce the area shown in Figure~\ref{AVmapregion3fig}.
\label{compareregion3fig}}
\end{figure*}

In the lower right panels of
Figures~\ref{compareregion1fig}---\ref{compareregion3fig}, we show
blue optical images\footnote{Courtesy of the superb astrophotographer
  Robert Gendler: {\tt{http://www.robgendlerastropics.com/}}} of the
same regions as the extinction maps in
Figures~\ref{AVmapregion1fig},~\ref{AVmapregion2fig},
and~\ref{AVmapregion3fig} (reproduced in the upper left of
Figures~\ref{compareregion1fig}--\ref{compareregion3fig}, for
reference).  The extinction features seen in our maps rival the
resolution of the equivalent features seen in the optical images.  The
morphology of the features are in excellent agreement as well.

As a further morphological comparison, the lower left panels of
Figures~\ref{compareregion1fig}---\ref{compareregion3fig} show the
publicly available Westerbork maps of H{\sc i} from
\citet{brinks1986}. Although these maps have significantly lower
resolution ($45\arcsec$) than the extinction maps, the broad
morphology is again in good agreement.

While the comparison between the optical image and the extinction map
show many of the same detailed structures, the new extinction maps
offer a far less ambiguous view.  First, they lack the complicated
foreground of young stars that confuses interpretation of extinction
features in broad-band images. But more importantly, the new maps
measure the total extinction of the dust column, regardless of where
it is with respect to the stars.  In contrast, there are extinction
features in the broad-band optical image which appear extremely
strong, when in fact the strength results only from the dust layer
being closer to the foreground, allowing it to block more of light due
to geometry alone.  A clear example of this effect can be seen in the
optical image in Figure~\ref{compareregion1fig}, where the structure
at ($10.92^\circ$, $41.42^\circ$) appears to be much higher column density
than the neighboring structure at ($10.99^\circ$,
$41.45^\circ$). Comparing to the adjacent map of $\widetilde{A_V}$,
however, reveals that the latter clump is actually the higher
extinction region. Instead, the high apparent extinction is due
primarily to the much higher reddening fraction in that region.  This
example points to the peril of using color and/or unsharp masking to
infer the extinction from maps of the attenuation
\citep[e.g.,][]{regan2011}, particularly in inclined galaxies with
large variations in $f_{red}$.

\subsection{Comparison with Emission-Based Maps of Dust Mass} \label{emissionsec}

Thanks to revolutionary new mid- and far-infrared facilities (e.g.,
{\emph{Spitzer}}, {\emph{Herschel}}), and sophisticated models
\citep[e.g.,][]{desert1990,draine2007,compiegne2011}, it has now
become standard to use long-wavelength spectral energy distributions
(SEDs) to infer the mass, composition, and temperature of the dust,
and the illuminating radiation field \citep[e.g.,][]{tuffs2004,
  boissier2004, walter2007, munozmateos2009, noll2009, montalto2009,
  dunne2011, boselli2010, smith2010, compiegne2010, tamura2010,
  popescu2011, galametz2011, skibba2011, galliano2011, dale2012,
  bendo2012, aniano2012, smith2012, galametz2012, groves2012}.  In
nearby, well-resolved galaxies, these techniques can generate maps of
the dust mass as a function of position, which, for a given dust
model, can be converted directly into predictions for the extinction.

Recently, \citet{draine2013} published a state-of-the-art analysis of
M31's dust, using emission maps from {\emph{Spitzer}}'s IRAC and MIPS
cameras \citep[][respectively]{barmby2006,gordon2006}, and from high
resolution HERSCHEL maps made with PACS and SPIRE
\citep[][]{groves2012}.  The SEDs cover from 3.6$\mu$m to
$\sim$500$\mu$m, using observations in 13 different bandpasses.  The
resulting maps of the dust column density were generated at two
resolutions: 24.9$\arcsec$ and 39$\arcsec$, corresponding to the
resolution of the SPIRE $350\mu$m and the MIPS $160\mu$m observations,
respectively, and then sampled with either 10$\arcsec$ or 16$\arcsec$
pixels. A high ($6\arcsec$) resolution emission map has also been
published by \citet{montalto2009}, but the level of contamination from
stellar sources makes it difficult to compare to the work here.

The emission-based dust map of \citet{draine2013} can be used to make
a direct prediction for the extinction.  \citet{draine2013} derive the
extinction-to-dust mass surface density ratio $A_V /
\Sigma_{Md}=7.394$ for their specific model of the dust composition,
where $\Sigma_{Md}$ is the dust mass surface density in $\msun/\pc^2$.

In the upper right of
Figures~\ref{compareregion1fig}---\ref{compareregion3fig}, we plot the
predicted distribution of $A_V$ derived from the 24.9$\arcsec$
resolution \citet{draine2013} dust map (i.e., $A_{V,emission}$).  The
morphological agreement between the $A_{V,emission}$ map and the
CMD-based $\widetilde{A_V}$ map is superb, particularly in regions
where $\widetilde{A_V}\gtrsim 1\mags$.  Even very small features are
clearly reproduced in both of the maps, in spite of their completely
independent derivations.  It is hard not to see the level of agreement
as a triumph for both techniques.

That said, the maps do present some differences. The CMD extinction
maps have nearly 4 times the spatial resolution ($93\pc$ vs $25\pc$),
which naturally produces clearer views of the smallest
features. However, the emission-based maps will be sensitive to any
dust in a given pixel, whereas the CMD extinction-based maps can miss
dust that is locked into small dense structures with negligible
filling factors.  The emission-based maps are also possibly more
accurate at low extinctions, where the CMD-mapping technique is
limited by our ability to: (1) detect broadening over the intrinsic
width of the RGB; and (2) generate accurate ``zero-reddening'' models
of the RGB. On the other hand, the emission-based maps face their own
difficulties at low extinctions, where the uncertainty in wide-field
background subtraction and calibration can easily produce large-scale
systematic offsets in the photometry. Emission maps may also have
difficulty disentangling emission from unresolved dust components with
different temperatures \citep[e.g.,][]{galliano2011}.

The most glaring discrepancy between the two techniques is the offset
in the normalizations.  Examination of the color scales in
Figures~\ref{compareregion1fig}---\ref{compareregion3fig} shows that
the emission-based maps predict extinctions that are more than twice
those that are observed\footnote{Note that this offset is not due the
  RGB only sampling half of the dust layer, because we explicitly fit
  for the fraction of reddened stars (eqn.\ \ref{pAVeqn}) when
  modeling the CMD.}. This difference is present even in low
extinction regions, suggesting that it cannot be due to simply missing
very dense dust clouds with very small filling factors, which would
only affect the highest column density regions.  Given the simplicity
of the CMD-based extinction technique compared to the emission based
technique \citep[which relies on assumptions about the radiation field
  as well as calibrations within the Milky Way;
  e.g.,][]{weingartner2001}, it is far more likely that this global
offset represents an issue with the current calibration of the dust
emission models.

We make a more quantitative assessment of this offset by comparing a
resolution-matched version of our CMD-based extinction map to the
emission-based map.  We use the mean extinction $\langle A_V \rangle$
for all comparisions, rather than the median $\widetilde{A_V}$; these
two quantities can be significantly different for a log-normal
distribution, and the mean is the better estimate of the total dust
column density and thus is a fairer test of the quantity tracked by
the emission-based map.  We extract the posterior probability
distribution function of $\langle A_V \rangle$ directly from the MCMC
samples, rather than deriving it from the combination of the best fit
$\widetilde{A_V}$ and $\sigma$ (i.e., $\langle A_V \rangle =
\widetilde{A_V} \exp{(\sigma^2/2)}$; Eqn.~\ref{meaneqn}); this choice
avoids any biases due to correlations between $\widetilde{A_V}$ and
$\sigma$.  We then degrade the resolution of the $\langle A_V \rangle$
map from ${\rm FWHM}=6.645\arcsec$ to ${\rm FWHM}=24.9\arcsec$ by
smoothing with a Gaussian kernel of standard deviation of
10.19$\arcsec$ ($= \sqrt{24.9^2 - 6.645^2}/2\sqrt{2\ln{2}}$); this
degrades the resolution from $25\pc$ to $94\pc$.  We use this
resolution-matched map of $\langle A_V \rangle$ in all subsequent
comparisons with results from \citet{draine2013}.

\begin{figure*}
\centerline{
\includegraphics[width=2.5in]{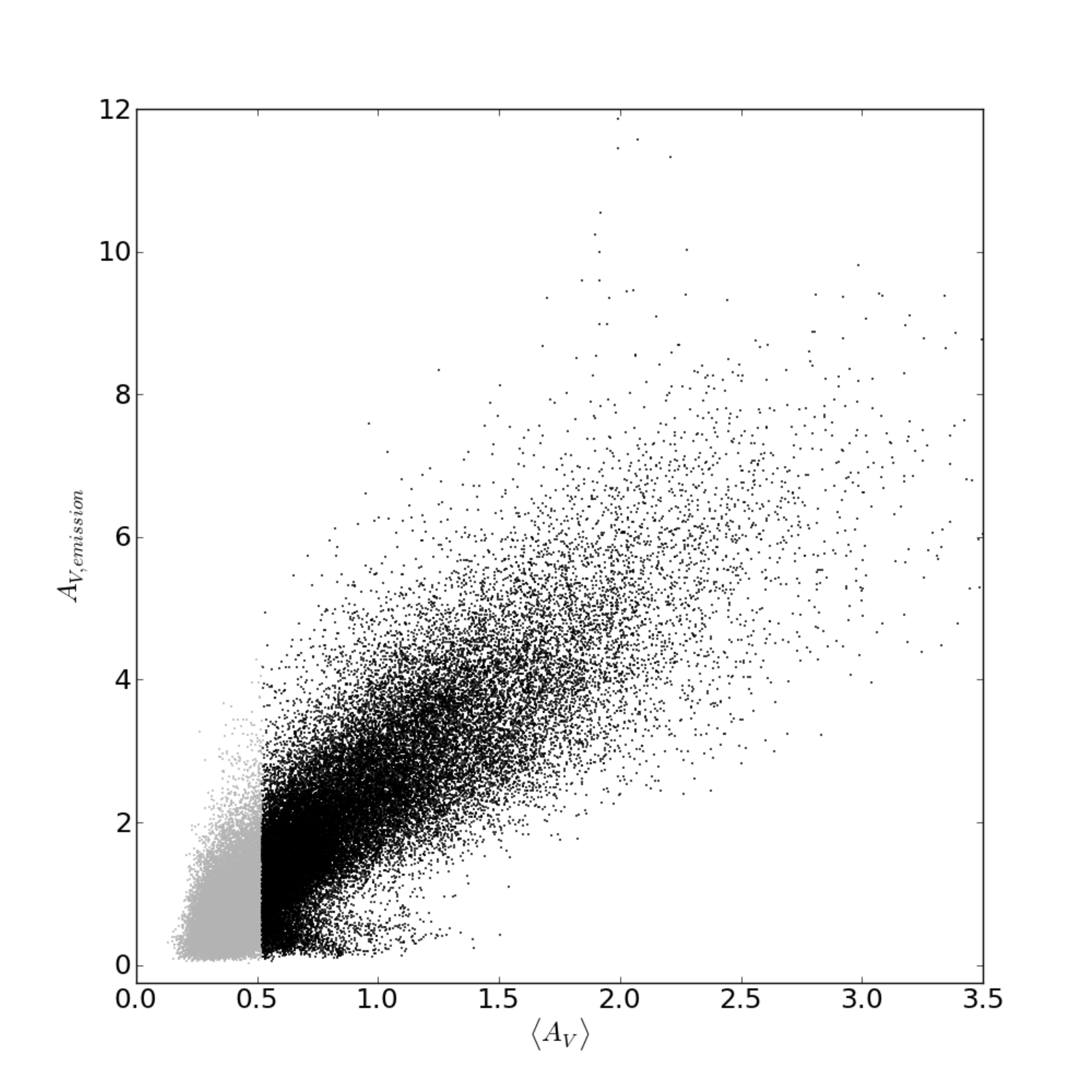}
\includegraphics[width=2.5in]{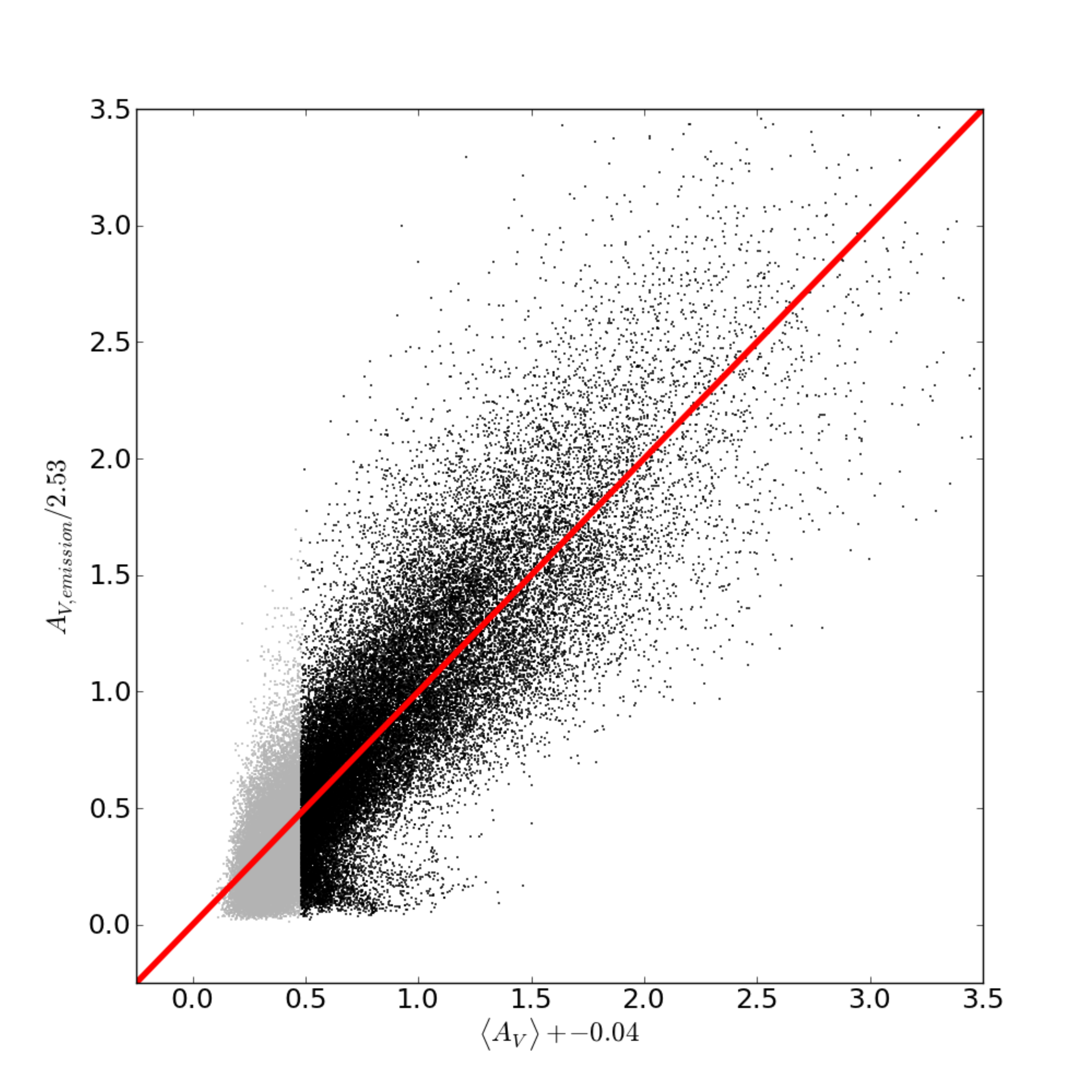}
}
\centerline{
\includegraphics[width=2.5in]{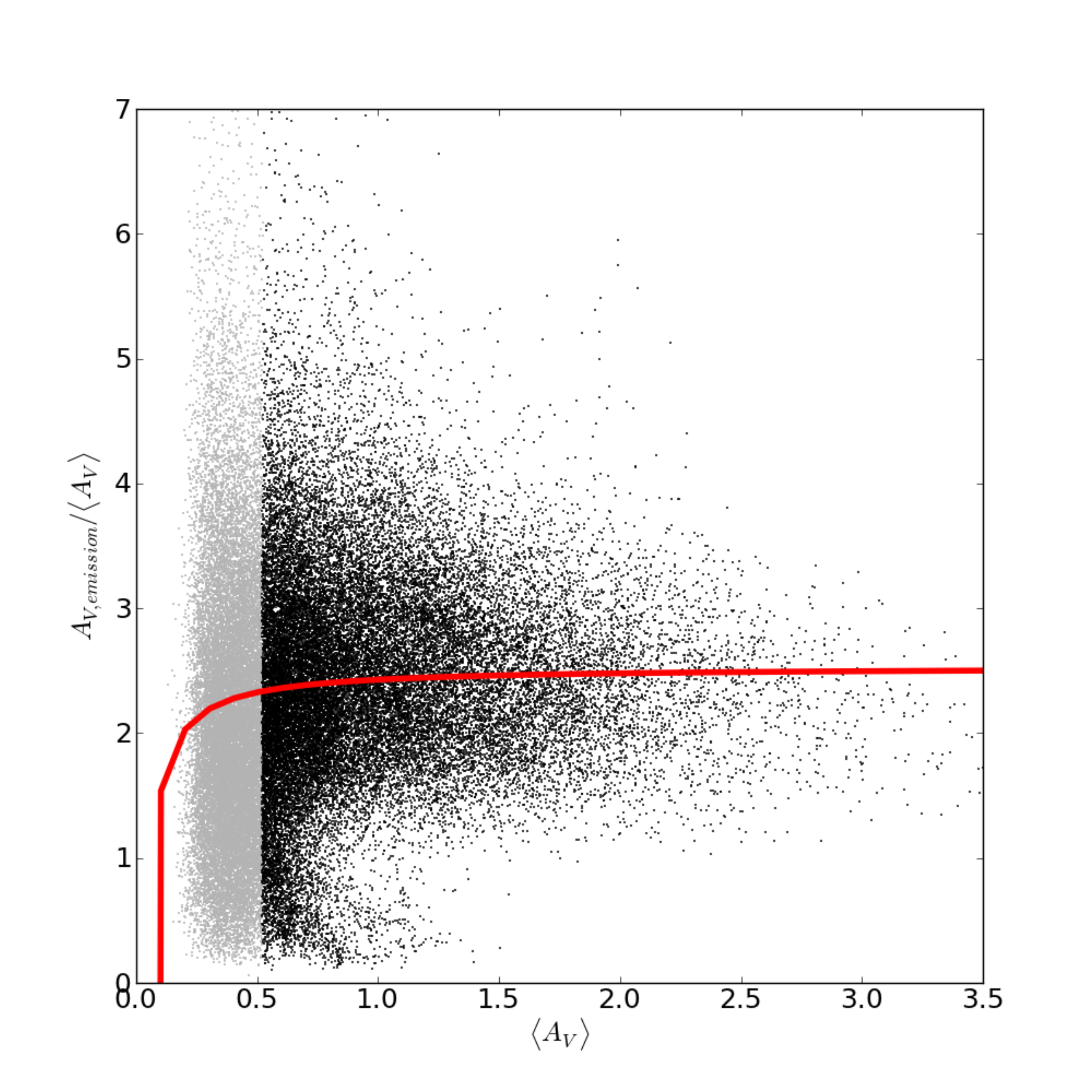}
\includegraphics[width=2.8in]{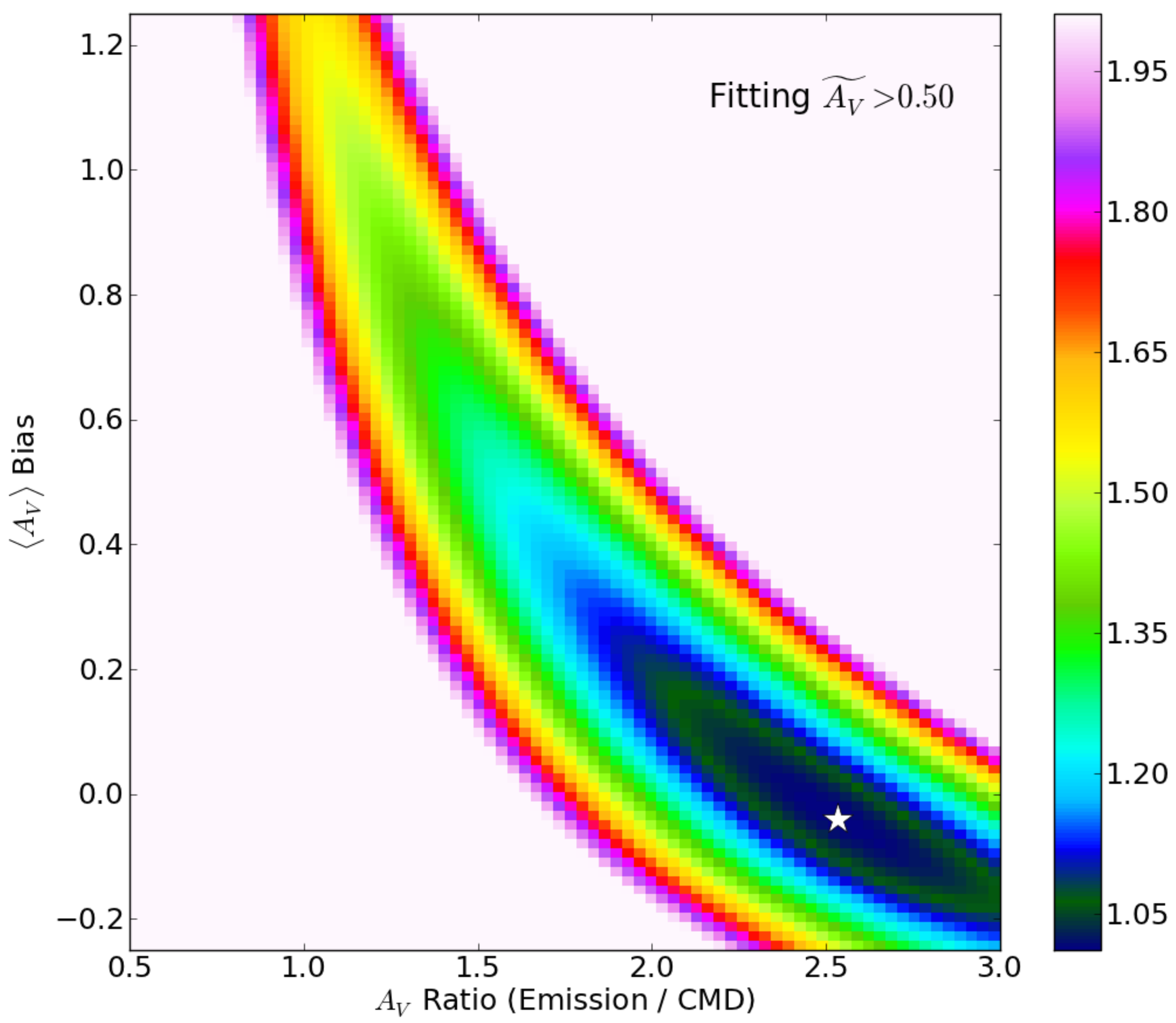}
}
\caption{(Top Row) Pixel-by-pixel comparisons between the
  \citet{draine2013} emission-based $A_V$ map and the
  resolution-matched map of mean extinction $\langle A_V \rangle$
  derived from the NIR CMD.  The original comparison (left) shows
  strong correlations, as expected from the morphological
  agreement. However, the steep slope and the vertical offset from the
  origin indicate issues with the overall scaling between the two
  methods, and a bias in the CMD-based extinction maps at low dust
  column. After correcting the \citet{draine2013} maps (right) by an
  overall scaling factor of $R=2.53$, and correcting the CMD-based
  extinction by an additive bias factor of $b=-0.04$, the agreement
  between the two maps is much improved. The scatter is still
  significant, however.
  (Bottom Left) The pixel-by-pixel ratio between the two maps and the
  resolution-matched map of mean extinction $\langle A_V \rangle$
  derived from the NIR CMD. The red line shows the predicted ratio
  using the scale$+$bias correction used in the center panel. The
  results are similar if the median extinction of the reddened
  component is used instead of the mean, although the scale factors
  are slightly different. (Bottom Right) The $\chi^2$ distribution of
  the values of the scale factor correction $R$ to the
  \citet{draine2013} map and the bias correction $b$ to the CMD-based
  map, derived where
  $\widetilde{A_V}>0.5$. \label{drainecorrelationfig}}
\end{figure*}

The top left panel in Figure~\ref{drainecorrelationfig} shows the
pixel-by-pixel correlation between the \citet{draine2013} extinction
maps and the mean extinction $\langle A_V \rangle$ of the reddened
component, derived from the NIR CMD. The pixels are highly correlated,
as expected from the excellent morphological agreement seen in
Figures~\ref{compareregion1fig}---\ref{compareregion3fig}.  However,
it is clear from the top left of Figure~\ref{drainecorrelationfig}
that the extinctions from the \citet{draine2013} maps are higher by a
factor of $\gtrsim$2. There is also a small tail of points where there
is little dust emission, but the CMD-based extinction is higher. These
points are all due to regions where the photometry is slightly biased
to redder colors near the chip edge in the very outer disk, where
there are other photometry errors (Fields 2 and 8 of Brick 22), or
where crowding is high and the RGB is intrinsically broad in the very
inner disk (Fields 3 and 4 of Brick 3).

\subsubsection{Correcting the Emission- and CMD-based Extinctions} \label{emissioncorrsec}

We quantify the differences between the emission-based and CMD-based
extinction maps by calculating a transformation that would bring the
maps into agreement.  We assume that the \citet{draine2013}
extinctions are too high, and must be divided by a factor $R$.  We
include an additional correction factor $b$ for possible systematic
biases in the CMD-based extinctions (see Sec.\ \ref{limitationsec} \&
\ref{systematicsec}).  We then estimate appropriate values for $R$ and
$b$ by calculating $\chi^2$ on a grid of $R$ and $b$, where $\chi^2$
is calculated using the observed ratio $A_{V,emission} / \langle A_V
\rangle$ compared to the prediction for the model as a function of
$\langle A_V \rangle$ (i.e., $A_{V,emission} / \langle A_V \rangle =
R(1 + b/\langle A_V \rangle )$; lower left of
Figure~\ref{drainecorrelationfig}). When calculating $\chi^2$, we use
an empirical estimate of the uncertainty in the measured ratio as a
function of $\langle A_V \rangle$, since we lack a realistic noise
model for the ratio $R$. We limit the calculation to only points with
$\widetilde{A_V} > 0.5\mags$ where measurements for both dust models
are expected to be more robust.

The resulting $\chi^2$ distribution is shown in the lower right panel
of Figure~\ref{drainecorrelationfig}. The best-fit model parameters
are $R=2.53$ and $b=-0.04\,{\rm mag}$ when using the mean CMD-based
extinction $\langle A_V \rangle$. If the CMD-based extinctions are
assumed to be unbiased ($b=0$) then the scaling for the
\citet{draine2013} extinctions shifts to slightly lower values of $R$.

In the upper right panel of Figure~\ref{drainecorrelationfig}, we show
the pixel-by-pixel correlation after applying these corrections, such
that $A_{V,emission}\rightarrow A_{V,emission}/R$ and $\langle A_V
\rangle \rightarrow \langle A_V \rangle + b$. The prediction for the
ratio using the same model is also shown as the red line on the lower
left panel.  The overall distribution now has the desired slope of 1
(red line) and is consistent with passing through (0,0).

The fits shown in Figure~\ref{drainecorrelationfig} suggest that (1)
there is a negligible global bias in the CMD-derived extinction values,
corresponding to a color shift of $\sim\!0.005$ magnitudes in
$F110W-F160W$, and (2) there is a large scaling
difference between the emission- and CMD-based extinctions.  We now
discuss the second of these conclusions in more detail, and defer the
less pressing issue of biases to Sec.\ \ref{spatialsystematicssec}.

\subsubsection{Are the \citet{draine2013} Extinctions Really Too High?}

Figure~\ref{drainecorrelationfig} strongly suggests that there is a
factor of $\sim\!2.5$ difference between the extinction inferred
from CMD fitting and that inferred from modeling the IR emission.  The
distribution of $\chi^2$ values in Figure~\ref{drainecorrelationfig}
shows there is no reasonable value of the bias for which the ratio $R$
becomes 1.  Thus, we are in the uncomfortable position of accepting a
large systematic problem in either the CMD-based extinction or in
widely used models of dust emission.

\begin{figure*}
\centerline{
\includegraphics[width=3.75in]{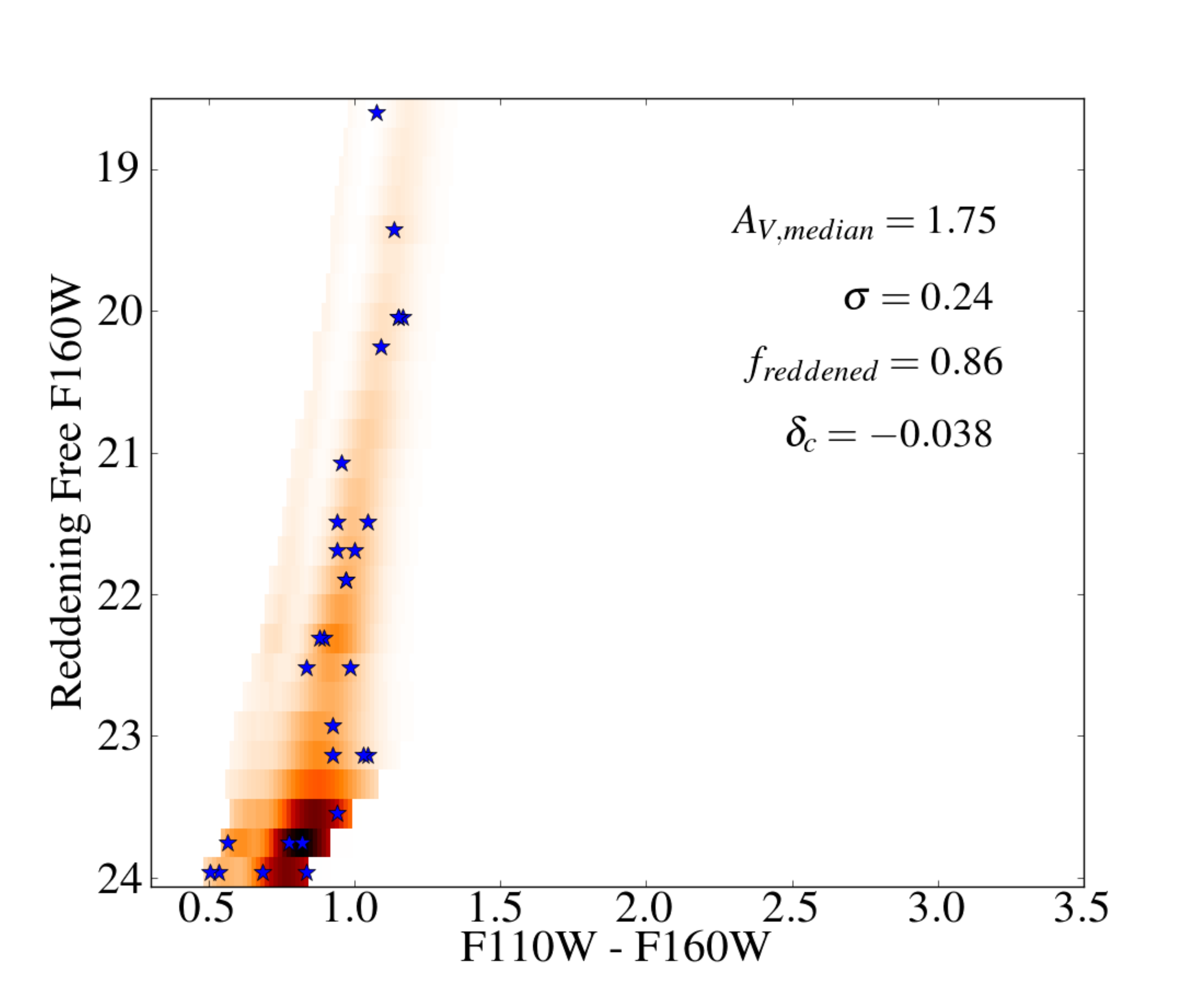}
\includegraphics[width=3.75in]{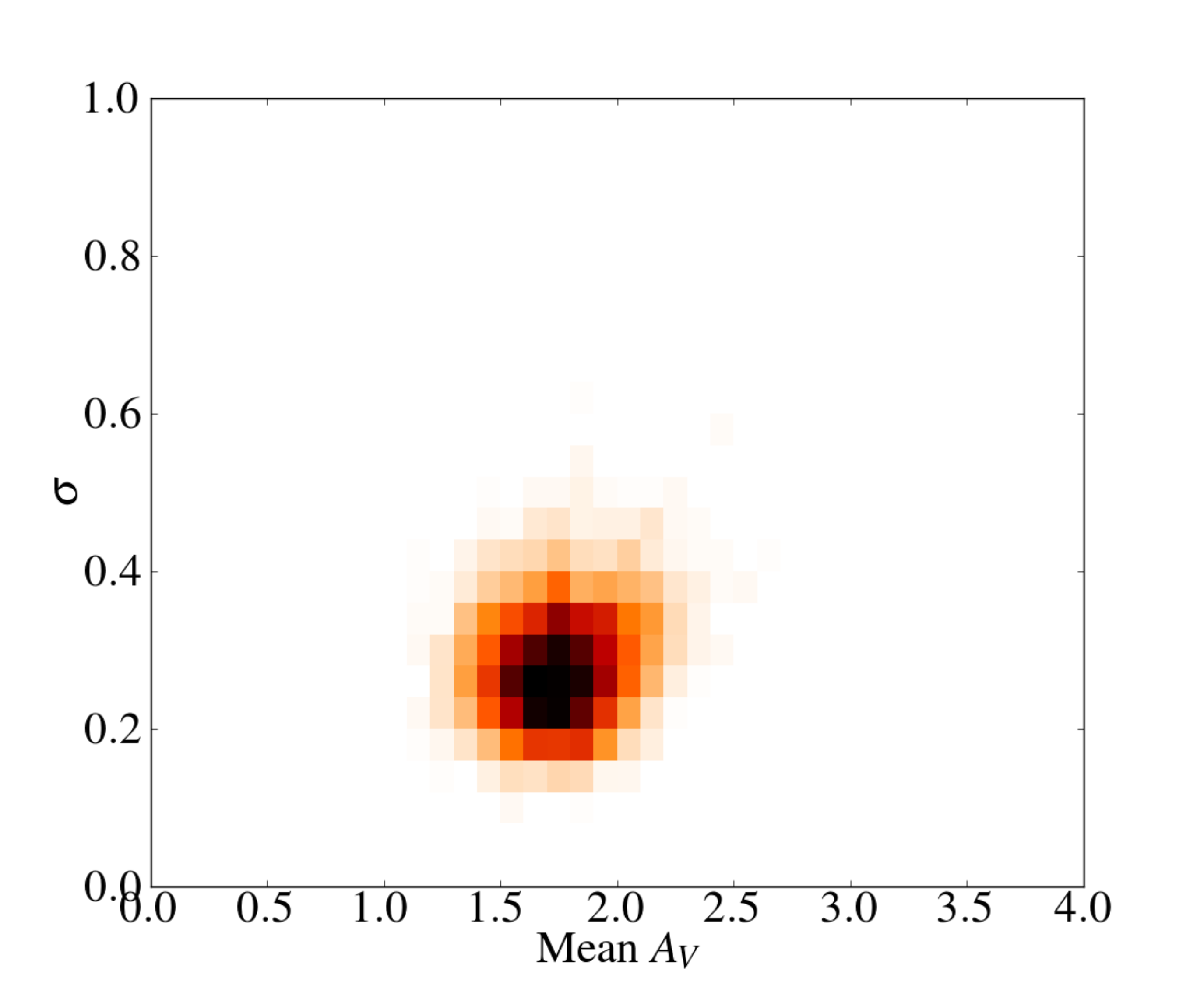}
}
\centerline{
\includegraphics[width=3.75in]{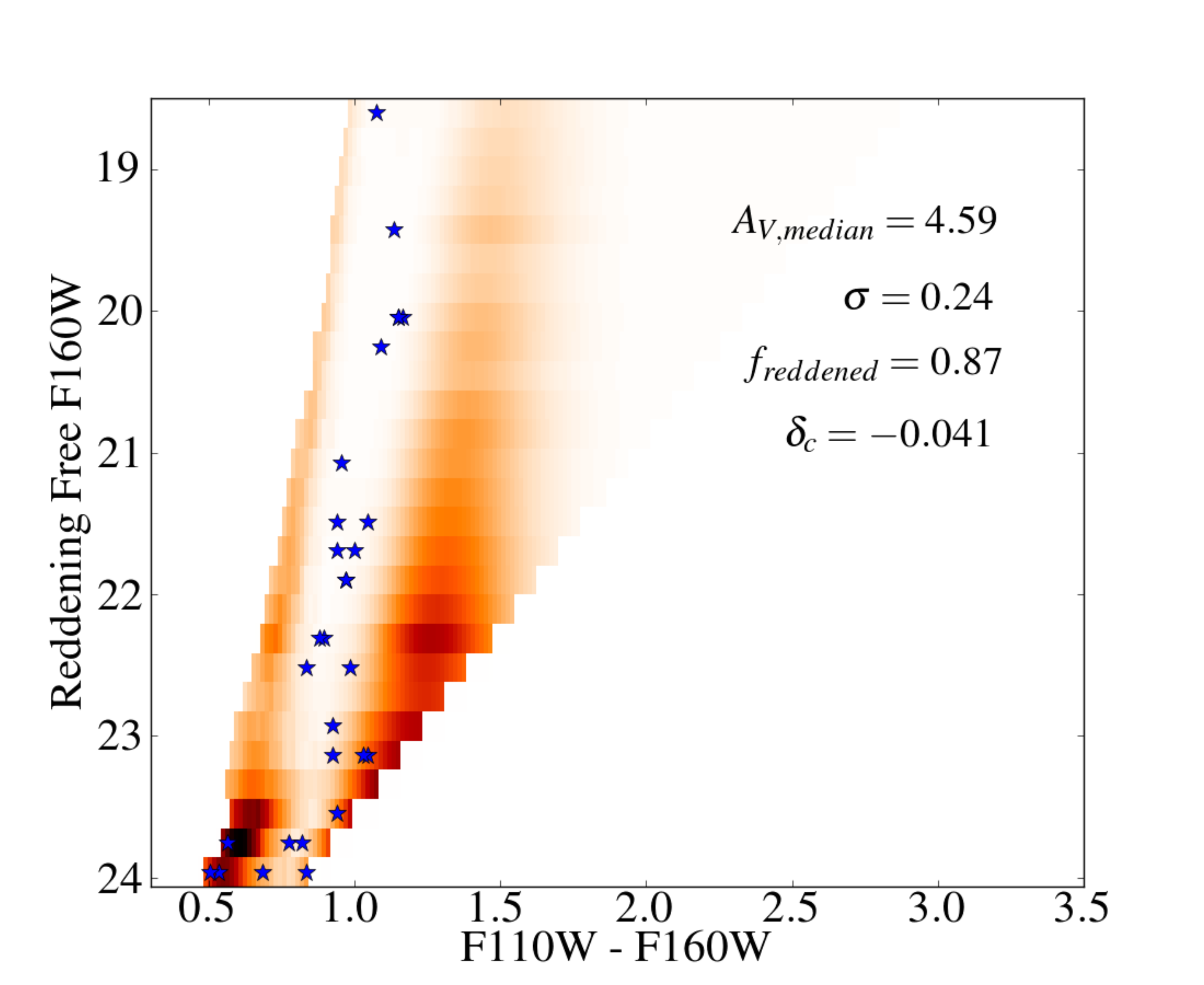}
\includegraphics[width=3.75in]{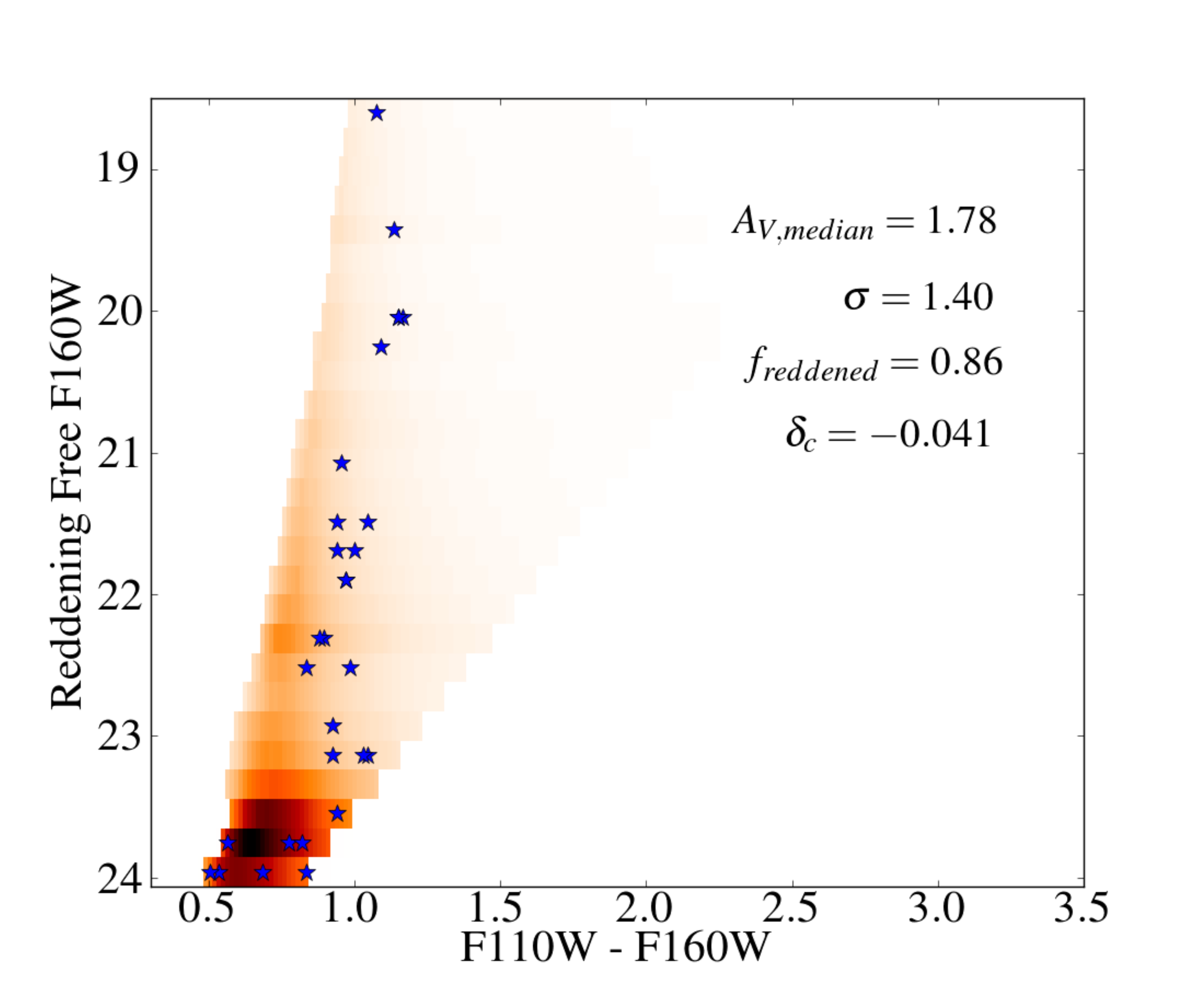}
}
\caption{Binned CMD showing the reddened model derived for stars found
  in a $25\pc$ pixel in Brick 15 (as in the right panel of
  Figure~\ref{modelcmdfig}). The upper row shows the actual best fit
  model derived from CMD fitting (left) and the joint probability
  distribution for $\langle A_V\rangle$ and $\sigma$ (right). The lower row
  shows example models corresponding to a factor of 2.53$\times$
  higher mean extinction, as inferred from the \citet{draine2013} dust
  emission model, made by rescaling the median extinction (left) or
  the width of the reddening distribution (right). Clearly, the data
  (shown as blue stars) are not consistent with the higher mean
  extinction characteristic of the emission-based dust model. Note:
  slight differences in the best fit parameter values given on each plot are due
  to difference instances of fitting the same pixel.
  \label{modelcmddrainecomparefig}}
\end{figure*}

We argue that the factor of 2.5 scaling difference is more likely to
have resulted from a problem with the \citet{draine2013}
emission-based maps.  Simply put, there is no obvious way that the
extinction inferred from the CMD can have mistaken the degree of
reddening by such a large factor.  We demonstrate this in
Figure~\ref{modelcmddrainecomparefig}, which shows the CMD of the same
region plotted in Figure~\ref{modelcmdfig}. The underlying density map
shows the CMD expectation for the true best fit model (upper left) and
for a model where the true reddening is $\sim$2.5 times higher than
what we derived (shown in the lower row as an adjustment to the median
extinction $\widetilde{A_V}$ or to the width of the distribution, on
the left and right, respectively).  A higher extinction is completely
at odds with the actual observations.

The only ways out of this conclusion are deeply unsatisfying. For the
CMD-based extinctions to be artificially low at essentially all
pixels, we would have had to (1) mistaken the true location of the
unreddened CMD by many tenths of a magnitude, or (2) invoked a
reddening law with a drastically different relationship between
attenuation and reddening in the NIR.  We can rule out the first
escape route because we have many regions where the vast majority of
stars are unreddened (i.e., small $f_{red}$); in these regions, the
NIR RGB is essentially exactly where our model places it, and is fully
consistent with theoretical isochrones. Moreover, such a color offset
would manifest as a constant bias factor, not a difference in
scaling. The second solution is equally difficult to accept. While
reddening laws are known to vary in the ultraviolet
\citep[e.g.,][]{gordon2003}, there is no evidence for significant
variations in the NIR at the required amplitude.

In contrast to the difficulty explaining a large scale error in the
CMD-based extinctions, there are many possible ways to explain scale
errors in the emission-based extinction maps.  Converting the spectral
energy distribution of the dust emission to a prediction for the
extinction requires calculating the dust mass in the context of a
specific model for the physical properties of the dust (composition,
grain size distribution, etc) and for the starlight radiation field.
The \citet{draine2013} dust map is derived using the
\citet{draine2007} models, which are calibrated to match the emission
of diffuse, dusty regions in the Milky Way
\citep[e.g.,][]{weingartner2001}, with additional adjustments in the
volume of grain material to give better consistency with depletion
measurements of the dust mass per hydrogen atom
\citep{draine2013}. Given the complexity of the models and the
calibration, it is possible that these locally calibrated models do
not apply in a different environment. Indeed, \citet{draine2013}
specifically raises the possibility of their systematically
overestimating the dust mass when discussing differences with
{\emph{Planck}} emission, although they argue that systematic errors
initially appear smaller than a factor of $\sim$1.3.

Additional complications arise from the degeneracy inherent in fitting
dust models to spectral energy distributions, and from the inability
to fully resolve complex, multi-temperature dust distributions
\citep[e.g.,][]{galliano2011}.  Dust emission is highly non-linear
with temperature, and thus the ensemble emission from a collection of
dust clouds with varying temperatures does not accurately reflect the
mean mass-weighted temperature.  There is also evidence that cold dust
may be much more emissive than expected \citep{paradis2009}.  If the
component of high-emissivity dust is pervasive in M31 as well as the
MW, then it may help to explain why the \citet{draine2013} map infers
more extinction than is actually seen.  However, this effect seems
restricted to the molecular phase, which is not dominant in M31.

Comparable mismatches between extinction and the \citet{draine2007}
models have been seen before in the literature. Both \citet{alton2004}
and \citet{dasyra2005} found that multiwavelength radiative transfer
modeling of edge-on galaxies underpredicts the observed sub-mm
emission at 850$\mu$m when using the \citet{draine2007} models.  In
other words, to be consistent with the observed amount of extinction,
either there must be another component of dust not accounted for in
their radiative transfer model, or the dust must produce a factor of
$\sim$3-4$\times$ greater emission at long wavelengths than predicted
by the \citet{draine2007} model. The inverse of this latter
possibility --- i.e., that a given amount emission can be produced by
1/3 the amount of dust -- would be consistent with our results, after
including the model revisions in \citet{draine2013}.  Recent
simulations, however, suggest that the mismatch may instead be due to
asymmetries and inhomogeneities in the disk that affect the outcome
of radiative transfer calculations \citep[e.g.,][]{saftly2015}.

Even more clearly, recent work by \citet{planck2015} showed a nearly
identical discrepancy to the one we identify here.  They measured the
optical extinction observed towards thousands of QSOs observed with
the Sloan Digital Sky Survey and compared it to the extinction
predicted by fitting \citep{draine2007} models to all-sky data from
{\it Planck}, {\it IRAS}, and {\it WISE}. This comparison found that
the \citet{draine2007} extinction estimates were too high by a factor
of $\sim\!2-2.4$. They found a comparable discrepancy when looking at
molecular clouds whose NIR extinctions were mapped using 2MASS by
\citet{schneider2011}.  We compare our results to those in \citet{planck2015}
in more detail in Sec.\ \ref{plancksec} below.

Finally, \citet{lombardi2014} find a factor of $\sim$2 difference between
observations in Orion and the \citet{weingartner2001} predictions for
the ratio of 2.2$\mu$m extinction to the 850$\mu$m opacity. They argue
that this discrepancy is reduced with recent models by
\citet{ormel2011}.  Interestingly, however, the ratio measured by
\citet{planck2015} is actually in much better agreement with
\citet{weingartner2001}, in spite of showing a siginificant difference
between the measured and the predicted extinctions from the similar
\citet{draine2007} models.

In summary, it appears that the most widely used \citet{draine2007}
dust models overestimate the mass of dust needed to produce a given
amount of emission in M31, even after revising the dust masses
downwards using the modifications in \citet{draine2013}. This
discrepancy is particularly troubling, given the ever growing
importance of dust modeling in the era of flagship mid- and far-IR
observatories.  Further application of CMD-based extinction
measurements outside the Milky Way will be invaluable in improving
existing dust models.

\subsection{Small-Scale Variations in Emission- and CMD-based Extinctions: The Impact of the Interstellar Radiation Field} \label{smallscalesec}

\begin{figure*}
\centerline{
\includegraphics[width=3.5in]{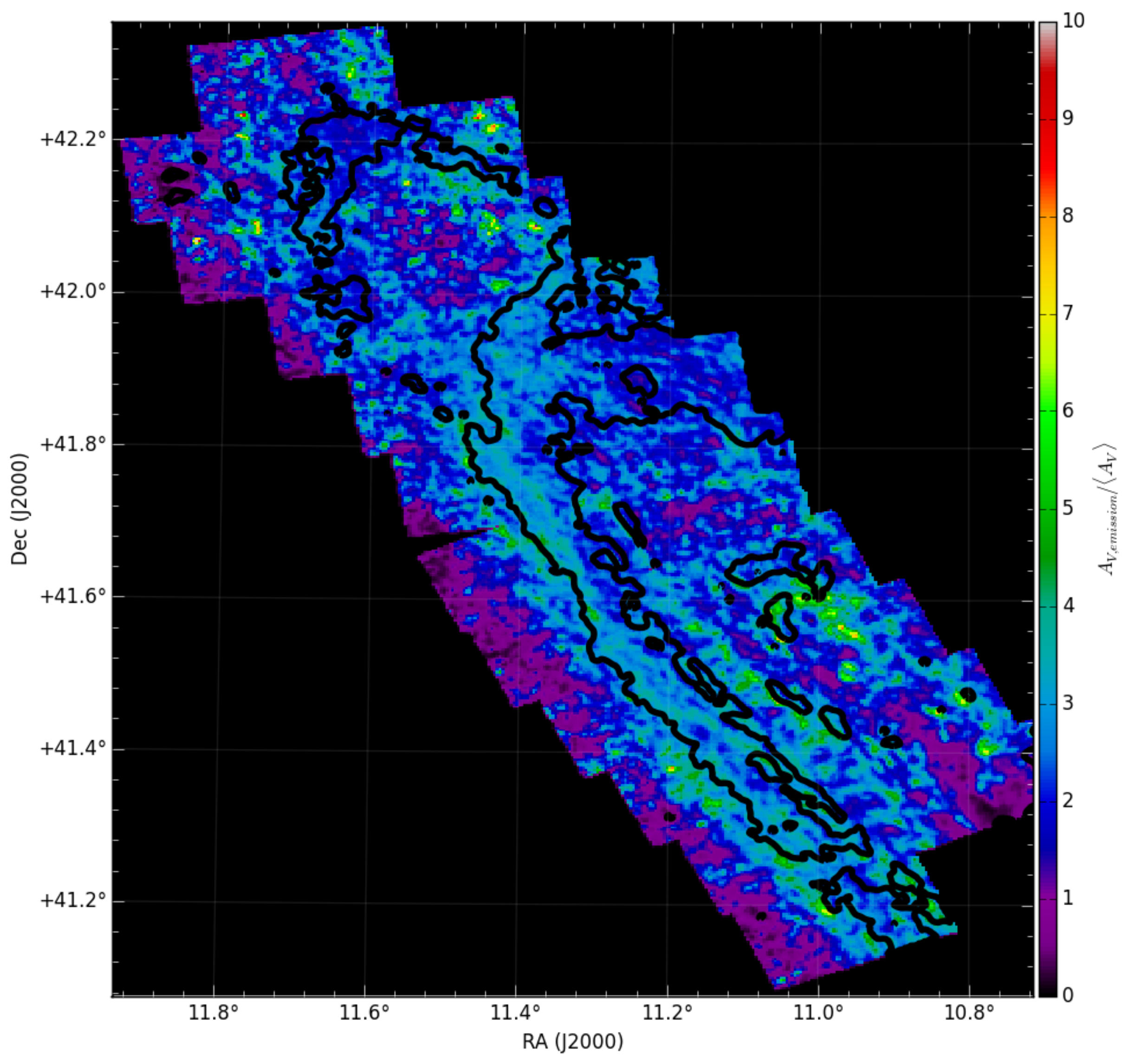}
\includegraphics[width=3.5in]{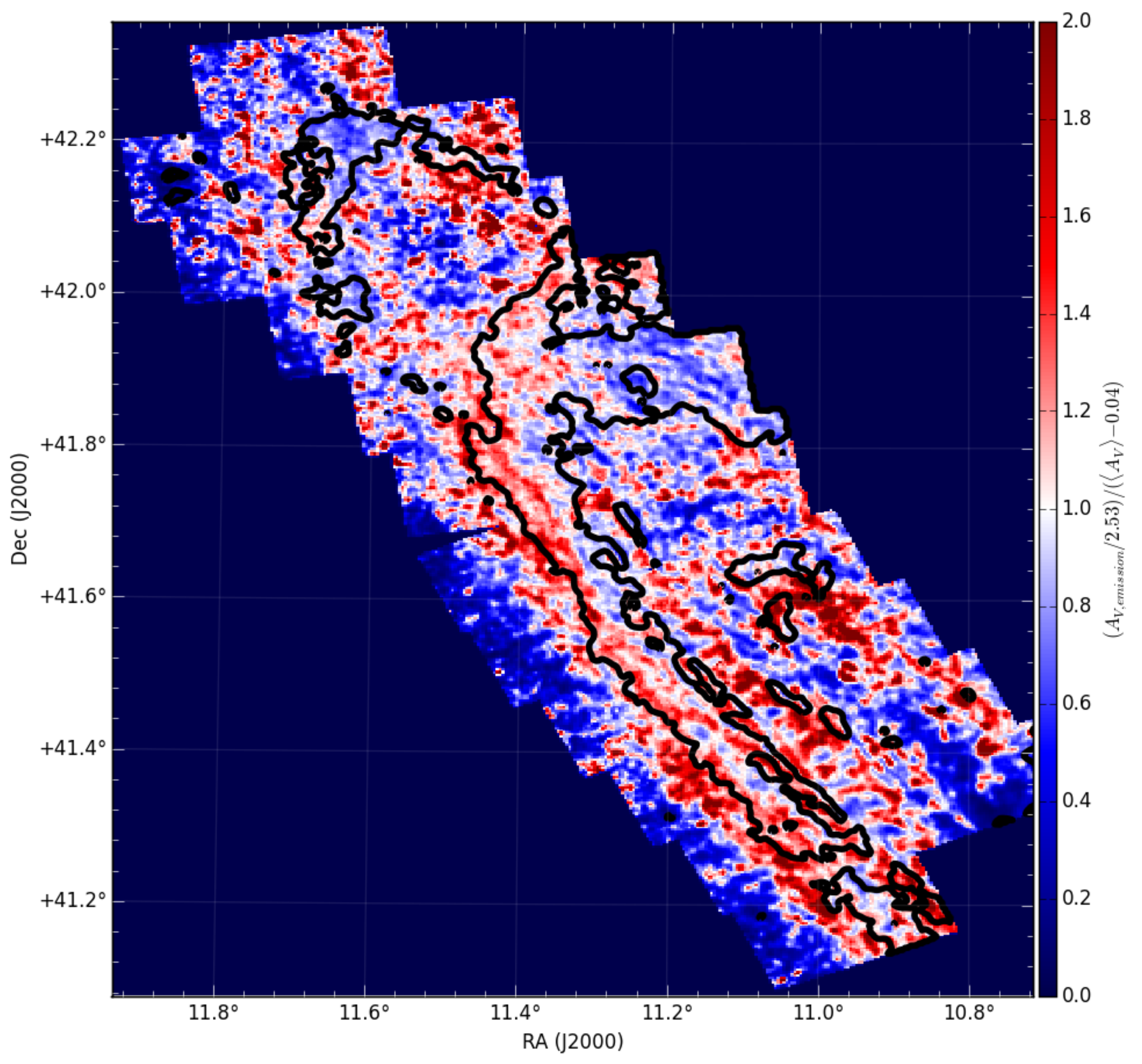}
}
\caption{Images of the pixel-by-pixel ratio of the \citet{draine2013}
  emission-based $A_V$ map divided by the resolution-matched map of
  mean extinction $\langle A_V \rangle$ derived from the NIR CMD,
  before (left) and after (right) applying the scale$+$bias correction
  used in the top-right panel of Figure~\ref{drainecorrelationfig}
  (Sec.\ \ref{emissioncorrsec}). The heavy dark contours indicate
  where $\langle A_V \rangle > 1$.  Inspection of the uncorrected map
  shows that the mismatch between the two maps is clearly spatially
  dependent. The overall difference in scale in the high extinction
  regions falls between 2-3, with some tendency towards higher values
  on the far side of the disk.  The scale correction brings the
  agreement to roughly $\pm30\%$ 
  in regions with $\langle A_V \rangle > 1$.  The
  ratio deviates most strongly from 1 at low to intermediate
  extinctions, due to the expected systematic underestimates in the
  CMD-based extinctions (discussed in Sec.\ \ref{accuracysec}); the
  agreement in these regions is greatly improved by the correction of
  an additive bias factor, although at the cost of worse agreement at
  the very lowest extinctions. The correction is not as effective as
  in the high extinction regions, most likely because the bias factor
  would be expected to vary radially.\label{draineratiomapfig}}
\end{figure*}

In Figure~\ref{draineratiomapfig} we plot the ratio between the
emission-based $A_V$ from \citet{draine2013} and the CMD-based mean
extinction $\langle A_V \rangle$, after smoothing the latter to match
the resolution of the former.  The left panel plots the raw ratio, and
the right panel plots the ratio after correcting the
\citet{draine2013} extinctions by a factor of $R=2.53$ and the
CMD-based mean extinctions by an offset $b=-0.04\mags$.  The dark contour
shows the locus within which $\langle A_V \rangle\!>\!1$ in the smoothed map;
CMD-based extinctions outside this locus will potentially be subject
to larger systematic uncertainties.  The redder regions of the right
hand panel indicate regions where the mismatch between the
\citet{draine2013} and the CMD-based extinctions are larger.  In the
bluer regions the agreement is better, but only approaches good
agreement ($A_{V,emission}/\langle A_V \rangle\!\sim\!1$ in the left
panel) at the very lowest extinctions, where both methods agree there
is very little dust.

Figure~\ref{draineratiomapfig} shows an amount of scatter consistent
with that seen in the pixel-by-pixel correlations shown in
Figure~\ref{drainecorrelationfig}. However, it also shows that the
deviations from the mean ratio are often spatially coherent. In the
following section (\ref{spatialsystematicssec}) we assess the possible
contribution from systematic errors in the CMD-based extinctions, but
before doing so we discuss how such behavior could also be produced by
emission-based extinction measurements.

\paragraph{Non-linearities in Smoothing} To produce
Figure~\ref{draineratiomapfig} we have smoothed the CMD-based dust
map. However, this smoothing is weighted by the column-density of the
dust.  In contrast, the smoothing that takes place during emission
observations is flux-weighted. The emission from dust is 
a strong power of dust temperature, and thus the resolution of a
mid- or far-IR telescope will effectively smooth the dust map
non-linearly, with more weighting given to regions with higher dust
temperatures. This can lead to emission maps missing 
cold dust \citep[e.g.,][]{galliano2011}, and producing systematic
mismatches with the observed extinction when there are strong, sharp
gradients in the dust temperature. This latter situation is likely to
be found on the edges of star-forming regions, such as just beyond the
boundaries of the star-forming ring \citep[see Figure 9
  of][]{draine2013}. Such an effect could be responsible for the more
significant deviations (redder colors) frequently seen just outside
the $\langle A_V \rangle\!=\!1$ contour.

\paragraph{Variations in Dust Composition} Models of dust emission typically
allow the dust composition, and in particular the polyaromatic
hydrocarbon (PAH) content, to vary spatially. \citet{draine2013} fit
for PAH abundance $q_{pah}$ and find strong spatial variations in and
out of M31's star-forming rings, as well as large gradients within the
galaxy. Some of the abundance variations are lessened when variations
in the interstellar radiation field are also included, particularly in
the inner galaxy, but the change in PAH abundance across major
star-forming features appears to persist in the more structured
northern half of the galaxy. Interestingly, the largest red region
outside the $\langle A_V \rangle\!=\!1$ contour is also a region where
the PAH abundance appears anomalously high for such a low column
density region \citep[Figure 11 of][]{draine2013}. It may be possible
that coherent variations in the inferred dust composition may be
propagating into coherent variations in the dust extinction. It would
not be surprising if M31's dust composition varied systematically
between the diffuse gas in the low column density regions and the
denser gas in high column density regions. It is also possible that
some of the bias factor $b$ that we have ascribed to the CMD-fitting
method may instead be a systematic in how the dust composition changes
in the more diffuse ISM. We return to the issue of possible PAH correlations
in Sec.\ \ref{plancksec}.

\paragraph{The Spectrum of the Interstellar Radiation Field} One of
the most interesting features in Figure~\ref{draineratiomapfig} is the
presence of two large regions at high extinction where the ratio
$A_{V,emission}/\langle A_V \rangle$ is much lower than typical -- one
in the $10\kpc$ star-forming ring just west of the major axis, and one
in the outermost spiral arm along the major axis. These two regions
are the locations of two well-known recent star formation events.
They can clearly be seen in Figure~13 of \citet{lewis2015}, which
plots the ``Scalo $b$'' parameter, defined as the current star
formation rate ($SFR$) divided by the past average SFR.  This
parameter is closely associated with the hardness of the underlying
spectrum. As the Scalo $b$ parameter increases, so does the fraction
of young stars, which will increase the contribution from hot stars to
the underlying interstellar radiation field, potentially changing the
efficiency of dust heating for a given bolometric energy density in
the interstellar radiation field.

\begin{figure}
\centerline{
\includegraphics[width=3.75in]{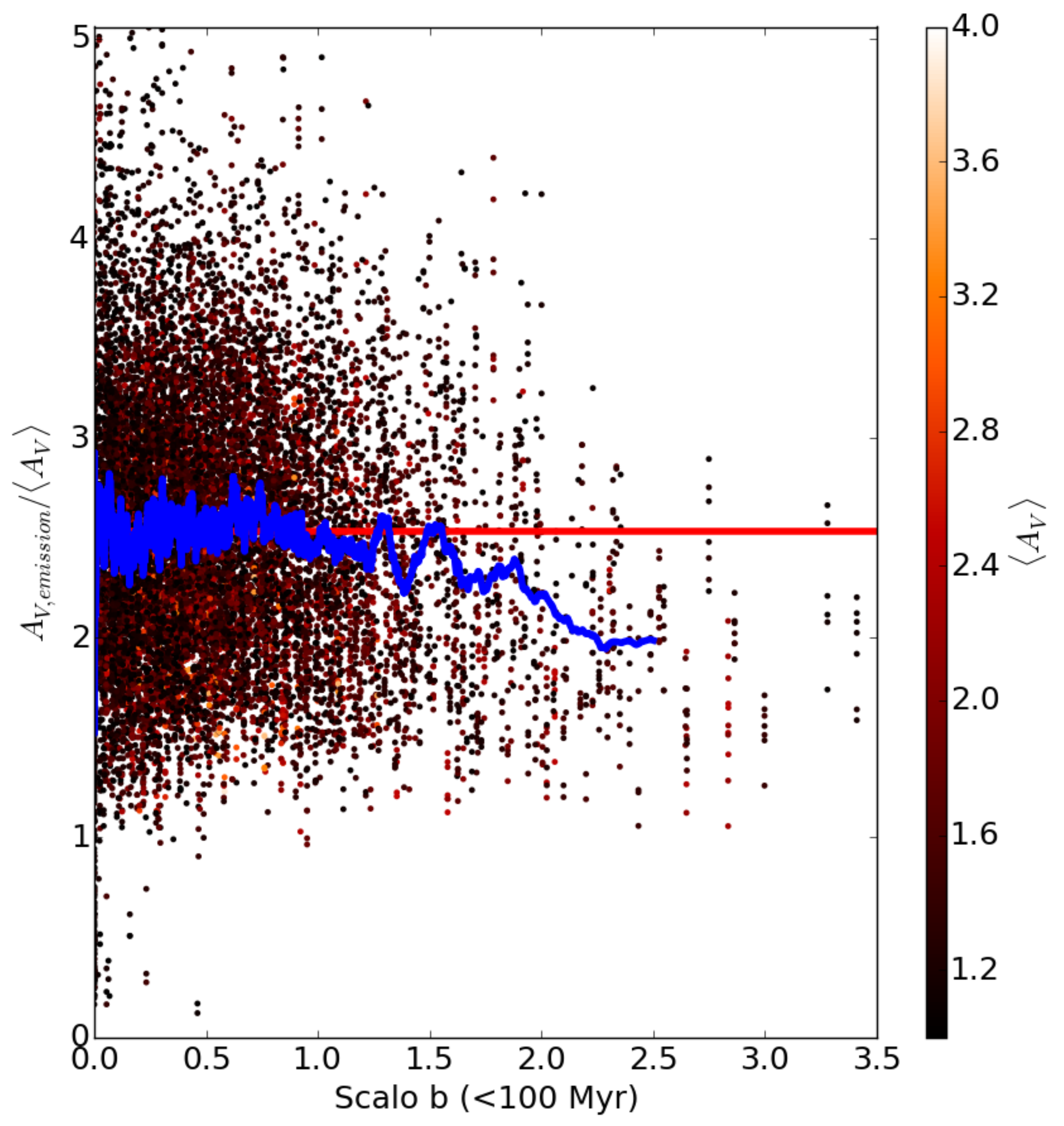}
}
\caption{The observed pixel-by-pixel ratio of the \citet{draine2013}
  emission-based $A_V$ map divided by the resolution-matched map of
  mean extinction $\langle A_V \rangle$ derived from the NIR CMD
  (i.e., Figure~\ref{draineratiomapfig}), as a function of the Scalo
  $b$ parameter ($\equiv\!SFR_{100}/\langle SFR \rangle$) from
  \citet{lewis2015}. The ratio is systematically smaller as the
  contribution from hot young stars increases, shown as the running
  average plotted in blue. For $SFR_{100}/\langle SFR \rangle \gtrsim
  2$, almost all of the pixels are below the globally measured value
  of $\sim\!2.5$ (Figure~\ref{drainecorrelationfig}), shown as the red
  horizontal line. Points are color-coded according to the mean
  CMD-based extinction.
  \label{scalobfig}}
\end{figure}

To explore a possible empirical correlation between the spectrum of
the underlying stars and the spatial variation of
$A_{V,emission}/\langle A_V \rangle$, in Figure~\ref{scalobfig} we
plot the observed ratio of dust extinctions as a function of the Scalo
$b$ parameter ($= SFR_{100}/\langle SFR \rangle$) where the
``current'' star formation rate $SFR_{100}$ is integrated over the
past $100\Myr$ using the spatially-resolved recent star formation
history derived in \citet{lewis2015}, derived on 100 pc bins.  We also
plot a rolling average constructed by taking the mean of 200
sequential points ranked by the Scalo $b$ parameter.  We restrict
this comparison to regions where $\langle A_V \rangle\!>\!1$, which
should better sample the denser neutral and molecular ISM.

Figure~\ref{scalobfig} reveals that there is a systematic drop in
$A_{V,emission}/\langle A_V \rangle$ that begins when the current star
formation rate becomes higher than the past average star formation
rate. At lower fractions of young stars, the mean ratio is
$\sim\!2.5$, but this drops to $\sim\!2$ in regions where the fraction
of young stars is very high ($SFR_{100}/\langle SFR \rangle \gtrsim
2$).  This trend is not nearly as strong when we limit the measurement
of the current star formation rate to only the past $25\Myr$ (not
shown), indicating that the cumulative effect of star formation over a
longer timescale is more important in shaping the dust emission than
just the short term impact of O- and massive B-stars. The observed
trend cannot be due to contamination from younger stars on the RGB
discussed earlier in Sec.\ \ref{systematicsec}, which would tend to
reduce the CMD-based extinction and increase the plotted ratio, which
is the opposite of what is seen in Figure~\ref{scalobfig}.

Figure~\ref{scalobfig} also points to an interesting path forward.
The recent star formation histories available in \citet{lewis2015} in
principle allow the spectrum of the interstellar radiation field to be
calculated indpendently of the dust emission. By assuming a vertical
distribution, the density of the interstellar radiation field could
also be calculated.  Including these independent measurements would
remove a significant uncertainty in modeling the dust emission, and
would potentially lead to improvements in modeling the dust itself.
This test would be particularly interesting in other higher star
formation rate intensity galaxies in the Local Group, including M33
and the Magellanic Clouds.

\subsection{Spatially-Dependent Systematics} \label{spatialsystematicssec}

The power of the CMD-based extinctions is that they present a direct,
independent route to constraining the dust mass, with an angular
resolution that is currently significantly higher than other
techniques. However, when comparing to other ISM tracers --- such as
one may want to do to derive gas-to-dust ratios, or to constrain the
physical models of the dust --- we have to be wary of possible
systematic errors.

We have previously discussed a number of possible systematic errors
that may affect the CMD-based extinctions (Sec.\ \ref{limitationsec}
\& \ref{systematicsec}).  The majority of these possible systematics
are due to uncertainties that are on the order of a few hundredths
of a magnitude in the NIR color distribution of the unreddened RGB.
While these are extremely small effects in terms of the CMD, they lead
to uncertainties of several tenths in the visual extinction, given
that $A_V = 7.56\,E({\rm F110W}-{\rm F160W})$.

One feature of the majority of the systematic errors, however, is that
they should only produce either smooth large-scale variations, or
variations that correlate with individual WFC3/IR chips, while leaving
the morphology of small scale pixel-to-pixel maps intact.
One class of possible large scale variations we may expect 
is one that correlates with the local stellar surface density.
The unreddened model we adopted was constructed in bins of local
surface density to simultaneously capture photometric errors,
photometric depth, and smooth stellar population changes with radius
(see Sec.\ \ref{unreddenedsec} and Figure~\ref{surfdensmapfig}).  We
therefore could have biases that depend on $\log_{10}\Sigma_{stars}$
if there are specific density bins in which the model of the
unreddened CMD is incorrect.  Possible origins of such a bias could
be: (1) the need to interpolate between neighboring bins when a
density range contains few unreddened regions; (2) failure of
$\log_{10}\Sigma_{stars}$ to correlate perfectly with radial changes
in the stellar population, as might be expected in the rapidly
changing, structurally complex inner regions; or (3) loss of
sensitivity to low extinction in the high surface density inner galaxy
where the RGB is broadened by crowding and/or a large metallicity
spread.

A second possible class of large scale systematic is one that varies
with distance from the major axis. The stellar disk of M31 is thick
(as we show in a companion paper), and as such there can be a range
of radii present in projection at any given position in the sky
(``projection mixing''; Sec.\ \ref{limitationsec}). This effect is
non-existent along the major axis, where the projected radius equals
the true radius.  However, the farther one moves from the major axis,
the larger the fractional range of radii that are projected onto the
same location. Any radial gradient in the stellar population will
affect the morphology of the RGB, and thus the superposition of a
large range in radii can lead to offsets and spreads in the apparent
RGB.  This effect would lead to values of $A_V$ that are biased in
regions that are further from the major axis\footnote{The amplitude of
  the effect may also have a secondary dependence on azimuth in the
  inner regions of the galaxy, where the fraction of stars that are
  due to M31's bulge and/or inner spheroid can become significant, and
  would not be properly captured by a model of a pure inclined disk.}.

We can assess these two large scale effects by looking at the
spatially-resolved correlations between the CMD-based extinctions and
other dust tracers.  Specifically, we break down our maps of the mean
extinction into bins of stellar surface density and reddening
fraction.  The latter is an excellent proxy for the spread of radii
present at a given position, because the same spread is directly
responsible for deviations from $f_{red}\!=\!0.5$.  We assign regions to
bins of $\log_{10}\Sigma_{stars}$ and $f_{red}$ using the models
developed when constructing the unreddened CMD and the prior on
$f_{red}$, respectively (Secs.\ \ref{unreddenedsec} \&
\ref{fittingsec}).

\begin{figure*}
\centerline{
\includegraphics[width=3.5in]{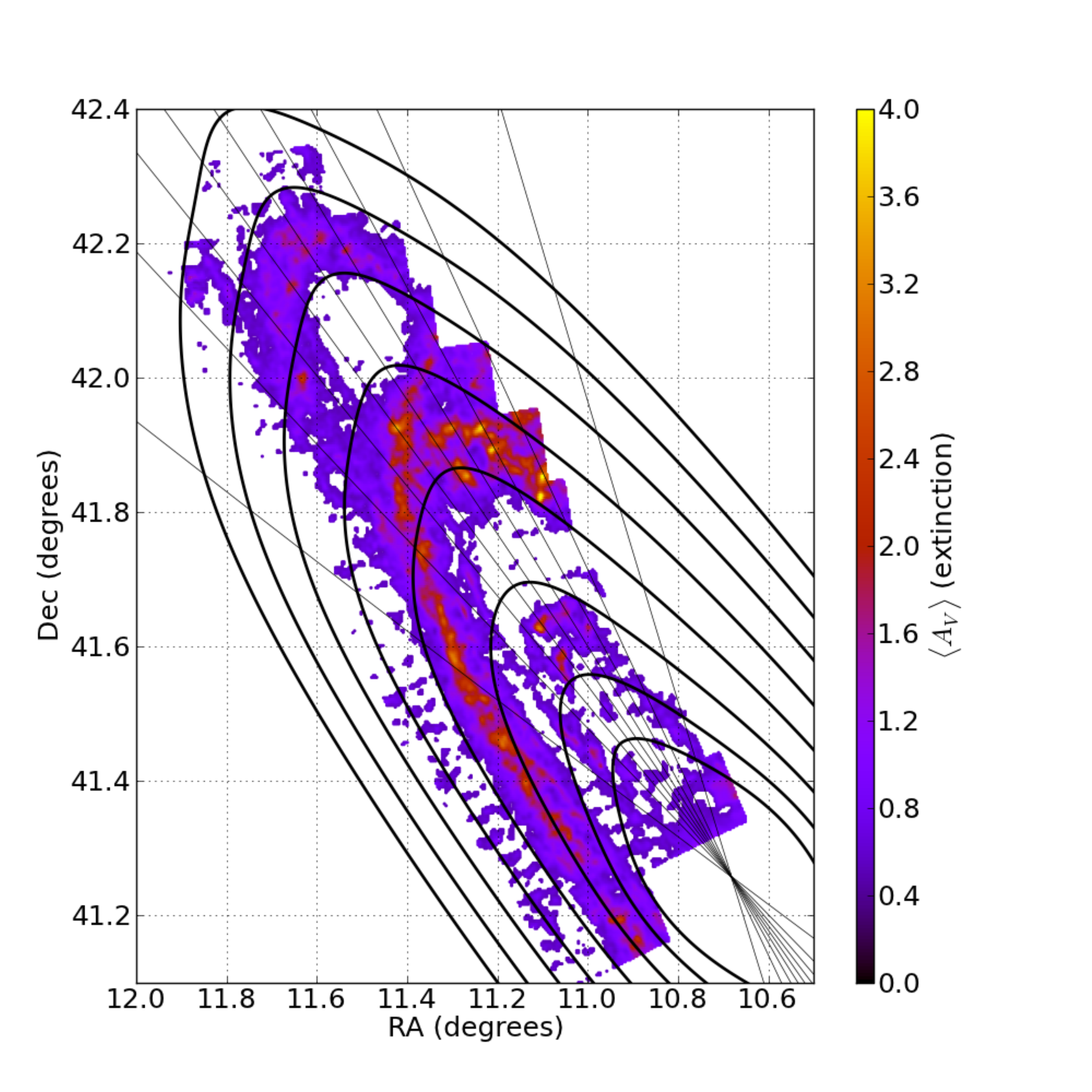}
\includegraphics[width=3.5in]{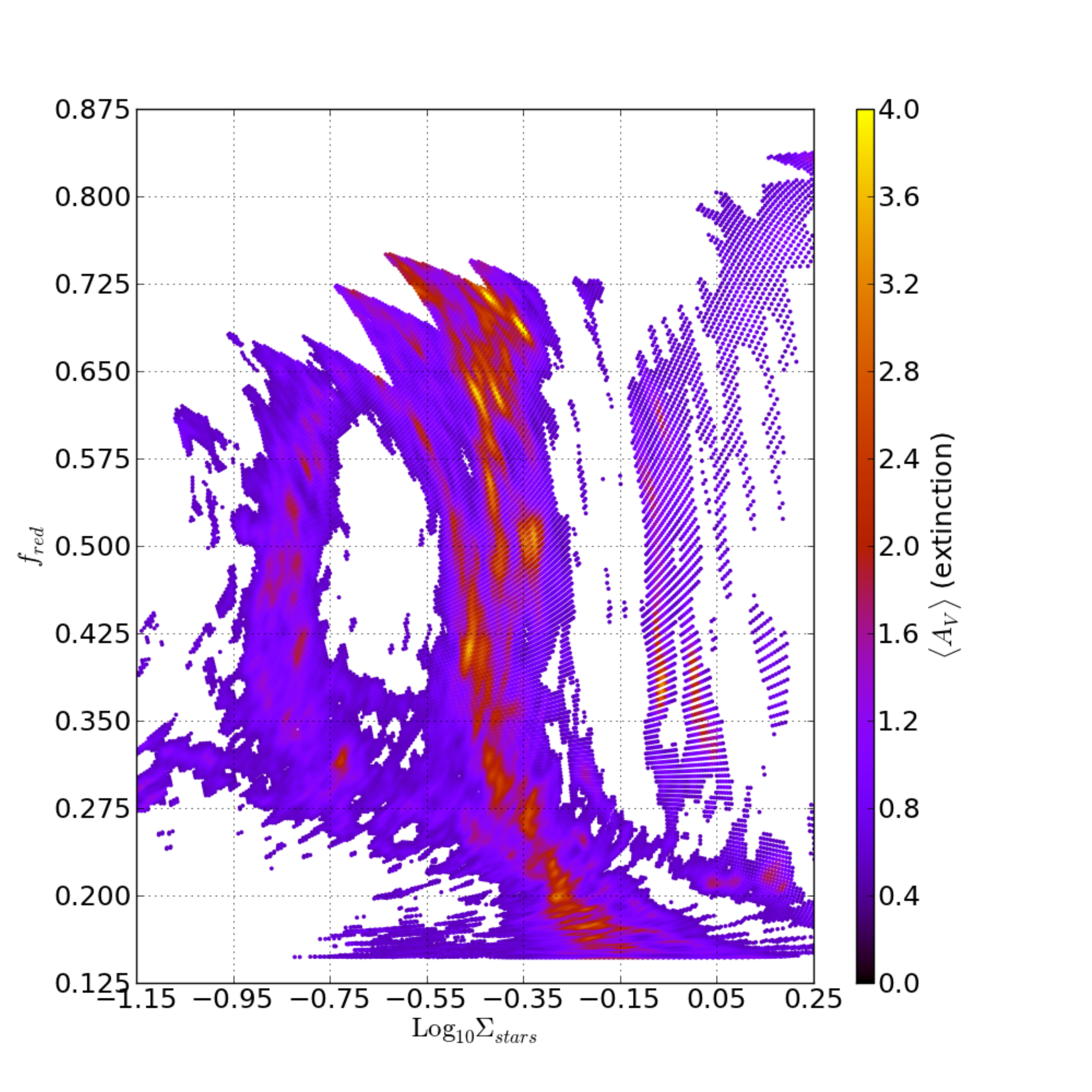}
}
\caption{Maps of the mean extinction $\langle A_V \rangle$ in terms of
  position on the sky (left), and transformed into coordinates
  tracking the local stellar surface density and the fraction of
  reddened stars (right); the latter quantity is a proxy for the range
  of radii that fall in a given line of sight. Only pixels with high
  extinction are plotted ($\langle A_V \rangle>0.5$). The heavy solid
  lines on the left panel show lines of constant
  $\log_{10}\Sigma_{stars}$, and the light diagonal lines indicate
  constant $f_{red}$ for the same coordinate grid used in the right
  panel.  \label{fredvslgnstarfig}}
\end{figure*}

The resulting transformation of the extinction map is shown in
Figure~\ref{fredvslgnstarfig}, restricted to regions where $A_V$ is
likely to well-measured ($\langle A_V \rangle>0.5\mags$).  The left
hand plot shows the more familar map, with the locii of constant
$\log_{10}\Sigma_{stars}$ shown as heavy contours, and the $f_{red}$
coordinate shown as ``spokes'' emanating from the center of the
galaxy. The right hand plot shows the extinction distribution remapped
into the $\log_{10}\Sigma_{stars}$ vs $f_{red}$ coordinate system.

\begin{figure*}
\centerline{
\includegraphics[width=7.25in]{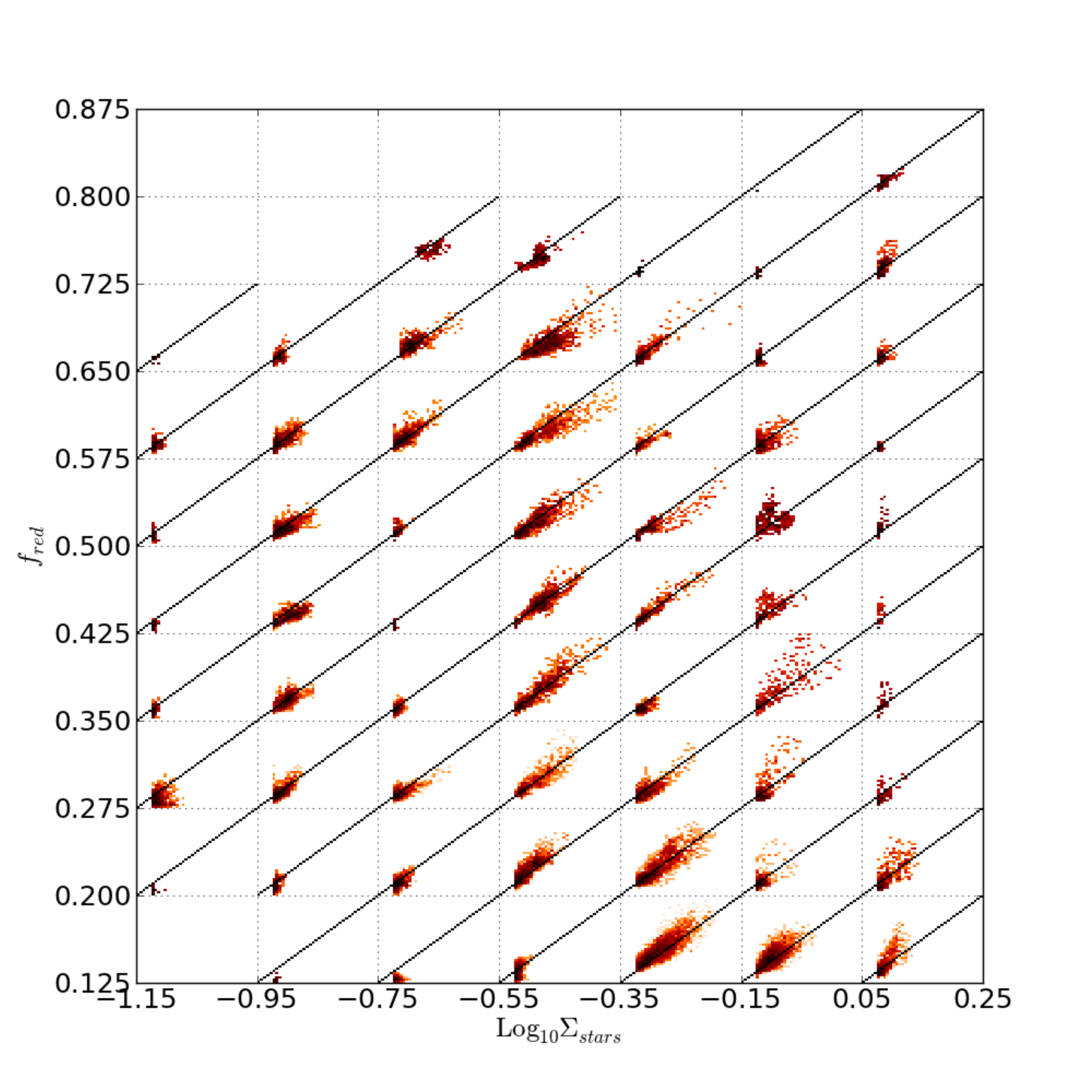}
}
\caption{Correlations between the CMD-based extinction and
  resolution-matched predictions for $A_V$ derived from emission
  (after correcting for $A_{V, emission}/\langle A_V\rangle =
  2.53$; see Sec.\ \ref{emissionsec}), calculated in bins of
  $\log_{10}\Sigma_{stars}$ and $f_{red}$, using the coordinate
  divisions shown in Figure~\ref{fredvslgnstarfig}. Each subpanel
  contains a plot equivalent to the upper right of
  Figure~\ref{drainecorrelationfig}, with the CMD-based extinction on
  the horizontal axis and the emission-based extinction on the
  vertical axis. In each subpanel, $A_V$ is plotted between 0 and
  $4\mags$, and data with $A_{V,emission}<0.5\mags$ is not
  shown. Perfect correlations would fall along the diagonal lines.  In
  every pixel, the correlations are strong. However, there are
  subtle differences in the slopes, depending on the specific
  values of $\log_{10}\Sigma_{stars}$ and $f_{red}$. These shifts are
  discussed in detail in
  Sec.\ \ref{spatialsystematicssec}. \label{AVcorrfig}}
\end{figure*}

In Figure~\ref{AVcorrfig} we plot the pixel-by-pixel correlations
between the mean CMD-based extinction $\langle A_V \rangle$
(restricted to where $\langle A_V \rangle > 0.5$) and the
scale-corrected emission-based extinction $A_{V,emission}/2.53$. Each
individual correlation sub-panel is calculated for pixels within one
of the bins of $\log_{10}\Sigma_{stars}$ versus $f_{red}$ shown in the
right panel of Figure~\ref{fredvslgnstarfig}. Thus, each sub-panel is
equivalent to the upper right panel of
Figure~\ref{drainecorrelationfig}, but restricted to a subregion of
the galaxy.

The correlations in Figure~\ref{AVcorrfig} can be used to look for
large-scale spatially-dependent variations in the systematic bias and
in the factor of 2.53 scale factor used to correct $A_{V,emission}$ to a
better estimate of $A_V$. In each
range of $\log_{10}\Sigma_{stars}$ and $f_{red}$, the diagonal line
shows the expectation for a perfect 1:1 correlation.  Data will fall
along this line wherever $A_{V,emission}/2.53$ is a good estimator of $A_V$
and there are no systematic biases in either extinction
measurement. If the scale factor is correct, but there is a
systematic bias, then the data would have the same slope as the 1:1
line, but would be shifted horizontally or vertically.

\subsubsection{Spatial Variations in $\langle A_V \rangle$ Correlations}

In general, Figure~\ref{AVcorrfig} suggests that the CMD- and
emission-based extinctions are well correlated throughout the galaxy,
with no strong positional-dependent biases like those discussed in
Sec.\ \ref{limitationsec} \& \ref{systematicsec}.  There are two
notable exceptions.

The first is seen at very high surface densities where the range of
emission-based extinctions is much larger than the CMD-based
extinctions.  This difference is most probably due to the CMD-based
extinction's lack of sensitivity in the very inner disk, where the CMD
is too broad to be sensitive to any but the highest extinctions
(Figure~\ref{RGBgradientfig}).  It may also be related to an
inappropriate model for the unreddened CMD, due to the inner region's
complex stellar populations, which lead to large spatial variations in
the fractions of M31's disk, bulge, bar, and inner spheroid.

The second place where the correlation between the two extinction
measures breaks down is at the very lowest surface densities. 
These regions are the origin of the small tail of stars
with spuriously high CMD-based extinctions seen in
Figure~\ref{drainecorrelationfig}; most of these are in the region of
bad photometry in Brick 22, or in the very outskirts of the disk where
there are both few reddened stars and low extinctions, which leads to
systematic errors (primarily due to systematic offsets in the
photometry across the WFC3/IR chip) being more noticeable.

\begin{figure}
\centerline{
\includegraphics[width=3.85in]{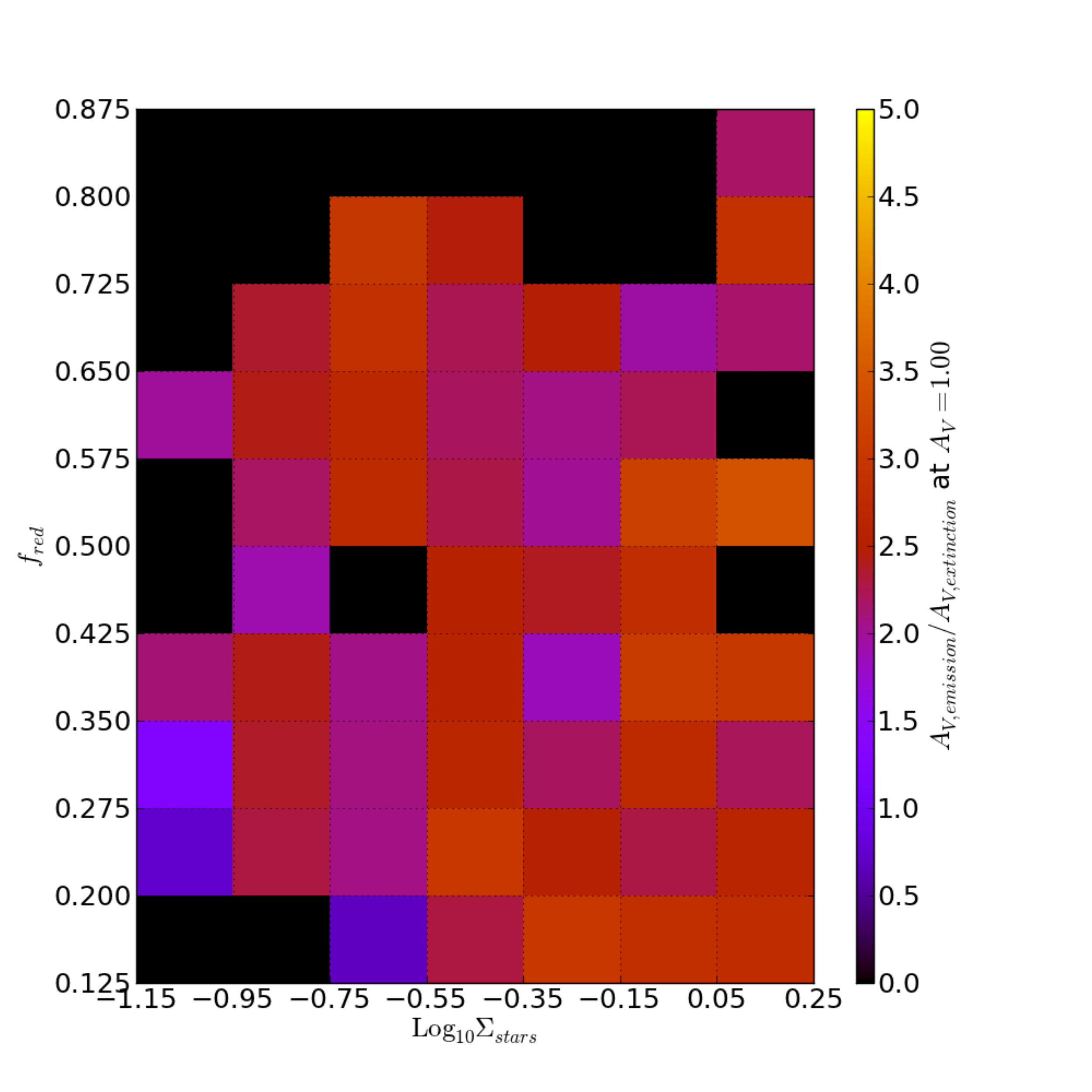}
}
\caption{Ratio between resolution-matched predictions for $A_V$
  derived from emission (before correcting $A_{V, emission}$; see
  Sec.\ \ref{emissionsec}) and the CMD-based extinction, calculated
  in bins of $\log_{10}\Sigma_{stars}$ and $f_{red}$, using the
  coordinate divisions shown in Figure~\ref{fredvslgnstarfig}. The
  ratio is inferred at $\langle A_V \rangle=1$ from fitting the
  correlations shown in Figure~\ref{fredvslgnstarfig}, but without
  applying the factor of 2.53 correction to the dust-emission based
  extinctions; only bins whre the data extends to at least
  $\langle A_V \rangle=0.75$ are shown. The color bar spans from zero to
  twice the expected ratio. The spatial variations in the ratio are
  modest for the dust-emission estimates. \label{AVratioat1fig}}
\end{figure}

To quantify the trends in Figure~\ref{AVcorrfig}, in
Figure~\ref{AVratioat1fig} we show the ratio
$A_{V,emission}/\langle A_V \rangle$ at
$\langle A_V \rangle=1$, as derived from fitting a linear function to the
correlation in each subpanel of Figure~\ref{AVcorrfig} (i.e., as a
function of $\log_{10}\Sigma_{stars}$ and $f_{red}$), after removing
the factor of $\sim\!2.5$ correction to the \citet{draine2013}
extinctions.  The plot has been scaled such that the overall ratio is
in the middle of the color bar, and the range goes from zero to twice
the observed ratio.

Figure~\ref{AVratioat1fig} confirms our impressions from
Figure~\ref{AVcorrfig}, namely that the ratio of $A_V$'s is relatively
uniform, except for the previously noted problems at very low surface
densities.  

\subsection{Testing the \citet{planck2015} Exctinction Estimator} \label{plancksec}

The \citet{planck2015} studies of reddening towards Milky Way
molecular clouds and towards background QSOs strongly suggest that
some of the spatial variation in $A_{V,emission}/\langle A_V \rangle$
that we see in Figures~\ref{draineratiomapfig}~\&~\ref{AVcorrfig} may
be correlated with a parameter related to the background radiation
field. In the physical dust models of \citet{draine2007}, the dust is
heated by a distribution of radiation field intensities, described by
a $\delta$-function at the minimum intensity $U_{min}$ and a power-law
distribution extending from $U_{min}$ to $U_{max}$ with slope
$\alpha$. \citet{planck2015} argue that the dependence of the observed
extinction on $U_{min}$ suggests that in practice $U_{min}$ is not
solely tracing the minimum radiation field intensity, but instead is
also compensating for variations in the properties of the underlying
dust, some of which may be manifesting in departures from the spectral
shape of the dust opacity expected in the {\it Planck} 857, 545, and
353 bands. Although these band are not used in the \citet{draine2013}
fits, the SPIRE 350$\mu$m and 500$\mu$m bands cover comparable
wavelengths as the first two of these {\it Planck} bandpasses and both
papers produce comparable dust mass estimates in M31 \citep[Appendix A
  of][]{planck2015}.  Thus, the \citet{draine2013} fits are likely to
be subject to the same effects.

\citet{planck2015} argue that $U_{min}$ can be used to make more
accurate estimates of the expected extinction using the results of
\citet{draine2007} fitting. They propose an empirical linear formula to
predict the extinction:

\begin{equation}
  A_{V,emission,Planck} = (a\,U_{min} + b) \times A_{V,emission}
\end{equation}

\noindent where $a=0.31$ and $b=0.35$ for the Milky Way's diffuse gas
and $a=0.33$ and $b=0.27$ for Milky Way molecular clouds. At a typical
value of $U_{min}\!=\!0.5$, $A_{V,emission}/A_{V,emission,Planck}$
would have values of 2.0 and 2.3 for the diffuse ISM and molecular
clouds, respectively.  These values are comparable to the overall
factor of 2.53 overestimate we derive here.

\begin{figure*}
\centerline{
\includegraphics[width=3.5in]{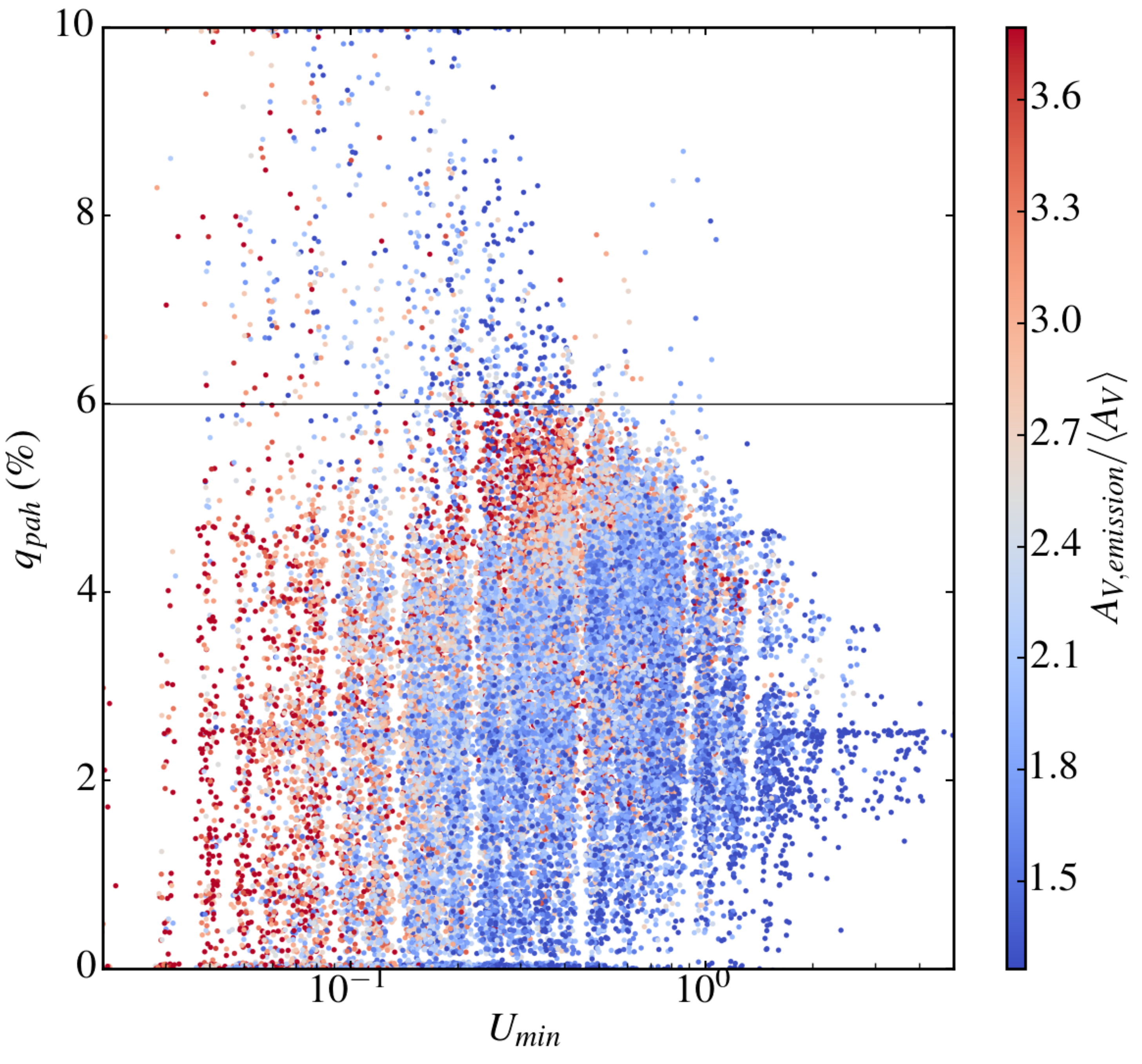}
\includegraphics[width=3.5in]{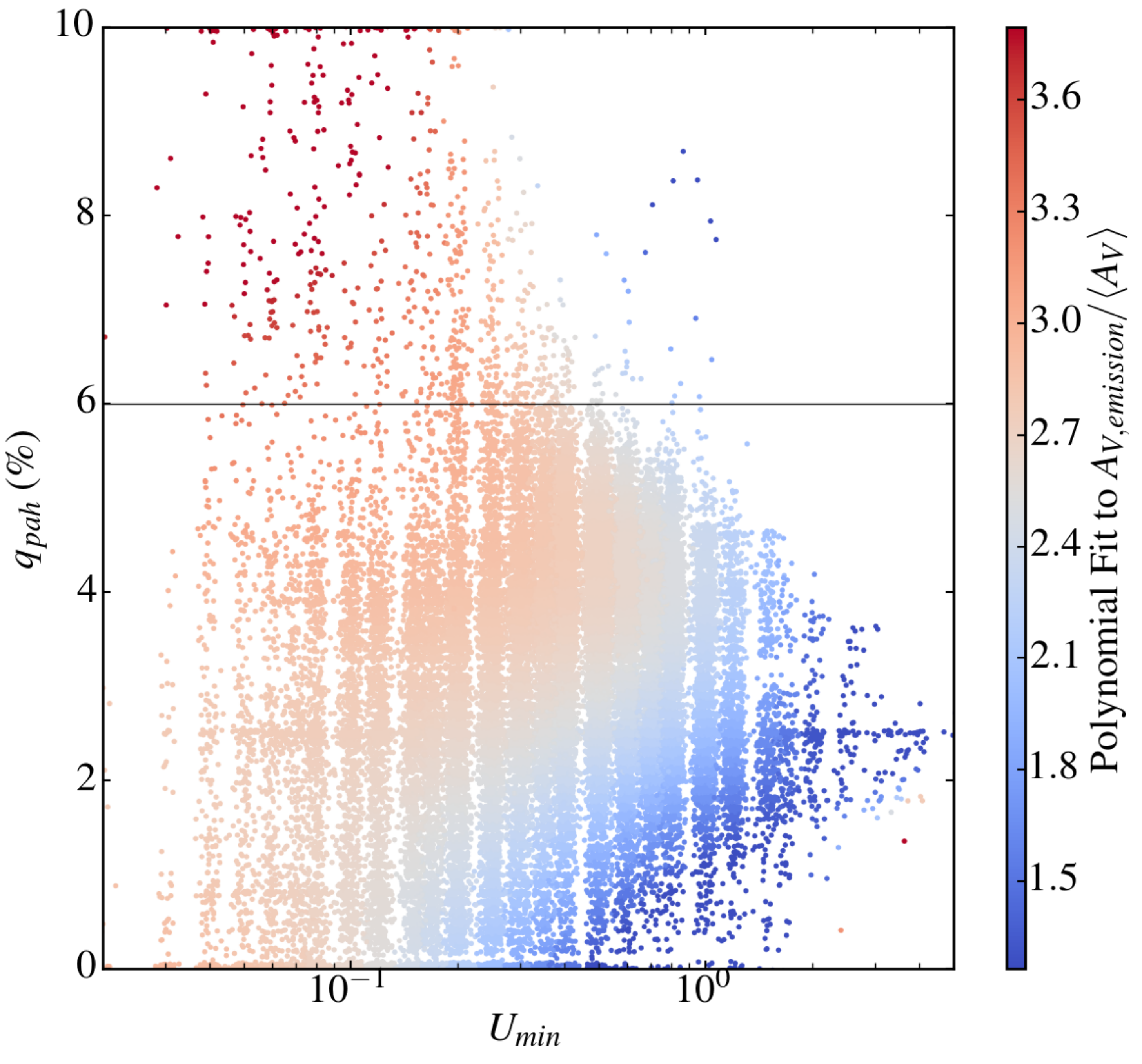}
}
\caption{The ratio between the emission-based extinction and the mean
  CMD-based extinction, as a function of $U_{min}$ and $q_{pah}$
  derived in \citet{draine2013}.  The left panel shows the data and
  the right panel shows the results of a 2nd order polynomial fit. The
  data show that the observed ratio depends on both $U_{min}$ and
  $q_{pah}$, not just on $U_{min}$ as was assumed in
  \citet{planck2015}. The range of the color scale is chosen such that
  the grey value indicates the mean ratio of $A_{V,emission}/\langle
  A_{V}\rangle=2.53$ found in Sec.\ \ref{emissioncorrsec}.  The
  data is only plotted for points with $A_{V,emission}>0.5$, where
  uncertainties and biases in both methods are smaller. The fit was
  restricted to regions where $q_{pah}<6$\%, because the regions with higher
  $q_{pah}$ are all found at the outer limits of where
  \citet{draine2013} considers their fits to be reliable.
  For clarity, small random offsets have been added to both $U_{min}$ and
  $q_{pah}$ to compensate for the quantization of the original data and to
  better show the individual values of $A_{V,emission}/\langle
  A_{V}\rangle$. \label{Uminvsqpahfig}}
\end{figure*}

We can look for a similar linear dependence of $A_{V,emission}$ on
$U_{min}$ within our data for M31.  In Figure~\ref{Uminvsqpahfig} we
plot $U_{min}$ versus the percentange of the dust mass in PAHs from
\citet{draine2013}, $q_{pah}$.  Each pixel overlapping PHAT is
color-coded by its measured value of $A_{V,emission}/\langle A_V
\rangle$.  We have excluded regions where $A_{V,emission}<0.5\mags$,
to eliminate piexls where the emission and CMD-based extinctions will
have the largest uncertainties.

Consistent with \citet{planck2015}, the left panel of
Figure~\ref{Uminvsqpahfig} shows a tendency for
$A_{V,emission}/\langle A_V \rangle$ to be overestimated where
$U_{min}$ is low.  However, the value of $A_{V,emission}/\langle A_V
\rangle$ also appears to depend on $q_{pah}$ as well.  The right panel
of Figure~\ref{Uminvsqpahfig} shows a 2nd order polynomial fit to the
data with $q_{pah}<6$\%; we exclude the highest values of $q_{pah}$
because they are all found in a region that is very close to the limit
where \citet{draine2013} consider their fits to no longer be robust to
uncertainties in background subtraction.  The resulting fit provides
an alternate way to estimate the extinction from the
\citet{draine2007} dust models:

\begin{equation}  \label{Uminqpahfiteqn}
\begin{split}
  A_{V,emission}/\langle A_V \rangle = 3.09 - 0.24 q_{pah} + 0.05 q_{pah}^2\\
  - 4.77 U_{min} - 1.26 U_{min}^2 \\
  + 2.45 U_{min} q_{pah} - 0.34 U_{min} q_{pah}^2 \\
  - 1.26 U_{min}^2 q_{pah} + -0.18 U_{min}^2 q_{pah}^2,
\end{split}
\end{equation}

\noindent where $q_{pah}$ is expressed as a percentage (rather than a fraction).
We also provide a simpler first-order linear correction that produces
an adequate fit to the data, but that leaves slightly larger systematic residuals:

\begin{equation} 
\begin{split}
  A_{V,emission}/\langle A_V \rangle = 2.57 + 0.08 q_{pah} - 1.24 U_{min} \\
                                        + 0.17 U_{min} q_{pah}.
\end{split}
\end{equation}

\begin{figure*}
\centerline{
\includegraphics[width=3.5in]{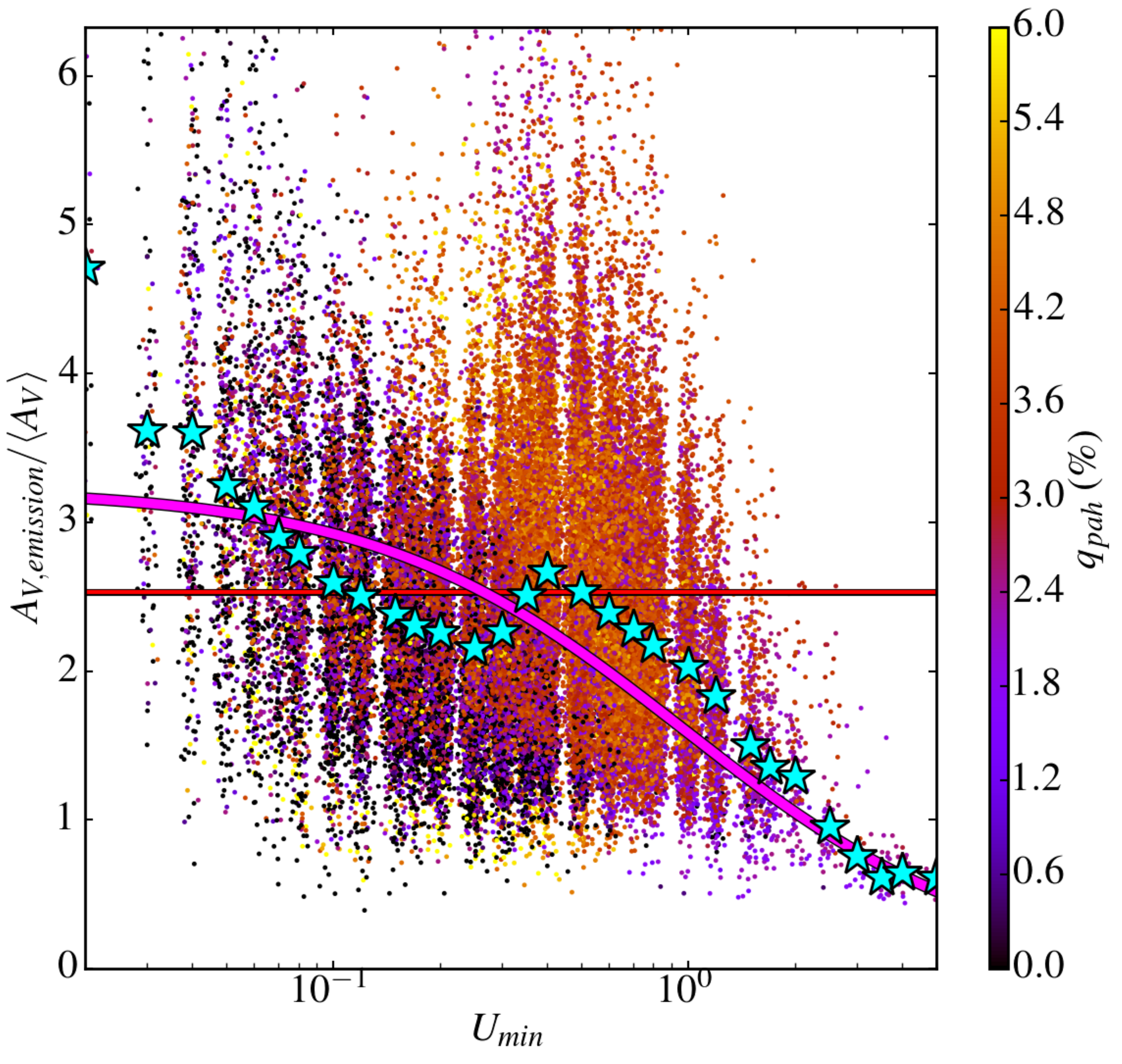}
\includegraphics[width=3.5in]{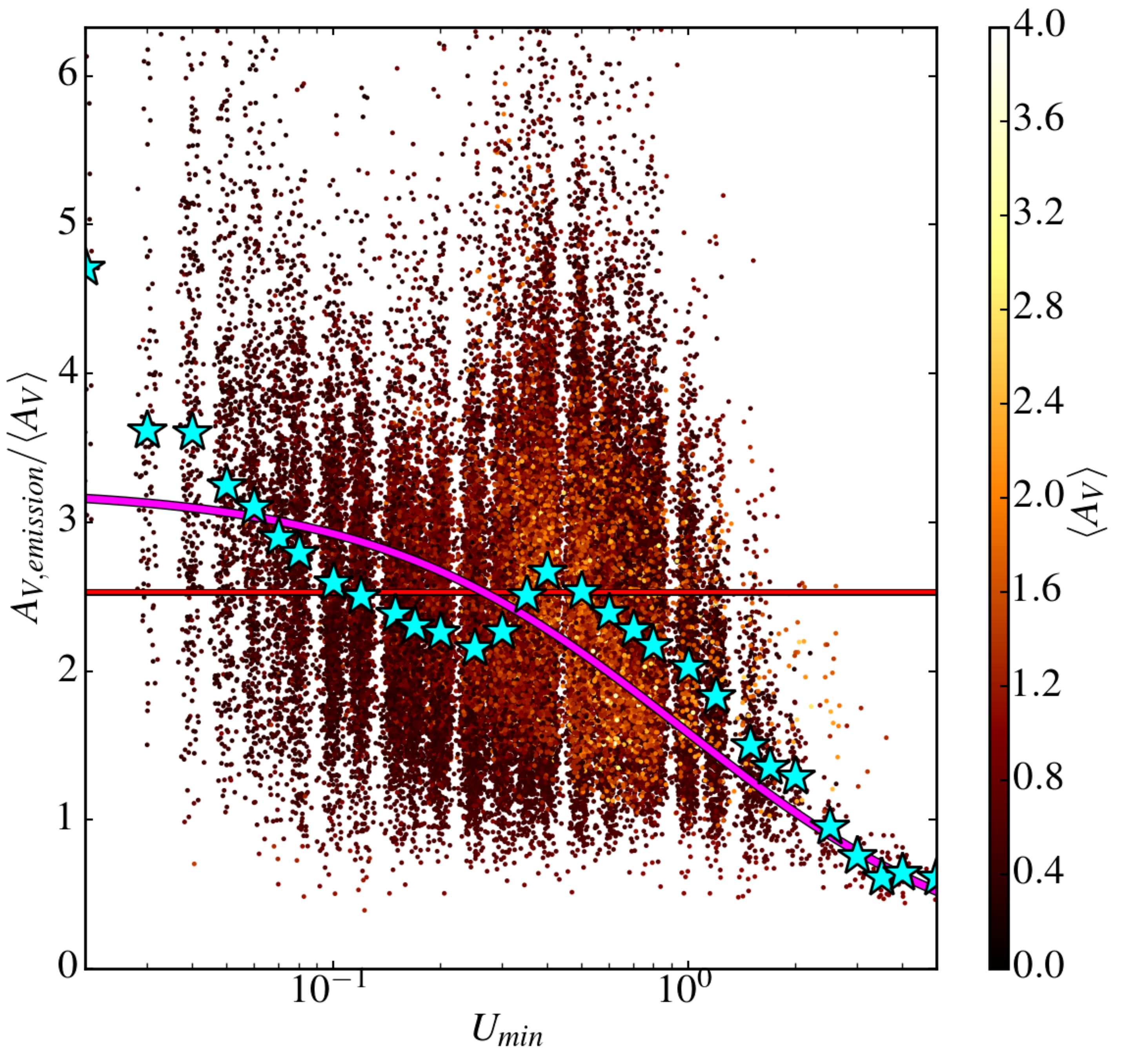}
}
\centerline{
\includegraphics[width=3.5in]{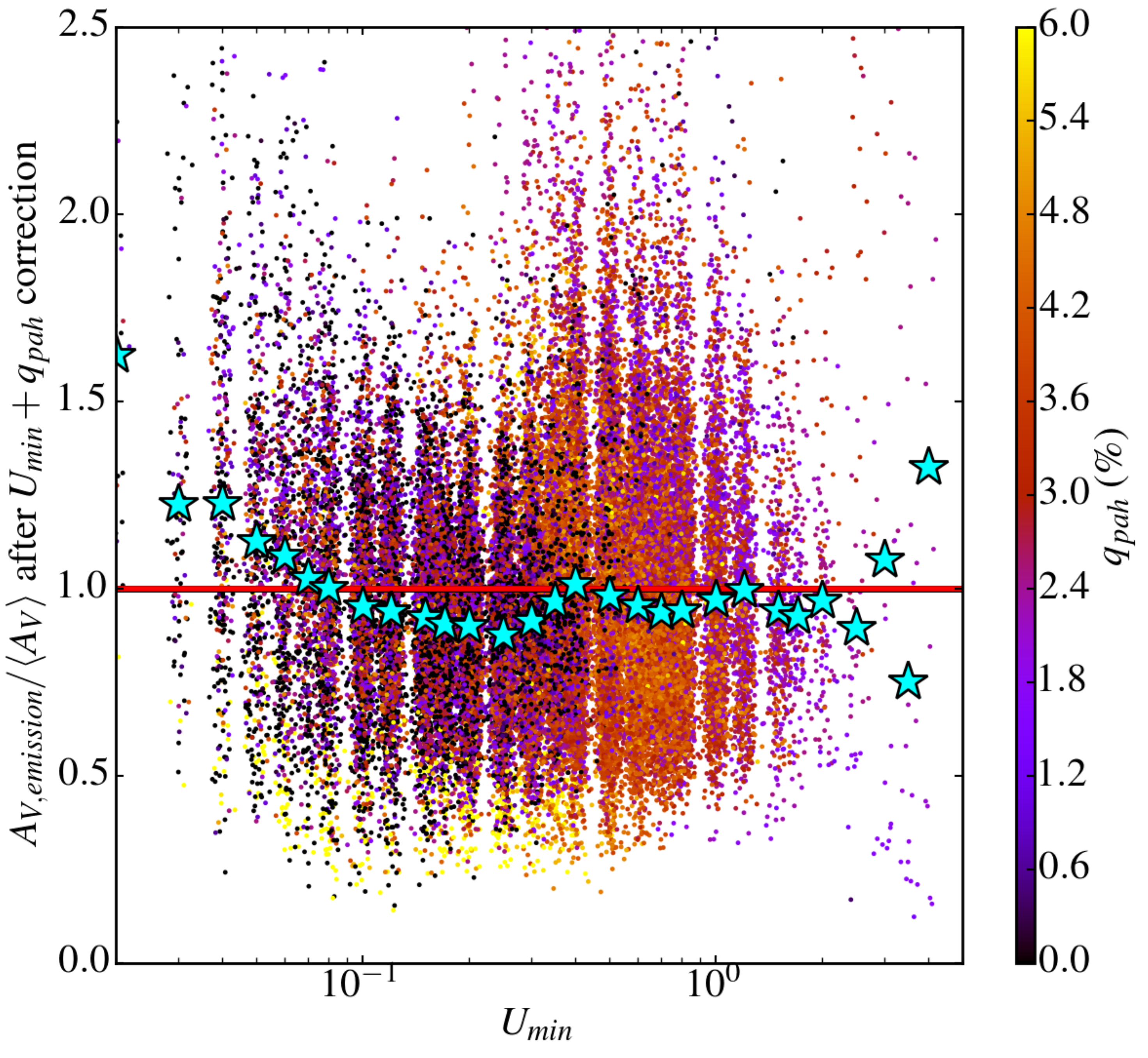}
\includegraphics[width=3.5in]{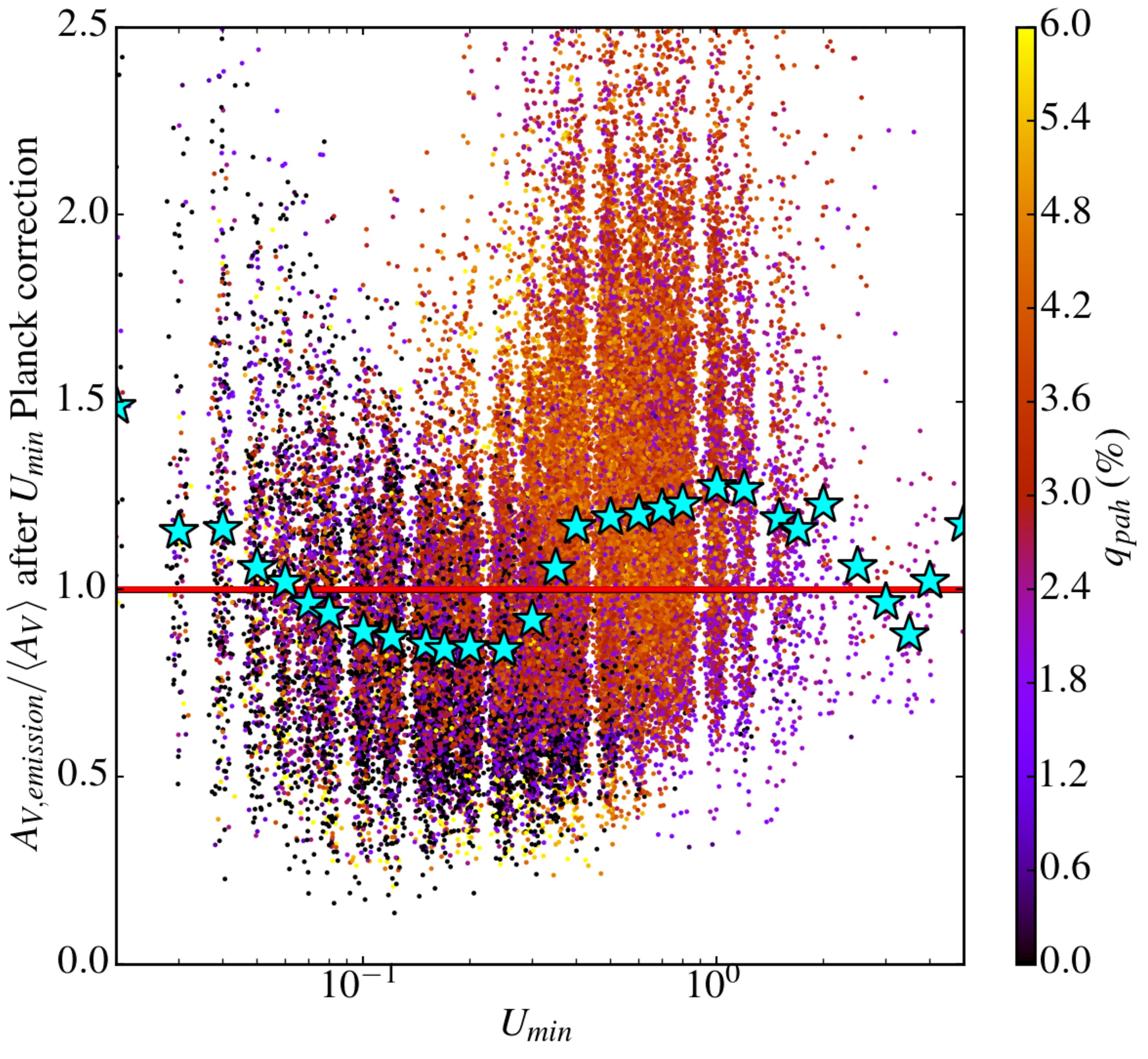}
}
\caption{The observed ratio between the emission-based extinction and
  the mean CMD-based extinction, as a function of $U_{min}$, color
  coded by $q_{pah}$ (top left, bottom row) or $\langle A_V \rangle$
  (top right). Large cyan stars indicate the median value of
  $A_{V,emission}/\langle A_{V}\rangle$ in bins of $U_{min}$. In
  the top row, the solid magnenta line is the expectation based on the
  \citet{planck2015} correction, and the horizontal red line is the
  mean correction derived in Sec.\ \ref{emissioncorrsec}. The bottom
  row shows the ratio after correcting $A_{V,emission}$ using the
  joint $U_{min}+q_{pah}$ correction shown in
  Figure~\ref{Uminvsqpahfig} (left) and the \citet{planck2015}
  correction based on $U_{min}$ (right). The correction based on both
  $U_{min}$ and $q_{pah}$ shows much weaker systematic residuals,
  except at the very lowest values of $U_{min}$.  The spatial
  correlation of the residuals is also reduced (see
  Figure~\ref{rationewcorrmapfig}).  For clarity, small random offsets
  have been added to $U_{min}$ to compensate for the quantization of
  the original data and to better show the individual values of
  $A_{V,emission}/\langle A_{V}\rangle$. \label{Uminvsratiofig}}
\end{figure*}

Figure~\ref{Uminvsratiofig} shows the result of applying either the
\citet{planck2015} correction or the $U_{min}+q_{pah}$ correction from
eqn.\ \ref{Uminqpahfiteqn}. Because our map of M31 includes both
diffuse and molecular cloud components, for the \citet{planck2015}
correction we adopt a slope and intercept midway between the two very
similar linear corrections recommended for these two regimes
($a=0.32$, $b=0.31$).

The upper panels of Figures~\ref{Uminvsratiofig} show the observed
ratio $A_{V,emission}/\langle A_V \rangle$ as a function of $U_{min}$,
color-coded by either $q_{pah}$ (upper left) or the mean extinction
($\langle A_V \rangle$; upper right).  The cyan stars show the median
ratio in small bins of $U_{min}$, and the magnenta line shows the
estimator derived in \citet{planck2015}.  The horizontal line is the
overall 2.53 scaling we derived in Sec.\ \ref{emissioncorrsec} (i.e.,
Figure~\ref{drainecorrelationfig}). The lower panels show the ratio
that results after correcting the emission-based extinction by the
2nd-order polynomial fit from eqn.\ \ref{Uminqpahfiteqn} (left panel)
or the \citet{planck2015} correction (right panel).

Figure~\ref{Uminvsratiofig} shows that the \citet{planck2015}
extinction estimator overall does a good job of correcting the
extinction predicted by the \citet{draine2007} models (lower
right). The overall scale of the correction agrees very well, such
that the ratio $A_{V,emission}/\langle A_V \rangle$ has a median
within 25\% of 1 after applying the $U_{min}$ correction to
$A_{V,emission}$.  However, when compared to the result of applying
the $U_{min}+q_{pah}$ correction (lower left), the \citet{planck2015}
correction leaves larger, systematic residuals. Both corrections leave
significant scatter, however, which may be due to some of the effects
discussed in Sec.\ \ref{smallscalesec}.

\begin{figure*}
\centerline{
\includegraphics[width=3.75in]{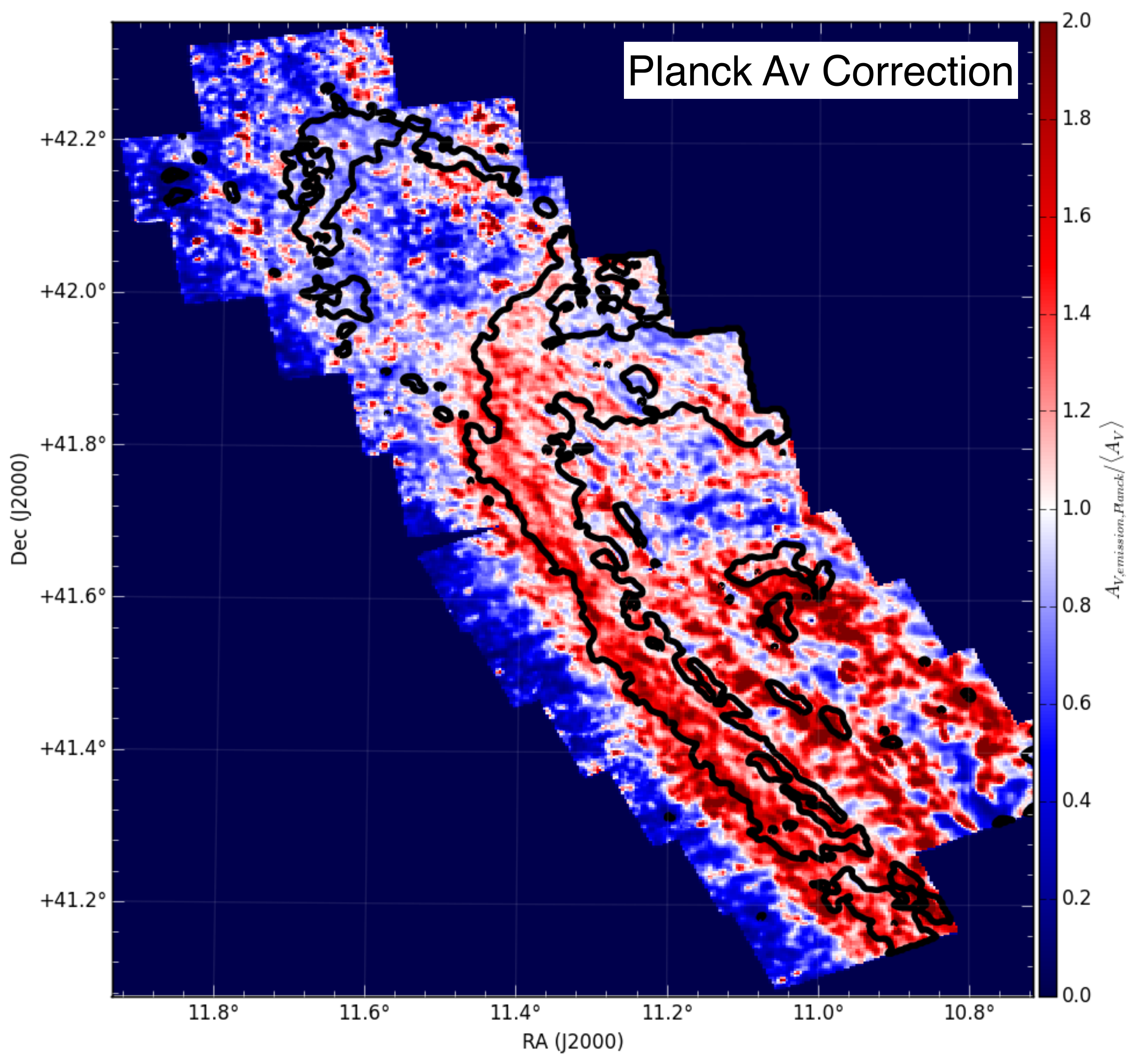}
\includegraphics[width=3.75in]{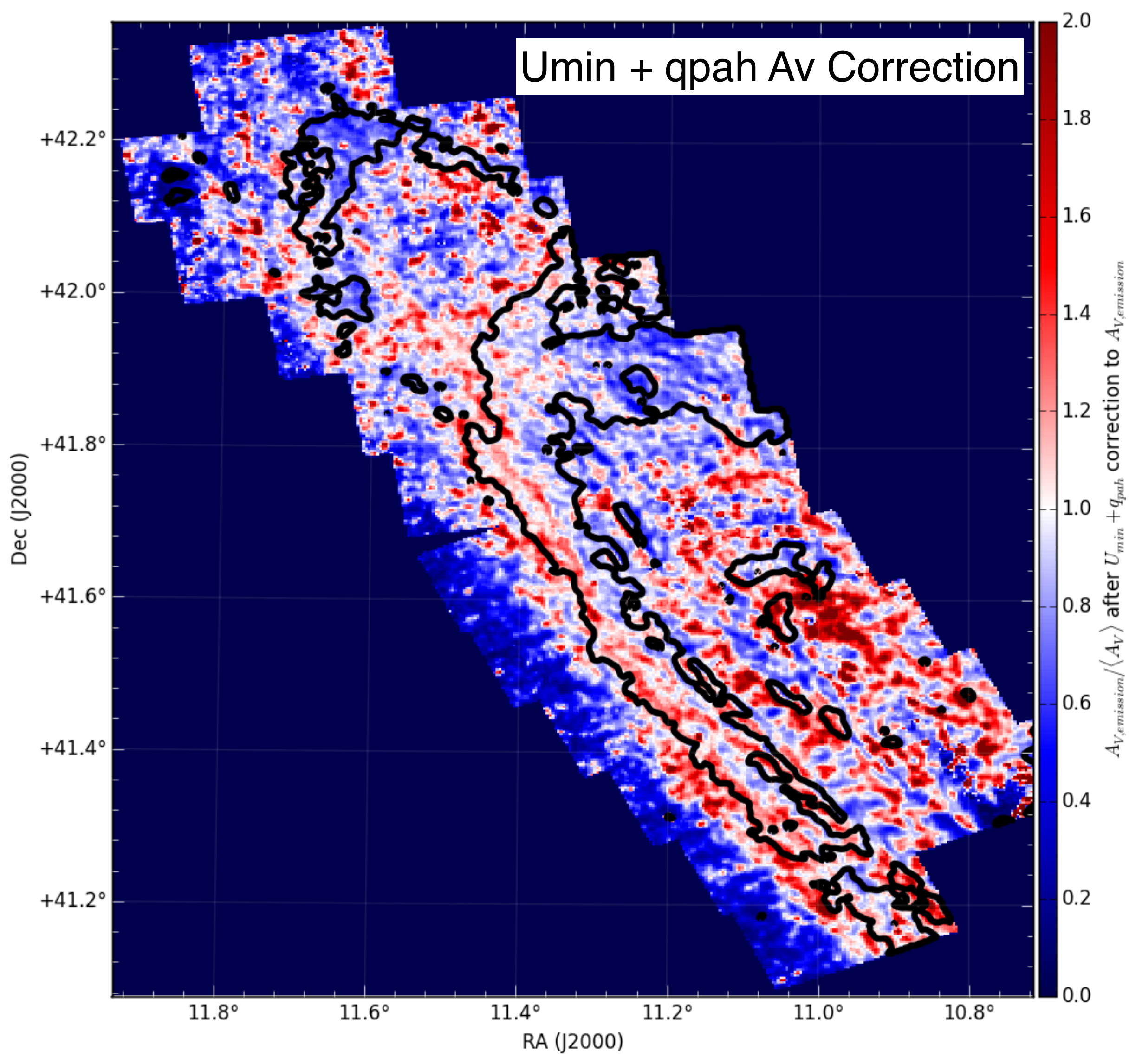}
}
\caption{Maps of the ratio between the emission-based extinction and
  the mean CMD-based extinction, after correcting $A_{V,emssion}$ by
  applying either the \cite{planck2015} correction that uses only
  $U_{min}$ (left) or a polynomial correction based on both $U_{min}$
  and $q_{pah}$ (right).  The \citep{planck2015} correction does a
  good job of correcting the overall normalization, but introduces a
  clear radial bias and does not correct some of the largest,
  spatially-coherent features in the map.  The correction based on
  both $U_{min}$ and $q_{pah}$ produces a much more uniform map. The
  notable discrepancies are in the inner disk where $\langle A_{V} \rangle$
  values are known to be more uncertain, and in the two most actively
  starforming complexes identified in
  Figure~\ref{scalobfig}. \label{rationewcorrmapfig}}
\end{figure*}

In Figure~\ref{rationewcorrmapfig} we compare the spatial distibution
of $A_{V,emission}/\langle A_V \rangle$ after making the
\citet{planck2015} correction (left) or the $U_{min}+q_{pah}$
correction from eqn.\ \ref{Uminqpahfiteqn} (right). The left panel
shows that the systematic residuals that the \citet{planck2015}
correction leaves with $U_{min}$ produce large-scale
spatially-correlated residuals in the predicted extinction map.  With
the \citet{planck2015} corrections, the extinction in the inner galaxy
and half of the major star forming ring is overestimated, and the
extinction in the outer galaxy is underestimated.  When compared to
Figure~\ref{draineratiomapfig}, the \citet{planck2015} correction appears
to produce larger, more spatially-correlated residuals than applying
a single $\sim$2.5 correction factor.

In contrast, the $U_{min}+q_{pah}$ correction from
eqn.\ \ref{Uminqpahfiteqn} produces a map that appears to be free from
the large scale systematics that were visible in
Figures~\ref{draineratiomapfig}~\&~\ref{AVcorrfig}.  The only
noticeable exceptions are the two regions of high star formation
intensity discussed in Sec.\ \ref{smallscalesec} (see
Figure~\ref{scalobfig}) where the corrected value of $A_{V,emission}$
is too low, and a region near the end of M31's long bar in the inner
disk, where the corrected value of $A_{V,emission}$ is too high.  This
latter mismatch may be due to an error in $\langle A_V \rangle$
associated with rapidly changing stellar poplations at this position.
Other than these regions, and the expected convergence of the
uncorrected value of $A_{V,emission}$ and $\langle A_V \rangle$ in
regions where $A_V\rightarrow 0$ in the outermost disk, the corrected
map looks quite uniform.  This suggests that our earlier suspicion
(Sec.\ \ref{smallscalesec}) that some of the observed systematic
variations were tracing the variation in $q_{pah}$ was correct.

\subsubsection{Interpreting the Dependence on $q_{pah}$}

As \citet{planck2015} noted, the dependence of $A_{V,emission}/\langle
A_V \rangle$ on other parameters of the \citet{draine2007} fits does
not necessarily suggest that the true values of those parameters are
actually changing.  Any fitting procedure will tend to produce
correlated residuals in the values of parameters when the underlying
model is not perfect. The mass contained in PAHs is quite small
(typically less than 5\%), and they are responsible for very little of
the extinction in the NIR. Thus, the fact that we see a correlation
between the $A_{V,emission}/\langle A_V \rangle$ and $q_{pah}$ is
unlikely to indicate that variations in the PAH fraction are actually
causing variations in the extinction.  Instead, it is likely that
both $U_{min}$ and $q_{pah}$ are compensating for subtle differences
in the spectral shape of the dust's opacity and emissivity, and for
complexities in how the interstellar radiation field relates to dust
emission.

One interesting feature in Figure~\ref{Uminvsratiofig} points to
possible connections to the role of dust heating and to the
differences between dust in the diffuse ISM and in star forming
regions.  The upper panels of Figure~\ref{Uminvsratiofig} appear to
show two broad diagonal sequences that have a similar variation of
$A_{V,emission}/\langle A_V \rangle$ with $U_{min}$, but that are
shifted horizontally from each other. At low values of $U_{min}<0.2$,
the majority of points have low PAH fractions and low extinctions. At
higher values of $U_{min}>0.2$, the sequence is dominated by regions
with higher PAH fractions and higher extinctions. This bifurcation may
reflect that the appropriate conversion between emission and
extinction largely separates into two different regimes -- one
associated with the diffuse ISM and one associated with higher star
formation rate intensities and larger typical gas columns.  Although
the \citet{draine2007} models attempt to include this effect by
allowing some fraction of the dust to be heated by harder radiation
fields associated with photo-dissociation regions (PDRs), it is
unlikely that a single physical dust model can fully capture the
complexity of dust heating and composition that is found in a typical
PDR.

In light of this, although
Figures~\ref{Uminvsqpahfig}--\ref{rationewcorrmapfig} show that the
correction to $A_{V,emission}$ given in eqn.\ \ref{Uminqpahfiteqn}
does an excellent job of predicting the extinction in M31, the
correction should not necessarily be expected to do an equally good
job in other circumstances.  The joint distribution of $U_{min}$ and
$q_{pah}$ is likely to reflect the dust heating and composition that
is particular to M31, to first order. The second order dependence of
$A_{V,emission}/\langle A_V \rangle$ on $U_{min}$ and $q_{pah}$ seems
likely to arise from deviations between the assumptions of the
\citet{draine2007} physical dust model and the real behavior of M31's
dusty ISM.  Neither of these situations will necessarily be reproduced
in galaxies with very different star formation rates, metallicities,
or dense gas fractions.  Likewise, if the variation of
$A_{V,emission}/\langle A_V \rangle$ with $U_{min}$ and $q_{pah}$ is
due to the fit compensating for subtle mismatches in the underlying
dust model, then fits that use data taken in different bandpasses
and/or with different physical resolutions may not compensate for
these effects in exactly the same way.  As a result, we do not expect
that the correction in eqn.\ \ref{Uminqpahfiteqn} will be one that
holds generically in all situations.  The overall scaling of a factor
of $\sim\!2-2.5$ seems robust, given that comparable corrections were
needed both in M31 and in the diffuse and the molecular components of
the Milky Way \citep{planck2015}, but the exact second-order
dependence of $A_{V,emission}/\langle A_V \rangle$ on $U_{min}$ and
$q_{pah}$ may be variable.

\section{Conclusions}  \label{conclusionsec}

In this paper we have developed a new method for mapping extinction in
external galaxies.  The method is based on using NIR observations of
RGB stars.  These stars have a very narrow distribution in NIR color as a
function of magnitude, allowing one to diagnose extinction as a
redward shift in the position of the RGB on a CMD.  We model the NIR
CMD as a combination of a foreground of unreddened stars, and a
background of stars that lie behind a thin layer of dust. The
background stars sample the distribution of extinctions within the
dust layer, which we characterize with a single log-normal
function. We fit for the fraction of stars behind the dust layer, and
the characteristic parameters of the log-normal (median extinction,
dimensionless width), from which we derive the mean and dispersion of
the distribution of $A_V$. We carry out this analysis using stars
from the PHAT survey of M31, selected in $25\pc$ ($6.6\arcsec$) bins
with $12.5\pc$ sampling.

The resulting map is the highest resolution tracer of extinction
available in M31 to date. More importantly, it offers a completely
independent assessment of the properties of the cold ISM, allowing it
to be compared with other widely-used tracers, including emission from
dust and from cool gas (H{\sc i} and CO).

Our maps reveal a factor of $\sim$2.5 offset between the extinction
directly measured from RGB stars and the extinction inferred from
recent maps of the dust emission \citep[e.g.,][]{draine2013}. This
difference is unlikely to be due to the NIR CMD fitting, and instead
suggests that there is significantly more emission associated with a
given amount of dust extinction than expected for the most recent
calibration of the \citet{draine2007} models. Given how important
these models have become to the study of external galaxies,
understanding the origin of this offset is vital.  Similar offsets
have been seen in the Milky Way \citep{planck2015}, where they were
found to have a second order dependence on the \citet{draine2007}
estimates of the strength of the background radiation field. We
show that the offset in M31 depends on the measured percentage of
PAH's as well. We provide a fitting formula to allow the extinction
to be estimated from the results of fitting to the \citet{draine2007}
models.

Given the success of this technique, it is natural to think of
possible extensions.  Within M31, one might consider including optical
data in the analysis to provide better sensitivity at low
extinction. However, unlike the NIR, the color of the optical RGB is
quite sensitive to age, leading to a much broader RGB for galaxies
with extended star formation histories, and greater sensitivity to
stellar population gradients. Given that the efficacy of the technique
depends on having sufficient reddening to move stars out of the
unreddened CMD, it is not immediately clear that optical data's greater
sensitivity to reddening would counteract the problem of an
intrinsically broader RGB.  We address these issues more quantitatively
in an appendix (Sec.\ \ref{opticalsec}).

Some of the difficulty in incorporating optical and UV data into an
RGB-only analysis can be sidestepped by carrying out joint
UV$+$optical$+$NIR fitting of every star's spectral energy
distribution (Gordon et al 2015, submitted). This approach was very
successful in \citet{zaritsky2002} and \citet{zaritsky2004}'s analyses
of the Magellanic Clouds, but used UV and optical data alone, as was
appropriate for the much lower extinctions in these galaxies. Fitting
PHAT's 6-filter photometry should offer an excellent opportunity to
probe down to lower extinctions and to assess the detailed connections
between dust and different ages of stellar populations, although with more
dependence on the accuracy of the underlying stellar models.

As another possible extension, one might also think of applying this
technique to other galaxies, with M33 being an obvious target. In more
distant galaxies, however, the apparent width of the RGB will be a
limiting factor when using current facilities; more distant galaxies
have higher levels of crowding that limit the photometric accuracy
possible with a given telescope+camera \citep[e.g.][]{olsen2003},
leading to a wider RGB. However, a next generation telescope (either
30m class from the ground, or an 10-12m in space) with a high
resolution NIR camera would make this technique viable throughout the
local volume.

\acknowledgements 

We are happy to acknowledge Jouni Kainulainen for discussions that
helped to crystallize our thinking during the writing of this
paper. Bruce Draine is warmly thanked for providing his dust map in
advance of publication, and for many illuminating discussions.  We
also are grateful to Robert Gendler
({\tt{http://www.robgendlerastropics.com/}}) for allowing us to use
his beautiful image of M31. We also thank L\'{e}o Girardi, Brent Groves, Dan
Foreman-Mackey, Hans-Walter Rix, Thomas Robitaille, and Sarah Kendrew
for help during the long incubation of this paper, and for
facilitating JJD's much-needed conversion to Python. We are also
extremely grateful to the referee for providing a truly excellent,
well-considered report. JJD gratefully acknowledges the hospitality of
the Max-Planck Institut f\"ur Astronomie and Caffe Vita during part of
this work. This work was supported by the Space Telescope Science
Institute through GO-12055.  Support for DRW is provided by NASA
through Hubble Fellowship grant HST-HF-51331.01 awarded by the Space
Telescope Science Institute.

This research made use of Astropy, a community-developed core Python
package for Astronomy \citep{astropy2013}, and APLpy, an open-source
plotting package for Python hosted at {\tt{http://aplpy.github.com}},
as well as numpy, scipy, and matplotlib
\citep{oliphant2007,hunter2007}.  This research has made use of NASA's
Astrophysics Data System Bibliographic Services.

\appendix

\section{Applying This Technique at Other Wavelengths} \label{opticalsec}

Although we have developed and applied this technique in the NIR, it
is natural to consider whether it might also be useful in other
wavelength regimes.  In particular, by incorporating optical data we
could potentially improve our sensitivity to low extinctions, because
of the greater impact of reddening and extinction at bluer wavelengths.
As we now show, the increased sensitivity to reddening comes at a cost,
because the underlying CMD is far more sensitive to population effects
in bluer filter.

The key issues that affect the sensitivity of our technique are: (1)
the amount of reddening produced by a given $A_V$; (2) the amount of
reddening needed to move a star clearly out of the RGB; (3) the
spatial uniformity of the unreddened RGB; (4) the maximum $A_V$ that
can be detected before losing a star below the photometric limit; and
(5) the total number of well-measured stars on the RGB.  Incorporating
data from bluer wavelengths improves the first issue, but its effect
on the remaining issues is unclear.

\begin{figure*}
\centerline{
\includegraphics[width=7.25in]{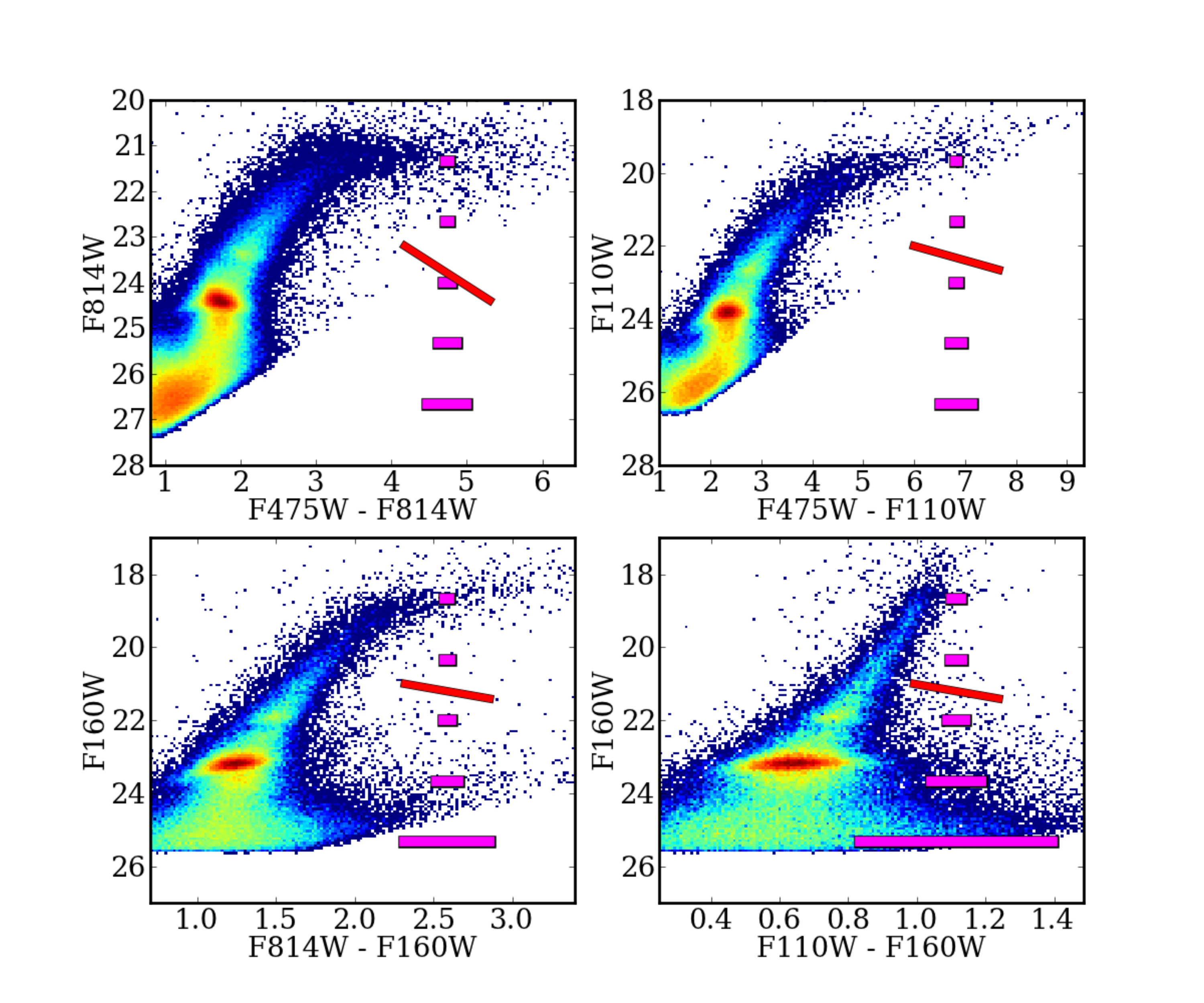}
}
\caption{Comparing the sensitivity of different filter combinations to
  reddening and extinction.  Plots show logarithmically-scaled Hess
  diagrams of the RGB region in 4 different filter combinations, for
  one of the lowest reddening regions of the PHAT survey (Brick 12,
  Field 1, plotting the {\tt{``gst''}} photometry). The horizontal
  magenta bars indicate the median reported photometric uncertainty
  at the plotted magnitude, including an additional systematic
  component of $\sigma_{color}=0.015$ \citep[see][]{dalcanton2012};
  these bars do not include errors due to crowding, which will
  increase significantly near the limiting magnitude of the data. In
  general, the width of the RGB is significantly larger than the
  photometric uncertainties, indicating that variations in the age and
  metallicity of the stellar population is primarily responsible for
  broadening the RGB. This effect is worse in the optical filters, but
  negligible in the NIR, where the RGB is intrinsically narrow and has
  a width almost entirely dominated by photometric uncertainties. The
  red bar shows the shift in color and magnitude that would be caused
  by $\Delta A_V=2$ magnitudes of extinction; and the range of plotted
  colors is adjusted so that it covers the amount of reddening caused
  by a constant $\Delta A_V=10$ magnitudes of extinction.  The
  intrinsic width of the RGB in terms of $\Delta A_V$ is much broader
  in the optical filters than the NIR, which cancels out any possible
  benefit which may have resulted from the optical's greater
  sensitivity to extinction, compared to the NIR. \label{optvsirfig}}
\end{figure*}

To assess the impact of these issues outside the NIR, we construct
CMDs of the unreddened RGB in a variety of filter combinations.  In
Figure~\ref{optvsirfig} we show Hess diagrams constructed within Brick
12, Field 7, which has extremely low reddening as judged from both the
NIR CMD and the \citet{draine2013} maps. To allow us to compare
arbitrary filter combinations, we use the 
simultaneous six filter {\tt{*.gst}} photometry described in Williams
et al (2014), which supersedes the first generation
photometry in \citet{dalcanton2012}; these photometry have not had
cuts applied to optimize rejection of spurious measurements redward of
the RGB, but are sufficient for our purposes here. The horizontal
purple bars indicate the median photometric uncertainty at 5 different
magnitudes; the uncertainties are a quadrature sum of the photon
counting uncertainties reported by DOLPHOT and a flat $0.015\mags$
uncertainty due to calibration \citep{dalcanton2012}. The diagonal red
line shows the effect of $\Delta A_V=2\mags$, using
$A_{F475W}/A_V=1.1591$ and $A_{F814W}/A_V=0.5962$, which we derived
analogously to $A_{F110W}/A_V$ and $A_{F160W}/A_V$ above in
Sec.\ \ref{cloudsec}.

The first, and most important, figure of merit needed to interpret
Figure~\ref{optvsirfig} is the relative width of the RGB compared to
the length of the reddening vector.  If the RGB is narrow compared to
the reddening vector, then even modest $A_V$ will move a star clearly
off the unreddened sequence, increasing the sensitivity to low dust
columns.  We facilitate this comparison by plotting each CMD with a
color range that corresponds to a constant $\Delta A_V$, such that the
reddening vector has the same width in each panel.  With this display
choice, CMDs that appear wider are indeed ``wider compared to a given
$\Delta A_V$''.

Inspection of Figure~\ref{optvsirfig} shows that the RGB is
intrinsically narrower in the NIR than in any other filter
combination.  In other words, although the reddening in $F110W-F160W$
is smallest for a given $A_V$, our ability to actually detect that
reddening is largest. In terms of $A_V$, the NIR RGB has a width of
$\Delta A_V=0.38\mags$ for stars with $19.5<F160W<20.5$, whereas the
optical CMD (upper left) has $\Delta A_V=0.54\mags$. 

The narrowness of the NIR CMD comes in large part because of the
cancellation of age and metallicity in RGB isochrones. Older stars
tends to have redder RGBs, but these stars also tend to have lower
metallicities, which drives their colors back blueward. In the NIR,
these effects are intrinsically small, and also largely cancel out.
In contrast, in the optical these effects are much larger and do not
effectively cancel out. This unfortunate fact leads to broader CMDs in
the optical. Moreover, it points to increased sensitivity of the
optical CMD to variations in the underlying stellar population's age
and/or metallicity.  While this is good for stellar population
studies, it means that one expects much larger spatial gradients in
the properties of the unreddened RGB, making the construction of an
accurate, smoothly varying ``unreddened'' CMD far more difficult. Thus,
by the criteria listed above, issues (2) and (3) are negatively affected
by including bluer filters. 

Issue (4) is also further compromised when moving to bluer
filters. Inspection of the optical CMDs in the top row shows that the
blue magnitude limit in $F475W$ cuts off stars with more than
$A_V\sim2$. In contrast, the redder CMDs in the bottom row allow a
much larger range of $A_V$ to be detected, giving better constraints
on the median $A_V$ and the shape of the reddening distribution.  This
limitation can not necessarily be overcome by a larger investment of
telescope time, given that the data is already crowding limited.

The one area where switching to optical filter may improve sensitivity
is in the number of stars (i.e., issue \#5).  Comparing photometric
errors (magenta bars) to the width of the RGB shows that bluer filters
can accurately track the RGB down to fainter CMD features, increasing
the number of stars that can be used in a given area. This difference
potentially allows one to increase the resolution of the extinction
maps.  However, the higher stellar density would have only modest
effects in high extinction regions, since most of the newly detected
stars are close to the magnitude limit.

Rather than switching entirely to the optical, it may be that some
gains could be achieved by using $F814W-F160W$ to measure the
reddening.  The RGB is less than 5\% wider than in the NIR (in terms
of $\Delta A_V$), so the sensitivity to small extinctions should be
comparable.  However, the photometric errors are proportionally
smaller, due to the wider color baseline and the better accuracy in
$F814W$ compared to F110W.  These improved errors should allow us to
use stars further down the RGB, which increases the spatial density of
stars.  One can then use smaller analysis pixels to increase the
spatial resolution of the extinction map.  Unlike in the optical,
these additional stars are comfortably above the magnitude limit, even
for significant reddening, and thus they should provide useful
leverage on the distribution of reddenings.  It is not clear, however,
whether these gains trump the likely increase in the susceptibility of
the RGB structure to age and metallicity variations.

\section{Smooth Models for the Dust Distribution}  \label{smoothsec}

During the course of this work, we also considered a model where the
reddening seen in the CMD was produced by a smooth layer of dust
embedded within the stellar disk.  As we show here, this assumption
produces distributions of extinction that are inconsistent with the
data, leading to the need for the alternative ``thin screen with a
log-normal distribution of extinction'' model adopted in this paper.

We calculate the expected distribution of stellar extinctions for
a smooth dust model as follows.  We assume that the dust is embedded
within a stellar disk with a scale height $z_\star$.  We model the
stellar disk using the probability $p_\star(z; z_\star)$ that a star is
found at height $z$ relative to the midplane of the stellar disk
(assuming that positive values of $z$ correspond to stars that are
closer to us than the midplane at $z=0$).  The stellar density
distribution can be assumed to be a Gaussian, an exponential, or any
other reasonable functional form.

We first assume that the dust is smoothly distributed within a
vertical distribution $p_d(z; z_d, z_0)$ with scale height $z_{d}$,
centered at a height $z_0$ relative to the midplane of the stellar
disk.  We assume that the total extinction of the integrated dust
column is $A_{V,tot}$.  The integrated extinction seen for a star at
height $z$ is then

\begin{equation}
A_V(z) = A_{V,tot}\,\int^{\infty}_z{p_d(z; z_d, z_0)\,dz}.
\end{equation}

\noindent If we assume that the vertical distribution of dust is an exponential
and let $z^\prime \equiv z-z_0$, then

\begin{equation}   \label{AVzeqn}
A_V(z=z^\prime+z_0) = \frac{A_{V,tot}}{2}\,\left[\frac{|z^\prime|-z^\prime}{|z^\prime|} + 
       \frac{z^\prime}{|z^\prime|}\,e^{-|z^\prime|/z_d} \right]
\end{equation}

The probability $p_A(A_V)\,dA_V$ of finding a star with an extinction between
$A_V$ and $A_V+dA_V$ is then

\begin{eqnarray}  \label{pAVeqnsmooth}
p_A(A_V)\,dA_V &=& p_\star(z(A_V); z_\star)\,\left|\frac{dz}{dA_V}\right|\,dA_V \\
         &=& p_\star(z(A_V); z_\star)\,\left(\frac{z_d}{A_{V,tot}/2}\right)\,
              \frac{1}{\frac{z^\prime}{|z^\prime|}\,\frac{A_V}{A_{V,tot}/2} - 
                       \frac{|z^\prime|-z^\prime}{|z^\prime|}}\,dA_V
\end{eqnarray}

\noindent where $z^\prime\equiv z(A_V)-z_0$ and $z(A_V)$ is the inverse of
equation~\ref{AVzeqn}:

\begin{equation}
z(A_V) = z_0 - z_d \, \frac{\Delta A_V}{|\Delta A_V|} \,
  \ln{\left(\frac{|\Delta A_V| - \Delta A_V}{|\Delta A_V|} + \frac{\Delta A_V}{|\Delta A_V|}\,\left(\frac{A_V}{A_{V,tot}/2}\right)\right)},
\end{equation}

\noindent assuming $\Delta A_V \equiv A_{V,tot}/2 - A_V$.

\begin{figure*}
\centerline{
\includegraphics[width=3.25in]{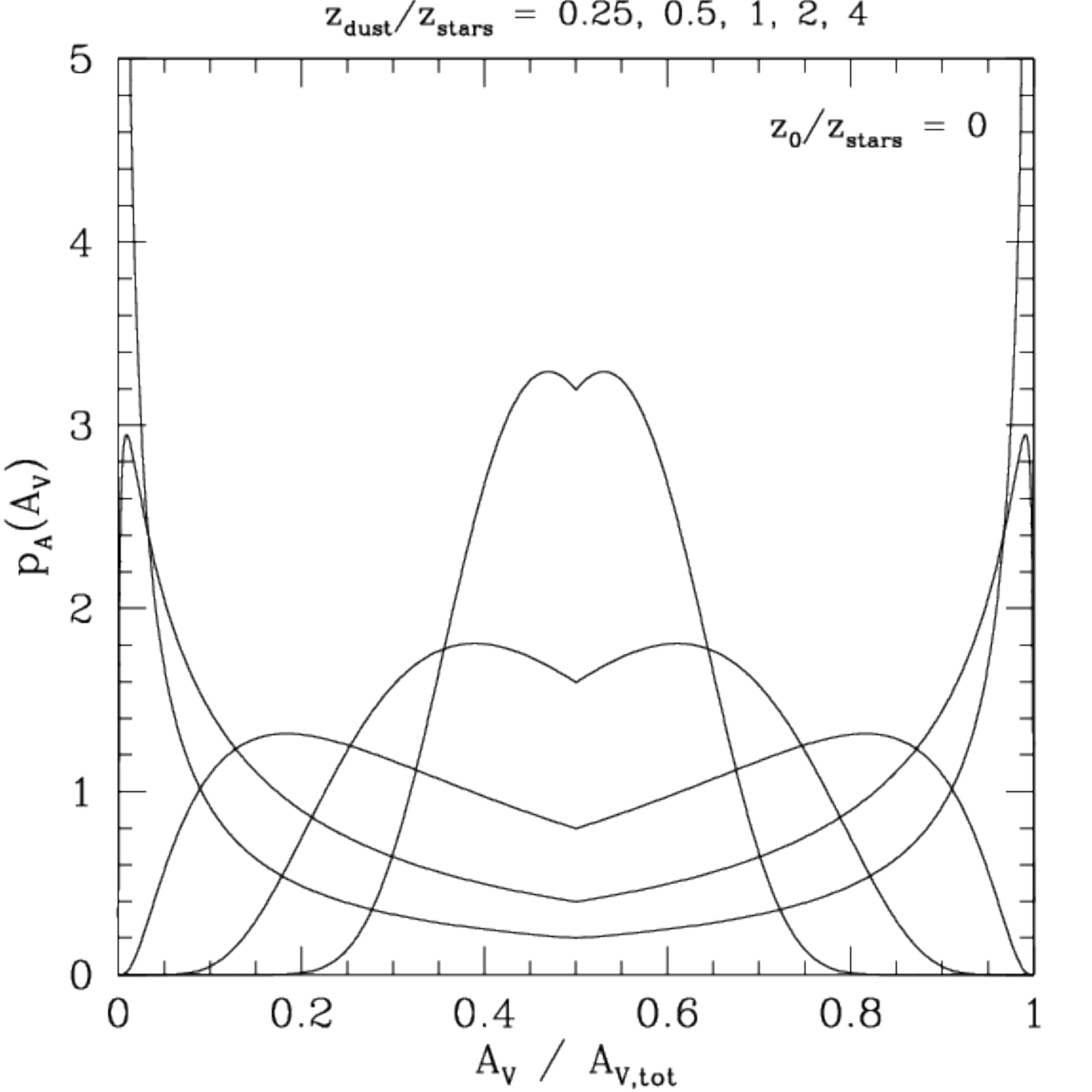}
\includegraphics[width=3.25in]{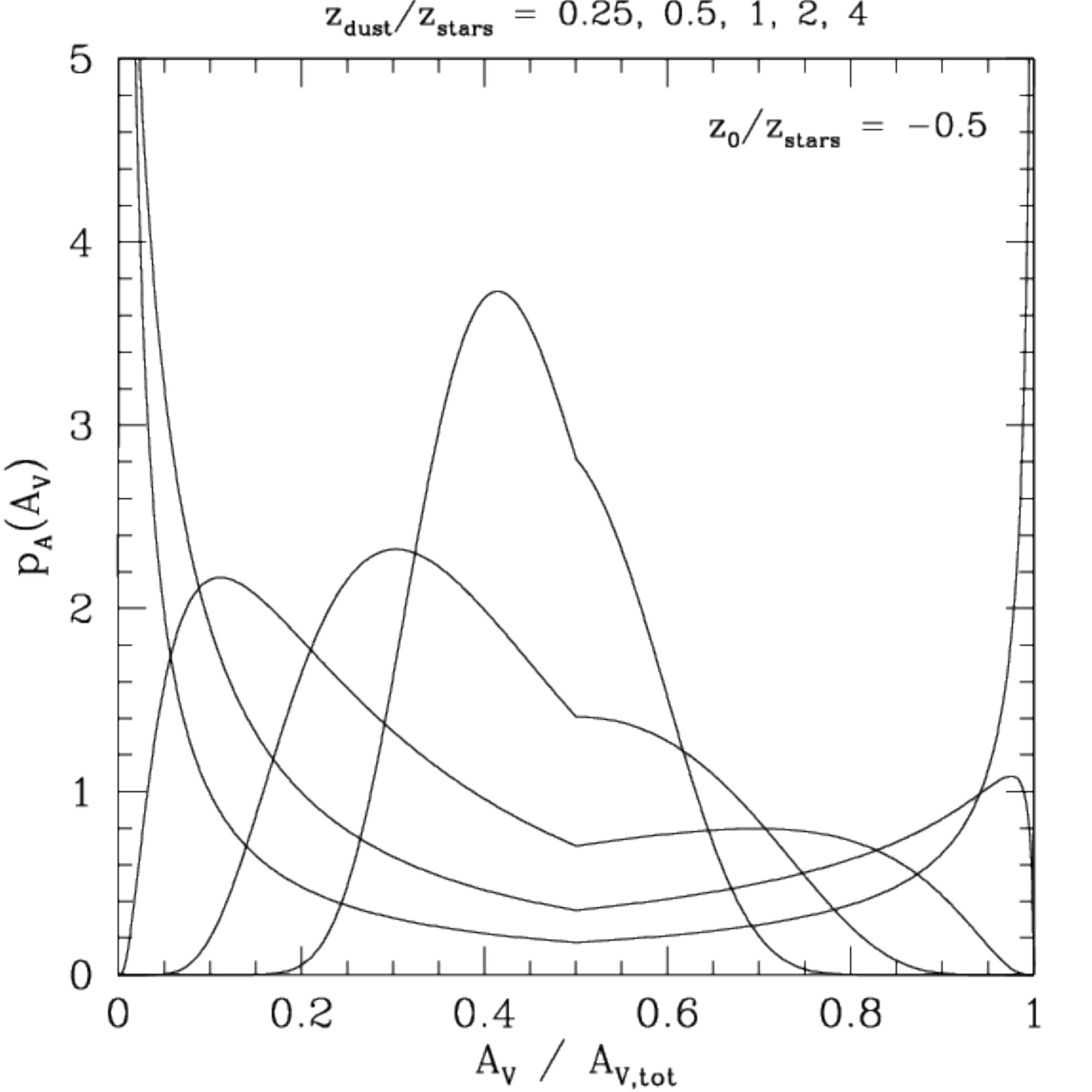}
}
\caption{
Probability distributions for a smooth exponential dust layer with
scale height $z_{dust}$ embedded within a Gaussian stellar disk of
scale height $z_{stars}$, for different ratios between the scale
heights of the dust and stars.  When the dust layer is thin compared
to the stars, the stars have a bimodal distribution of $A_V$, falling
in an either unreddened or reddened peak.  This bimodality becomes
less distinct as the thickness of the dust and stars become
comparable.  The left hand panel shows a dust layer centered at the
stellar midplane and the right hand panel shows a dust layer that is
shifted to the far side of the stellar midplane by one half the scale
height of the stars.  When the dust lies on the far side of the
midplane, a larger fraction of stars have low extinctions.
\label{pAVfig}}
\end{figure*}

The resulting distribution of $p_A(A_V)$ is
shown in Figure~\ref{pAVfig} for the case of a
Gaussian distribution of stars.  When the scale height of the dust is
less than that of the stars, the probability distribution has two
strong peaks --- one at $A_V=0$, corresponding to unreddened stars seen
on the near side of the dust layer, and one at $A_V=A_{V,tot}$,
corresponding to stars seen on the far side of the dust layer, after
they have experienced the entire dust column.  The relative height of the two
peaks depends on the position of the dust layer relative to the stars,
with the unreddened peak becoming stronger when the dust layer is on
the far side of the stellar midplane.  As the scale height of the dust
increases, there are fewer stars with either zero reddening or the
maximum reddening (i.e., all stars are embedded within the dust), and
the two peaks move together towards a typical extinction of half the
total integrated extinction.

To transfer $p_A(A_V)$ to the observed reddening distribution for the
RGB, one must convolve equation~\ref{pAVeqnsmooth} with the intrinsic
distribution of RGB colors along the reddening vector. For an
intrinsic Gaussian distribution of RGB colors and a thin dust layer,
one would then expect to see two skewed Gaussian peaks -- an
unreddened Gaussian skewed towards redder colors, and a highly
reddened Gaussian skewed toward bluer colors -- provided that the
expected total reddening from dust is significantly larger than the
intrinsic width of the RGB colors.  The reddened and unreddened
Gaussians should have identical widths when the dust is centered on
the midplane, but can have somewhat different widths if the dust lane
is not aligned with the stars (right hand panel of
Figure~\ref{pAVfig}), in the sense that the peak containing a higher
fraction of stars should also be wider.  However, there are many
regions in our data which appear to violate this condition.
Frequently the reddened distribution is nearly an order of magnitude
broader than the unreddened peak, which cannot be accounted for in
models with a smooth dust layer.  We therefore reject a smooth dust layer
as being an appropriate model for these observations.

\vfill \clearpage

\end{document}